\documentclass[11pt,letterpaper,twoside,reqno,nosumlimits]{amsart}

\allowdisplaybreaks

\usepackage[usenames,dvipsnames]{xcolor}
\usepackage{fancyhdr}
\usepackage{amsmath,amsfonts,amsbsy,amsgen,amscd,amssymb,amsthm}
\usepackage{url}

\usepackage{mathtools}

\usepackage[font=small,margin=0.25in,labelfont={bf},labelsep={colon}]{caption}

\usepackage{subcaption}

\usepackage{tikz}
\usepackage{microtype}
\usepackage{enumitem}

\definecolor{dark-gray}{gray}{0.3}
\definecolor{dkgray}{rgb}{.4,.4,.4}
\definecolor{dkblue}{rgb}{0,0,.5}
\definecolor{medblue}{rgb}{0,0,.75}
\definecolor{rust}{rgb}{0.5,0.1,0.1}

\usepackage[colorlinks=true]{hyperref}

\hypersetup{urlcolor=rust}
\hypersetup{citecolor=dkblue}
\hypersetup{linkcolor=dkblue}

\usepackage{setspace}

\usepackage{graphicx}
\usepackage{booktabs,longtable,tabu} 
\usepackage{multirow} 

\usepackage{float}

\usepackage[full]{textcomp}

\usepackage[scaled=.98,sups,osf]{XCharter}
\usepackage[scaled=1.04,varqu,varl]{inconsolata}
\usepackage[type1]{cabin}
\usepackage[charter,vvarbb,scaled=1.07]{newtxmath}
\usepackage[cal=boondoxo]{mathalfa}
\usepackage[T1]{fontenc}

\usepackage{bm} 

\graphicspath{{figures/}}

\theoremstyle{definition}

\numberwithin{equation}{section} 
\numberwithin{figure}{section}
\numberwithin{table}{section}

\floatstyle{plaintop}
\newfloat{recipe}{thp}{lor}
\floatname{recipe}{Recipe}
\numberwithin{recipe}{section}

\providecommand{\mathbold}[1]{\bm{#1}}  
                                
\newcommand{\diff}[1]{\mathrm{d}{#1}}
\newcommand{\idiff}[1]{\, \diff{#1}}

\newcommand{\vct}[1]{\mathbold{#1}}

\newcommand{\triplenorm}[1]{{\left\vert\kern-0.25ex\left\vert\kern-0.25ex\left\vert #1
    \right\vert\kern-0.25ex\right\vert\kern-0.25ex\right\vert}}

\usepackage[numbers]{natbib}

\newcommand{\om}{\omega}
\newcommand{\vom}{\vct{\omega}}
\newcommand{\vu}{\vct{u}}

\begin{document}

\title[Potentially singular behavior of the $3$D Navier-Stokes equations]{Potentially singular behavior of the $3$D Navier--Stokes equations}

\author[T. Y. Hou]{Thomas Y. Hou}
\address{Applied and Computational Mathematics, California Institute of Technology, Pasadena, CA 91125, USA}
\email{hou@cms.caltech.edu}

\date{\today}

\begin{abstract}
Whether the $3$D incompressible Navier--Stokes equations can develop a finite time singularity from smooth initial data is one of the most challenging problems in nonlinear PDEs. In this paper, we present some new numerical evidence that the incompressible axisymmetric Navier--Stokes equations with smooth initial data of finite energy seem to develop potentially singular behavior at the origin. 
This potentially singular behavior is induced by a potential finite time singularity of the $3$D Euler equations that we reported in the companion paper \cite{Hou-euler-2021}.
We present numerical evidence that the $3$D Navier--Stokes equations develop nearly self-similar singular scaling properties with 
maximum vorticity increased by a factor of $10^7$. 
We have applied several blow-up criteria to study the potentially singular behavior of the Navier--Stokes equations. 
The Beale-Kato-Majda blow-up criterion and the blow-up criteria based on the growth of enstrophy and negative pressure seem to imply that the Navier--Stokes equations using our initial data develop a potential finite time singularity. We have also examined the Ladyzhenskaya-Prodi-Serrin regularity criteria \cite{ladyzhenskaya1957,prodi1959,serrin1962} that are based on the growth rate of $L_t^q L_x^p$ norm of the velocity with $3/p + 2/q \leq 1$. Our numerical results for the cases of $(p,q) = (4,8),\; (6,4),\; (9,3)$ and $(p,q)=(\infty,2)$ provide strong evidence for the potentially singular behavior of the Navier--Stokes equations. The critical case of $(p,q)=(3,\infty )$ is more difficult to verify numerically due to the extremely slow growth rate in the $L^3$ norm of the velocity field and the significant contribution from the far field where we have a relatively coarse grid. Our numerical study shows that while the global $L^3$ norm of the velocity grows very slowly, the localized version of the $L^3$ norm of the velocity experiences rapid dynamic growth relative to the localized $L^3$ norm of the initial velocity. This provides further evidence for the potentially singular behavior of the Navier--Stokes equations.

\end{abstract}

\maketitle

\section{Introduction}

The three-dimensional ($3$D) incompressible Navier--Stokes equations in fluid dynamics govern the motion of viscous incompressible flows. 
They have been used to model ocean currents, weather patterns, and other fluids related phenomena. The question regarding the global regularity of the $3$D Navier--Stokes equations with smooth initial data of finite energy is one of the most important fundamental questions in nonlinear partial differential equations and is one of the seven Millennium Problems posted by the Clay Mathematics Institute \cite{fefferman2006existence}. 
The main difficulty associated with the global regularity of the $3$D Navier--Stokes equations is that the nonlinearity due to vortex stretching is super-critical. A closely related problem is the global regularity of the $3$D incompressible Euler equations \cite{majda2002vorticity}. It is generally believed that the $3$D Euler equations are more likely to develop a finite time singularity than the $3$D Navier--Stokes equations due to the lack of viscous regularization.

In this paper, we present numerical evidence that the $3$D incompressible axisymmetric Navier--Stokes equations equations with smooth initial data of finite energy seem to develop potentially singular solutions at the origin. The potentially singular behavior of the Navier--Stokes equations is induced by the potential finite time singularity of the $3$D Euler equations that we reported in a companion paper  \cite{Hou-euler-2021}. This singularity scenario is different from the Hou-Luo scenario \cite{luo2014potentially,luo2014toward}, which occurs on the boundary.
An important feature of our new blowup scenario is that the solution of the $3$D Euler equations develops nearly self-similar scaling properties that are compatible with those of the $3$D Navier--Stokes equations. To investigate whether the $3$D Navier--Stokes equations would develop potential singularity, we solve the Navier--Stokes equations with a relatively large viscosity $\nu=5\cdot 10^{-3}$.
Surprisingly, this relatively strong viscous regularization enhances nonlinear alignment of vortex stretching and the $3$D Navier--Stokes equations develop nearly self-similar singular scaling properties with maximum vorticity increased by a factor of $10^7$. 
To the best of our knowledge, such a large growth rate of maximum vorticity has not been reported in the literature for the $3$D Navier--Stokes equations. 

We consider the $3$D axisymmetric  Navier--Stokes equations in a periodic cylindrical domain. We impose a no-slip no-flow boundary condition at $r=1$ for the Navier--Stokes equations. We use a periodic boundary condition in the axial variable $z$ with period $1$.
Let $u^\theta$, $\omega^\theta$, and $\psi^\theta$ be the angular components of the velocity, the vorticity, and the vector stream function, respectively. 
In \cite{hou2008dynamic}, Hou and Li introduced the following change of variables: 
\[
u_1= u^\theta/r, \quad \omega_1=\omega^\theta/r, \quad \psi_1 = \psi^\theta/r,
\]
and transformed the Navier--Stokes equations into the form 
\begin{subequations}\label{eq:axisymmetric_NSE_0}
\begin{align}
u_{1,t}+u^ru_{1,r}+u^zu_{1,z} &=2u_1\psi_{1,z} + \nu\left(u_{1,rr} + \frac{3}{r}u_{1,r}\right) + \nu u_{1,zz},\label{eq:as_NSE_0_a}\\
\om_{1,t}+u^r\om_{1,r}+u^z\om_{1,z} &=2u_1u_{1,z} + \nu\left(\om_{1,rr} + \frac{3}{r}\om_{1,r}\right) + \nu \om_{1,zz}\label{eq:as_NSE_0_b}\\
 -\left(\partial_r^2+\frac{3}{r}\partial_r+\partial_z^2\right)\psi_1 &= \om_1,\label{eq:as_NSE_0_c}
\end{align}
\end{subequations}
where $u^r=-r\psi_{1,z}, \; u^z =2\psi_1 + r\psi_{1,r}$.

\vspace{-0.1in}
\subsection{The main features of the potentially singular solution}

The solution of the Navier--Stokes equations with our initial condition shares several attractive features of the two-scale traveling wave solution studied by Hou-Huang in \cite{Hou-Huang-2021,Hou-Huang-2022}. In particular, the oddness of angular vorticity induce two antisymmetric (with respect to $z$) vortex dipoles, which generate a hyperbolic flow structure near $r=0$. The antisymmetric vortex dipoles produce a strong shear layer for the axial velocity, which transports the solution toward $z=0$. Moreover, the $2$D velocity field $(u^r(t),u^z(t))$ in the $rz$-plane forms a closed circle right above $(R(t),Z(t))$, the location at which $u_1$ achieves its maximum. The corresponding streamlines are trapped in the circle region in the $rz$-plane and a significant portion of $u_1$ and $\omega_1$ are retained in this circle region instead of being transported upward along the $z$-direction. This is an important mechanism that leads to a sustainable growth of the solution. The induced antisymmetric local convective circulation further pushes the solution near $z=0$ toward the symmetry axis $r=0$. 

One interesting feature of our initial condition is that $\psi_{1z}$ becomes large, positive and relatively flat in a local region near the origin. This local region is characterized by a rectangular domain with $ 0 \leq r \leq 0.9 R(t)$ and $0 \leq z \leq 0.5 Z(t)$.
Beyond this local domain, $\psi_{1,z}(r,z,t)$  decays rapidly and becomes negative near the tail region. The large value of $\psi_{1,z}$ in this local region near the origin generates a large growth of $u_1$ through the vortex stretching term and the small or negative value of $\psi_{1,z}$ in the tail region generates a slower growth rate of $u_1$. This difference in the growth rates in the local region and the tail region produces a one-scale traveling wave solution approaching the origin. The traveling wave induced by the strong nonlinear alignment of vortex stretching is so strong that it overcomes the stabilizing effect of advection in the upward $z$-direction (see e.g. \cite{hou2008dynamic,lei2009stabilizing}).

Although the angular vorticity is set to zero initially,
the oddness of $u_1$ in $z$ generates a large positive gradient $u_{1z}$, which contributes positively to the rapid growth of $\omega_1$ through the vortex stretching term $2 u_1 u_{1z}$ in \eqref{eq:as_NSE_0_b}. The rapid growth of $\omega_1$ in turn feeds back to the rapid growth of $\psi_{1,z}$, leading to even faster growth of $u_1^2$. The whole coupling mechanism described above forms a positive feedback loop.

Another interesting feature is that the flow spins rapidly around the symmetry axis due to the rapid growth of $u_1$ dynamically (recall $u^\theta = r u_1$). Near the symmetry axis, the upward axial velocity dominates the angular and radial velocities. The streamlines induced by the velocity field travel upward along the vertical direction and then move outward along the radial direction. The local solution structure resembles the structure of a tornado. However, we do not observe the formation of a vacuum region in our blow-up scenario as in the two-scale traveling singularity in \cite{Hou-Huang-2021,Hou-Huang-2022}. 

\vspace{-0.1in}
\subsection{The potentially singular behavior of the Navier--Stokes equations}

After a short transition time, the solution of the $3$D Navier--Stokes equations develops potentially singular scaling properties.
If we introduce $\xi = (r-R(t))/Z(t)$ and $\zeta=z/Z(t)$ as the dynamically rescaled variables, we observe that the rescaled solutions of the $3$D Navier-Stokes equations seem to develop nearly self-similar profiles in a local region of $(\xi, \zeta)$ .  
Moreover, the nearly self-similar profile is very stable with respect to the small perturbation of the initial data. 

During the rapid growth period, the solution of the $3$D Navier--Stokes equations develops nearly self-similar scaling properties. Both $R(t)$ and $Z(t)$ seem to scale like $O((T-t)^{1/2})$.
The maximum vorticity grows like $O((T-t)^{-1})$. Moreover, we show that $\int_0^t \| \vom (s) \|_{L^\infty} ds $ seems to grow without bound. According to the well-known Beale-Kato-Majda blow-up criterion \cite{beale1984remarks}, this seems to imply that the Navier--Stokes equations with our initial data would develop a finite time singularity.
Another quantity of interest is pressure $p$. According to \cite{sverak2002}, if there is a blowup for the Navier--Stokes equations, the pressure must tend to negative infinity at the singularity time. Our study indicates that $\|-p\|_{L^\infty}$ seems to grow like $O((T-t)^{-1})$ and $\|\frac{1}{2} |{\bf u}|^2 + p\|_{L^\infty} \sim O((T-t)^{-1})$.  Moreover, we observe that both $\int_0^t \|-p (s)\|_{L^\infty} ds$ and $\int_0^t \|\frac{1}{2} |{\bf u}(s)|^2 + p (s)\|_{L^\infty} ds$ experience rapid dynamic growth. 
This provides support for the potentially singular behavior of the $3$D Navier--Stokes equations.

Another quantity of interest is the growth rate of enstrophy $\|\vom (t)\|_{L^2}^2$. We observe a very rapid dynamic growth of the enstrophy. A linear fitting suggests that the enstrophy grows roughly like $O((T-t)^{-1/2})$. Moreover, $\int_0^t\|\vom (s)\|_{L^2}^4 ds $ seems to grow without bound. A standard energy estimate implies that $\int_0^t\|\vom (s)\|_{L^2}^4 ds $ must blow up if the solution of the Navier--Stokes equations develops a finite time singularity. The rapid growth of $\int_0^t\|\vom (s)\|_{L^2}^4 ds $ provides additional support that the $3$D Navier--Stokes equations seem to develop a finite time singularity.

We have also examined the Ladyzhenskaya-Prodi-Serrin regularity criteria \cite{ladyzhenskaya1957,prodi1959,serrin1962} that are based on the estimate of the $L_t^q L_x^p$ norm of the velocity with $3/p + 2/q \leq 1$. We study the cases of $(p,q) = (4,8),\; (6,4),\; (9,3)$, and $(\infty,2)$ respectively. Denote by $\|{\bf u}(t)\|_{L^{p,q}} = \left (\int_0^t \|{\bf u(s)}\|_{L^p(\Omega)}^q ds \right )^{1/q}$. We further define a localized version of $\| {\bf u}(t)\|_{L^{p,q}_{loc}}$ computed over $\Omega_{loc} = \{(r,z) \;|\: r \leq 0.001\}$. The partial regularity results due to Caffarelli-Kohn-Nirenberg\cite{caffarelli1982partial} imply that there is no finite time singularity for the axisymmetric Navier--Stokes equations away from the symmetry axis $r=0$. Thus, it is sufficient to check the dynamic growth of $\| {\bf u}(t)\|_{L^{p,q}_{loc}}$.  Our numerical results show that $\|{\bf u}(t)\|_{L^{p,q}_{loc}}/\|{\bf u}(0)\|_{L^{p,q}_{loc}}$ develops rapid growth dynamically. This provides strong evidence for the development of a potential finite time singularity of the Navier--Stokes equations. 

We have further investigated the endpoint case of $(p,q)=(3,\infty)$ due to Escauriaza-Seregin-Sverak \cite{sverak2003}.  The analysis of the critical case of $(p,q)=(3,\infty)$ is due to a compactness argument. According to a 
recent result by Tao \cite{tao2020}, as one approaches a finite blowup time $T$, the critical $L^3$ norm of the velocity may blow up as slowly as $\left (\log \log \log \frac{1}{T-t}\right )^c$ for some absolute constant $c$. 
If the $L^3$ norm of the velocity indeed grows in a triple logarithmic rate, it would be almost impossible to capture such slow growth with our current computational capacity. We indeed observe very slow growth of $\|{\bf u}(t) \|_{L^3}$ in the late stage. Moreover, our results show that the far field velocity has a significant contribution to the $L^3$ norm of the velocity. Since our current adaptive mesh strategy allocates a majority of the grid points to the most singular region, we have a relatively coarse grid in the far field. Moreover, the frequent changes of adaptive mesh in the late stage introduce a relatively large numerical dissipation in the far field. This makes it very difficult to capture the extremely slow growth of $\|{\bf u}(t) \|_{L^3}$ in the late stage. 

In order to capture such mild dynamic growth rate, we compute the relative growth of the localized version of the $L^3$ norm of the velocity over a localized domain $\Omega_{loc}^* = \{ (r,z) \in [0,0.001]^2\}$. We observe rapid dynamic growth of $\| {\bf u}(t)\|_{L^3(\Omega_{loc}^*)}/\| {\bf u}(0)\|_{L^3(\Omega_{loc}^*)}$. This provides further evidence for the potentially singular behavior of the Navier--Stokes equations using our initial data.

\vspace{-0.05in}
\subsection{Comparison with the two-scale traveling wave singularity }

Despite some similarity in the solution behavior between the new initial condition considered in this paper and the initial condition considered in 
\cite{Hou-Huang-2021,Hou-Huang-2022},
there are some important differences between the two potential blowup scenarios. One important difference is that our solutions essentially have a one-scale structure instead of a two-scale structure observed in \cite{Hou-Huang-2021,Hou-Huang-2022}. Moreover, our solutions of the $3$D Euler equations seem to develop nearly self-similar scaling properties that are compatible with those of the $3$D Navier--Stokes equations. This scaling property is crucial for the potentially singular behavior of the $3$D Navier-Stokes equations. Due to the two-scale solution structure, the relative growth of the maximum vorticity of the $3$D Navier--Stokes solution using a constant viscosity $\nu = 10^{-5}$ reported in \cite{Hou-Huang-2021} is less than $2$. 

Another important difference is that
 the solution of the $3$D Euler equations considered in \cite{Hou-Huang-2021,Hou-Huang-2022} develops a three-scale structure and the thickness of the sharp front does not seem to settle down to a stable scale. 
Moreover, the Navier--Stokes equations with degenerate viscosity coefficients reported in \cite{Hou-Huang-2021,Hou-Huang-2022} develop strong shearing instability in the tail region. In comparison, our solutions have fast decay in the far field and do not suffer from this shearing instability in the tail region. As a result, there is no need to apply any numerical filter in the tail region.

\vspace{-0.075in}
\subsection{Numerical Methods}
We use a similar adaptive mesh strategy developed in \cite{Hou-Huang-2021} by constructing two adaptive mesh maps for $r$ and $z$ explicitly. The solutions of the Navier--Stokes equations with our initial data are much smoother than those considered in \cite{Hou-Huang-2021,Hou-Huang-2022} due to the relative large viscous regularization. More details on how to construct the adaptive mesh will be provided in the Appendix.
We use a second order finite difference method to discretize the spatial derivatives, and a second order explicit Runge--Kutta method to discretize in time. An adaptive time-step size is used according to the standard time-stepping stability constraint with the smallest time-step size of order $O(10^{-15})$.  
The overall method is second order accurate. We have performed careful resolution study and confirm that our method indeed gives at least second order accuracy in the maximum norm.

\vspace{-0.05in}
\subsection{Review of previous works}

For the $3$D Navier--Stokes equations, the partial regularity result due to Caffarelli--Kohn--Nirenberg \cite{caffarelli1982partial} is one of the best known results (see a simplified proof by Lin \cite{lin1998new}). This result implies that any potential singularity of the axisymmetric Navier--Stokes equations must occur on the symmetry axis. There have been some very interesting theoretical developments regarding the lower bound on the blow-up rate for axisymmetric Navier-Stokes equations \cite{chen2008lower,chen2009lower,sverak2009}. Another interesting development is a result due to Tao \cite{tao2016nse} who proposed an averaged three-dimensional Navier--Stokes equation that preserves the energy identity, but blows up in finite time. 

There have been a number of theoretical developments for the $3$D incompressible Euler equations, including the Beale--Kato--Majda blow-up criterion \cite{majda2002vorticity}, the geometric non-blow-up criterion due to Constantin--Fefferman--Majda \cite{cfm1996} and its Lagrangian analog due to Deng-Hou-Yu \cite{dhy2005}. In 2019, Elgindi \cite{elgindi2019finite} (see also \cite{Elg19}) proved an exciting result: the $3$D axisymmetric Euler equations develop a finite time singularity for a class of $C^{1,\alpha}$ initial velocity with no swirl. There have been a number of interesting theoretical results inspired by the Hou--Lou blowup scenario \cite{luo2014potentially,luo2014toward}, see e.g. \cite{kiselev2014small,choi2014on,choi2015,kryz2016,chen2019finite2,chen2020finite,chen2021finite3}
and the excellent survey article \cite{kiselev2018}.

There have been relatively few papers on the numerical study regarding the potential blow-up of the $3$D Navier--Stokes equations, although there were a number of attempts to look for potential Euler singularities numerically, see \cite{gs1991,es1994,kerr1993,hl2006,luo2014potentially,luo2014toward,brenner2016euler}. The work by Boratav and Pelz using Kida's high-symmetry initial data in \cite{bp1994} has generated some interests (see also \cite{hl2008} and a recent result \cite{elgindi2021}). We refer to a review article \cite{gibbon2008} for more discussions on potential Euler singularities.

The rest of the paper is organized as follows. In Section \ref{sec:setup}, we describe the setup of the problem. In Section \ref{sec:nse}, we describe the potentially singular behavior of the $3$D Navier--Stokes equations using a relatively large constant viscosity. 
Some concluding remarks are made in Section \ref{sec:conclude}. Some technical details regarding the construction of our adaptive mesh for the $3$D Navier--Stokes equations will be deferred to the Appendix. 


\vspace{-0.075in}
\section{Description of the Problem}\label{sec:setup}
In this paper, we study the $3$D axisymmetric incompressible Navier--Stokes equations.
Let ${\bf u}$ be the velocity field and define $\vom = \nabla\times \vu$ as the $3$D vorticity vector. 
To introduce the axisymmetric Navier--Stokes equations, 
we decompose the radially symmetric velocity field as follows 
\[\vct{u}(t,r,z) = u^r(t,r,z)\vct{e}_r + u^\theta(t,r,z)\vct{e}_\theta + u^z(t,r,z)\vct{e}_z,\]
\[\vct{e}_r = \frac{1}{r}(x,y,0)^T,\quad \vct{e}_\theta = \frac{1}{r}(-y,x,0)^T,\quad \vct{e}_z = (0,0,1)^T.\]
%
The vorticity can be represented in cylindrical coordinates as follows:
\[\vom (t,r,z) = -(u^\theta)_z\vct{e}_r + \omega^\theta(t,r,z)\vct{e}_\theta + \frac{1}{r}(r u^\theta)_r\vct{e}_z.\]
Let $\psi^\theta$ be the angular stream function.
By making the change of variables, 
$u_1 = u^\theta/r,\; \om_1=\om^\theta/r,\; \psi_1 = \psi^\theta/r$,
Hou and Li \cite{hou2008dynamic} derived the following equivalent axisymmetric Navier--Stokes equations:
\begin{subequations}\label{eq:axisymmetric_NSE_1}
\begin{align}
u_{1,t}+u^ru_{1,r}+u^zu_{1,z} &=2u_1\psi_{1,z} + \nu\left(u_{1,rr} + \frac{3}{r}u_{1,r}\right) + \nu u_{1,zz},\label{eq:as_NSE_1_a}\\
\om_{1,t}+u^r\om_{1,r}+u^z\om_{1,z} &=2u_1u_{1,z} + \nu\left(\om_{1,rr} + \frac{3}{r}\om_{1,r}\right) + \nu \om_{1,zz}\label{eq:as_NSE_1_b}\\
 -\left(\partial_r^2+\frac{3}{r}\partial_r+\partial_z^2\right)\psi_1 &= \om_1,\label{eq:as_NSE_1_c}\\
 u^r=-r\psi_{1,z},\quad u^z&=2\psi_1 + r\psi_{1,r}. \label{eq:as_NSE_1_d}
\end{align}
\end{subequations}
This reformulation has the advantage of removing the $1/r$ singularity from the cylindrical coordinates. 

Our smooth initial condition has a very simple form and is given below:
\begin{equation}
\label{eq:initial-data}
u_1 (0,r,z) =\frac{12000(1-r^2)^{18}
\sin(2 \pi z)}{1+12.5(\sin(\pi z))^2}, \quad \om_1(0,r,z)=0.
\end{equation}
The flow is completely driven by large swirl initially.
The other two velocity components are set to zero initially. 
Note that $u_1$ is an odd and periodic function of $z$ with period $!$. The oddness of $u_1$ induces the oddness of $\omega_1$ dynamically through the vortex stretching term in the $\omega_1$-equation. 
It is worth emphasizing that $u_1$ decays rapidly as $r$ approaches the boundary $r=1$. The specific form of the denominator is also important. It breaks the even symmetry of $\sin(2 \pi z)$ with respect to $z=1/4$ along the $z$ direction with a bias toward $z=0$.
This initial condition generates a solution that has comparable scales along the $r$ and $z$ directions, leading to a one-scale traveling solution moving toward the origin. We will show in Section \ref{IC:stability} that the solution behavior is very stable to a small perturbation of the initial data. 

Our initial condition does not seem to lead to potentially singular behavior if we judge the solution behavior in the very early stage. The maximum of $u_1$ actually decreases in the very early stage.  After a short transition time, the solution develops favorable structure dynamically and we observe strong nonlinear alignment of vortex stretching. Beyond this short transition time, we observe rapid dynamic growth of the maximum vorticity throughout the computation.

We will impose a periodic boundary condition in $z$ with period $1$ and the odd symmetry of $u$. Since $u^\theta,\om^\theta,\psi^\theta$ is an odd function of $r$ \cite{liuwang2006},  $u_1,\om_1,\psi_1$ is an even function of $r$. Thus, we impose the following pole conditions:
\begin{equation}\label{eq:even_r}
u_{1,r}(t,0,z) = \om_{1,r}(t,0,z) = \psi_{1,r}(t,0,z) = 0.
\end{equation}
For the Navier--Stokes equations, the velocity satisfies a no-slip no-flow boundary condition on the solid boundary $r=1$. The no-flow boundary condition is given by
\begin{equation}\label{eq:no-flow}
\psi_1(t,1,z) = 0\quad \text{for all $z$},
\end{equation}
and the no-slip boundary condition is given by
$u^\theta(t,1,z) = u^z(t,1,z) = 0$ for all $z$.
In view of \eqref{eq:as_NSE_1_d} and \eqref{eq:no-flow}, this further leads to $\psi_{1,r}(t,1,z) = 0$. Therefore, the no-slip boundary in terms of the new variables $u_1,\om_1,\psi_1$ reads
\begin{equation}\label{eq:no-slip1}
u_1(t,1,z) = 0,\quad \om_1(t,1,z) = -\psi_{1,rr}(t,1,z),\quad \text{for all $z$}.
\end{equation}
We will enforce the no-slip boundary condition for $\omega_1$ as a vorticity boundary condition by discretizing $\om_1(t,1,z) = -\psi_{1,rr}(t,1,z)$ and imposing $\psi_{1,r}(t,1,z) = 0$.
The periodicity and the odd symmetry of the solution imply that we only need to solve equations~\eqref{eq:axisymmetric_NSE_1} in the half-period domain 
\[\mathcal{D}_1 = \{(r,z): 0\leq r\leq 1, 0\leq z\leq 1/2\},\] 
and $u^r$ and $u^z$ satisfy the following conditions 
\[u^r = -r \psi_{1,z} = 0 \quad \text{on $r=0,1$}\quad \text{and}\quad u^z = 2\psi_1+r\psi_{1,r} = 0\quad \text{on $z=0,1/2$}.\]
Thus the boundaries of $\mathcal{D}_1$ behave like ``impermeable walls''. 
To numerically compute the potential singularity formation of the equations \eqref{eq:axisymmetric_NSE_1}-\eqref{eq:initial-data}, we adopt the numerical methods developed in my recent joint work with Dr. De Huang \cite{Hou-Huang-2021}. In particular, we design an adaptive mesh by constructing two adaptive mesh maps for $r$ and $z$ explicitly. The computation is performed in the transformed domain using a uniform mesh. When we map back to the physical domain, we obtain a highly adaptive mesh with the smallest mesh size of order $O(10^{-8})$. We will provide more details how to construct the adaptive mesh for the $3$D Euler and Navier--Stokes equations in the Appendix.
The detailed descriptions of the overall numerical methods can be found in Appendix A in \cite{Hou-Huang-2021}. 


\section{Potentially singular behavior of the 3D Navier--Stokes equations}
\label{sec:nse}
In this section, we will investigate the potentially singular behavior of the $3$D Navier--Stokes equations. 
In \cite{Hou-euler-2021}, we investigate the potential finite time singularity for the $3$D Euler equations. Our study shows that the $3$D Euler equations develop a sharp front in the late stage and it is extremely difficult to resolve the sharp front numerically. Since the solution of the $3$D Euler equations has scaling properties compatible with those of the $3$D Navier--Stokes equations, it is natural to consider whether the $3$D Navier--Stokes equations may develop a finite time singularity using the same initial data. Moreover, the viscous effect would regularize the sharp front, making it easier to resolve numerically.

It turns out that the choice of the viscosity coefficient plays a crucial role in generating a stable and sustainable growth of the maximum vorticity. On one hand, if the viscosity is too large, it would destroy the mechanism that leads to the potential Euler singularity. On the other hand, if the viscosity is too small, then it is not strong enough to stabilize the shearing instability generated by the $3$D Euler equations. After performing many experiments, we find that first solving the Navier--Stokes equations with viscosity $\nu=5\cdot 10^{-4}$ up to a short time $t_0 = 0.00227375$ and then increasing $\nu$ to $5\cdot 10^{-3}$ seem to give the nearly optimal growth rate. Using $\nu =5\cdot 10^{-4}$ for the early stage from $t=0$ to $t_0$ enables us to preserve the main mechanism leading to the potential Euler singularity. Using a larger viscosity $\nu =5\cdot 10^{-3}$ beyond $t_0$ enables us to stabilize the fluid dynamic instability induced by the $3$D Euler equations. This choice of viscosity coefficient produces a relatively long stable phase of nonlinear alignment of vortex stretching and nearly self-similar scaling properties.

\begin{figure}[!ht]
\centering
    \includegraphics[width=0.32\textwidth]{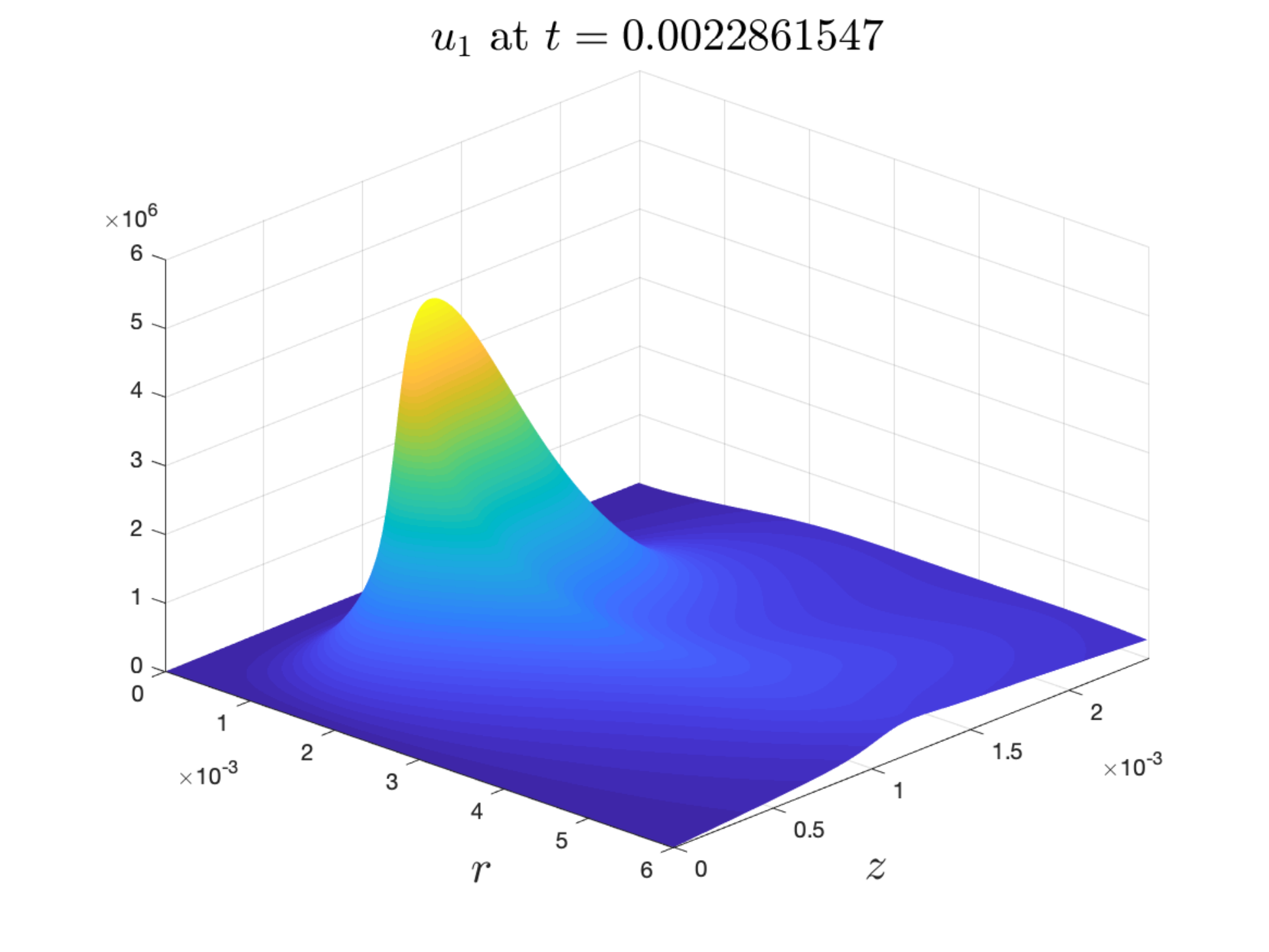}
    \includegraphics[width=0.32\textwidth]{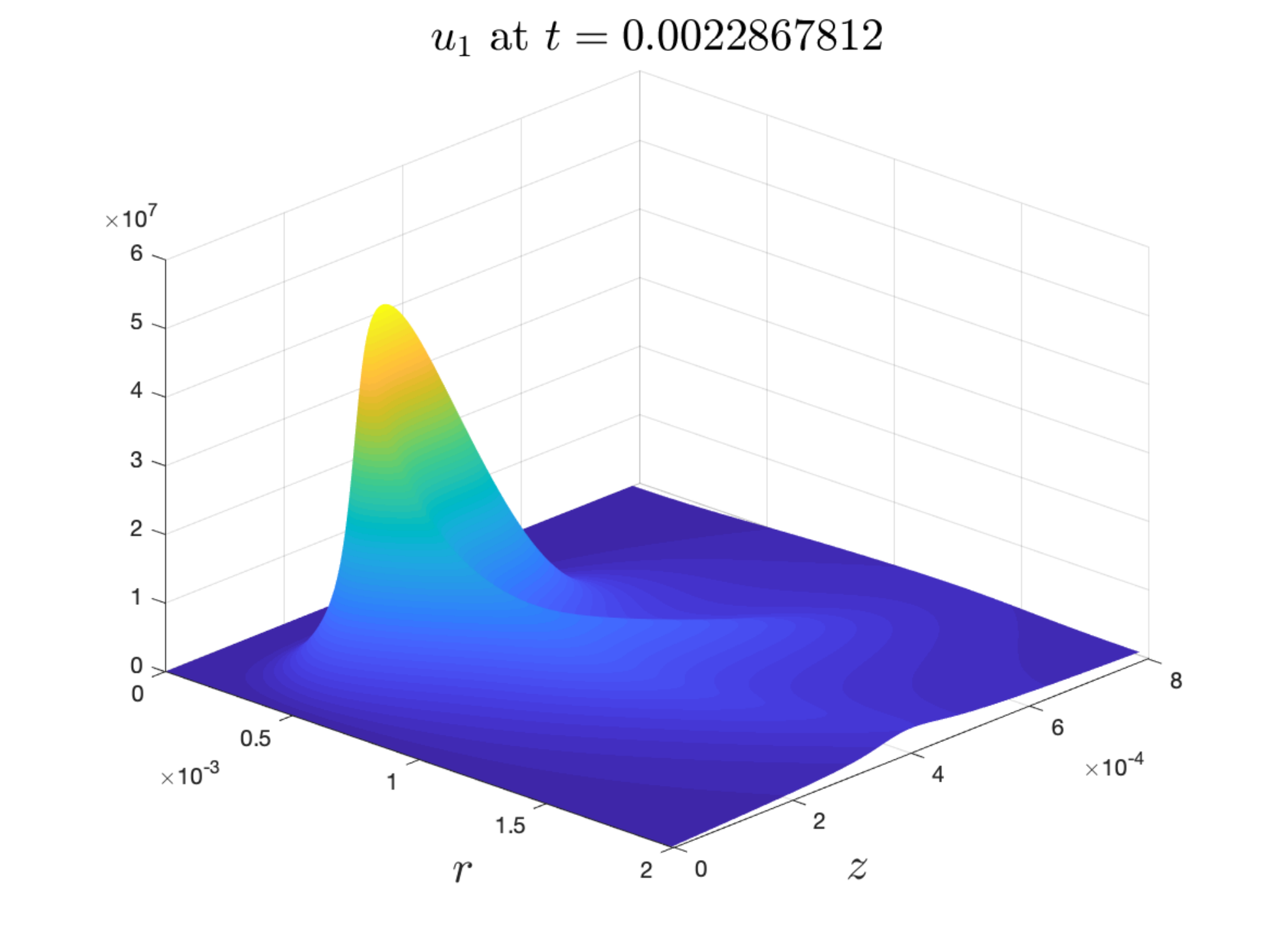}
    \includegraphics[width=0.32\textwidth]{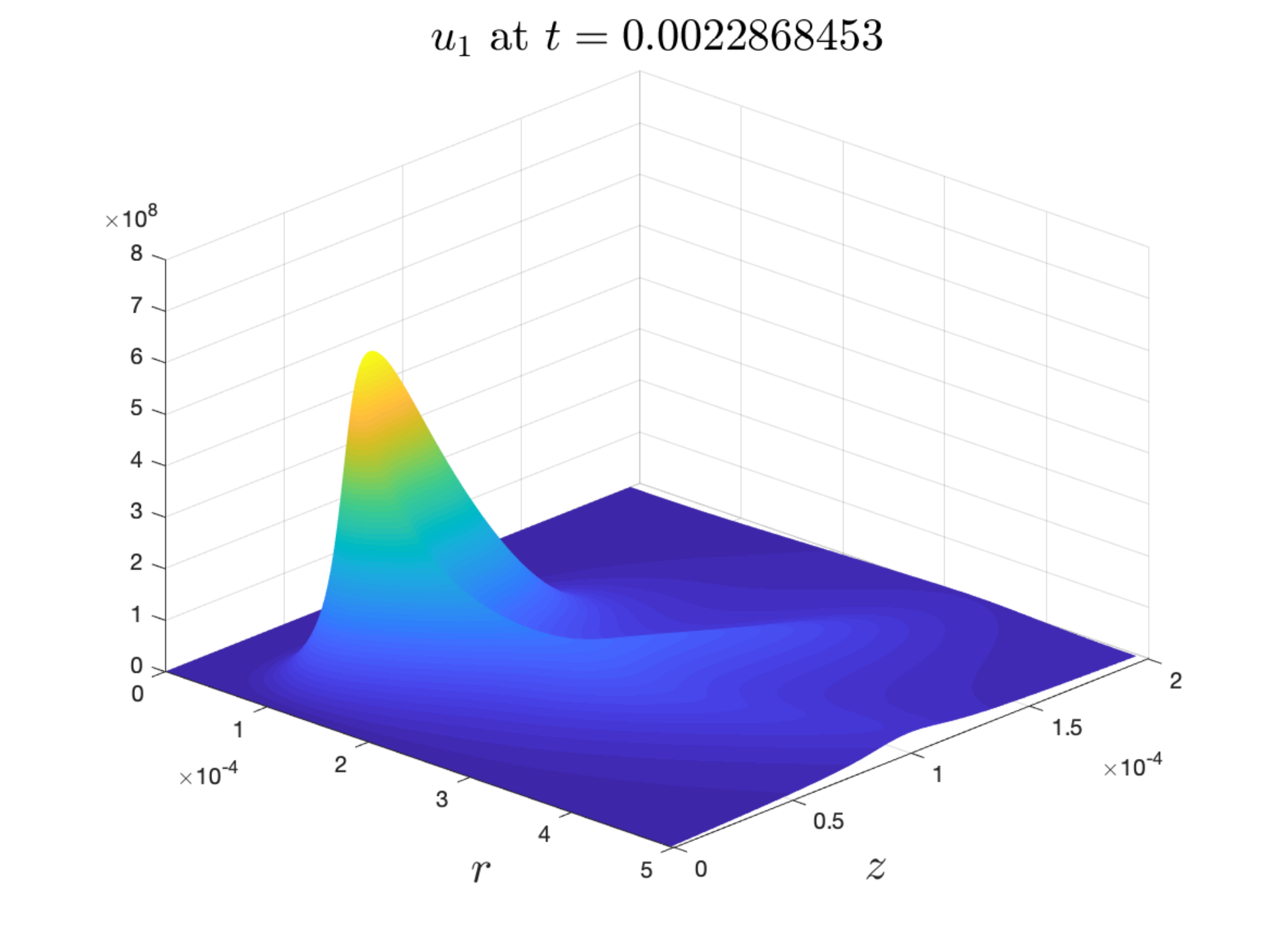}
    \includegraphics[width=0.32\textwidth]{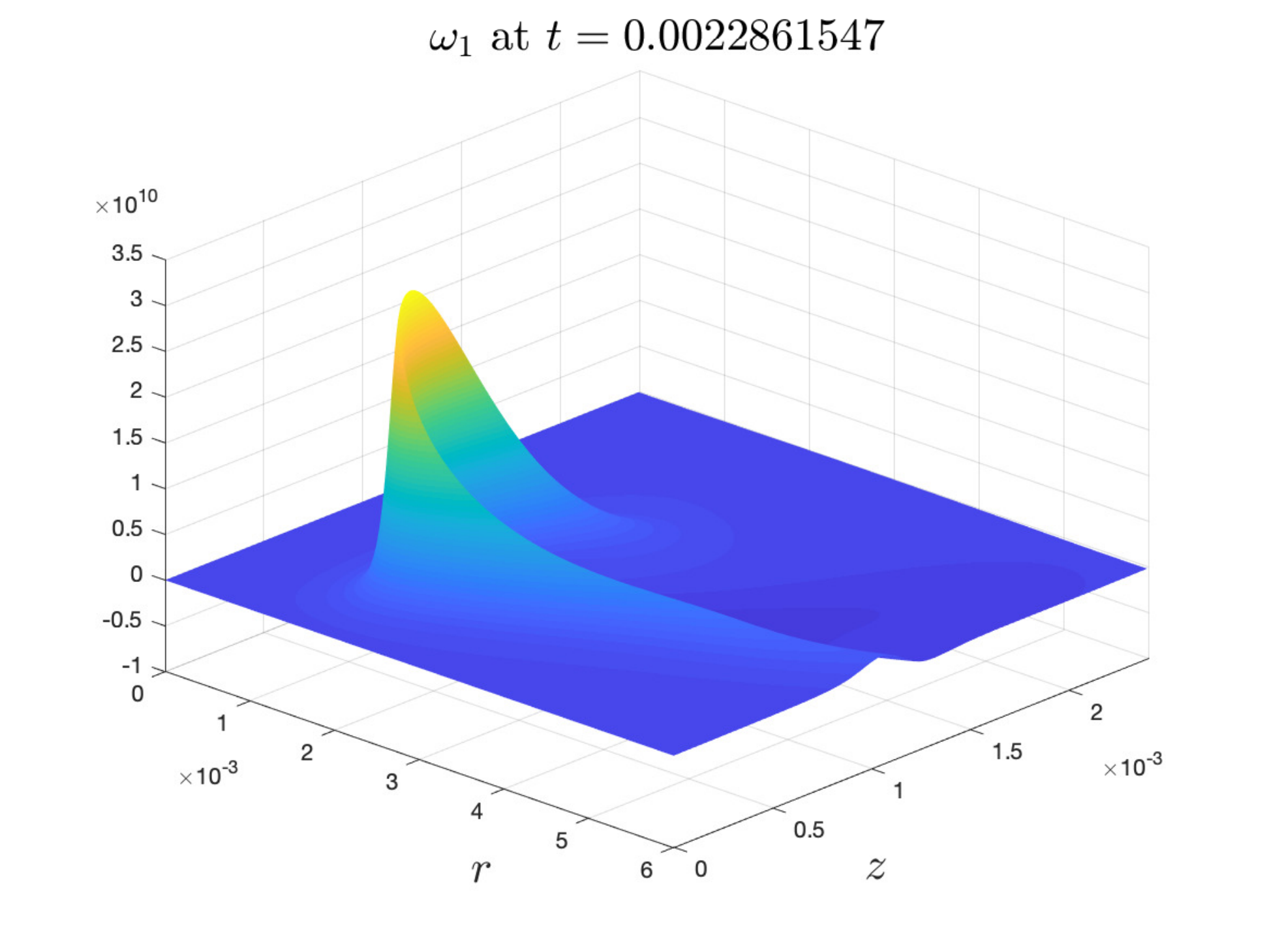}
    \includegraphics[width=0.32\textwidth]{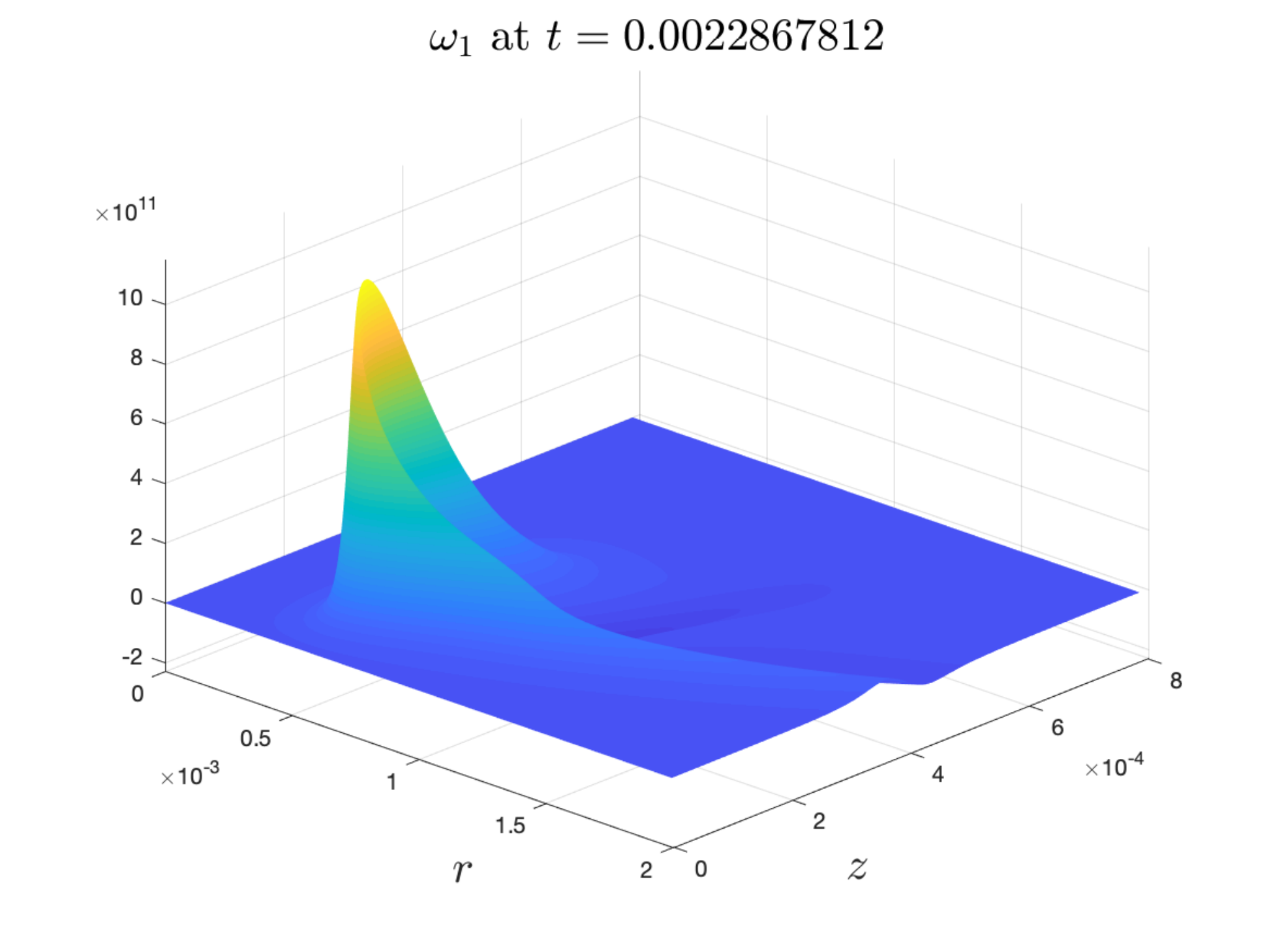}
    \includegraphics[width=0.32\textwidth]{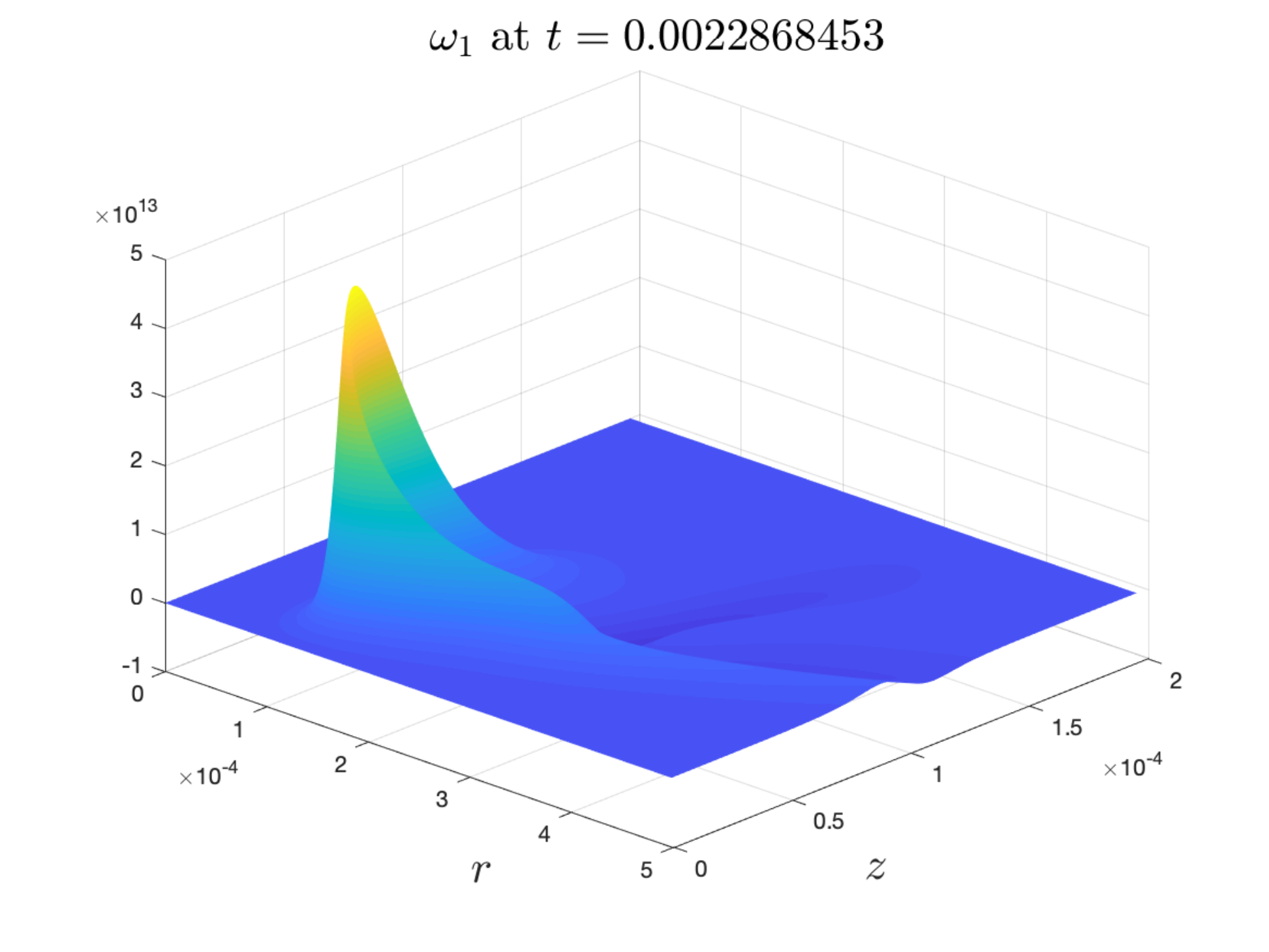}
    \caption[Profile evolution]{The evolution of the profiles of $u_1$ (row $1$) and $\om_1$ (row $2$) for the $3$D Navier--Stokes equations at three different times, $t_1=0.0022861547$, $t_2=0.0022867812$, $t_3=0.0022868453$, respectively. }  
    \label{fig:profile_evolution_nse}
\end{figure}

\subsection{Numerical Results}
\label{sec:first_sign_nse}
We have numerically solved the $3$D axisymmetric Navier--Stokes equations \eqref{eq:axisymmetric_NSE_1}-\eqref{eq:initial-data} on the half-period cylinder $\mathcal{D}_1=\{(r,z):0\leq r\leq 1, 0\leq z\leq 1/2\}$ using meshes of size $(n_1,n_2) = (256p,256p)$ for $p = 2, 3, \dots, 6$. In this subsection, we first present the major features of the potentially singular solution of the Navier--Stokes equations using our initial condition. In Section \ref{sec:performance_study_nse}, we carry out a careful resolution study of the numerical solutions. Then we investigate the nearly self-similar scaling properties in Section \ref{sec:scaling_study_nse}.

\subsubsection{Profile evolution}
\label{sec:profile_evolution_nse}
In this subsection, we investigate how the profiles of the solution evolve in time. We will use the numerical results computed on the adaptive mesh of size $(n_1,n_2) = (1536,1536)$. We have computed the numerical solution up to time $t_3=0.0022768453$ when it is still well resolved. 

In Figure \ref{fig:profile_evolution_nse}, we  present the $3$D solution profiles of $u_1,\om_1$ at $3$ different times $t_1 = 0.0022861547$, $t_2=0.0022867812$ and $t_3 = 0.0022868453$. We can see that the magnitudes of $u_1,\om_1$ grow very rapidly in time. The maximum vorticity has increased by a factor of $10^4$, $10^5$ and $10^6$, respectively at these three instants. We observe that the singular support of the profiles travels toward the origin (note that the domain size and the amplitude have been rescaled). Due to the strong viscous regularization, the profile of $u_1$ remains relatively smooth near $(R(t),Z(t))$, the maximum location of $u_1$. Moreover, the thin structure for $\om_1$ that we observed for the $3$D Euler equations in \cite{Hou-euler-2021} becomes much smoother. The tail part of $u_1$ and $\om_1$ is quite smooth and decays rapidly into the far field. This is quite different from the long tail that we observed for the two-scale traveling wave singularity reported in \cite{Hou-Huang-2021,Hou-Huang-2022}.

\begin{figure}[!ht]
\centering
    \begin{subfigure}[b]{0.35\textwidth}
        \centering
        \includegraphics[width=1\textwidth]{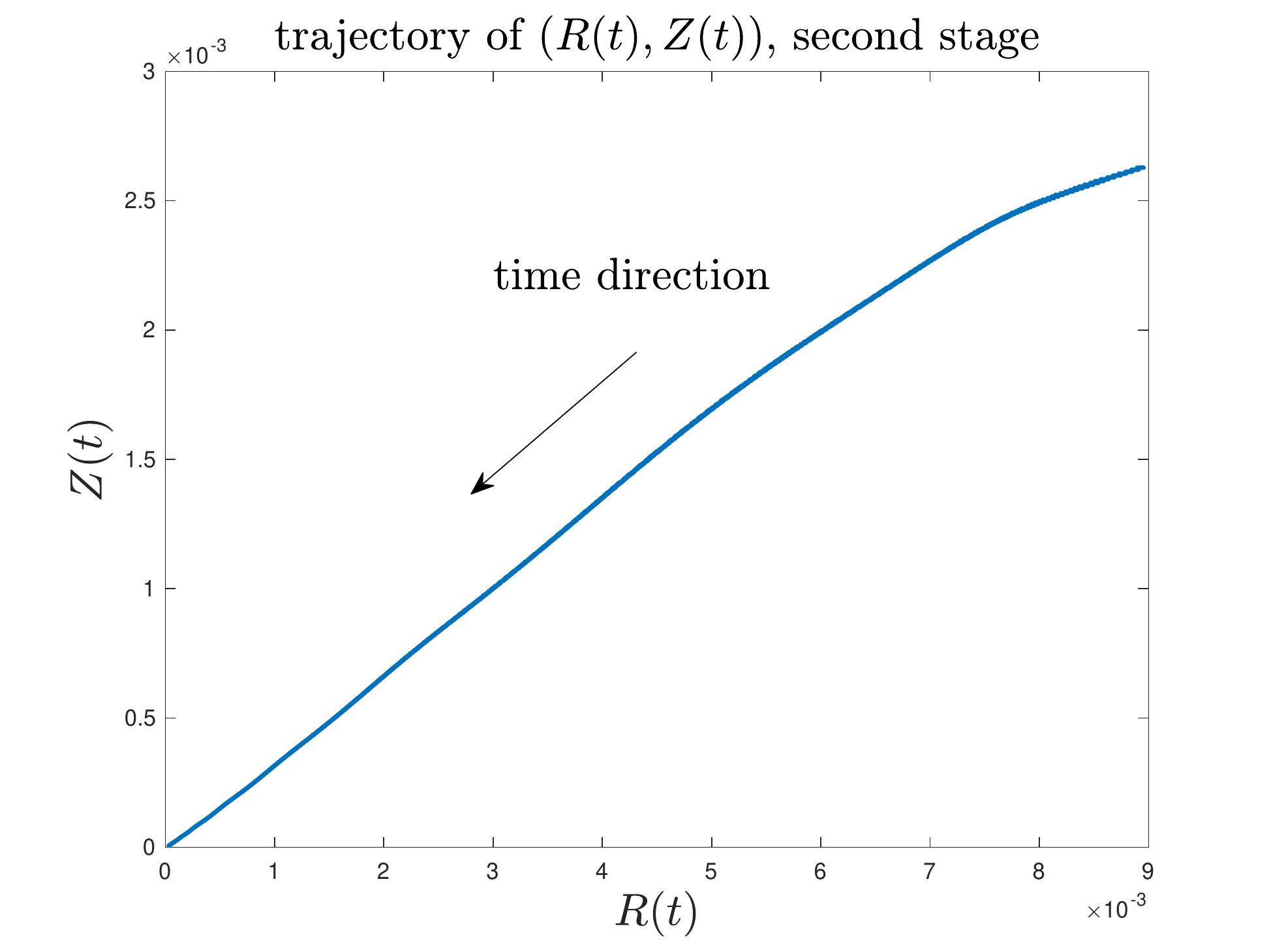}
        \caption{Trajectory $(R(t),Z(t))$}
    \end{subfigure}
    \begin{subfigure}[b]{0.35\textwidth}
        \centering
        \includegraphics[width=1\textwidth]{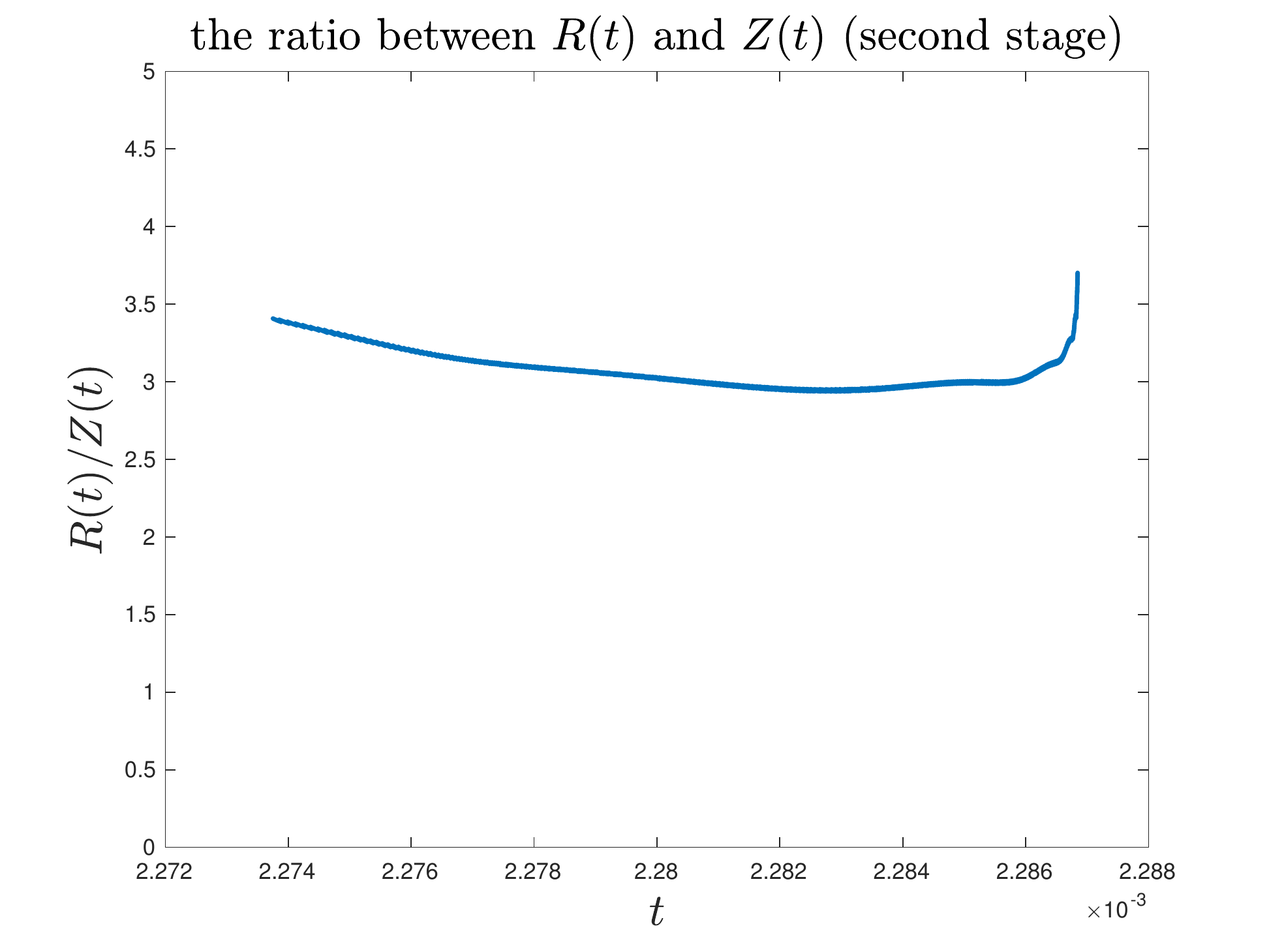}
        \caption{The ratio $R(t)/Z(t)$}
    \end{subfigure}
    \vspace{0.1in}
  \begin{subfigure}[b]{0.35\textwidth}
        \centering
        \includegraphics[width=1\textwidth]{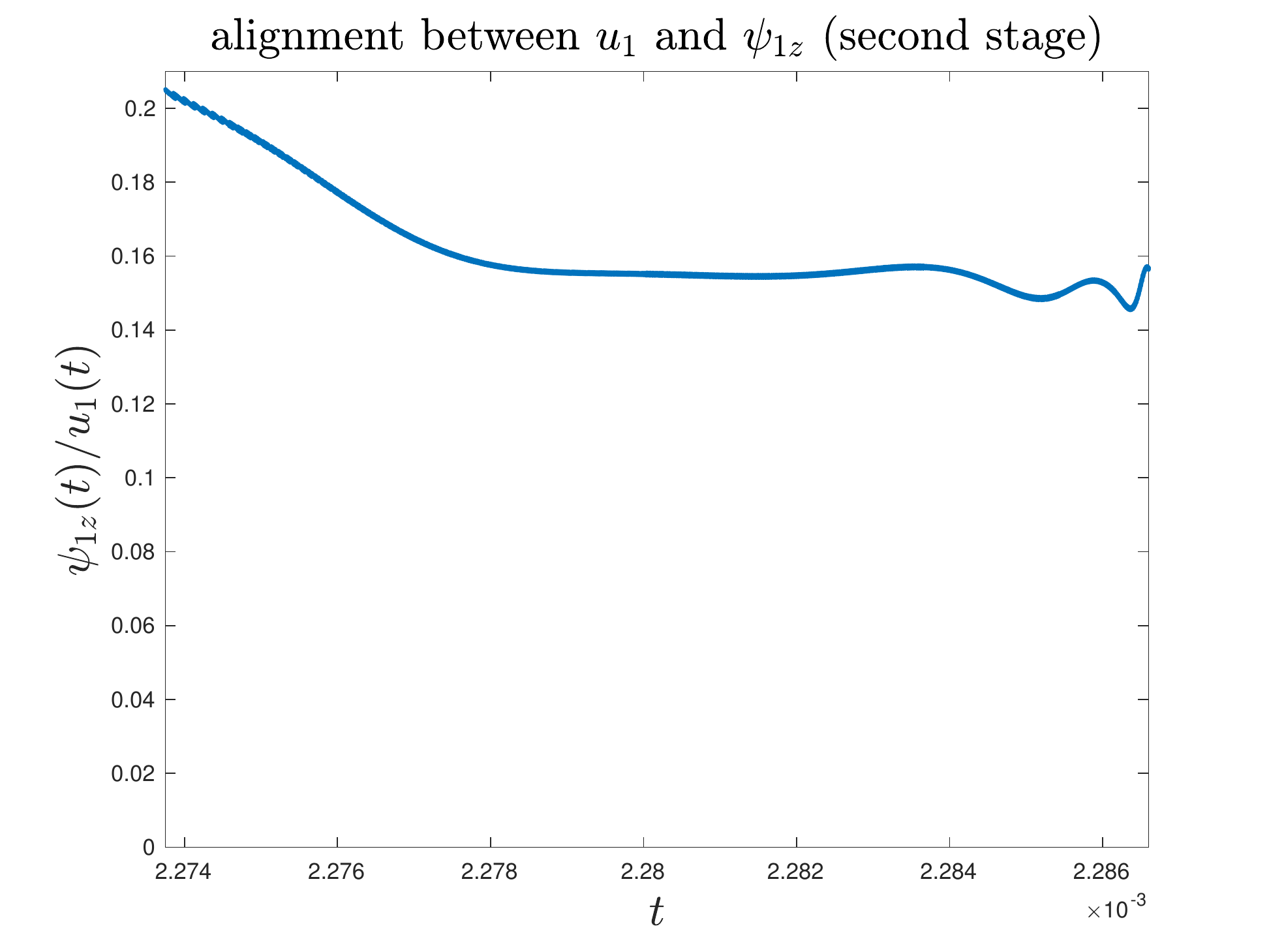}
        \caption{alignment $\psi_{1z}/u_1$ second stage}
    \end{subfigure}
    \begin{subfigure}[b]{0.35\textwidth}
        \centering
        \includegraphics[width=1\textwidth]{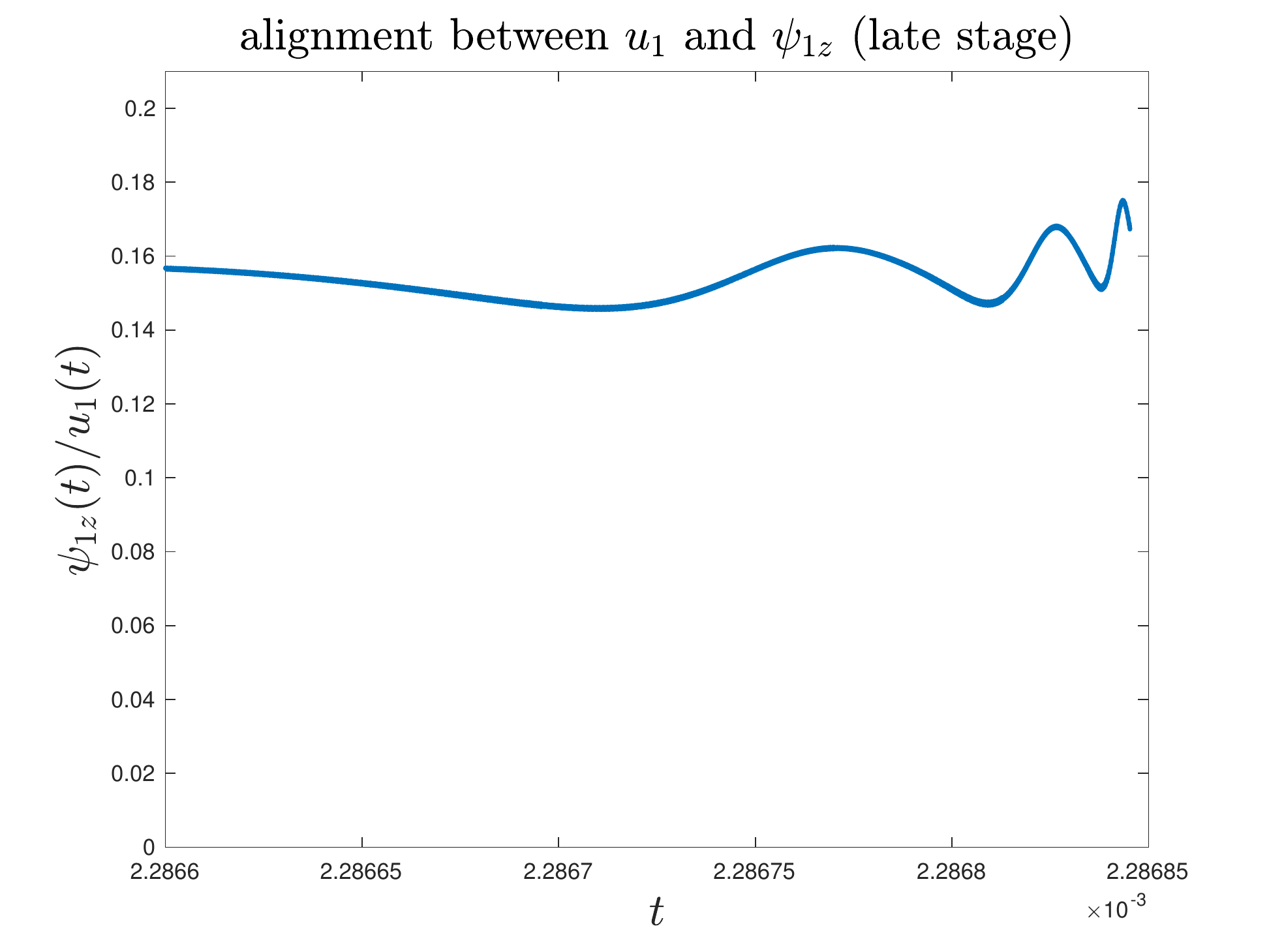}   
          \caption{alignment $\psi_{1z}/u_1$ late stage}
    \end{subfigure}
    \caption[Trajectory]{(a) the trajectory of $(R(t),Z(t))$ approaching to the origin, (b) the ratio $R(t)/Z(t)$,  (c) the alignment between $\psi_{1z}$ and $u_1$ at $R(t),Z(t))$ in the second stage with $ 0.00227375\leq t \leq 0.0022866$. (d) the alignment $\psi_{1z}$ and $u_1$ at $R(t),Z(t))$  in the late stage with $ 0.0022866 \leq t \leq 0.0022868453$.}  
     \label{fig:trajectory_nse}
        \vspace{-0.05in}
\end{figure}

\subsection{Trajectory and alignment}\label{sec:two_scale_nse}  The maximum location $(R(t),Z(t))$ of $u_1$ lies on $r=0$ initially and travels downward. Then it moves outward almost horizontally. After a short time, $(R(t),Z(t))$ turns around and propagates toward the origin, see Figure \ref{fig:trajectory_nse} (a). 
Due to the viscous regularization, the solution becomes smoother and is more stable. We are able to compute up to a time when $(R(t),Z(t))$ is very close to the origin. This is something we could not achieve for the $3$D Euler equations \cite{Hou-euler-2021}. We also observe a positive alignment between $\psi_{1z}$ and $u_1$ and the alignment becomes stronger and stronger in time, see Figure \ref{fig:trajectory_nse} (c)-(d). It is worth noting that there is a relatively long stable phase in which the alignment between $u_1$ and $\psi_{1z}$ is relatively flat. This indicates that the $3$D Navier--Stokes equations enter a relatively stable scaling relationship period. This is also something that we did not observe for the $3$D Euler equations \cite{Hou-euler-2021}. 

In the Figure \ref{fig:trajectory_nse} (b), we observe that the ratio $R(t)/Z(t)$ decays most of the time and has a very mild growth in the late stage of our computation. This property is very different from the two-scale traveling wave singularity reported in \cite{Hou-Huang-2021}. In our case, we have an essentially one-scale solution structure for the majority of the time. The mild growth of $R(t)/Z(t)$ in the late stage seems to be due to the numerical dissipation introduced by the frequent changes of adaptive mesh in the late stage. We will further discuss this issue in Section \ref{stretching-diffusion}.

\subsubsection{Rapid growth}\label{sec:rapid_growth_nse} 
\begin{figure}[!ht]
\centering
    \includegraphics[width=0.32\textwidth]{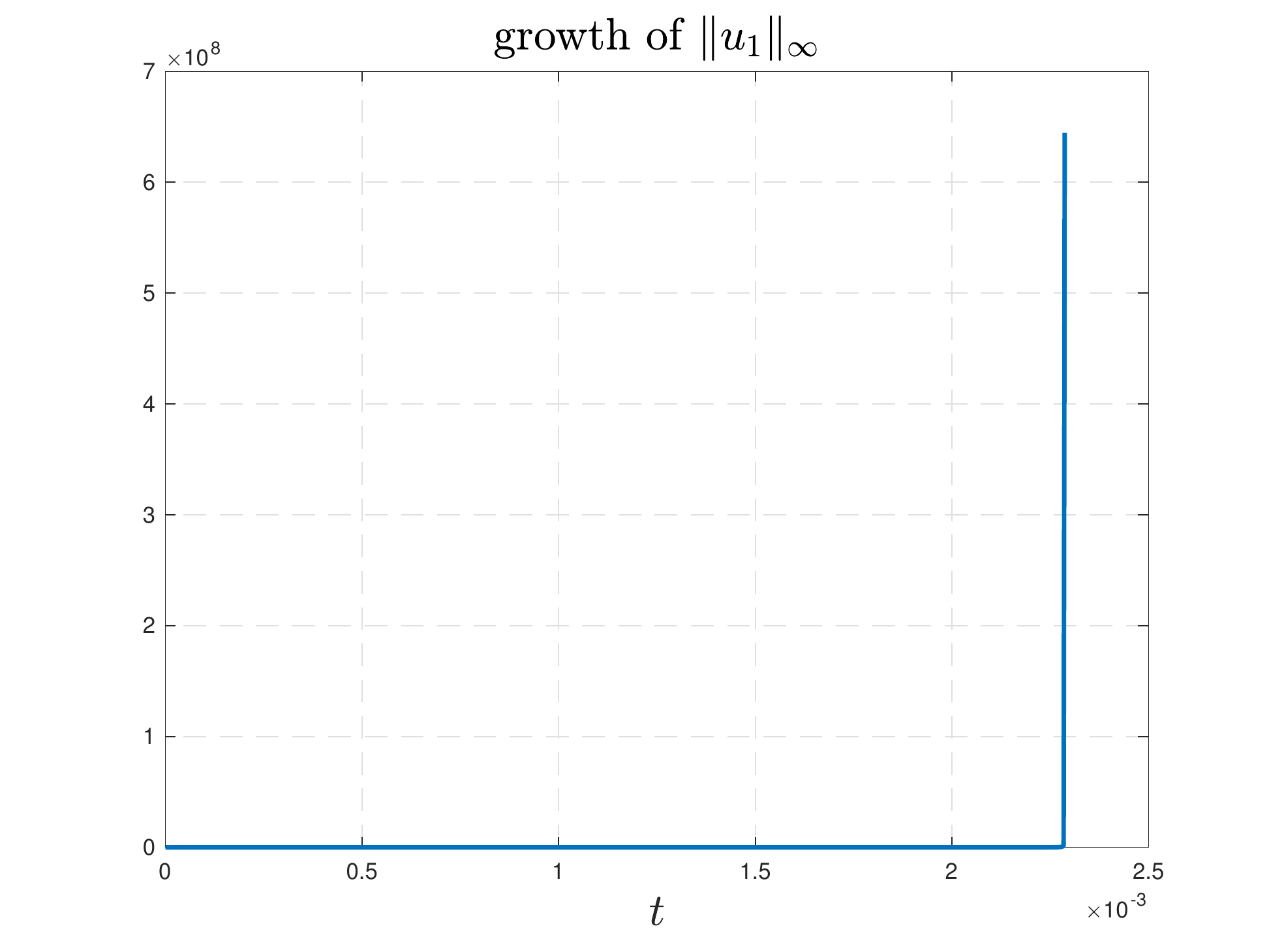}
    \includegraphics[width=0.32\textwidth]{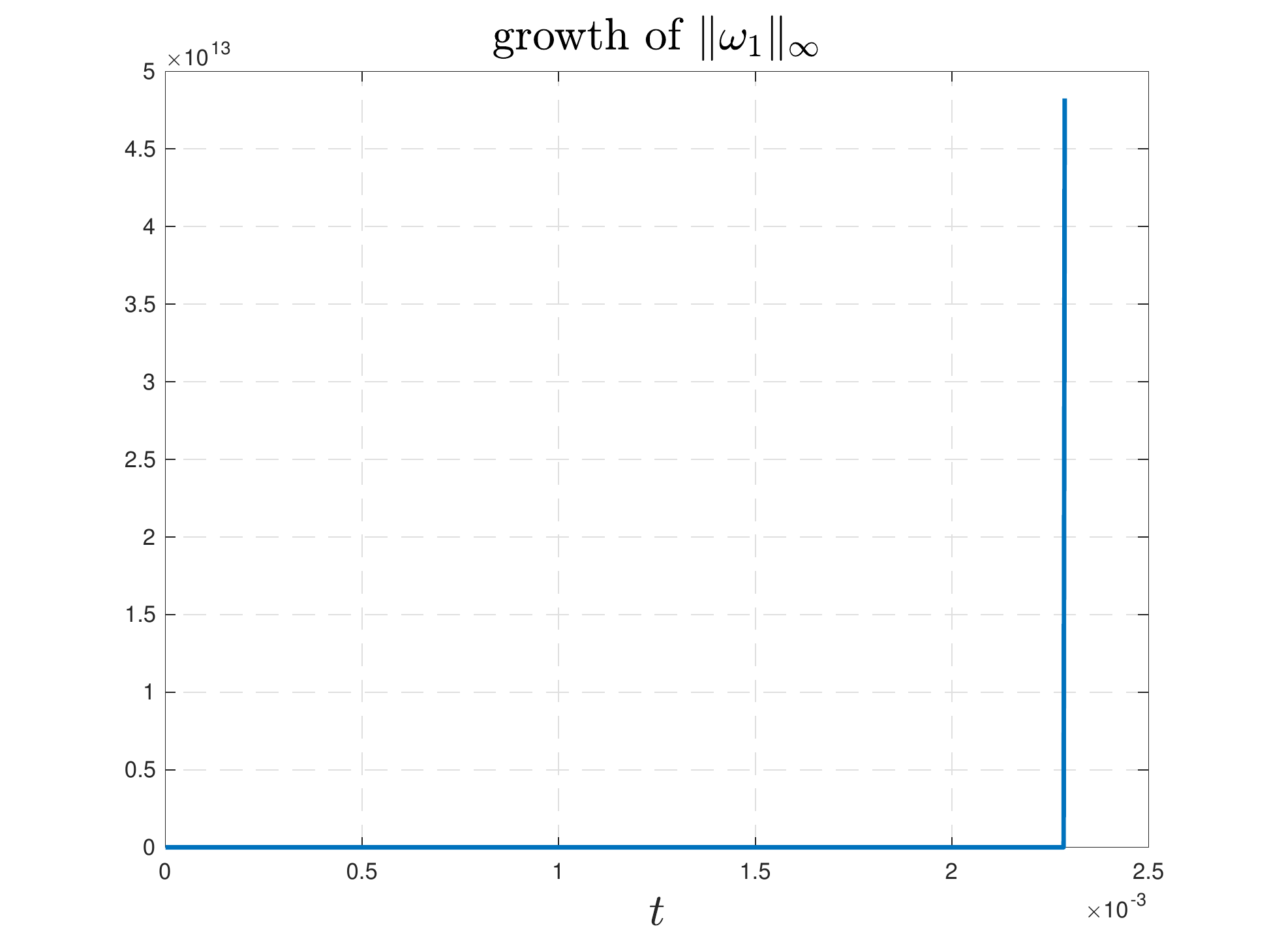} 
    \includegraphics[width=0.32\textwidth]{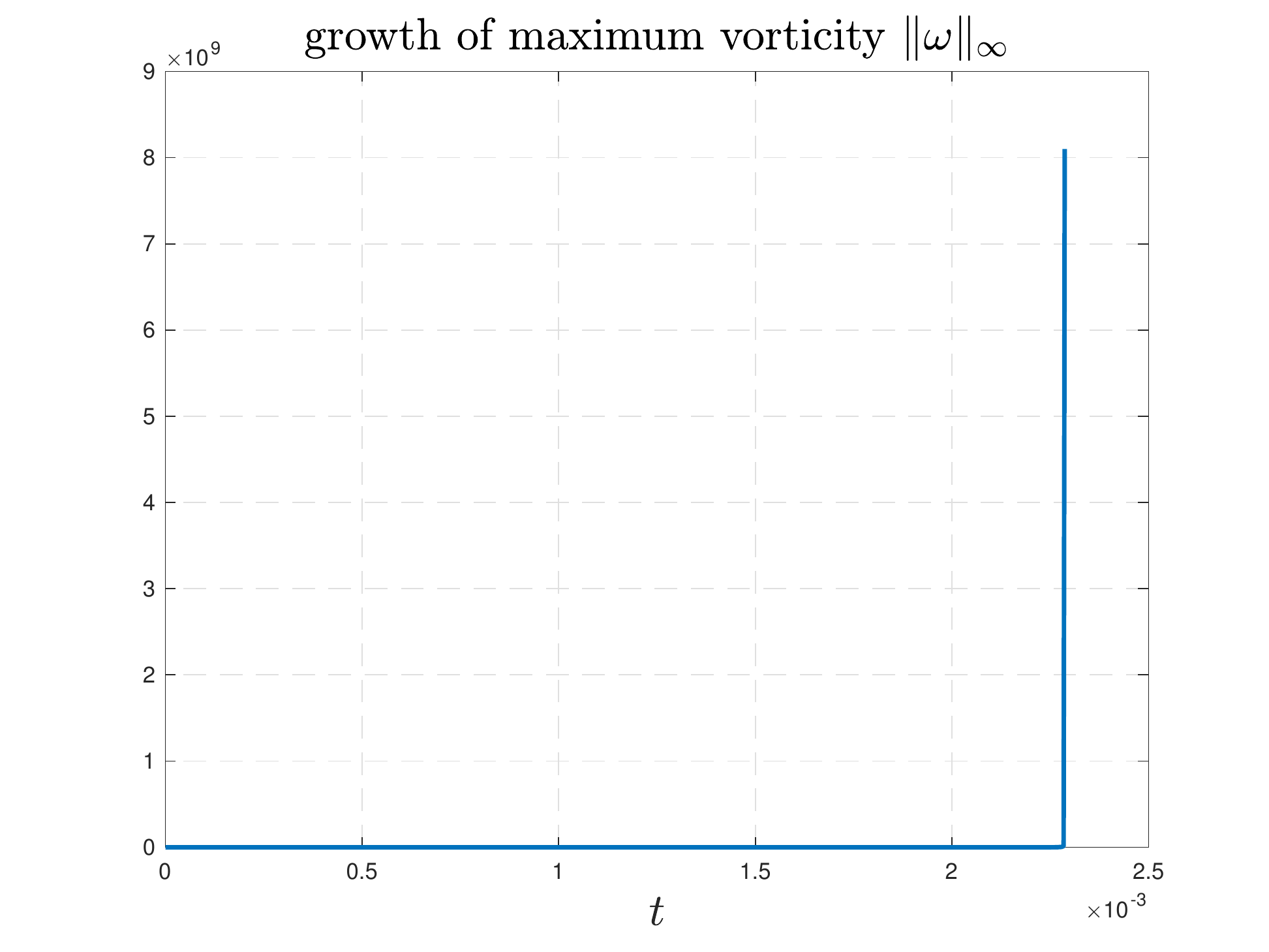}
    \includegraphics[width=0.32\textwidth]{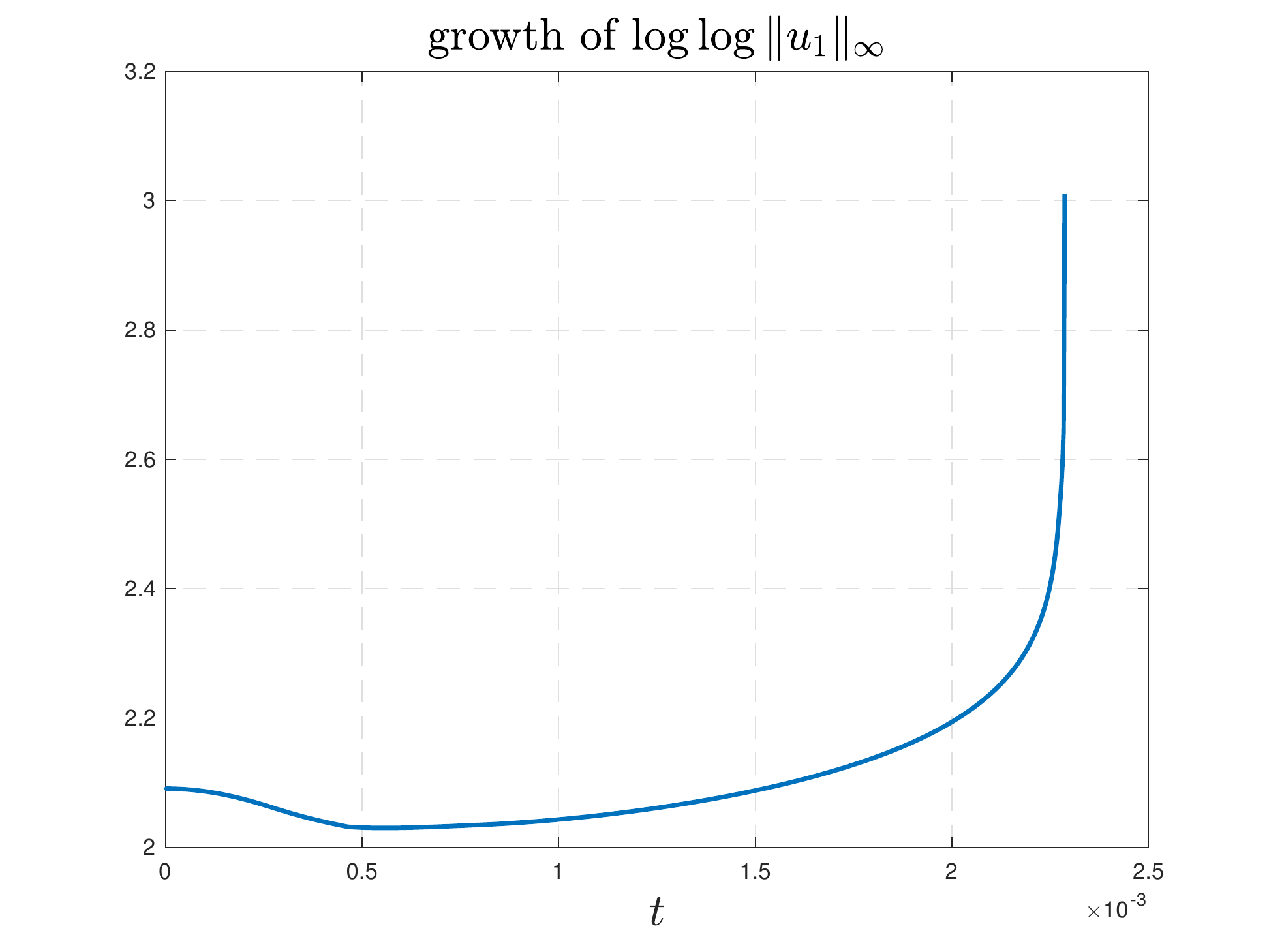}
    \includegraphics[width=0.32\textwidth]{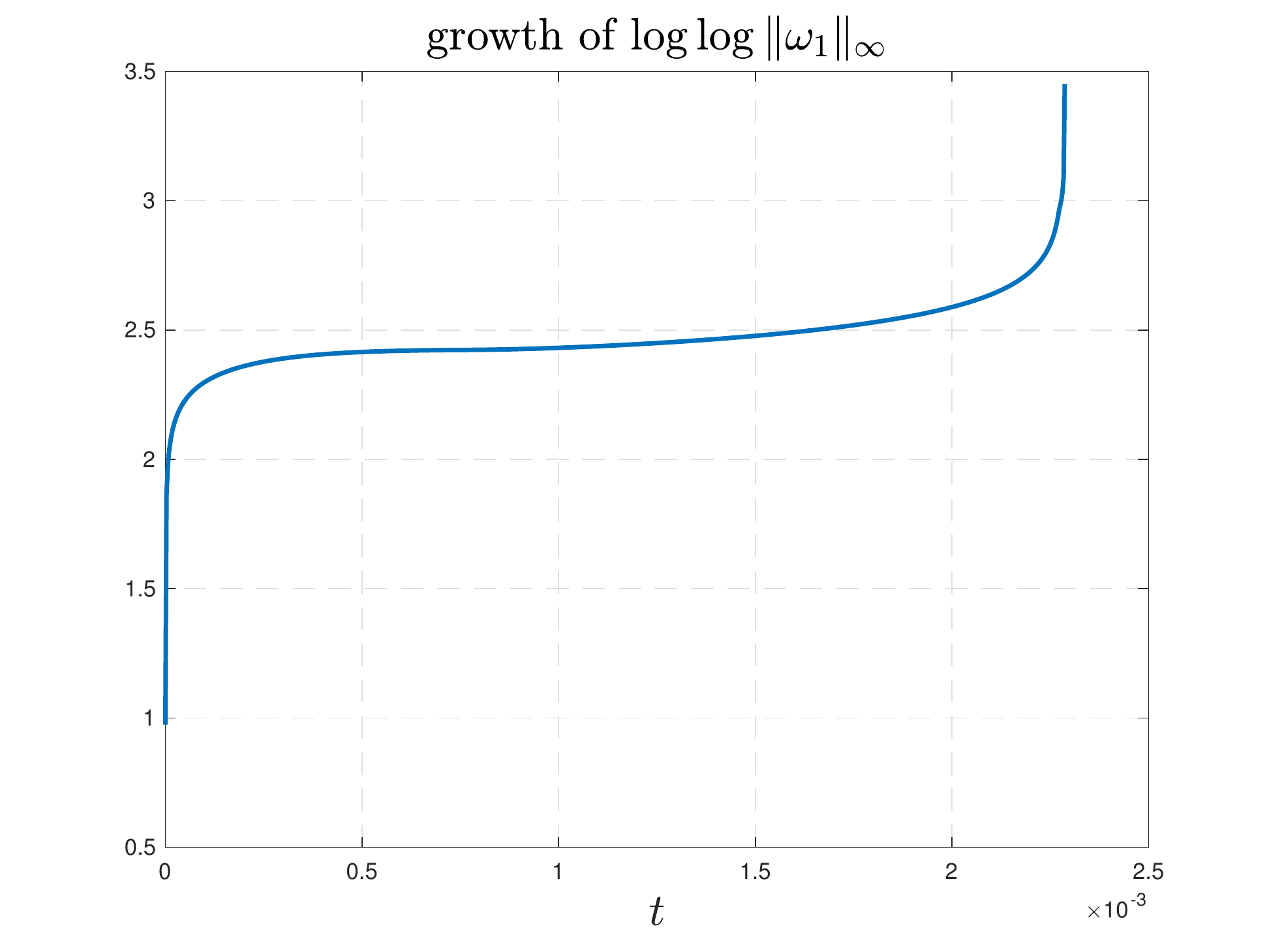}
    \includegraphics[width=0.32\textwidth]{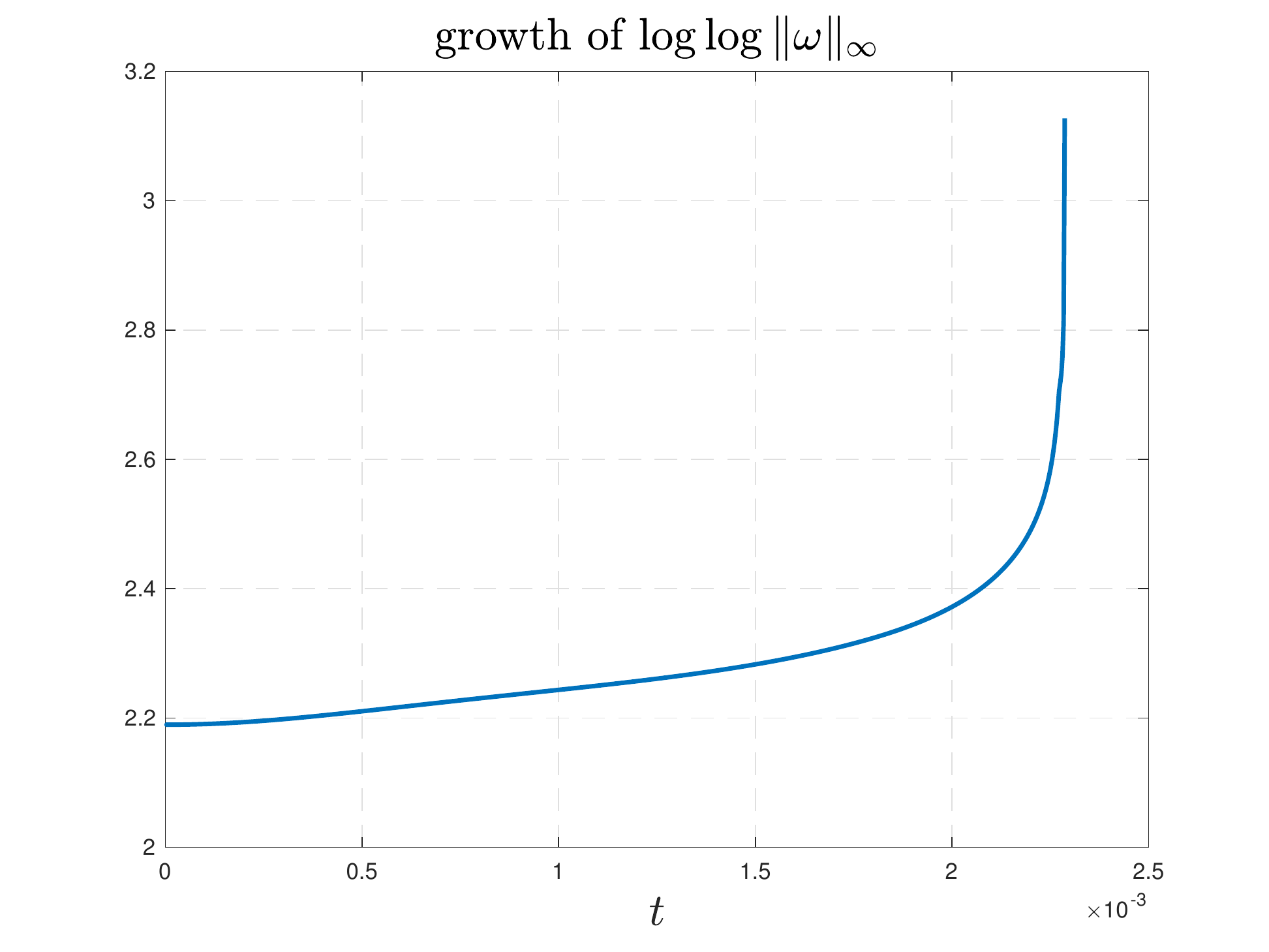}
    \caption[Rapid growth]{First row: the growth of $\|u_1\|_{L^\infty}$, $\|\om_1\|_{L^\infty}$ and $\|\vom\|_{L^\infty}$ as functions of time. Second row: $\log\log\|u_1\|_{L^\infty}$, $\log\log\|\om_1\|_{L^\infty}$ and $\log\log\|\vom\|_{L^\infty}$.} 
    \label{fig:rapid_growth_nse}
\end{figure}

We observe that the solution develops rapid growth dynamically. In the first row of Figure \ref{fig:rapid_growth_nse}, we report the maximum of $|u_1|,|\om_1|$ and $|\vom|$ as a function of time. 
We can see that these variables grow extremely rapidly in time. In the second row of Figure \ref{fig:rapid_growth_nse}, we plot that the double logarithm of $\|u_1\|_{L^\infty}$,  $\|\psi_{1z}\|_{L^\infty}$ and  $\|\vom\|_{L^\infty}$ as a function of time. We observe that the growth rate is much faster than double-exponential. Compared with the growth of the corresponding quantities for the $3$D Euler equations, the growth rate of these quantities for the $3$D Navier--Stokes equations is much larger in magnitude and is more stable. 

\begin{figure}[!ht]
\centering
    \includegraphics[width=0.38\textwidth]{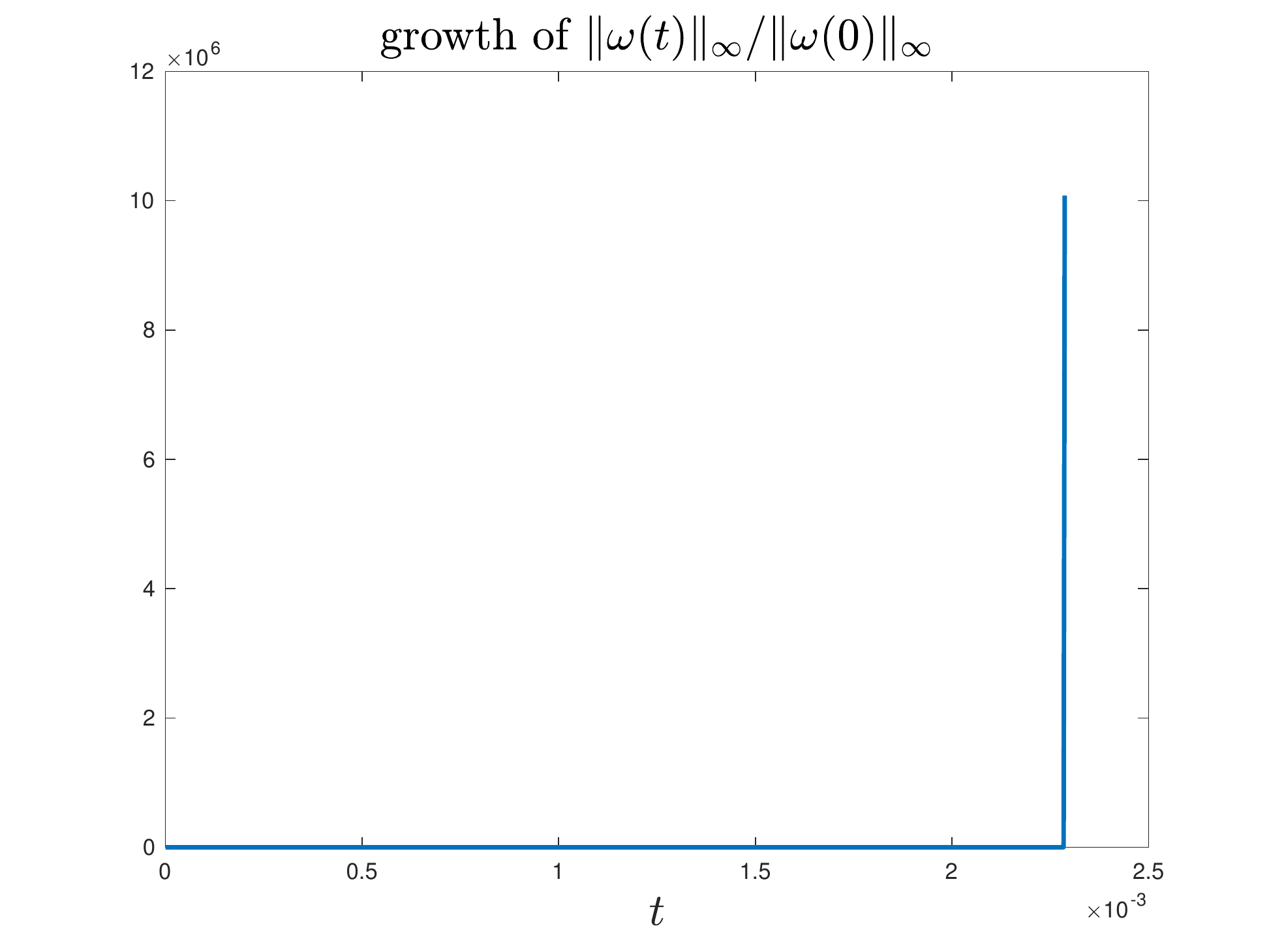}
    \includegraphics[width=0.38\textwidth]{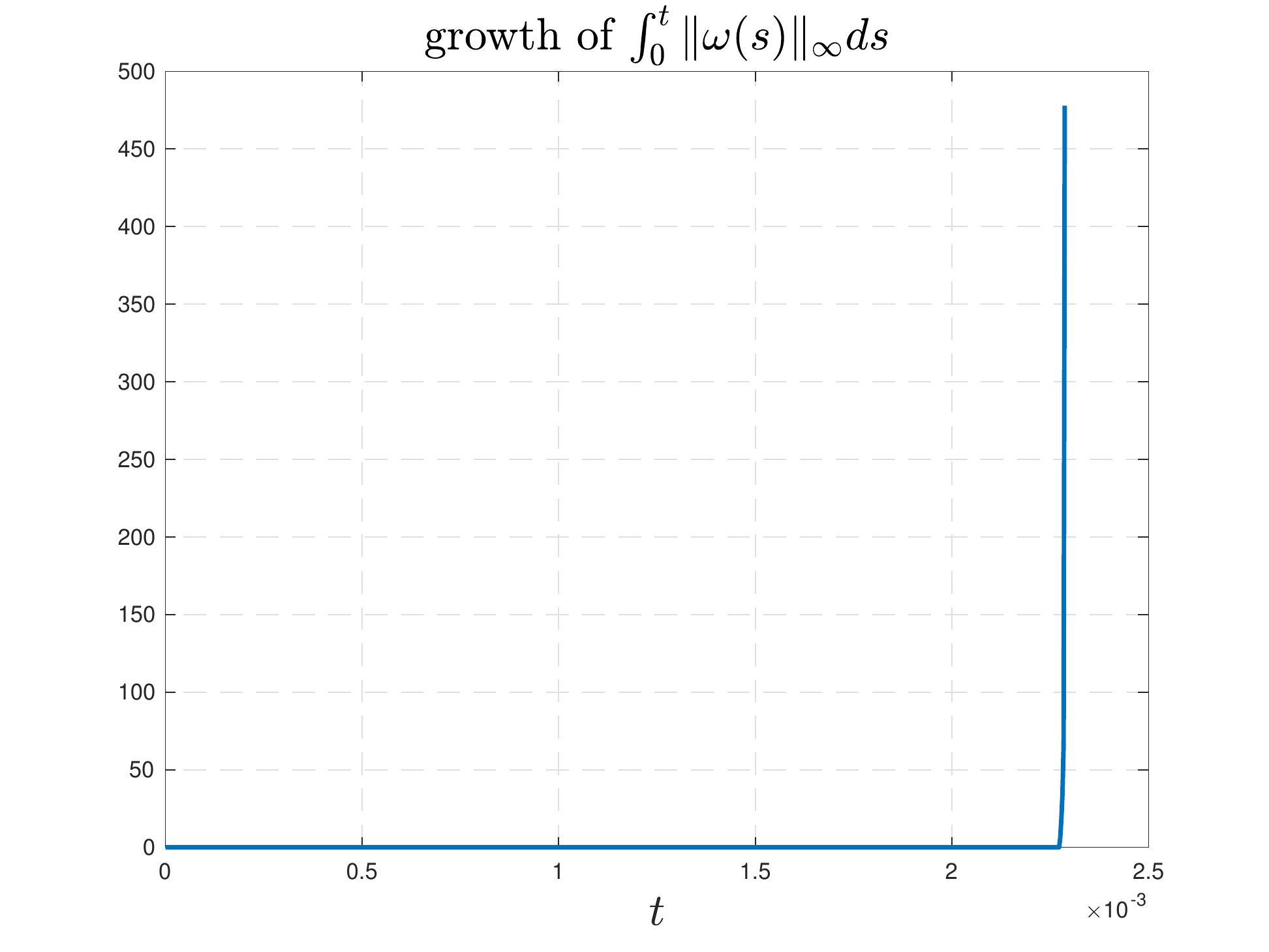} 
    \caption[Rapid growth]{ Left plot: the amplification of maximum vorticity relative to its initial maximum vorticity, $\|\vom (t)\|_{L^\infty}/\|\vom (0)\|_{L^\infty}$ as a function of time. Right plot: the time integral of maximum vorticity,  $\int_0^t \|\omega (s)\|_{L^\infty}ds$ as a function of time. The solution is computed using $1536\times 1536$ grid. The final time instant is $t_4=0.0022868502$.} 
    \label{fig:rapid_growth2_nse}
\end{figure}

We also compute the relative growth of maximum vorticity $\|\vom (t)\|_{L^\infty}/\|\vom (0)\|_{L^\infty}$ and $\int_0^{t}\|\vom(s)\|_{L^\infty}\idiff s$ in Figure \ref{fig:rapid_growth2_nse}. The final time of this computation is at $t_4=0.0022868502$. We observe that $\|\omega (t)\|_{L^\infty}/\|\omega (0)\|_{L^\infty}$ has increased by a factor of $10^7$ by the end of the computation. To best of our knowledge, such a large growth rate of the maximum vorticity has not been reported for the $3$D incompressible Navier--Stokes equations in the literature. The rapid growth of $\int_0^{t}\|\vom(s)\|_{L^\infty}\idiff s$ seems to suggest that the $3$D Navier--Stokes equations develop a potential finite time singularity according to the well-known Beale-Kato-Majda blow-up criterion \cite{beale1984remarks}.

\begin{figure}[!ht]
\centering
\vspace{-0.7in}
    \begin{subfigure}[b]{0.40\textwidth}
        \centering
        \includegraphics[width=1\textwidth]{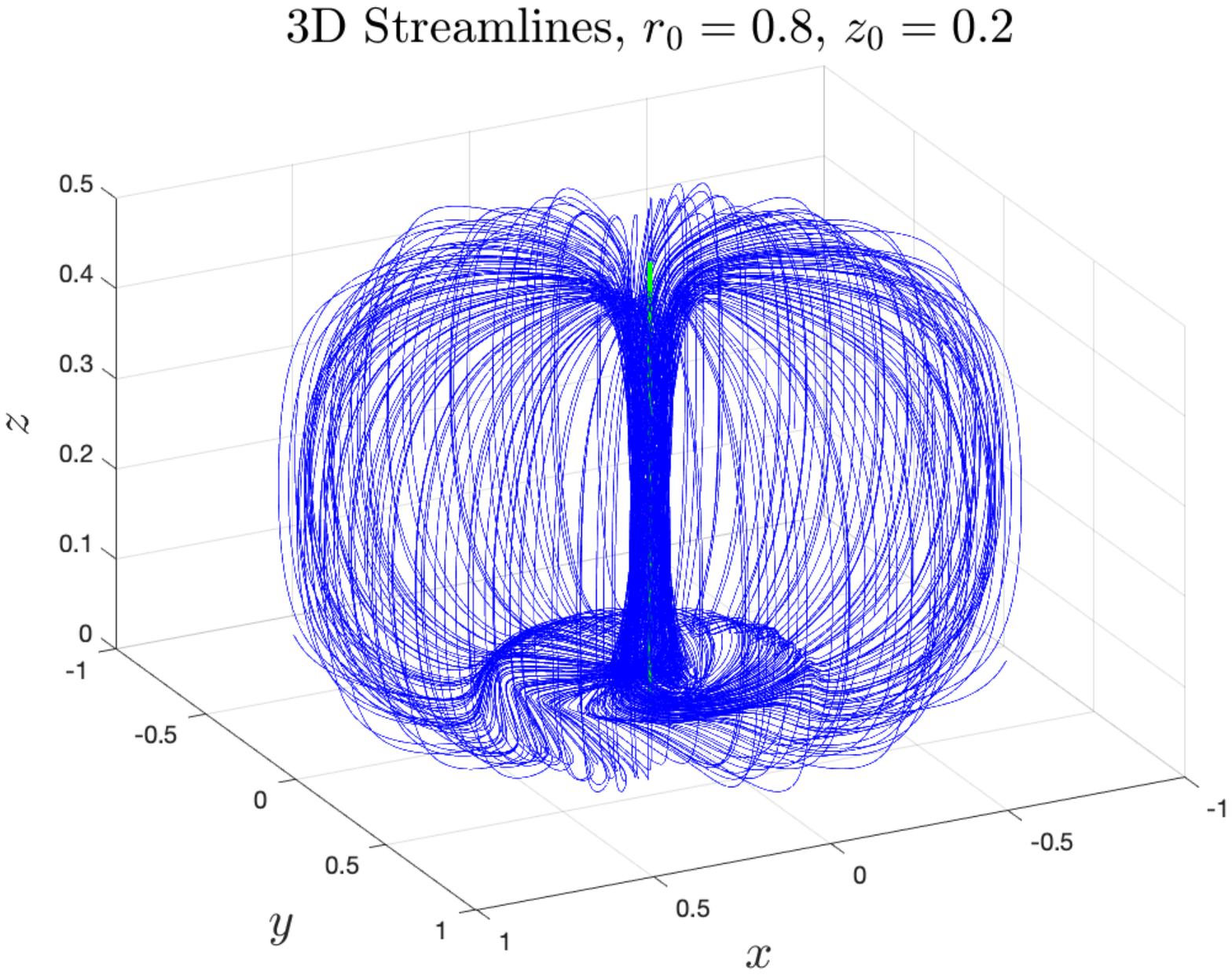}
        \vspace{-0.9in}
        \caption{$r_0 = 0.8$, $z_0 = 0.2$}
    \end{subfigure}
    \begin{subfigure}[b]{0.40\textwidth}
        \centering
        \includegraphics[width=1\textwidth]{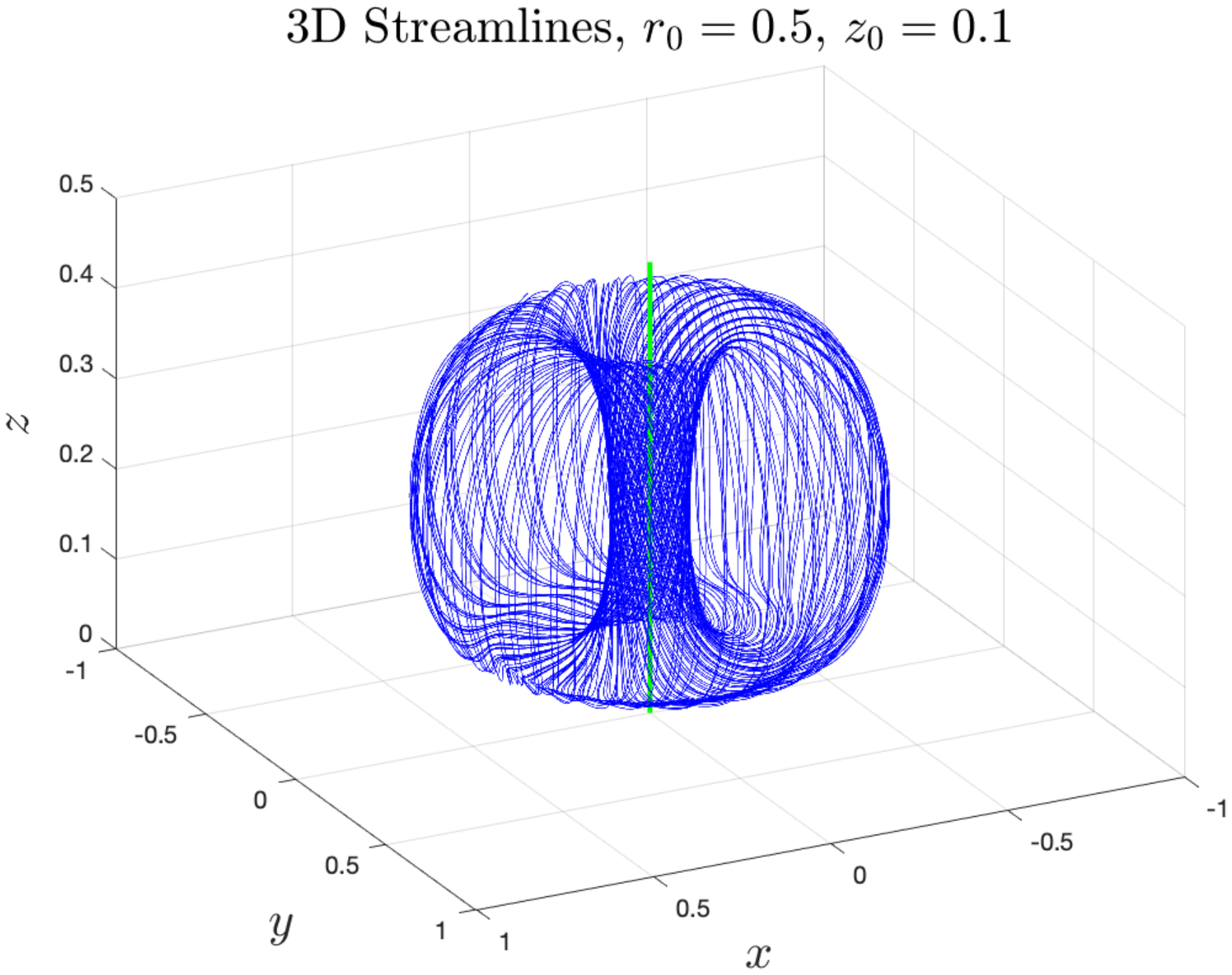}
        \vspace{-0.9in}
        \caption{$r_0 = 0.5$, $z_0 = 0.1$}
    \end{subfigure}
    \begin{subfigure}[c]{0.40\textwidth}
        \centering
        \vspace{-0.7in}
        \includegraphics[width=1\textwidth]{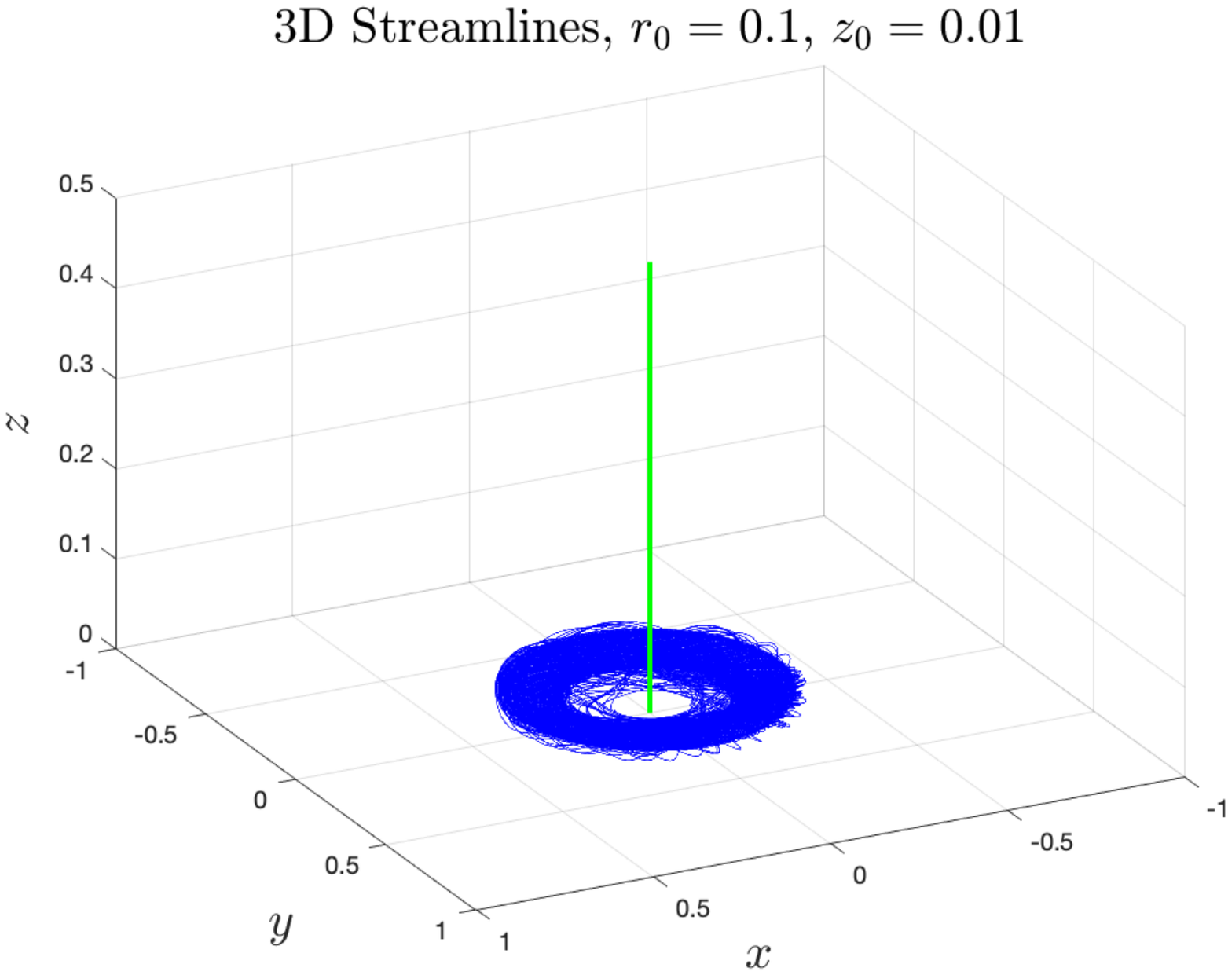}
        \vspace{-0.9in}
        \caption{$r_0 = 0.1$, $z_0 = 0.01$}
    \end{subfigure}
    \begin{subfigure}[d]{0.40\textwidth}
        \centering
         \vspace{-0.7in}
        \includegraphics[width=1\textwidth]{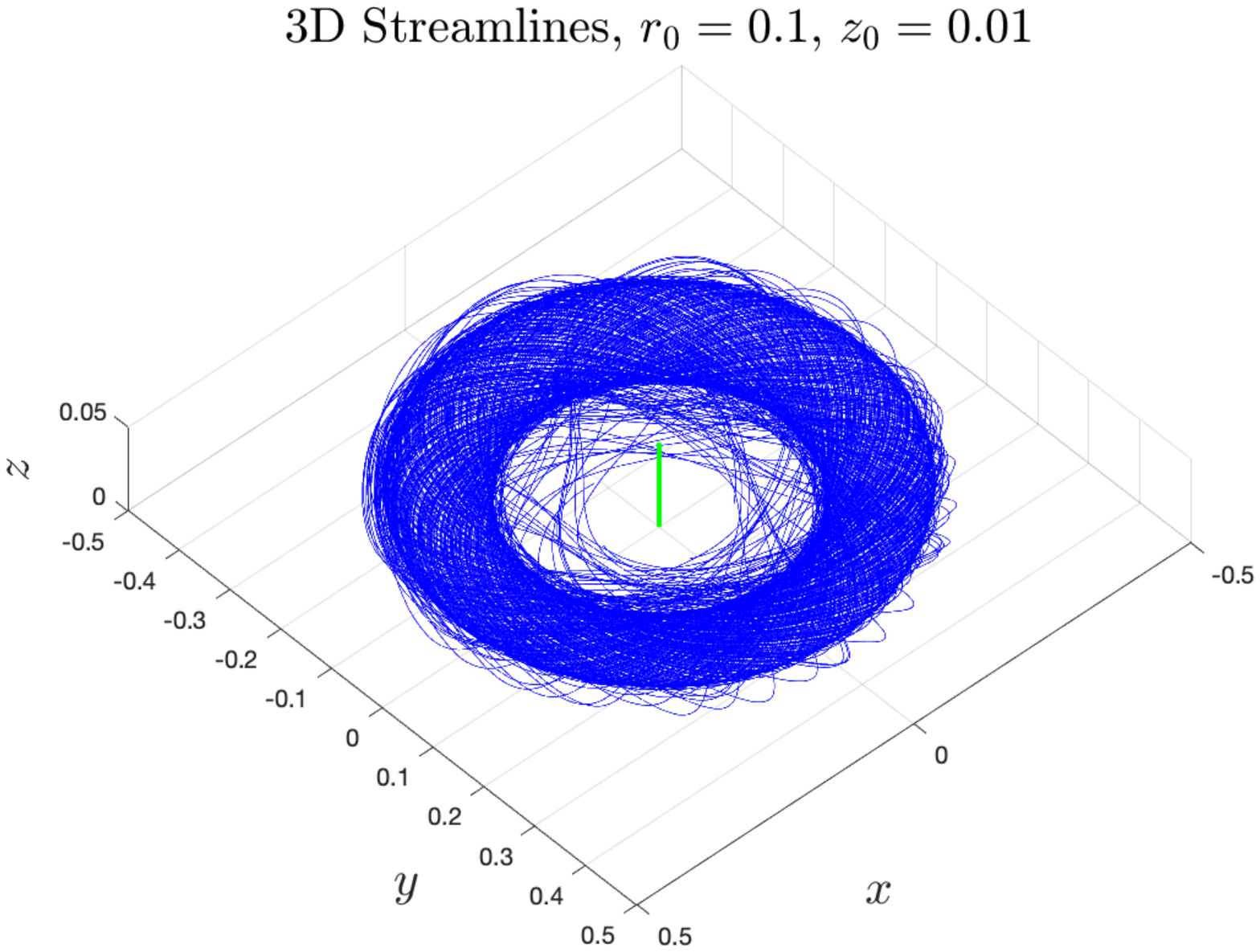}
        \vspace{-0.9in}
        \caption{same as (c), a zoom view}
    \end{subfigure}
    \caption[Local streamline]{The streamlines of $(u^r(t),u^\theta(t),u^z(t))$ at time $t_0=0.00227375$ with initial points given by (a) $(r_0,z_0) = (0.8,0.2)$, (b) $(r_0,z_0) = (0.5,0.1)$, (c) $(r_0,z_0) = (0.1,0.01)$ ($3$D view), (d) $(r_0,z_0)$ is the same as (c), a zoom view. The green pole is the symmetry axis $r=0$.}  
     \label{fig:streamline_3D_global_nse}
        \vspace{-0.05in}
\end{figure}

\subsubsection{Velocity field} In this subsection, we investigate the feature of the velocity field. We first study the $3$D velocity field by looking at the induced streamlines. In Figure~\ref{fig:streamline_3D_global_nse}, we plot the streamlines induced by the velocity field $\vu(t)$ at a relatively early time $t_0 = 0.00227375$. This is the time by which we end our computation using viscosity $\nu=5\cdot 10^{-4}$ and switch to the larger viscosity $\nu=5 \cdot 10^{-3}$. By this time, the ratio between the maximum vorticity and the initial maximum vorticity, i.e. $\|\vom (t)\|_{L^\infty}/\|\vom (0)\|_{L^\infty}$, has increased by a factor of $498.42$ only. We can also regard the solution at $t_0 = 0.00227375$ as the new initial condition for the subsequent computation using the larger viscosity $\nu=5 \cdot 10^{-3}$.

\begin{figure}[!ht]
\centering
\vspace{-0.7in}
    \begin{subfigure}[b]{0.40\textwidth}
        \centering
        \includegraphics[width=1\textwidth]{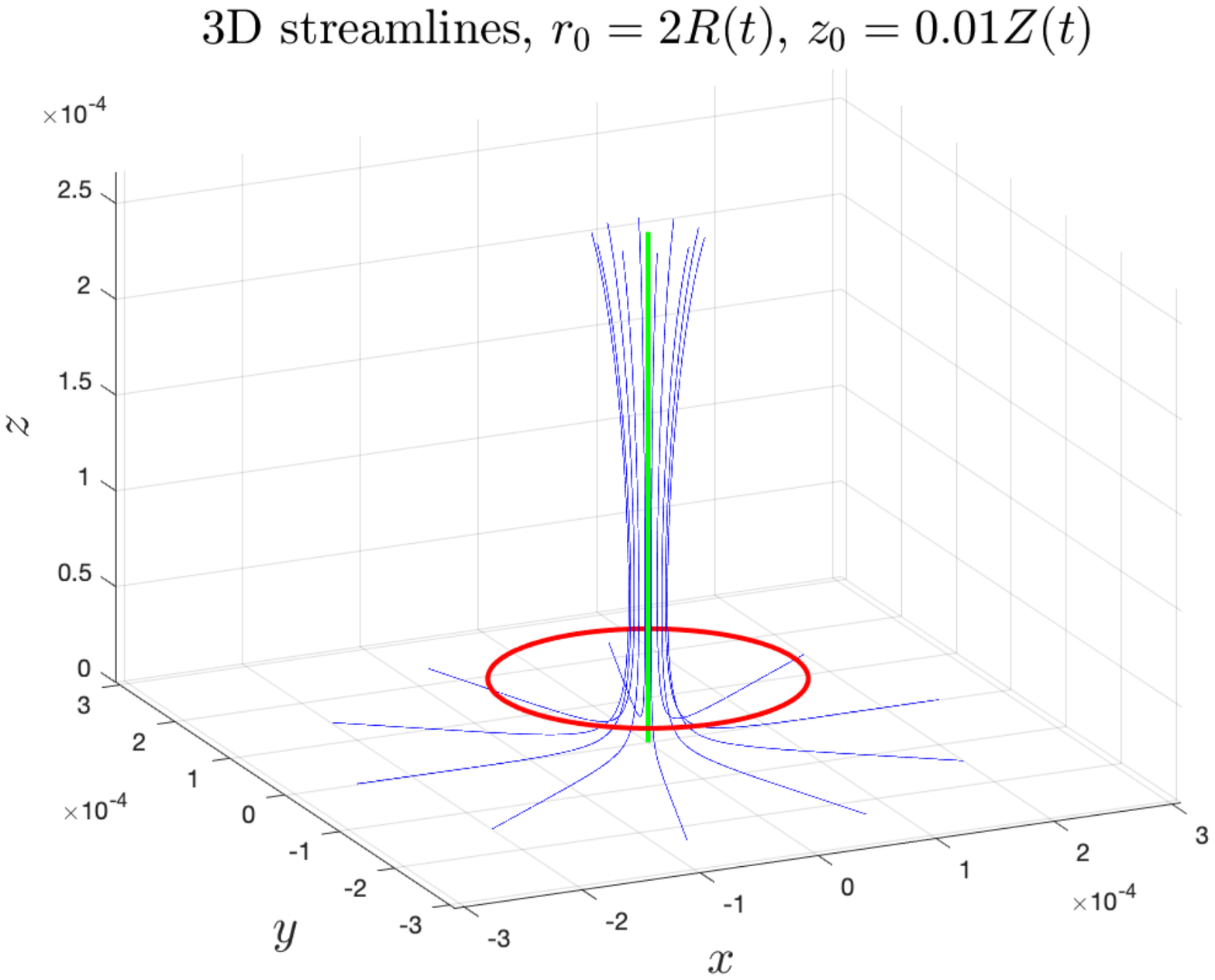}
        \vspace{-0.9in}
        \caption{$r_0 = 2R(t)$, $z_0 = 0.01Z(t)$}
    \end{subfigure}
    \begin{subfigure}[b]{0.40\textwidth}
        \centering
        \includegraphics[width=1\textwidth]{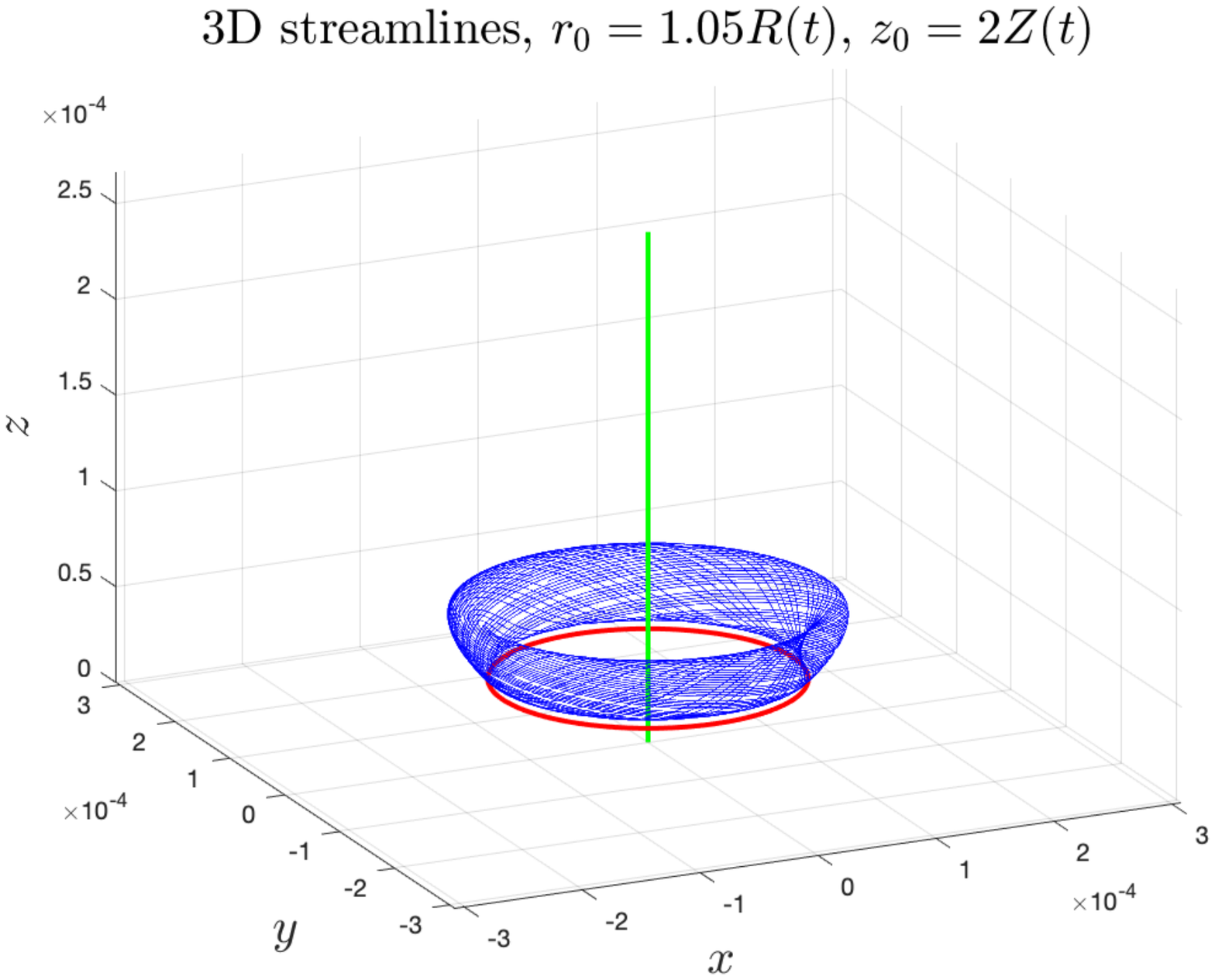}
        \vspace{-0.9in}
        \caption{$r_0 = 1.05R(t)$, $z_0 = 2Z(t)$}
    \end{subfigure}
    \begin{subfigure}[c]{0.40\textwidth}
        \centering
        \vspace{-0.7in}
        \includegraphics[width=1\textwidth]{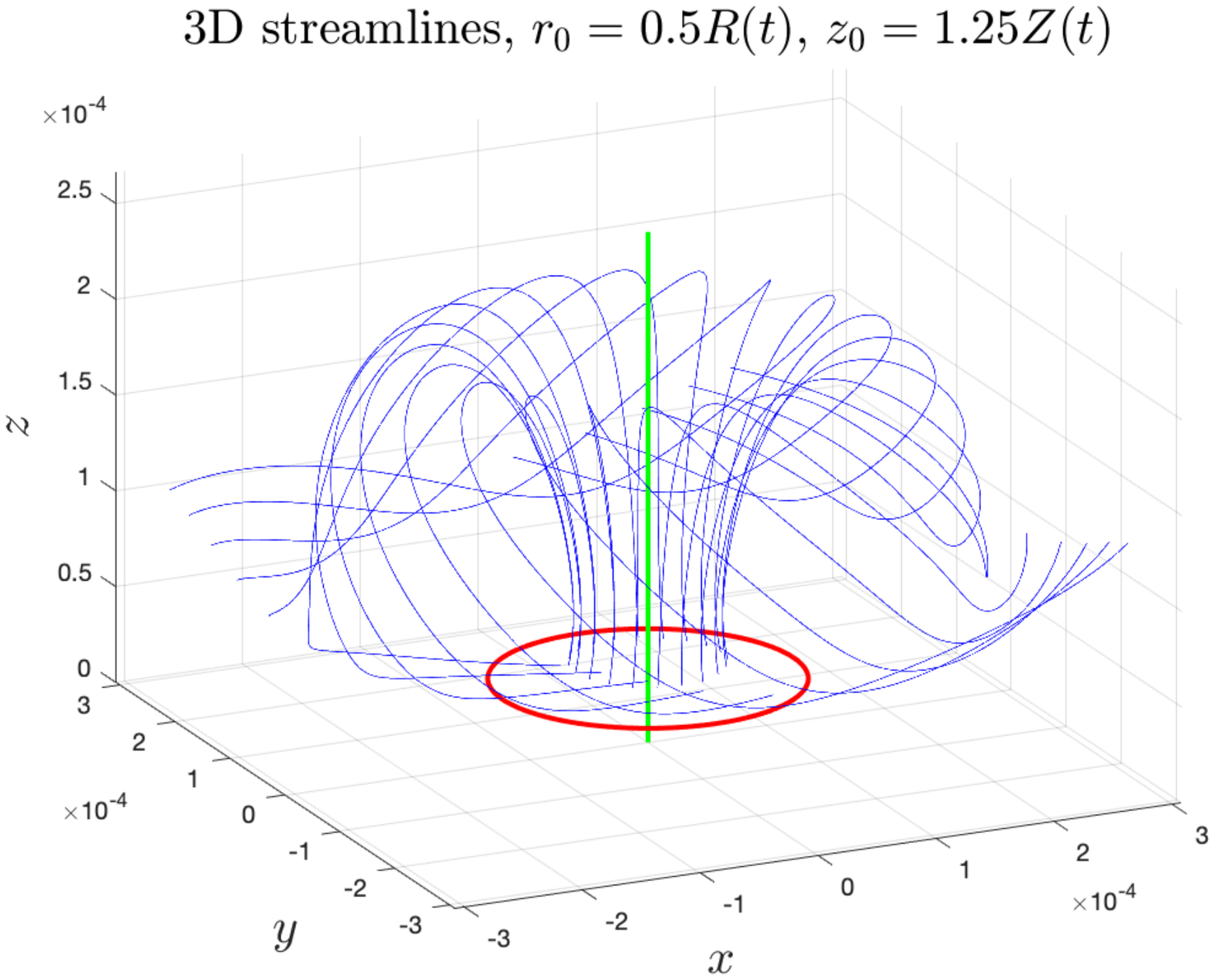}
        \vspace{-0.9in}
        \caption{$r_0 = 0.5R(t)$, $z_0 = 1.25Z(t)$}
    \end{subfigure}
    \begin{subfigure}[d]{0.40\textwidth}
        \centering
         \vspace{-0.7in}
        \includegraphics[width=1\textwidth]{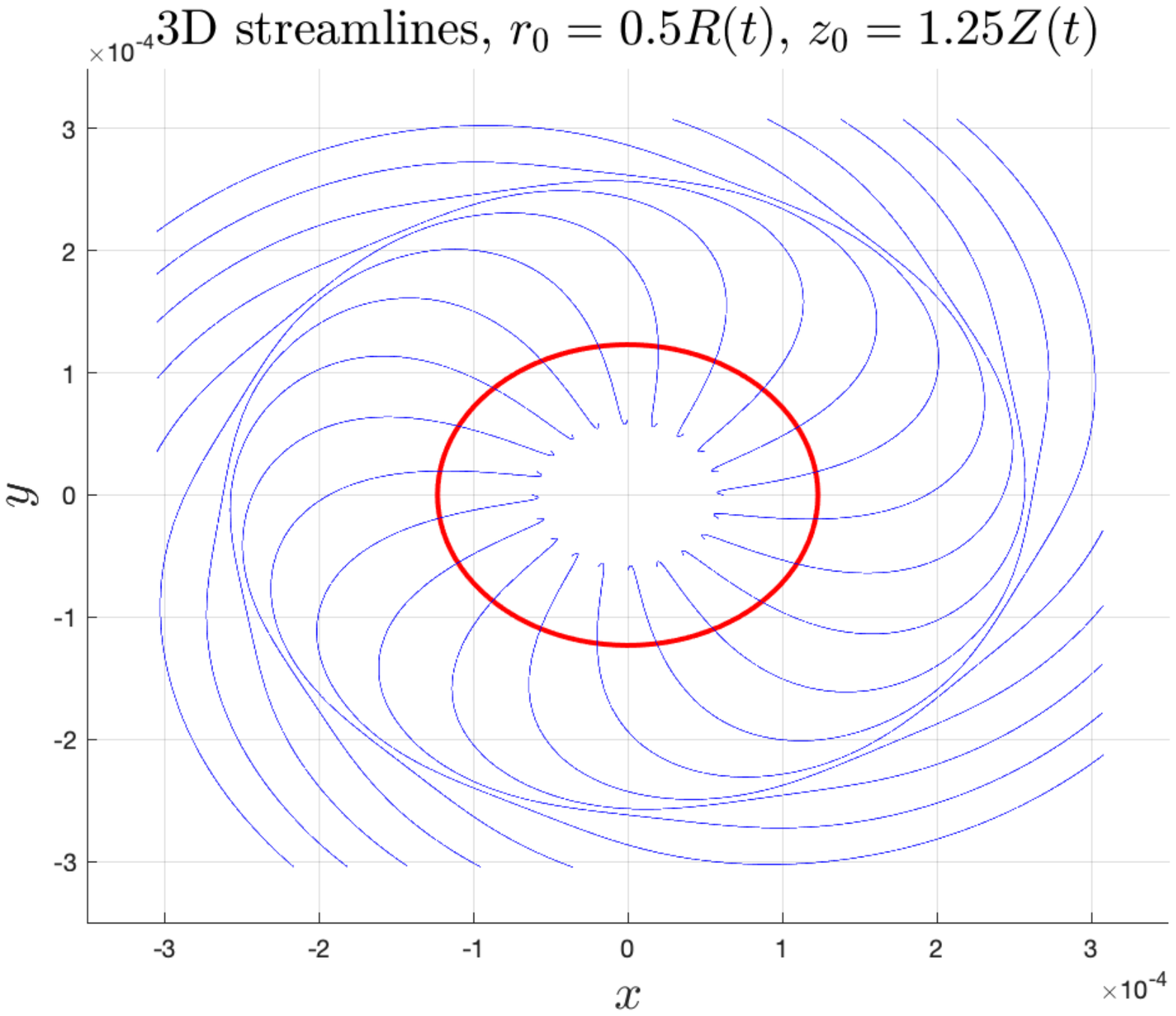}
        \vspace{-0.9in}
        \caption{same as (c), a top view}
    \end{subfigure}
    \caption[Local streamline]{The streamlines of $(u^r(t),u^\theta(t),u^z(t))$ at time $t_3=0.0022868453$ with initial points given by (a) $(r_0,z_0) = (2R(t),0.01Z(t))$, (b) $(r_0,z_0) = (1.05R(t),2Z(t))$, (c) $(r_0,z_0) = (0.5R(t),1.25Z(t))$ ($3$D view), (d) $(r_0,z_0)$ is the same as (c), a top view. $(R(t),Z(t))$ is the maximum location of $u_1(t)$, indicated by the red ring. The green pole is the symmetry axis $r=0$.}  
     \label{fig:streamline_3D_zoomin_nse}
        \vspace{-0.05in}
\end{figure}

Interestingly the induced streamlines look qualitatively the same as those obtained for the $3$D Euler equations \cite{Hou-euler-2021}. In Figure~\ref{fig:streamline_3D_global_nse}, we plot the streamlines induced by the velocity field $\vu(t)$ at $t_0 = 0.00227375$ in a macroscopic scale (the cylinder domain $\mathcal{D}_1\times [0,2\pi]$) for different initial points with (a) $(r_0,z_0) = (0.8,0.2)$, (b) $(r_0,z_0) = (0.5,0.1)$, and (c)-(d) $(r_0,z_0) = (0.1,0.01)$. The velocity field resembles that of a tornado spinning around the symmetry axis (the green pole). In  Figure~\ref{fig:streamline_3D_global_nse}(a), we observe that the streamlines first travel toward the symmetry axis, spin around the symmetry axis near $z=0$. As they get close enough to the symmetry axis, they move upward toward $z=1/2$ and then move away from the symmetry axis. For a slightly smaller $z_0$, we observe that the streamlines do not spin as much as they approach the symmetry axis. They seem to travel in a circular cycle in the $rz$-coordinates, see Figure~\ref{fig:streamline_3D_global_nse}(b). On the other hand, the behavior is quite different if the initial point is very close to $z=0$.  The streamlines will spin around the symmetry axis, forming a close torus, but do not travel upward, see Figure~\ref{fig:streamline_3D_global_nse}(c)-(d).

Next, we study the streamlines at the late stage of the computation. In Figure~\ref{fig:streamline_3D_zoomin_nse}, we plot the streamlines at time $t_3=0.0022868453$ for different initial points near the maximum location $(R(t),Z(t))$ of $u_1(t)$. By this time, $\|\omega (t_3)\|_{L^\infty}/\|\omega (0)\|_{L^\infty}$ has increased by a factor of $10^6$. The red ring represents the location of $(R(t),Z(t))$, and the green pole is the symmetry axis $r=0$. 
The first two settings of $(r_0,z_0)$ are exactly the same as the zoom-in case for the $3$D Euler equations \cite{Hou-euler-2021}. We make a small modification for the last case. 
More specifically, we have
(a) $(r_0,z_0) = (2R(t),0.01Z(t))$, (b) $(r_0,z_0) = (1.05R(t),2Z(t))$ and (c)-(d) $(r_0,z_0) = (0.5R(t),1.25Z(t))$.  Surprisingly, we observe the same qualitative behavior as we did for the $3$D Euler equations \cite{Hou-euler-2021}. In some sense, the Navier--Stokes equations preserve the same qualitative features of the $3$D Euler equations. The relatively large viscosity enables us to compute the solution for much longer time and we observe a much more rapid growth of the maximum vorticity.

\subsubsection{The $2$D flow} 
\begin{figure}[!ht]
\centering
    \includegraphics[width=0.40\textwidth]{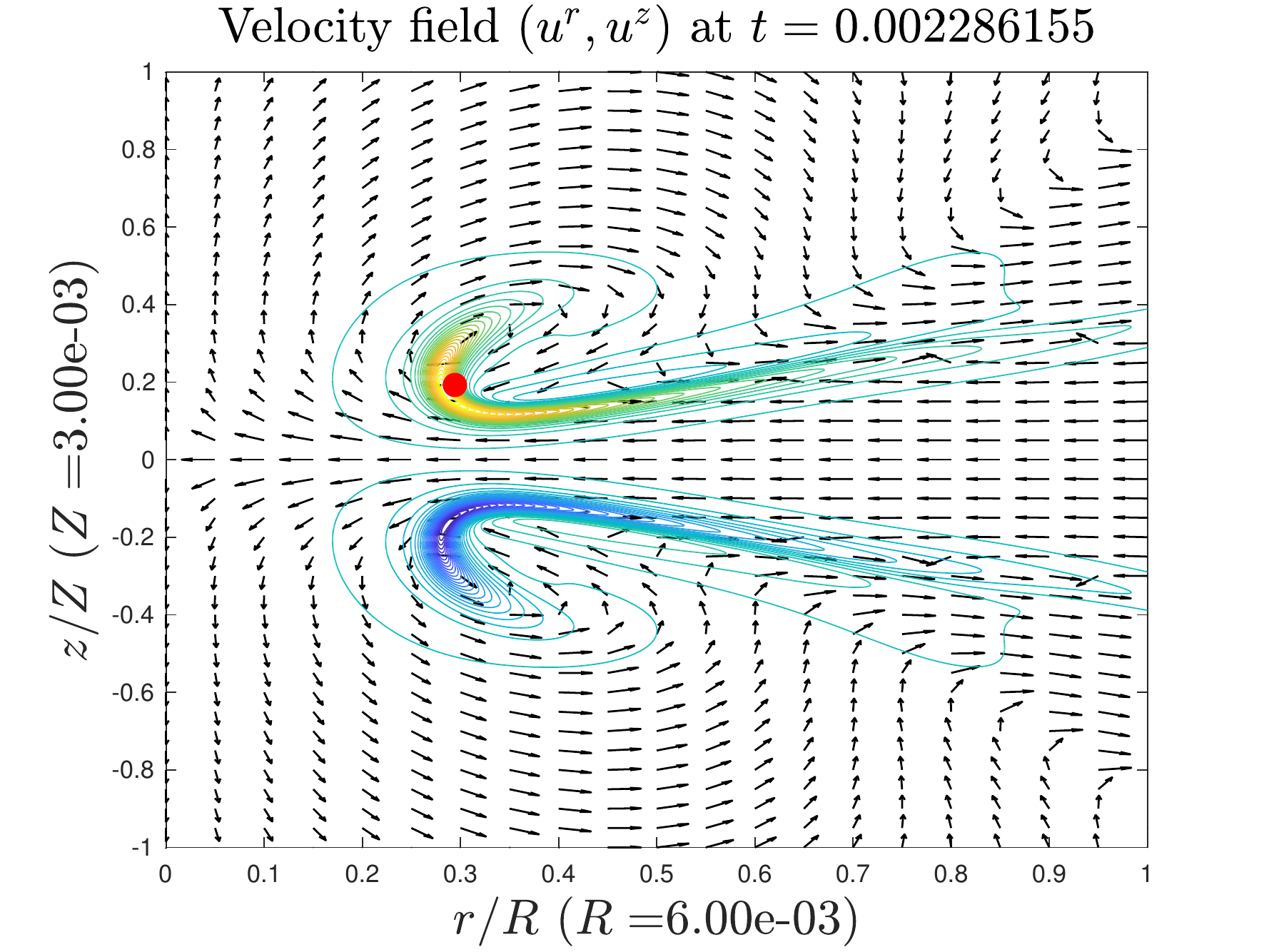}
    \includegraphics[width=0.40\textwidth]{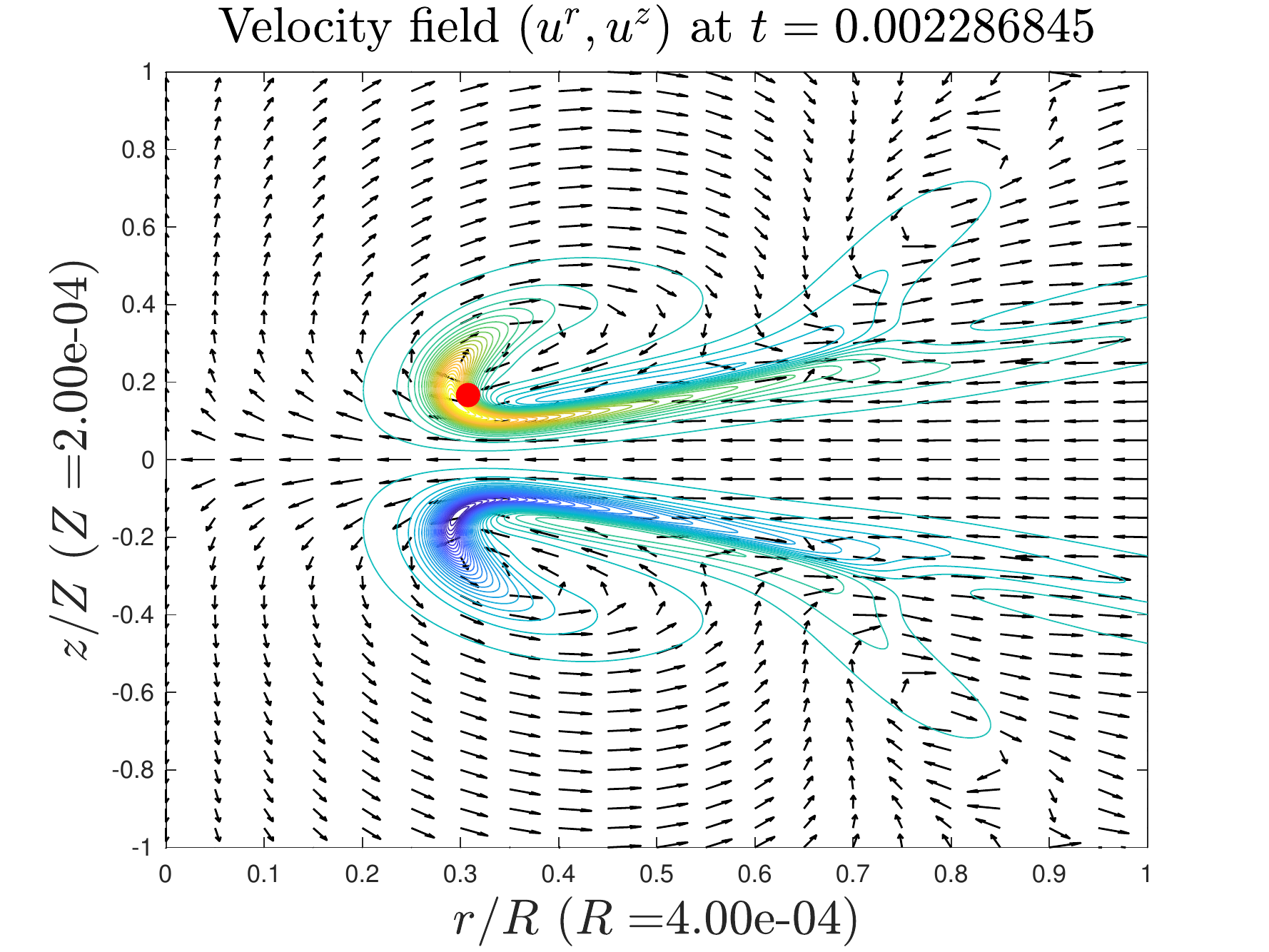} 
    \caption[Dipole]{The dipole structure of $\omega_1$  and the induced local velocity field at two different times, $t_1=0.0022861547$ (left plot) and $t_3=0.002286845$ (right plot). The red point is the maximum location $(R(t),Z(t))$ of $u_1(t)$.}  
     \label{fig:dipole_nse}
        \vspace{-0.05in}
\end{figure}

To understand the phenomena in the most singular region as shown in Figure~\ref{fig:streamline_3D_zoomin_nse}, we study the $2$D velocity field $(u^r,u^z)$. In Figure~\ref{fig:dipole_nse}(a)-(b), we plot the dipole structure of $\om_1$ in a local symmetric region and the hyperbolic velocity field induced by the dipole structure in a local microscopic domain $[0,R_b]\times [0,Z_b]$ at two different times, $t_1=0.0022861547$ and $t_3=0.0022868453$.
The dipole structure for the $3$D Navier--Stokes equations look qualitatively similar to that of the Euler equations except that the dipole structure for the Navier--Stokes equations is much smoother than that for the Euler equations \cite{Hou-euler-2021}.
By the time $t=t_1$, $\|\vom (t)\|_{L^\infty}/\|\vom (0)\|_{L^\infty}$ has increased by a factor of $10^4$ and by time $t=t_3$, $\|\vom (t)\|_{L^\infty}/\|\vom (0)\|_{L^\infty}$ has increased by a factor of $10^6$.

The $2$D velocity field shares the same qualitative features as those of the $3$D Euler equations. As in the case of the Euler equations, the negative radial velocity near $z=0$ induced by the antisymmetric vortex dipoles pushes the solution toward $r=0$, then move upward away from $z=0$. This is one of the driving mechanisms for a potential singularity on the symmetry axis. Since the value of $u_1$ becomes very small near the symmetry axis $r=0$, the streamlines almost do not spin around the symmetry axis, as illustrated in Figure~\ref{fig:streamline_3D_zoomin_nse}(a). 

Moreover, the velocity field $(u^r(t),u^z(t))$ also forms a closed circle right above $(R(t),Z(t))$. The corresponding streamlines are trapped in the circle region in the $rz$-plane. The fluid spins fast around the symmetry axis $r=0$, see Figure~\ref{fig:streamline_3D_zoomin_nse}(b). As in the case of the $3$D Euler equations, this local circle structure of the $2$D velocity field is critical in stabilizing the blow-up process.

%

\subsubsection{The effect of viscous regularization}
\begin{figure}[!ht]
\centering
\vspace{-0.7in}
    \includegraphics[width=0.40\textwidth]{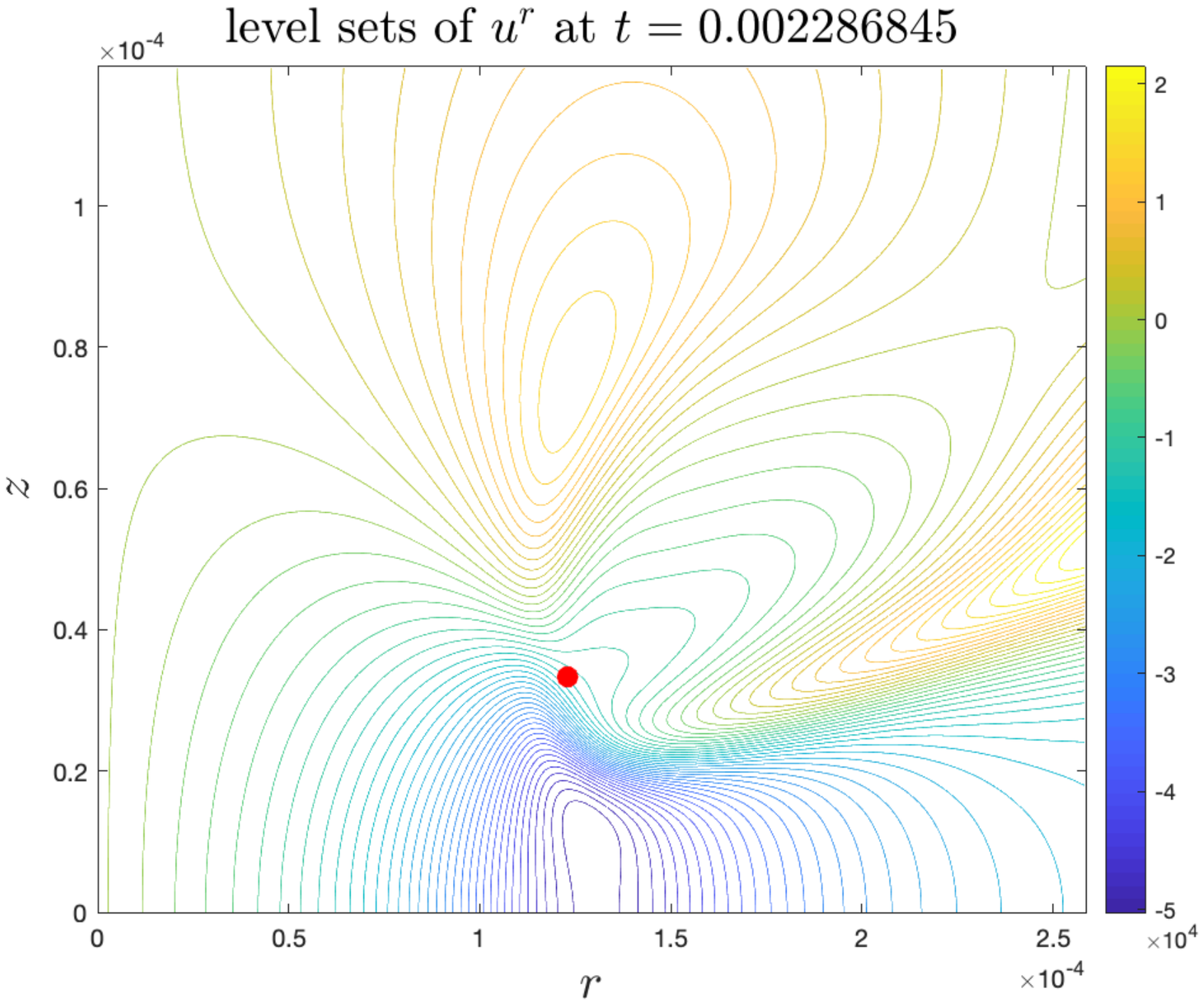}
    \vspace{-0.4in}
    \includegraphics[width=0.40\textwidth]{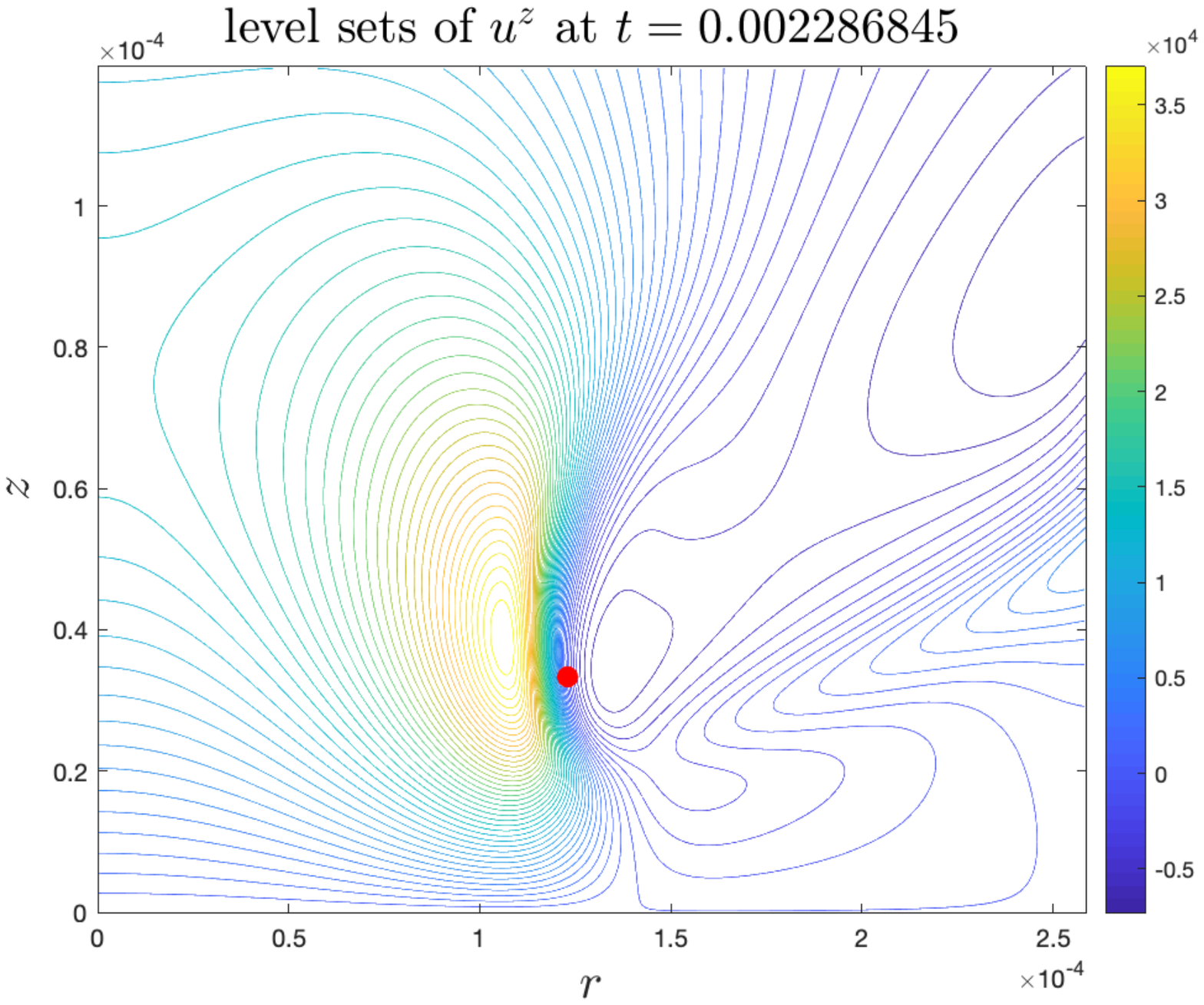} 
    \vspace{-0.4in}
    \caption[Velocity level sets]{The level sets of $u^r$ (left) and $u^z$ (right) at $t_3 = 0.002286845$. The red point is the maximum location $(R(t),Z(t))$ of $u_1(t)$.}  
     \label{fig:velocity_levelset_nse}
        \vspace{-0.05in}
\end{figure}

To study the effect of viscous regularization, 
we plot the velocity contours in Figure~\ref{fig:velocity_levelset_nse}. While they share the same qualitative features as those of the $3$D Euler equations, we notice that the velocity field becomes much smoother due to the strong viscous regularization. Thus, the sharp front does not have a rapidly decreasing thickness as in the case of the $3$D Euler equations. We have a stable scaling relationship for a much longer time.

\begin{figure}[!ht]
\centering
	\begin{subfigure}[b]{0.35\textwidth}
    \includegraphics[width=1\textwidth]{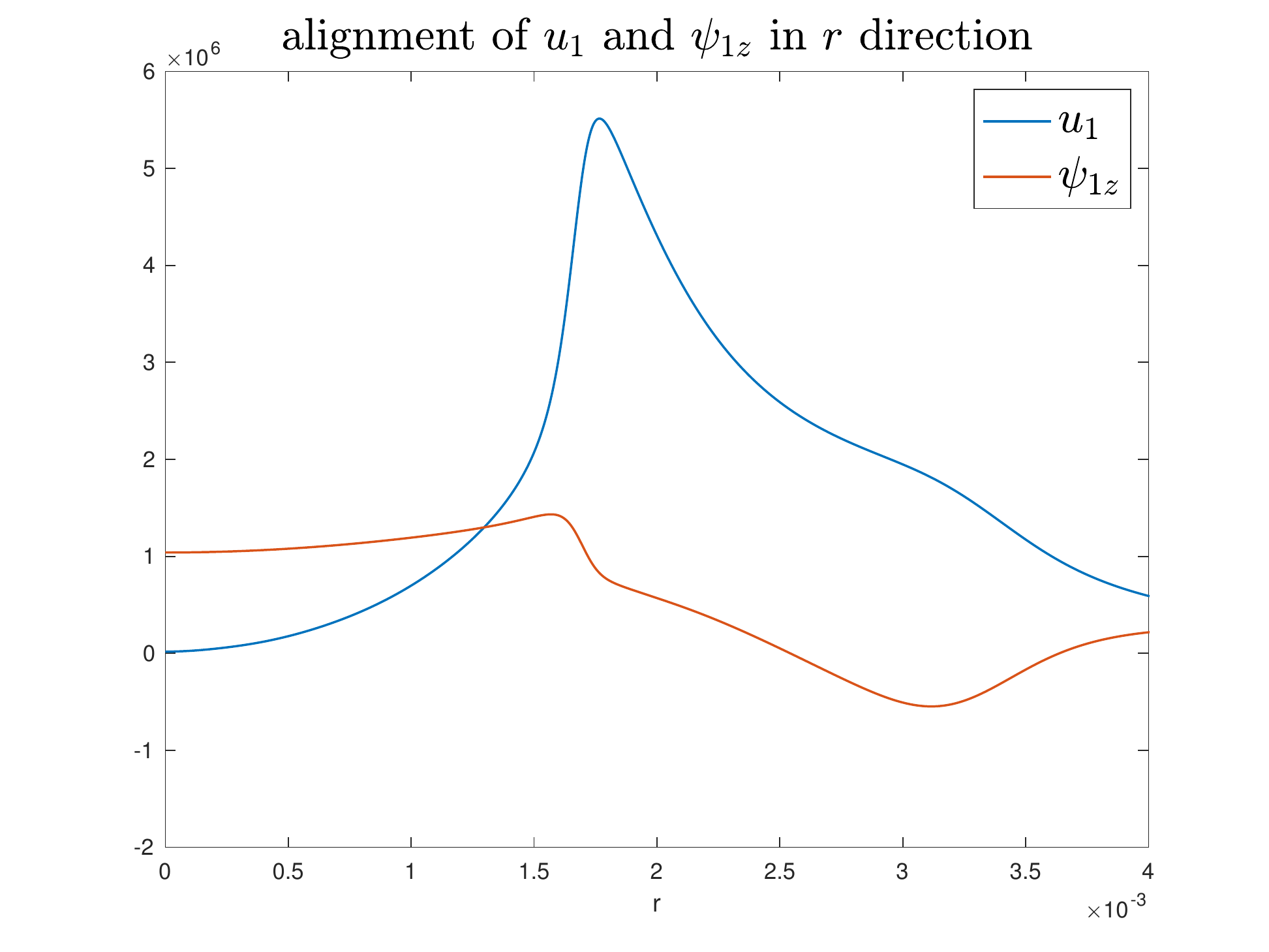}
    \caption{$r$ cross sections of $u_1,\psi_{1,z}$}
    \end{subfigure}
    \begin{subfigure}[b]{0.35\textwidth}
    \includegraphics[width=1\textwidth]{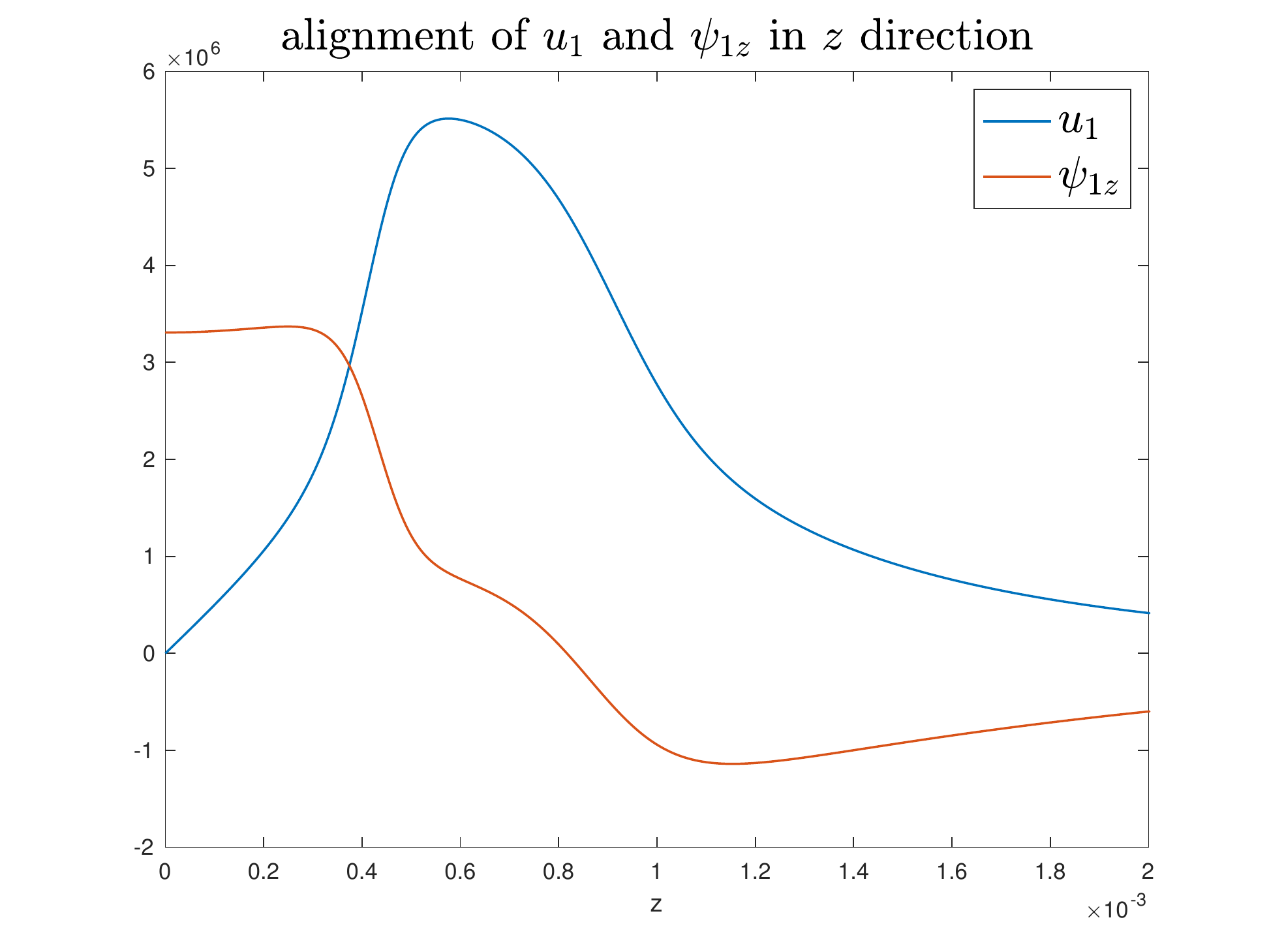}
    \caption{$z$ cross sections of $u_1,\psi_{1,z}$}
    \end{subfigure}
  	\begin{subfigure}[b]{0.35\textwidth}
    \includegraphics[width=1\textwidth]{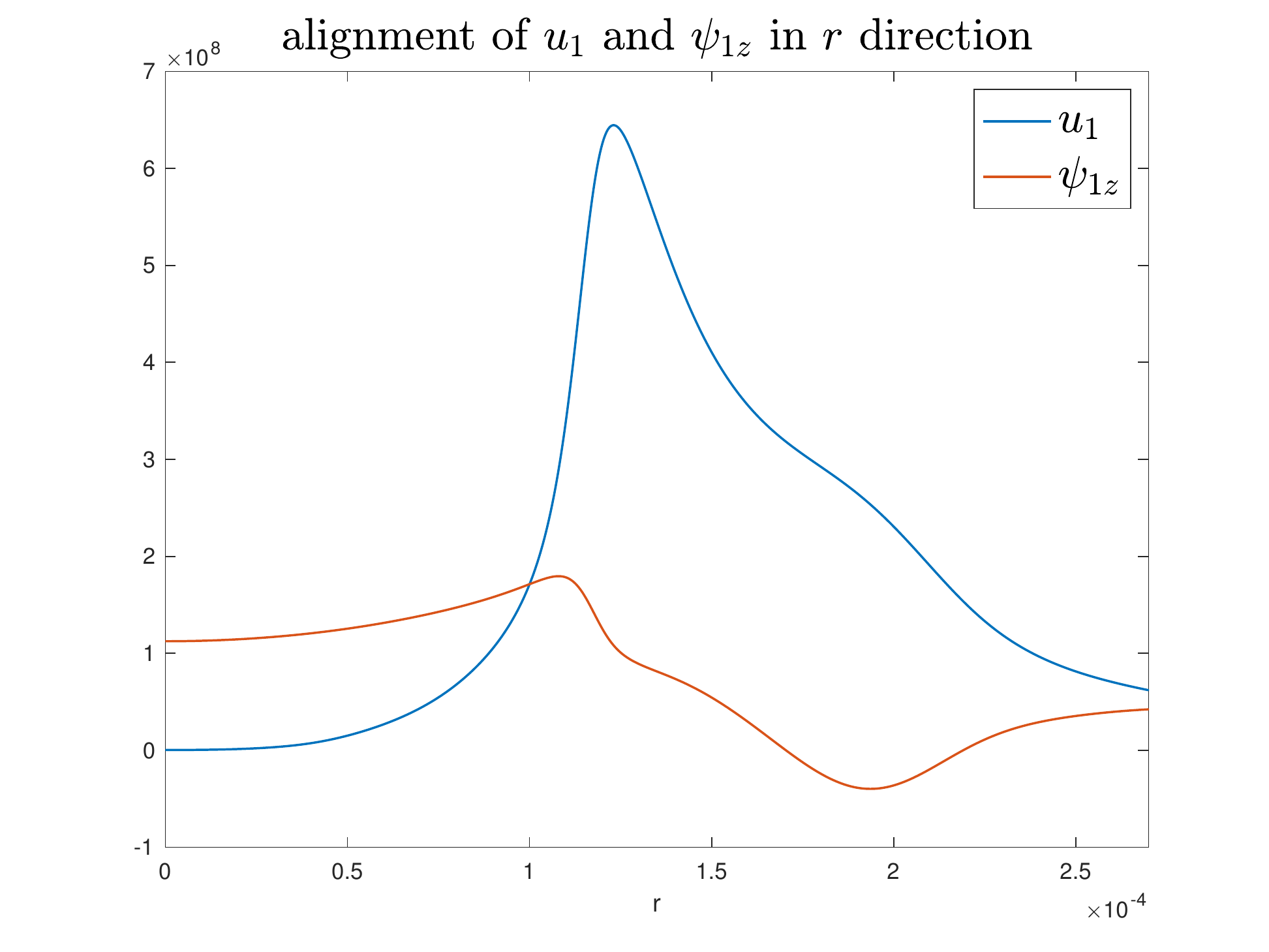}
    \caption{$r$ cross sections of $u_1,\psi_{1,z}$}
    \end{subfigure}
     \begin{subfigure}[b]{0.35\textwidth}
    \includegraphics[width=1\textwidth]{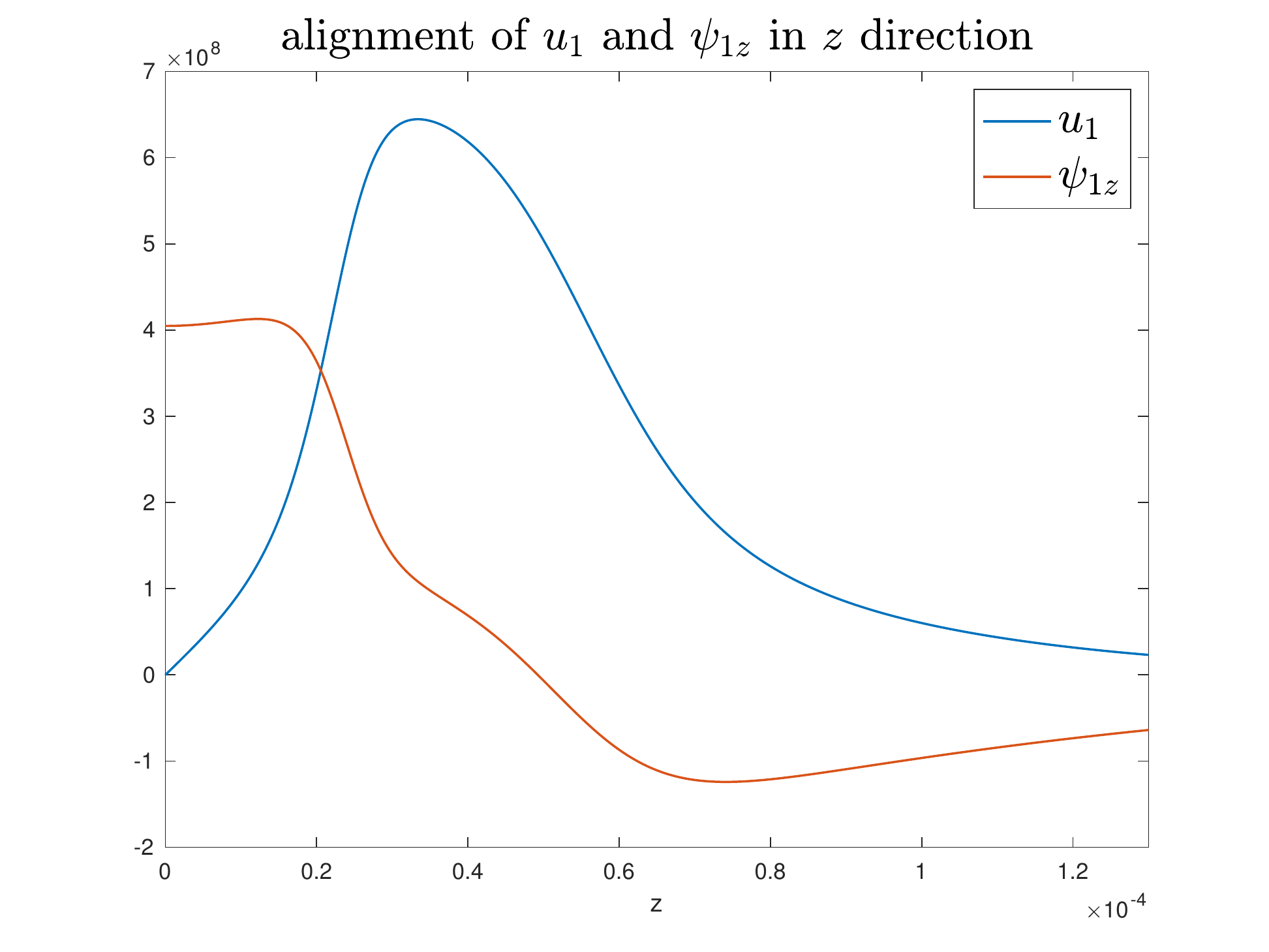}
    \caption{$z$ cross sections of $u_1,\psi_{1,z}$}
    \end{subfigure}
    \caption[Alignment]{The alignment between $u_1$ and $\psi_{1,z}$. (a) and (b): cross sections of $u_1$ and $\psi_{1,z}$ through the point $(R(t),Z(t))$ at $t_1=0.0022861547$. (c) and (d): cross sections of $u_1$ and $\psi_{1,z}$ through the point $(R(t),Z(t))$ at $t_3=0.0022868453$.}  
     \label{fig:alignment_nse}
        \vspace{-0.05in}
\end{figure}

Although we apply a relatively large viscosity to stabilize the potential Euler singularity, the main mechanism that leads to the potential Euler singularity is still preserved by the Navier--Stokes equations during the rapid growth period.
The relatively strong viscous regularization enhances nonlinear alignment of vortex stretching.  
In Figure \ref{fig:alignment_nse}(a)-(b), we demonstrate the alignment between $\psi_{1,z}$ and $u_1$ at two different times $t_1=0.0022861547$ and $t_3=0.0022868453$. The maximum vorticity has increased by a factor of $10^4$ at $t_1$ while the maximum vorticity has increased by a factor of $10^6$ at $t_3$. Although the maximum vorticity has grown so much by these times, the local solution structures have remained qualitatively the same. In particular, we do not observe the sharp drop in the $\psi_{1z}(t, r,Z(t))$ as a function of $r$ near $r=R(t)$ as we observed for the $3$D Euler equation in the late stage \cite{Hou-euler-2021}. This shows that the viscous effect has a strong stabilizing effect that enhances the nonlinear alignment of vortex stretching. We also observe that $\psi_{1z}$ is relatively flat in the region $\{(r,z) | 0 \leq r \leq 0.9R(t), \; 0 \leq z \leq 0.5Z(t)\}$. This property is critical for $u_1$ to remain large between the sharp front and $r=0$, thus avoiding the formation of a vacuum region and a two-scale structure.

We observe that the large, positive, and relative flat $\psi_{1z}(t,R(t),z)$ near $z=0$ induces a large growth of $u_1$ through the vortex stretching term $2\psi_{1,z}u_1$ in the $u_1$-equation \eqref{eq:as_NSE_1_a}. Due to the oddness of $u_1$ as a function of $z$, the large growth of $u_1$ near $z=0$ generates a large positive gradient of $u_1^2$ in the $z$-direction between $z=0$ and $z = Z(t)$. The vortex stretching term $2(u_1^2)_z$ in the $\omega_1$-equation \eqref{eq:as_NSE_1_b} then induces
a rapid growth of $\omega_1$. Moreover, we observe that the antisymmetric dipole structure generated by $\om_1$ produces a strong negative radial velocity in between the dipole, see Figure \ref{fig:dipole_nse}. This in turn generates rapid growth of $\psi_{1,z}$ near $z=0$ (recall $\psi_{1,z} = -u^r/r$). The rapid growth of $\psi_{1,z}$ in turn generates an even faster growth of $u_1$ through the vortex stretching term in the $u_1$-equation. The larger value of $u_1$ and larger positive gradient of $u_1^2$ in the $z$-direction then lead to faster growth of $\omega_1$ through the vortex stretching term in the $\omega_1$-equation. The whole coupling mechanism forms a positive feedback loop. 

\subsection{Numerical Results: Resolution Study}\label{sec:performance_study_nse}
In this subsection, we perform resolution study and investigate the convergence property of our numerical methods. In particular, we will study 
the effectiveness of the adaptive mesh in Section \ref{sec:mesh_effectiveness_nse}, and 
the convergence of the solutions as $h_\rho,h_\eta\rightarrow 0$ in Section \ref{sec:resolution_study_nse}.

\subsubsection{Effectiveness of the adaptive mesh}\label{sec:mesh_effectiveness_nse} 
\begin{figure}[!ht]
\centering
	\begin{subfigure}[b]{0.40\textwidth}
    \includegraphics[width=1\textwidth]{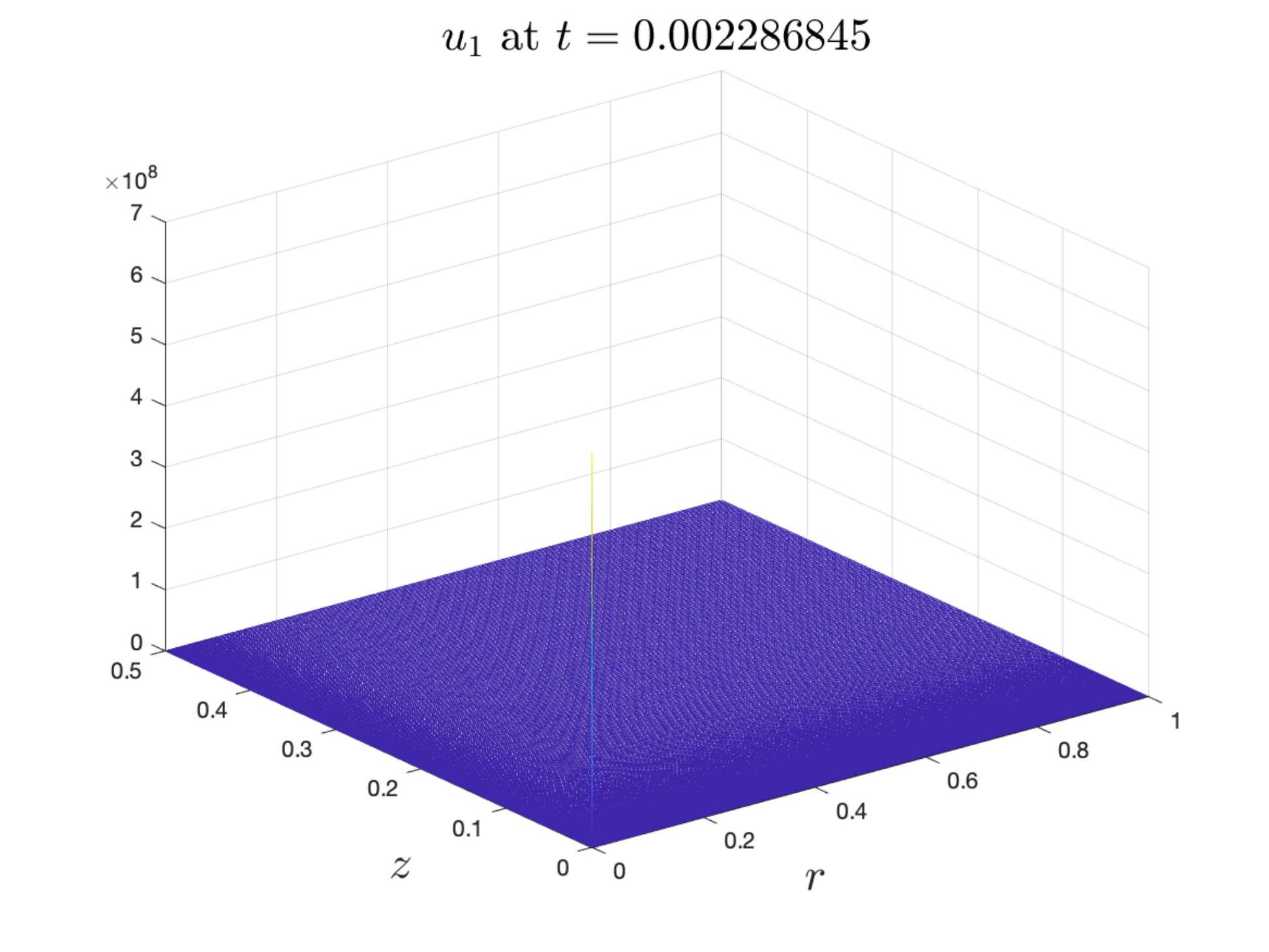}
    \caption{$u_1$ in the $rz$-plane}
    \end{subfigure}
  	\begin{subfigure}[b]{0.40\textwidth}
    \includegraphics[width=1\textwidth]{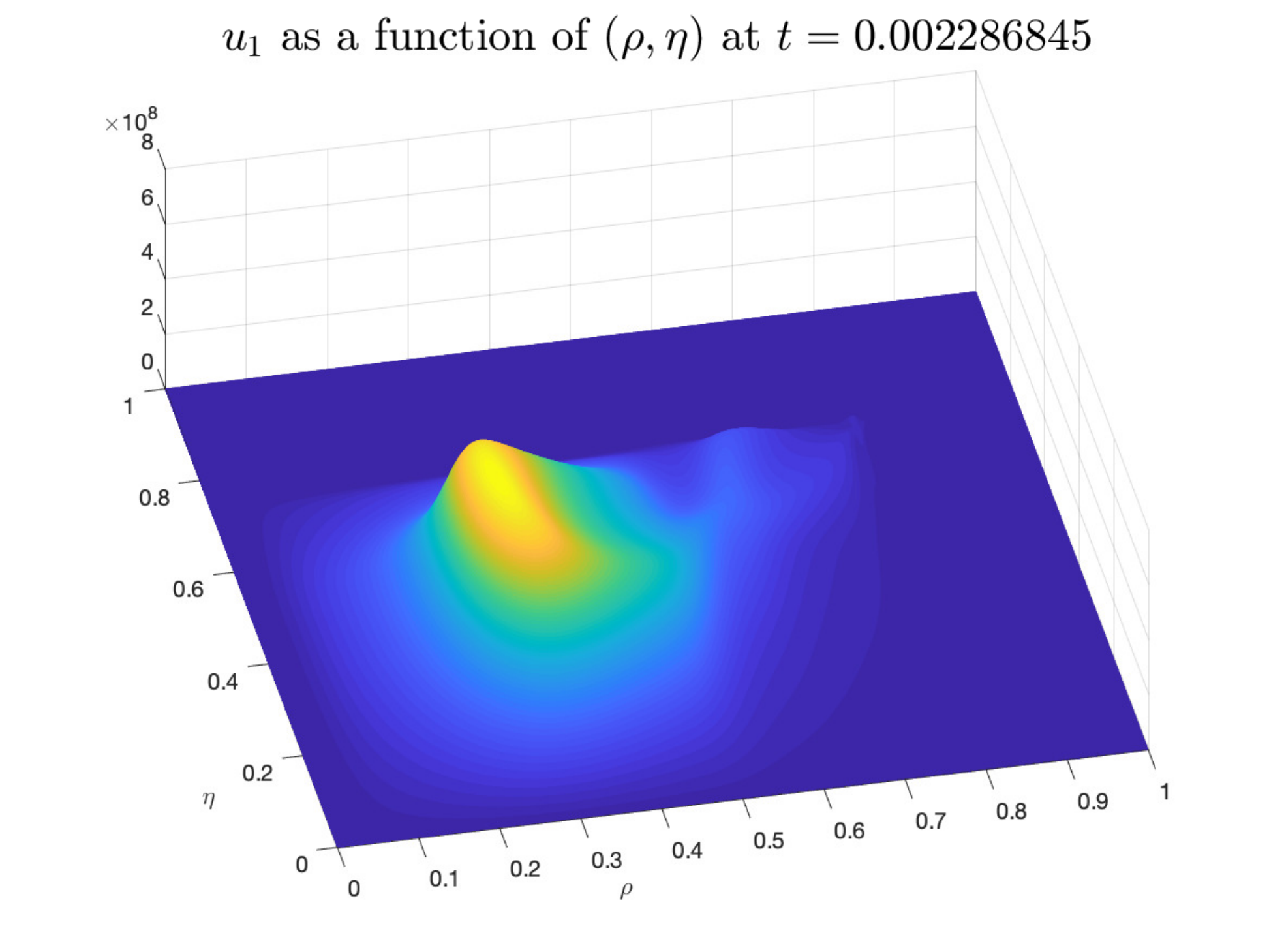}
    \caption{$u_1$ in the $\rho\eta$-plane}
    \end{subfigure}
    \caption[Mesh effectiveness]{The adaptive mesh resolves the solution in the $\rho\eta$-plane. (a) shows the focusing nearly singular profile of $u_1$ at $t_3=0.0022868453$ in the $rz$-plane on the whole computational domain $\mathcal{D}_1$. (b) plots the profile of $u_1$ in the $\rho\eta$-plane.}  
     \label{fig:mesh_effective_nse}
        \vspace{-0.05in}
\end{figure}

Since we solve the Navier--Stokes equations in the transformed $(\rho,\eta)$ coordinates, we would like to see if the solution remains smooth in the $(\rho,\eta)$ coordinates. In Figure~\ref{fig:mesh_effective_nse}(a), we plot the $3$D profile of $u_1$ at $t_3=0.0022868453$ in the original $rz$-plane. This plot suggests that the solution seems to develop a focusing and potentially singular solution at the origin. In Figure~\ref{fig:mesh_effective_nse}(b), we plot the profile of $u_1$ at the same time in the $\rho\eta$-plane. We can see clearly that the solution is quite smooth in the $(\rho,\eta)$ coordinates and our adaptive mesh resolves the potentially singular  solution in the $(\rho,\eta)$ coordinates.

\begin{figure}[!ht]
\centering
	\begin{subfigure}[b]{0.40\textwidth}
    \includegraphics[width=1\textwidth]{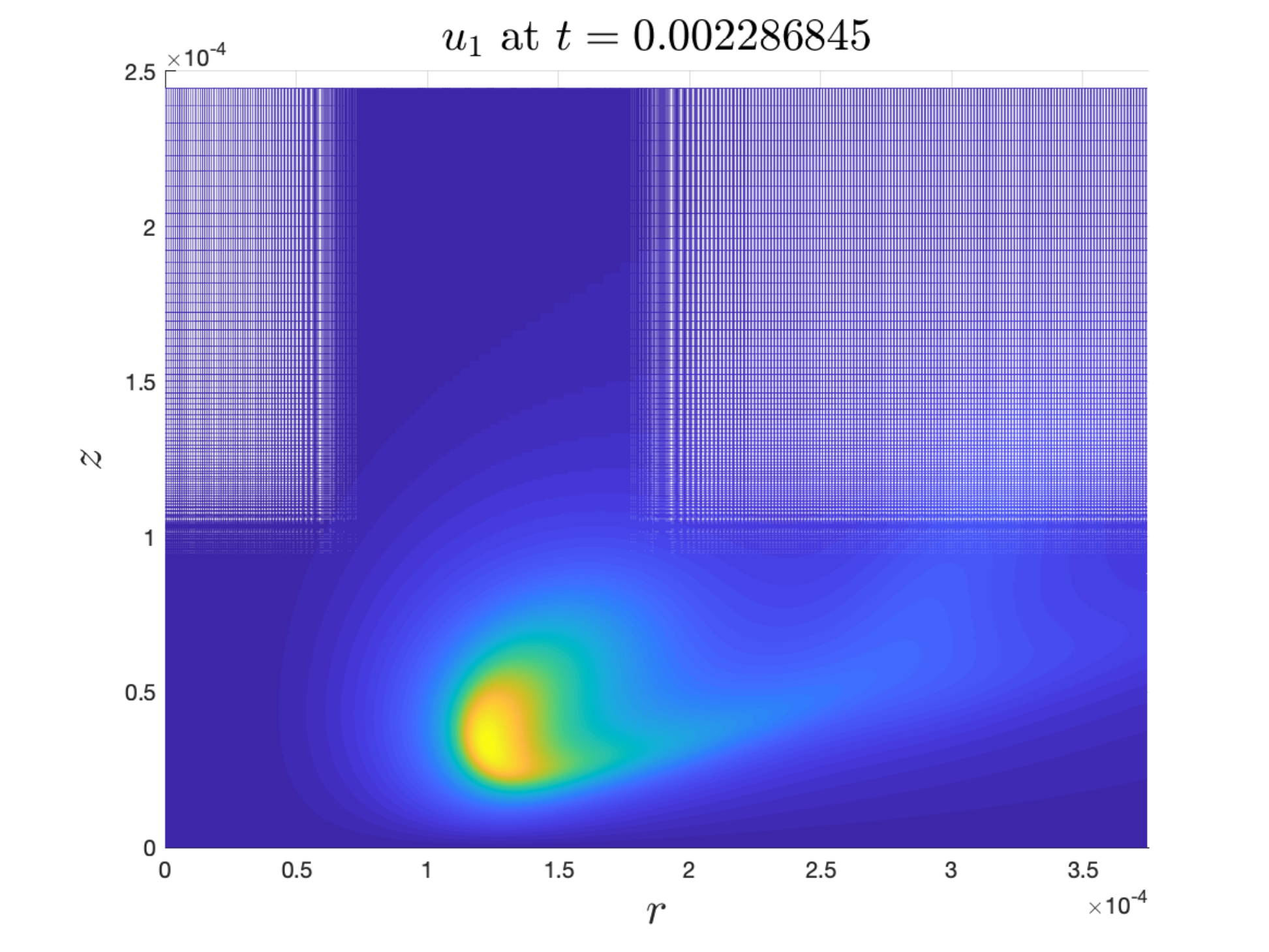}
    \caption{$u_1$ at $t_3=0.0022868453$}
    \end{subfigure}
  	\begin{subfigure}[b]{0.40\textwidth}
    \includegraphics[width=1\textwidth]{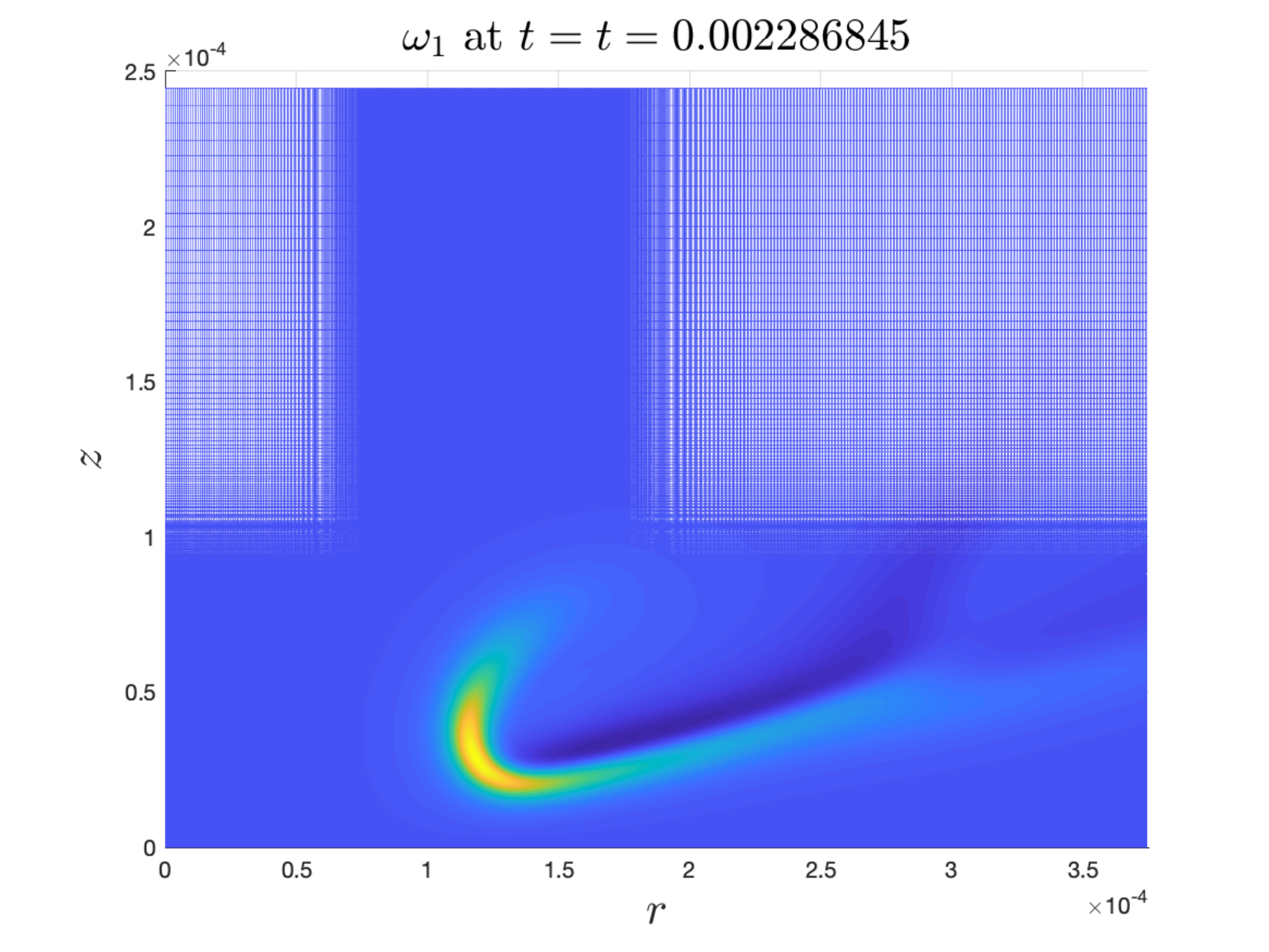}
    \caption{$\om_1$ at $t_3=0.0022868453$}
    \end{subfigure} 
    \caption[Mesh phases]{The adaptive mesh has different densities in different regions. }  
     \label{fig:mesh_phases_nse}
        \vspace{-0.05in}
\end{figure}

In Figure~\ref{fig:mesh_phases_nse}, we show the top views of the profiles of $u_1,\om_1$ in a local domain at $t_3=0.0022868453$. This figure demonstrates how the mesh points are distributed in different phases of the adaptive mesh. As we can see, we have the most mesh points in phase $1$ in both directions, and the adaptive mesh resolves the most singular part of the solution.

Inspired by my recent joint work with De Huang in \cite{Hou-Huang-2021}, we define the mesh effectiveness functions $ME_\rho(v),ME_\eta$ with respect to some solution variable $v$ as follows:
\[ME_\rho(v) = \frac{h_\rho v_\rho}{\|v\|_{L^\infty}} = \frac{h_\rho r_\rho v_r}{\|v\|_{L^\infty}},\quad ME_\eta(v) = \frac{h_\eta v_\eta}{\|v\|_{L^\infty}} = \frac{h_\eta r_\eta v_z}{\|v\|_{L^\infty} }.\]
We further define the corresponding mesh effectiveness measures (MEMs) as follows:
\[ME_{\rho,\infty}(v) = \|ME_\rho(v)\|_{L^\infty},\quad ME_{\eta,\infty}(v) = \|ME_\eta(v)\|_{L^\infty}.\]
As indicated in \cite{Hou-Huang-2021}, the MEMs quantify the the largest relative growth of a function $v$ in one single mesh cell. If the MEMs are small, the adaptive mesh has done a good job in resolving the function $v$ effectively. Thus, the MEMs provide a good measure on the effectiveness of our adaptive mesh.

%

Table~\ref{tab:MEM_mesh_nse} reports the MEMs of $u_1,\om_1$ at $t_3=0.0022868453$ on meshes of different sizes. We observe that the MEMs decrease as the grid sizes $h_\rho,h_\eta$ decrease. This is reasonable since  the MEMs are proportional to $h_\rho,h_\eta$. Table~\ref{tab:MEM_time_nse} reports the MEMs of $u_1,\om_1$ at different times using the same mesh size $(n_1,n_2) = (1536,1536)$. We can see that the MEMs remain relatively small throughout this time interval. This study implies that our adaptive mesh strategy is effective in resolving the potentially singular solution of the $3$D Navier--Stokes equations over the entire computational domain $\mathcal{D}_1$. 

\begin{table}[!ht]
\centering
\footnotesize
\renewcommand{\arraystretch}{1.5}
    \begin{tabular}{|c|c|c|c|c|}
    \hline
    \multirow{2}{*}{Mesh size} & \multicolumn{4}{c|}{MEMs on mesh at $t_3=0.0022868453$} \\ \cline{2-5} 
    						   & $ME_{\rho,\infty}(u_1)$ & $ME_{\eta,\infty}(u_1)$ & $ME_{\rho,\infty}(\om_1)$ & $ME_{\eta,\infty}(\om_1)$  \\ \hline 
    $512\times 512$ & $0.025$ & $0.015$ & $0.050$ & $0.027$ \\ \hline 
    $768\times 768$ & $0.016$ & $0.010$ & $0.035$ & $0.023$ \\ \hline 
    $1024\times 1024$ & $0.013$ & $0.008$ & $0.027$ & $0.018$  \\ \hline 
    $1280\times 1280$ & $0.011$ & $0.005$ & $0.022$ & $0.011$ \\ \hline 
    $1536\times 1536$ & $0.009$ & $0.004$ & $0.017$ & $0.009$  \\ \hline 
    \end{tabular}
    \caption{\small MEMs of $u_1,\om_1$ at $t=0.002286845$ on the meshes of different sizes.}
    \label{tab:MEM_mesh_nse}
\end{table}

\begin{table}[!ht]
\centering
\footnotesize
\renewcommand{\arraystretch}{1.5}
    \begin{tabular}{|c|c|c|c|c|}
    \hline
    \multirow{2}{*}{Time} & \multicolumn{4}{c|}{MEMs on mesh $(n_1,n_2) = (1536,1536)$} \\ \cline{2-5} 
    						   & $ME_{\rho,\infty}(u_1)$ & $ME_{\eta,\infty}(u_1)$ & $ME_{\rho,\infty}(\om_1)$ & $ME_{\eta,\infty}(\om_1)$  \\ \hline 
    $0.00227375$ & $0.006$ & $0.006$ & $0.013$ & $0.029$ \\ \hline 
    $0.0022861547$ & $0.007$ & $0.006$ & $0.018$ & $0.017$  \\ \hline 
    $0.0022867812$ & $0.008$ & $0.005$ & $0.018$ & $0.011$ \\ \hline 
    $0.0022868453$ & $0.009$ & $0.004$ & $0.017$ & $0.009$ \\ \hline 
    \end{tabular}
    \caption{\small MEMs of $u_1,\om_1$ at different times on the mesh of size $(n_1,n_2) = (1536,1536)$.}
    \label{tab:MEM_time_nse}
\end{table}

\begin{figure}[!ht]
\centering  
    \includegraphics[width=0.35\textwidth]{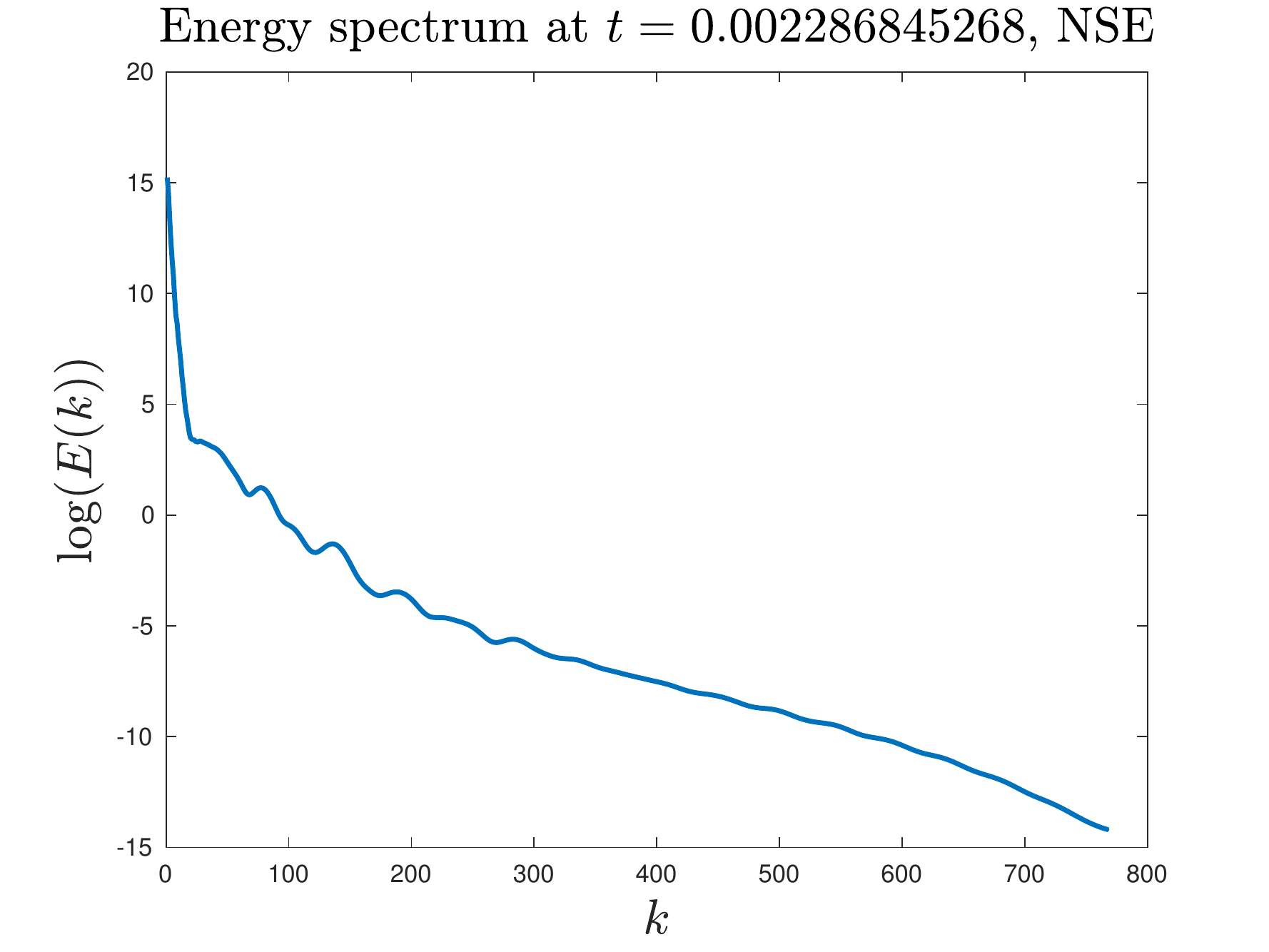}
   \includegraphics[width=0.35\textwidth]{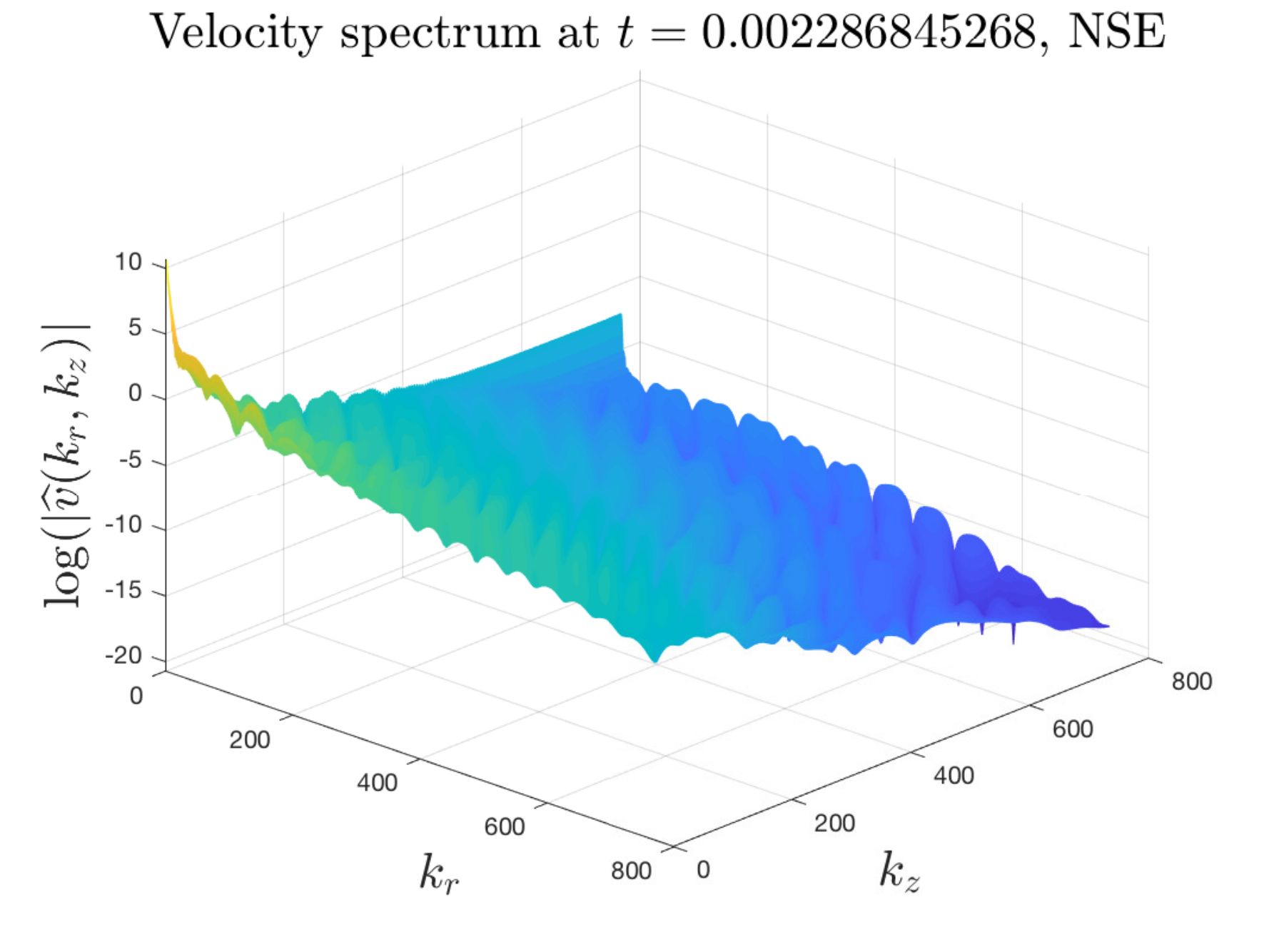}   
    \includegraphics[width=0.35\textwidth]{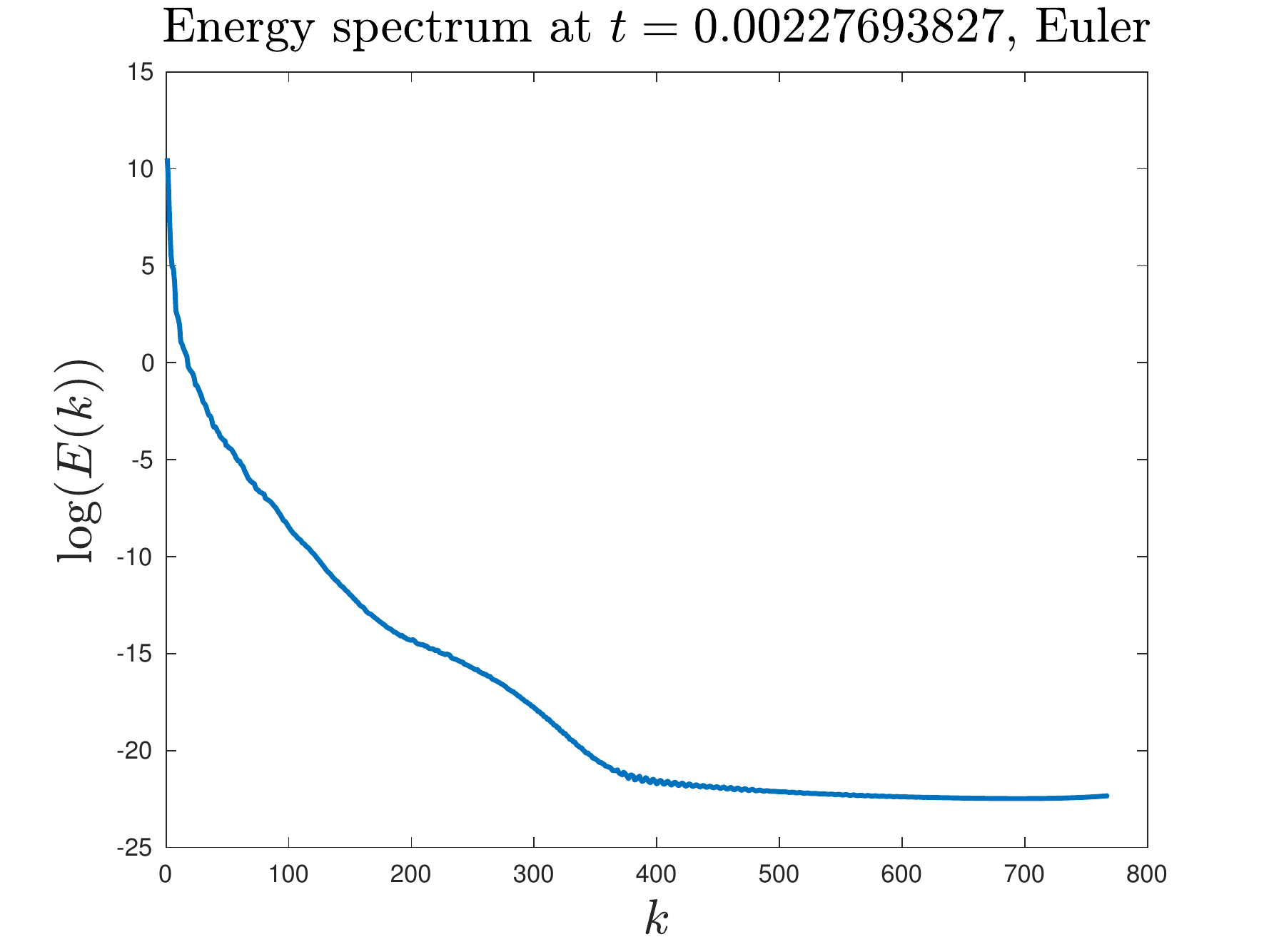} 
     \includegraphics[width=0.35\textwidth]{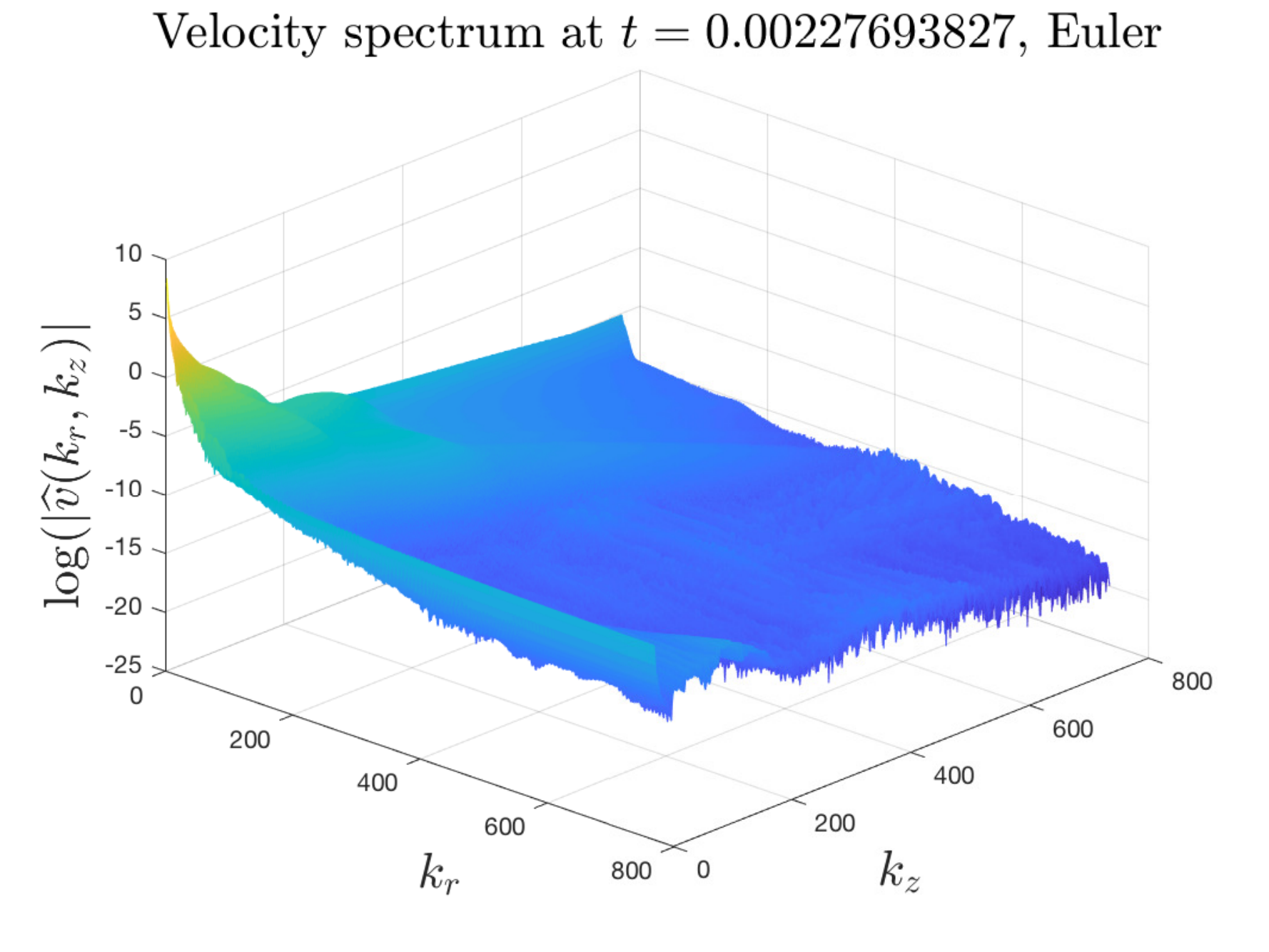}   
    \caption[Energy spectrum]{Top row. Left plot: Energy spectrum for the Navier-Stokes equations computed in the $(\rho,\eta)$ coordinates at $t_3=0.0022868453$ using $1536\times 1536$ grid. Right plot: Velocity spectrum in the $(\rho,\eta)$ coordinates.
  Bottom row. Left plot: Energy spectrum for the Euler equations computed in the $(\rho,\eta)$ coordinates at $t=0.00227693827$ using $1536\times 1536$ grid. Right plot: Velocity spectrum in the $(\rho,\eta)$ coordinates. $E(k)$ is the sum of the contributions from the square of the velocity spectrum, $|\widehat{\bf u}_{k_r,k_z}|^2$, from the shell centered at $k<|(k_r,k_z)|\leq k+1$ normalized by the total energy at time $t$. }
    \label{fig:energy_spectrum_4_nse}
       \vspace{-0.10in}
\end{figure}

In an effort to understand how well our adaptive mesh resolves the solution of the Navier--Stokes equations, we also study the energy spectrum and the velocity spectrum of the solution. Since the smallest mesh size of our adaptive mesh is of order $10^{-8}$, it would be prohibitively expensive to generate the Fourier spectrum in the $(r,z)$ coordinates. On the other hand, our computation is carried out in the transformed $(\rho,\eta)$ coordinates. So it makes sense to plot the energy spectrum and the velocity spectrum in the $(\rho,\eta)$ coordinates. To reduce the boundary effect, we have applied a soft cut-off $f_c(\rho,\eta)$ that is approximately equal to $1$ for $(\rho,\eta) \in (0.1,0.9)\times (0.13,0.9)$ and goes to zero smoothly at the boundary $\rho=0,1$ and $\eta=0,1$. In the top row of Figure \ref{fig:energy_spectrum_4_nse}, we plot the energy spectrum and the velocity spectrum for the solution of the Navier-Stokes equations using grid $1536\times1536$ at $t_3=0.0022868453$. The Navier-Stokes solution is in a late stage with maximum vorticity increased by a factor of $10^6$ by this time. Both the energy spectrum $E(k)$ and the velocity spectrum $\widehat{\bf u}_{k_r,k_z}$ indicate that our adaptive mesh resolves the high frequency solution accurately.
We can see a power law like structure developed in the energy spectrum for the Navier-Stokes solution with a slightly faster decay rate in the very high frequency regime. The velocity spectrum decays exponentially fast from low frequencies to high frequencies with no sign of high frequency instability.

For comparison, we also plot the corresponding energy spectrum and velocity spectrum for the Euler equations in the bottom row of Figure \ref{fig:energy_spectrum_4_nse}. We use grid $1536\times1536$ and a second order numerical viscosity to solve the Euler equations up to $t=0.00227693827$. Again, the energy spectrum and velocity spectrum show that the Euler solution is well resolved in the high frequency regime. We should emphasize that the Euler solution is in an early stage with maximum vorticity increased by a factor of more than $5000$ by this time. We do not observe a similar power law like structure for the energy spectrum since we are not close enough to the potential finite time singularity. Without using a relatively large viscous regularization, the solution of the Euler equations develops a sharp front dynamically, which makes it difficult for us to get sufficiently close to the potential singularity time.

\subsubsection{Resolution study}\label{sec:resolution_study_nse} 
In this subsection, we perform resolution study on the numerical solutions of the initial-boundary value problem \eqref{eq:axisymmetric_NSE_1} at various time instants $t$. We will estimate the relative error of a solution variable $f_p$ computed on the $256p\times 256p$ mesh by comparing it to a reference variable $\hat{f}$ that is computed at the same time instant on the finest mesh of size $1536\times 1536$.

We first study the sup-norm error of the solution, which is the most important measure on the accuracy of our numerical method. Tables~\ref{tab:sup-norm_error_1-1_nse}--\ref{tab:sup-norm_error_1-4_nse} report the sup-norm relative errors and numerical orders of different solution variables at times $t_0 = 0.00227375$ and $t_2 = 0.0022867812$, respectively.
The first time instant $t_0 = 0.00227375$ corresponds to the time at which we stop the computation using viscosity $\nu=5\cdot 10^{-4}$. After this time, we increase the viscosity to $\nu=5\cdot 10^{-3}$. We will use the solution computed by our finest resolution $1536\times 1536$ at this time as a new initial condition for the coarser meshes. Thus it is important to make sure that we have resolved the Navier--Stokes equations accurately up to this time.
The convergence results at both $t_0 = 0.00227375$ and $t_2 = 0.0022867812$ look qualitatively the same although the errors at $t_2 = 0.0022867812$ have increased modestly. This seems to be quite reasonable since the maximum vorticity has increased by a factor of $10^5$ by $t_2 = 0.0022867812$. In comparison, the maximum vorticity has increased only by a factor of $498.42$ by $t_0 = 0.00227375$. These results confirm that our method is at least $2$nd-order accurate. 

\begin{table}[!ht]
\centering
\footnotesize
\renewcommand{\arraystretch}{1.5}
    \begin{tabular}{|c|c|c|c|c|c|c|}
    \hline
    \multirow{2}{*}{Mesh size} & \multicolumn{6}{c|}{Sup-norm relative error at $t_0=0.00227375$ for $3$D Navier--Stokes equations} \\ \cline{2-7} 
    						   & $u_1$ & Order & $\omega_1$ & Order & $\psi_1$ & Order \\ \hline 
    $512\times512$ & $1.6181\times10^{-1}$ & -- & $4.9486\times10^{-1}$ & -- & $4.2245\times10^{-2}$ & -- \\ \hline 
    $768\times768$ & $6.3639\times10^{-2}$ & $2.302$ & $2.1788\times10^{-1}$ & $2.023$ & $1.5929\times10^{-2}$ & $2.405$ \\ \hline 
    $1024\times1024$ & $2.7017\times10^{-2}$ & $2.978$ & $9.4386\times10^{-2}$ & $2.908$ & $6.6958\times10^{-3}$ & $3.013$ \\ \hline 
    $1280\times1280$ & $9.6517\times10^{-3}$ & $4.613$ & $3.3906\times10^{-2}$ & $4.588$ & $2.3826\times10^{-3}$ & $4.631$ \\ \hline 
    \end{tabular}
    \caption{\small Sup-norm relative errors and numerical orders of $u_1,\om_1,\psi_1$ at $t_0 = 0.00227375$ for the $3$D Navier--Stokes equations.}
    \label{tab:sup-norm_error_1-1_nse}
    \vspace{-0.2in}
\end{table}

\begin{table}[!ht]
\centering
\footnotesize
\renewcommand{\arraystretch}{1.5}
    \begin{tabular}{|c|c|c|c|c|c|c|}
    \hline
    \multirow{2}{*}{Mesh size} & \multicolumn{6}{c|}{Sup-norm relative error at $t_0=0.00227375$ for $3$D Navier--Stokes  equations} \\ \cline{2-7} 
    						   & $u^r$ & Order & $u^z$ & Order & $\vom=(\om^\theta,\om^r,\om^z)$ & Order \\ \hline 
    $512\times512$ & $1.6238\times10^{-1}$ & -- & $3.6640\times10^{-1}$& -- & $5.0751\times10^{-1}$ & -- \\ \hline 
    $768\times768$ & $6.4154\times10^{-2}$ & $2.290$ &
    $1.4711\times10^{-1}$ & $2.251$ & $2.2110\times10^{-1}$ & $2.049$ \\ \hline 
    $1024\times1024$ & $2.7213\times10^{-2}$ & $2.981$ & $6.2551\times10^{-2}$ & $2.973$ &  $9.5467\times10^{-2}$ & $2.919$ \\ \hline 
    $1280\times1280$ & $9.7121\times10^{-3}$ & $4.617$ &
    $2.2333\times10^{-2}$ & $4.615$ & $3.4244\times10^{-2}$ & $4.595$ \\ 
    \hline
    \end{tabular}
    \caption{\small Sup-norm relative errors and numerical orders of $u^r,u^z,\vom$ at $t_0 = 0.00227375$ for $3$D Navier--Stokes equations.}
    \label{tab:sup-norm_error_1-2_nse}
    \vspace{-0.2in}
\end{table}

\begin{table}[!ht]
\centering
\footnotesize
\renewcommand{\arraystretch}{1.5}
    \begin{tabular}{|c|c|c|c|c|c|c|}
    \hline
    \multirow{2}{*}{Mesh size} & \multicolumn{6}{c|}{Sup-norm relative error at $t_2=0.0022867812$ for $3$D Navier--Stokes equations} \\ \cline{2-7} 
    						   & $u_1$ & Order & $\omega_1$ & Order & $\psi_1$ & Order \\ \hline 
    $512\times512$ & $2.9842\times10^{-1}$ & -- & $5.4093\times10^{-1}$ & -- & $7.4719\times10^{-2}$ & -- \\ \hline 
    $768\times768$ & $1.1109\times10^{-1}$ & $2.437$ & $2.1295\times10^{-1}$ & $2.299$ & $2.6937\times10^{-2}$ & $2.516$ \\ \hline 
    $1024\times1024$ & $5.4326\times10^{-2}$ & $2.486$ & $1.04621\times10^{-1}$ & $2.471$ & $1.3168\times10^{-3}$ & $2.488$ \\ \hline 
    $1280\times1280$ & $2.2040\times10^{-2}$ & $4.043$ & $4.2516\times10^{-2}$ & $4.035$ & $5.3224\times10^{-3}$ & $4.060$ \\ \hline 
    \end{tabular}
    \caption{\small Sup-norm relative errors and numerical orders of $u_1,\om_1,\psi_1$ at $t=0.002286781$ for the $3$D Navier--Stokes equations.}
    \label{tab:sup-norm_error_1-3_nse}
    \vspace{-0.2in}
\end{table}

\begin{table}[!ht]
\centering
\footnotesize
\renewcommand{\arraystretch}{1.5}
    \begin{tabular}{|c|c|c|c|c|c|c|}
    \hline
    \multirow{2}{*}{Mesh size} & \multicolumn{6}{c|}{Sup-norm relative error at $t_2=0.0022867812$ for $3$D Navier--Stokes equations} \\ \cline{2-7} 
    						   & $u^r$ & Order & $u^z$ & Order & $\vom=(\om^\theta,\om^r,\om^z)$ & Order \\ \hline 
    $512\times512$ & $2.5750\times10^{-1}$ & -- & $4.3600\times10^{-1}$& -- & $5.6599\times10^{-1}$ & -- \\ \hline 
    $768\times768$ & $9.4107\times10^{-2}$ & $2.482$ &
    $1.6258\times10^{-1}$ & $2.433$ & $2.1966\times10^{-1}$ & $2.334$ \\ \hline 
    $1024\times1024$ & $4.5528\times10^{-2}$ & $2.524$ & $7.9726\times10^{-2}$ & $2.477$ &  $1.0734\times10^{-1}$ & $2.489$ \\ \hline 
    $1280\times1280$ & $1.8425\times10^{-2}$ & $4.054$ &
    $3.2303\times10^{-2}$ & $4.049$ & $4.3522\times10^{-2}$ & $4.045$ \\ 
    \hline
    \end{tabular}
    \caption{\small Sup-norm relative errors and numerical orders of $u^r,u^z,\vom$ at $t_2=0.0022867812$ for the $3$D Navier--Stokes equations.}
    \label{tab:sup-norm_error_1-4_nse}
    \vspace{-0.2in}
\end{table}

%
%
%

\begin{figure}[!ht]
\centering  
    \includegraphics[width=0.35\textwidth]{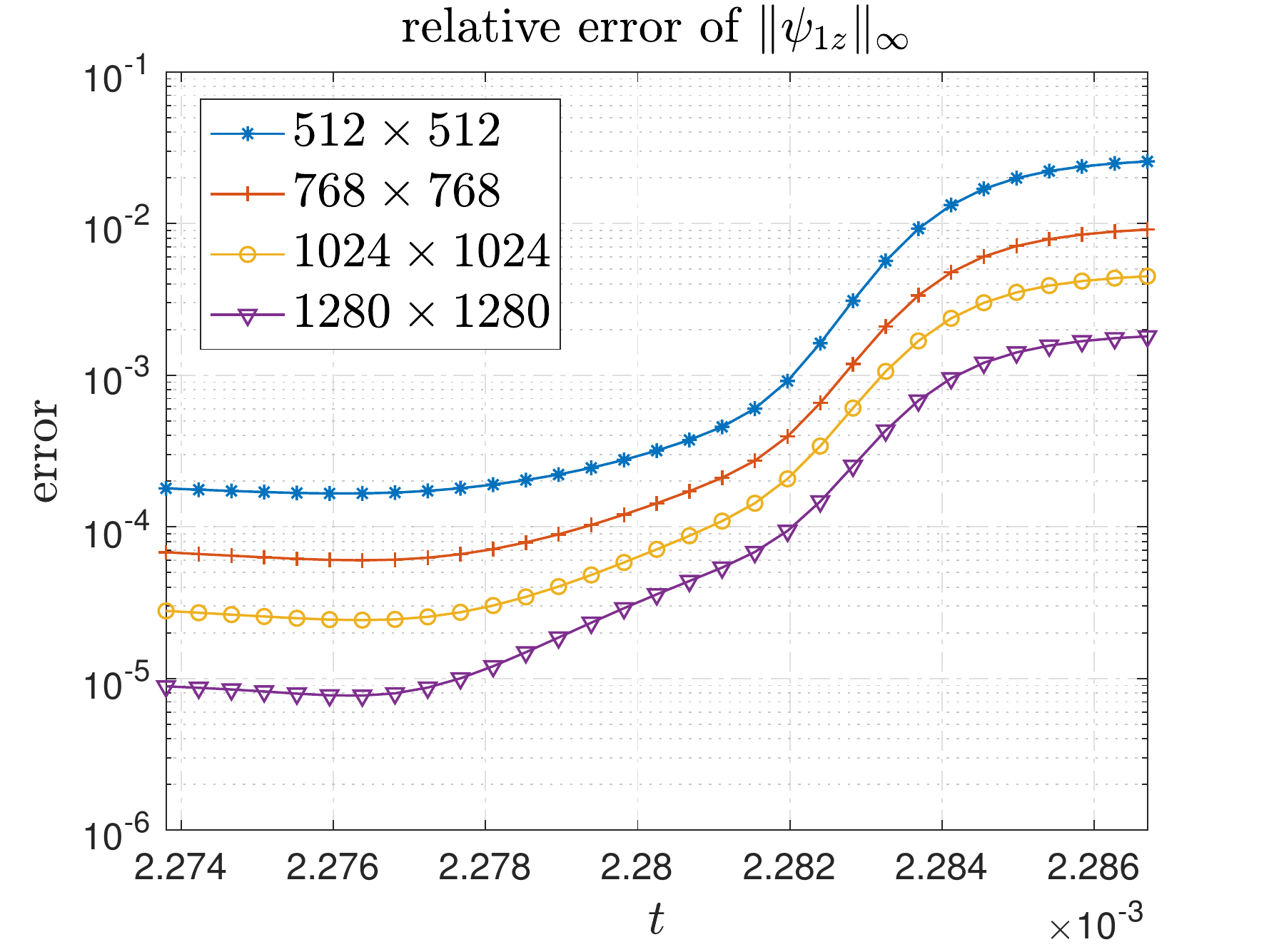}
    \includegraphics[width=0.35\textwidth]{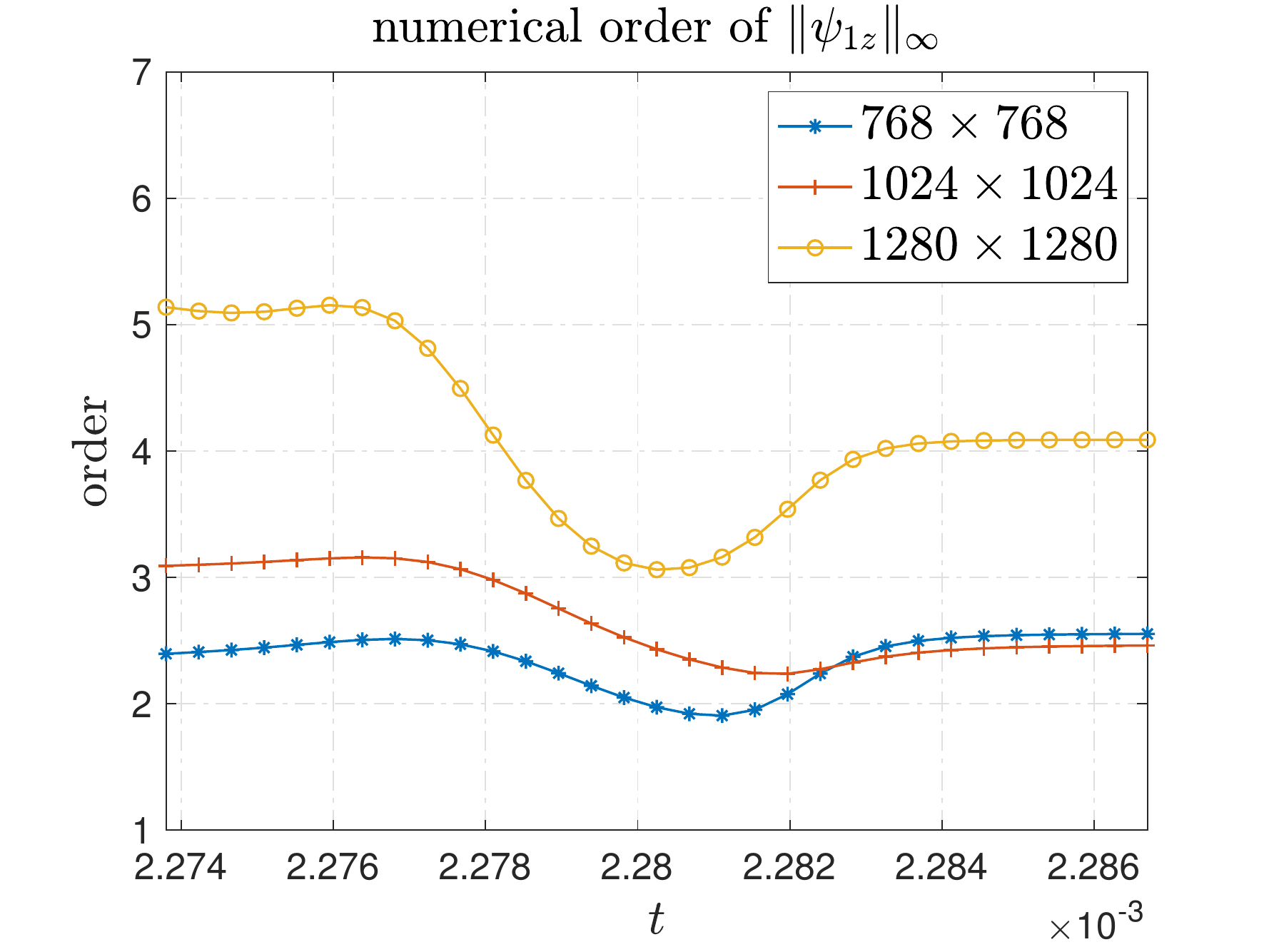} 
      \includegraphics[width=0.35\textwidth]{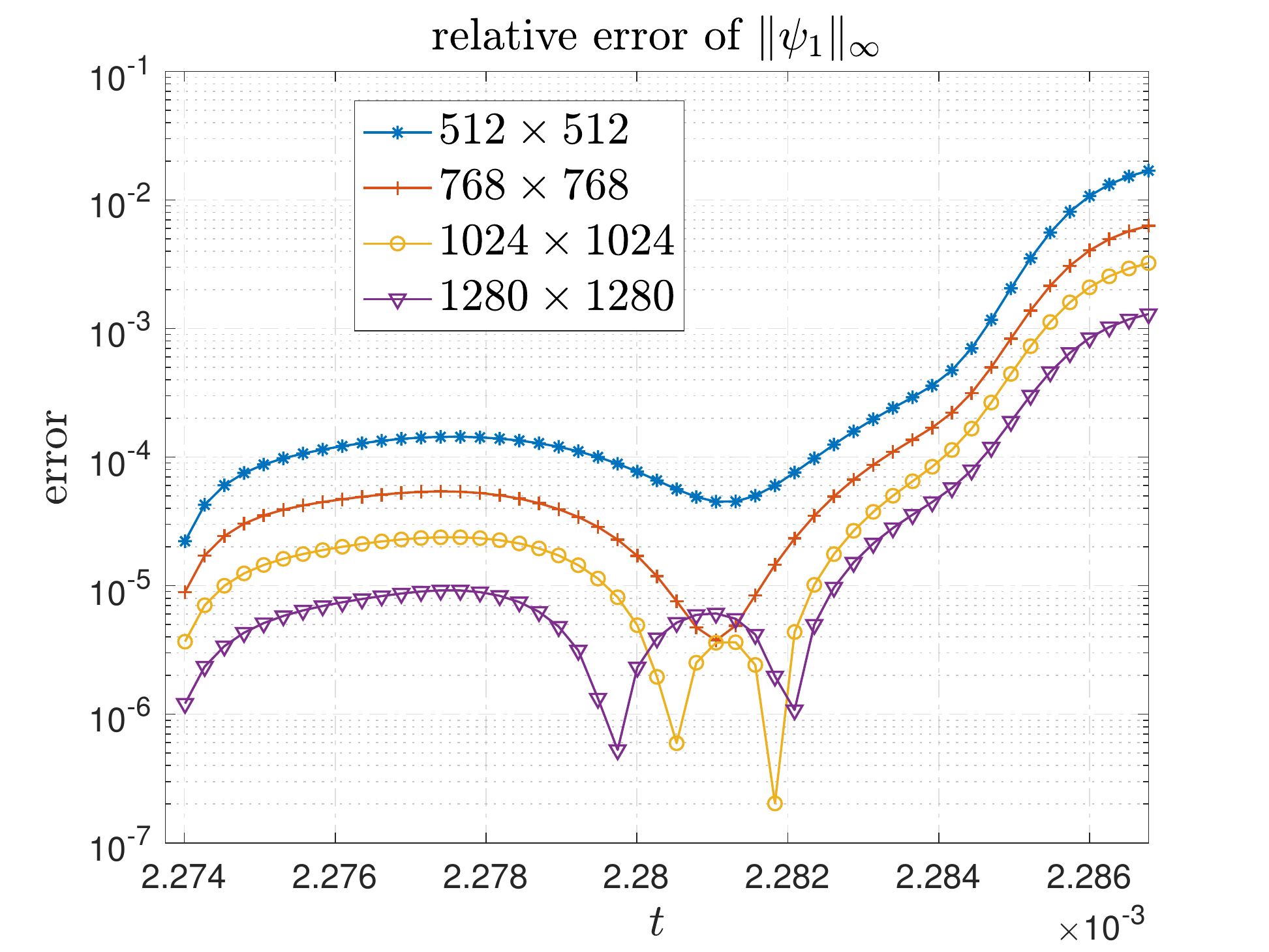}
    \includegraphics[width=0.35\textwidth]{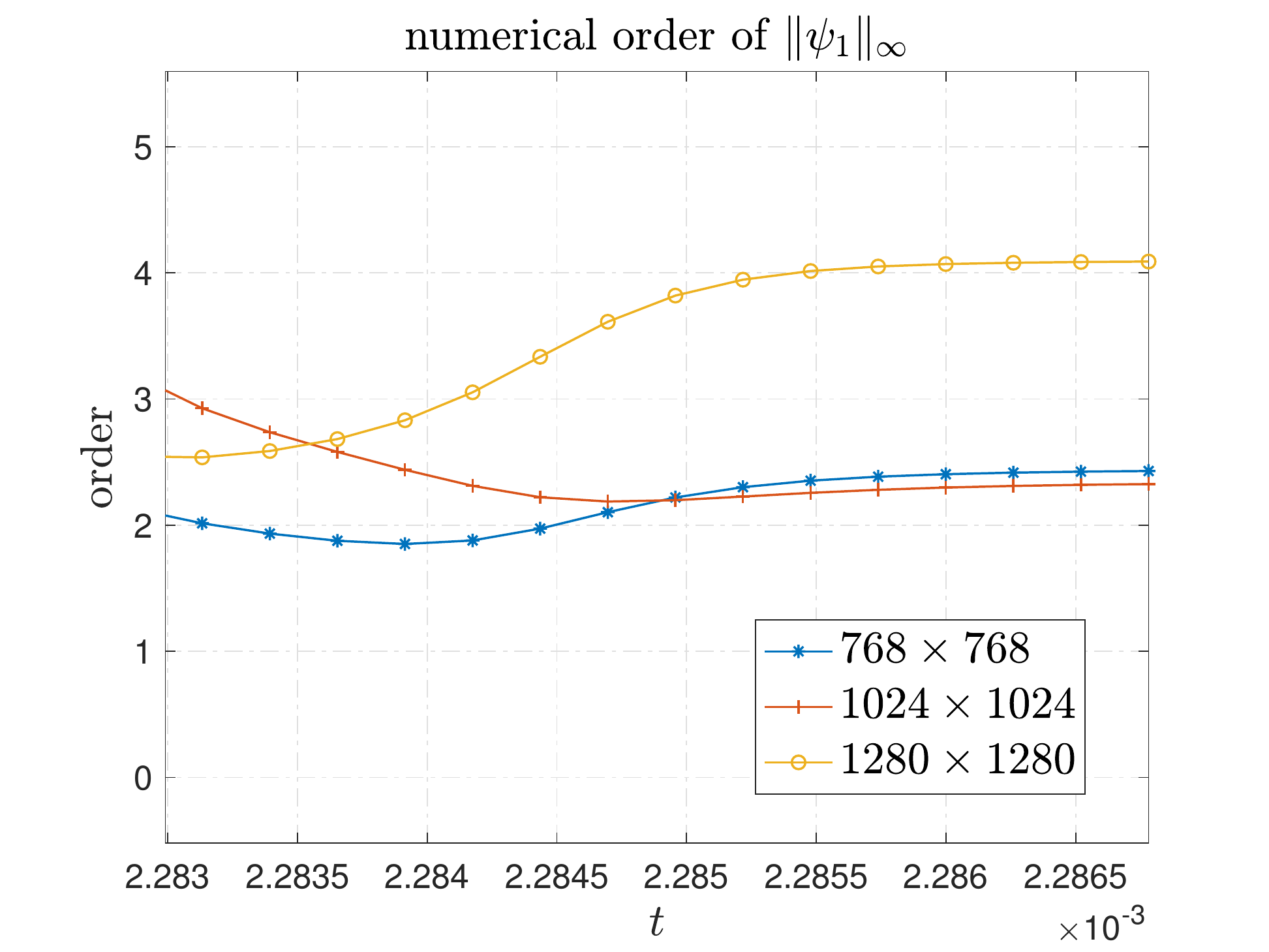} 
    \includegraphics[width=0.35\textwidth]{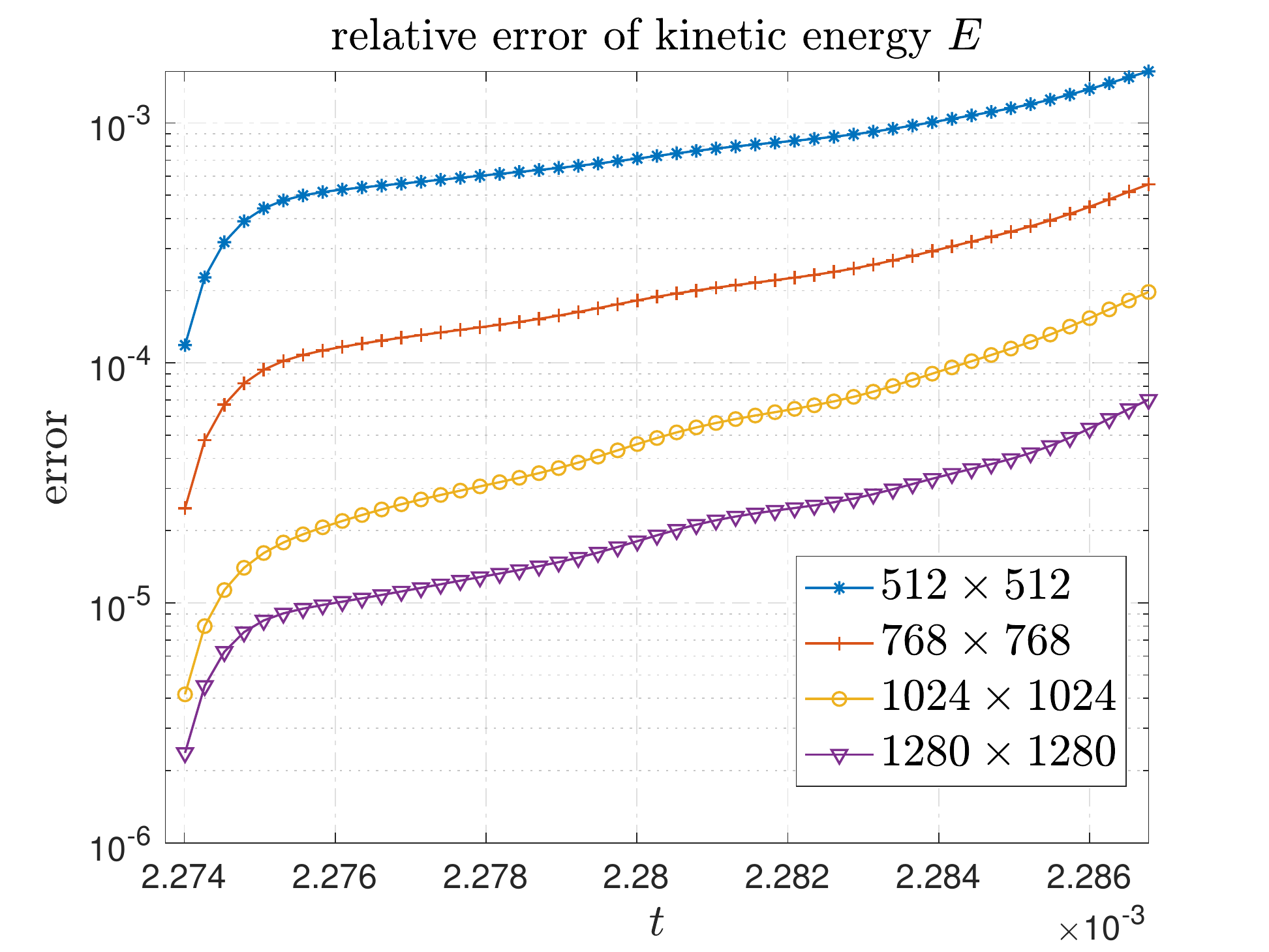}
    \includegraphics[width=0.35\textwidth]{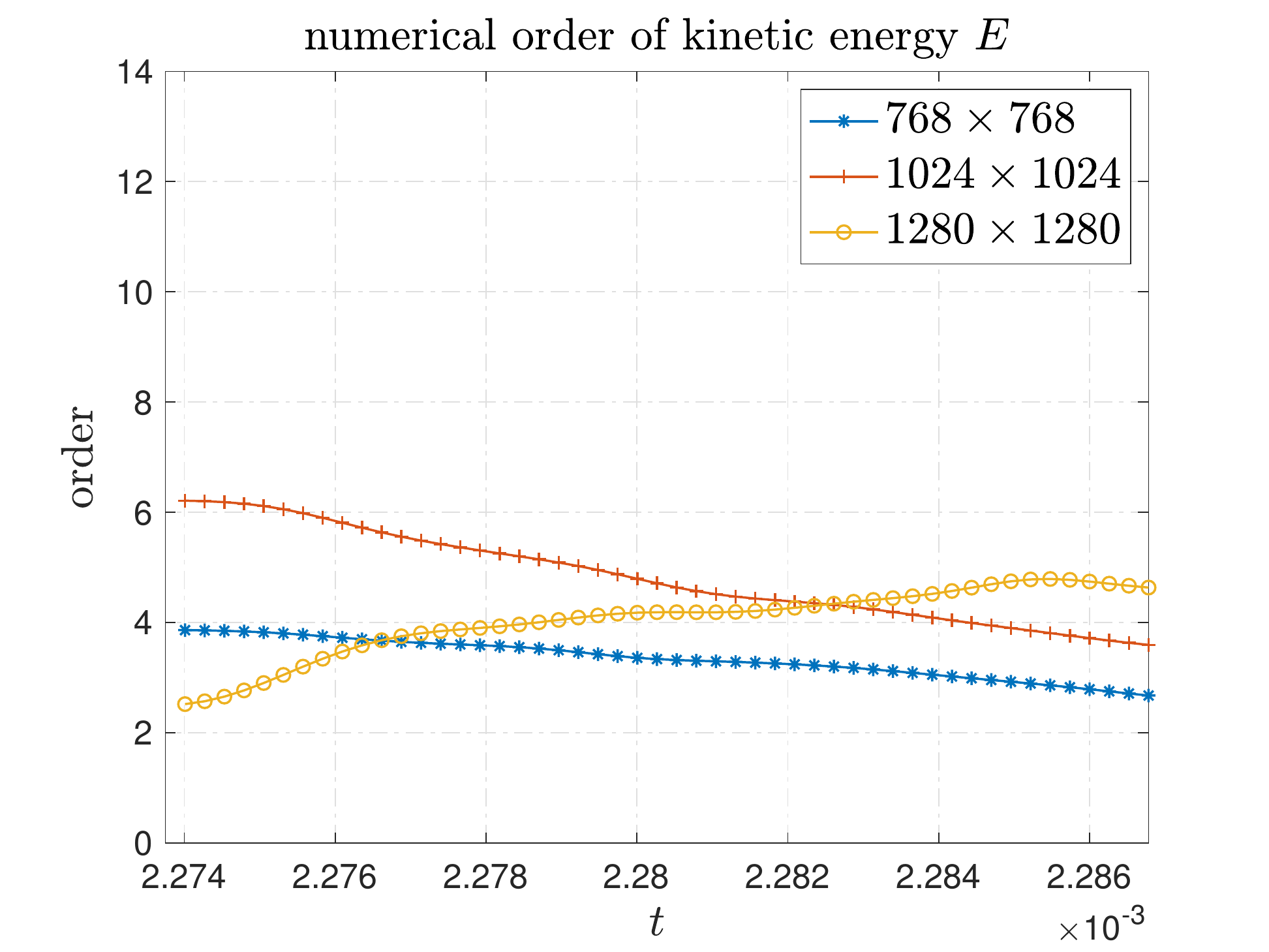}
    \caption[Relative error II]{First row: relative error and numerical order of $\|\psi_{1z}(t)\|_{L^\infty}$. Second row: relative error and numerical order of $\|\psi_1(t)\|_{L^\infty}$. Third row: relative error and numerical order of the kinetic energy, $E(t)$. The computation is between $t_0=0.00227375$ and $t_2=0.0022867812$.}  
    \label{fig:relative_error_2_nse}
       \vspace{-0.05in}
\end{figure}


\begin{figure}[!ht]
\centering  
    \includegraphics[width=0.35\textwidth]{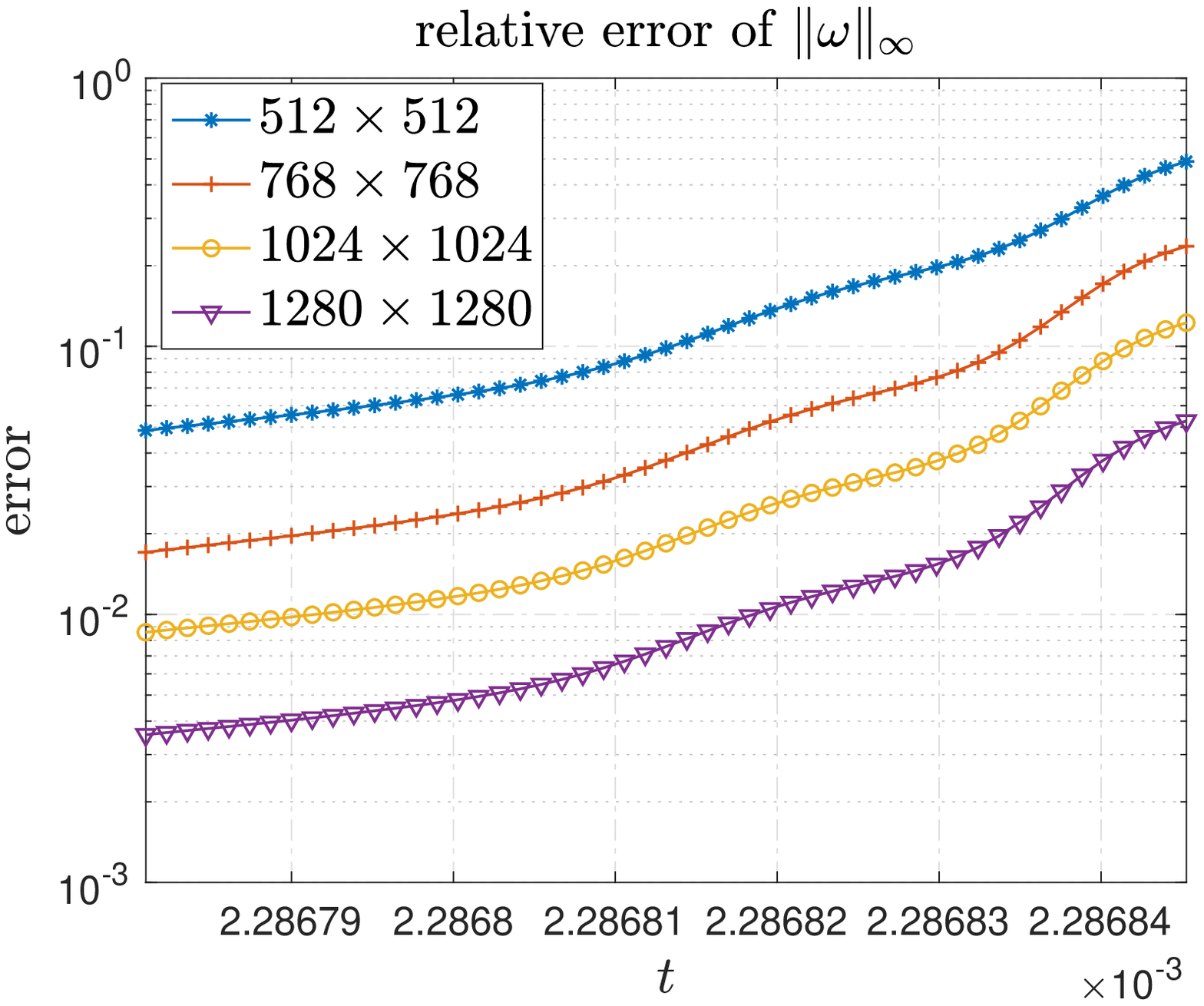}
    \includegraphics[width=0.35\textwidth]{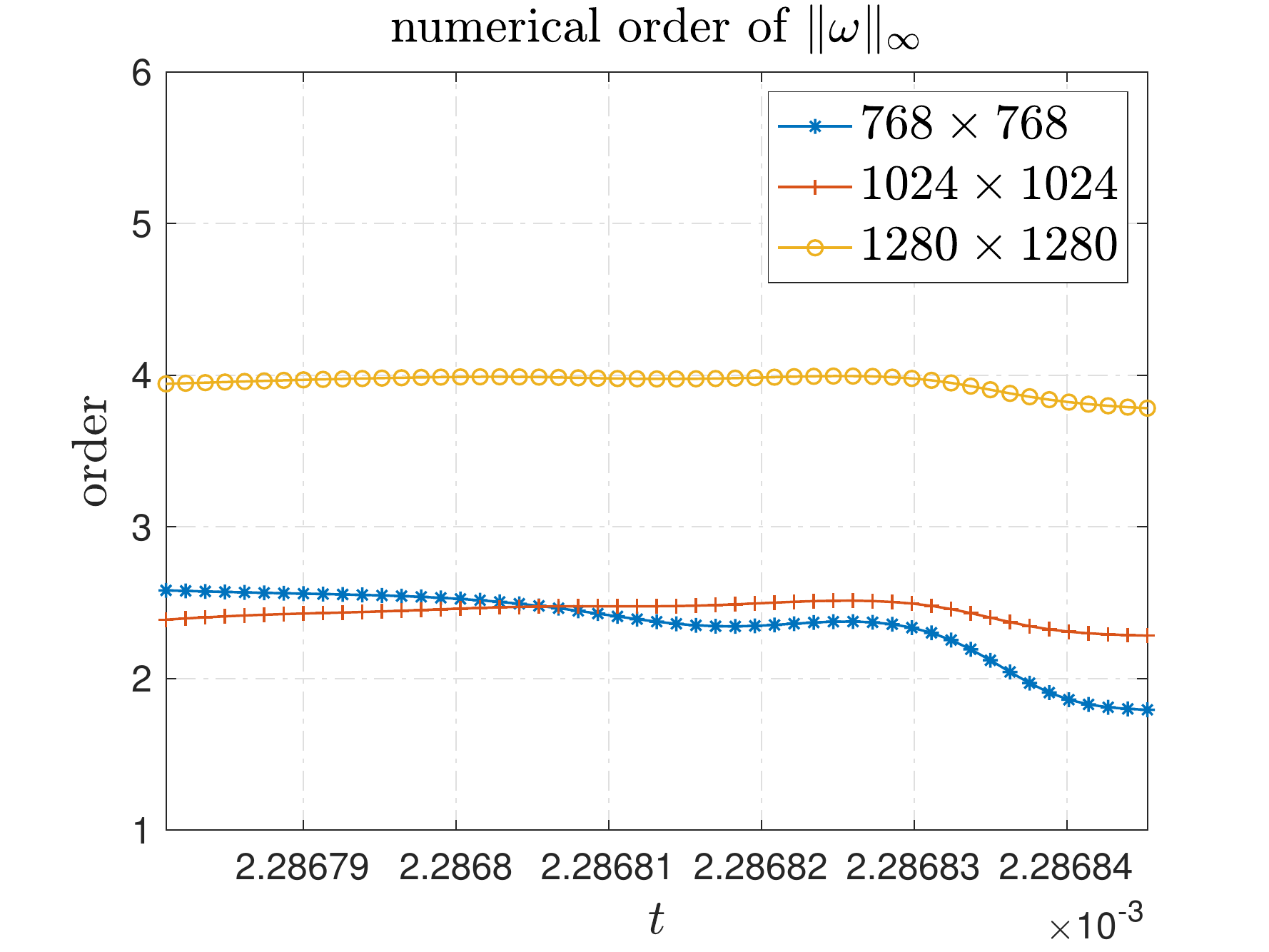} 
    \includegraphics[width=0.35\textwidth]{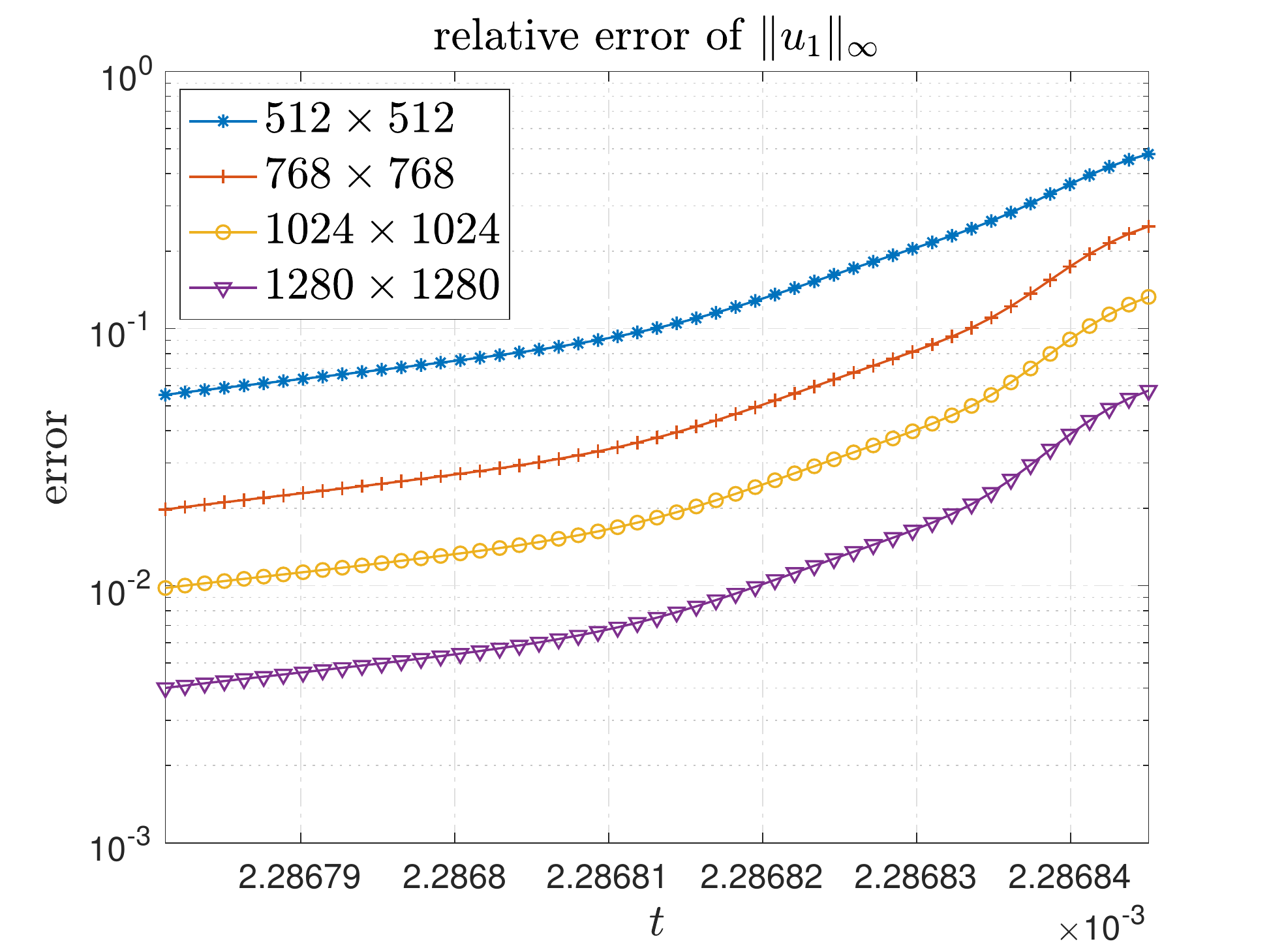}
    \includegraphics[width=0.35\textwidth]{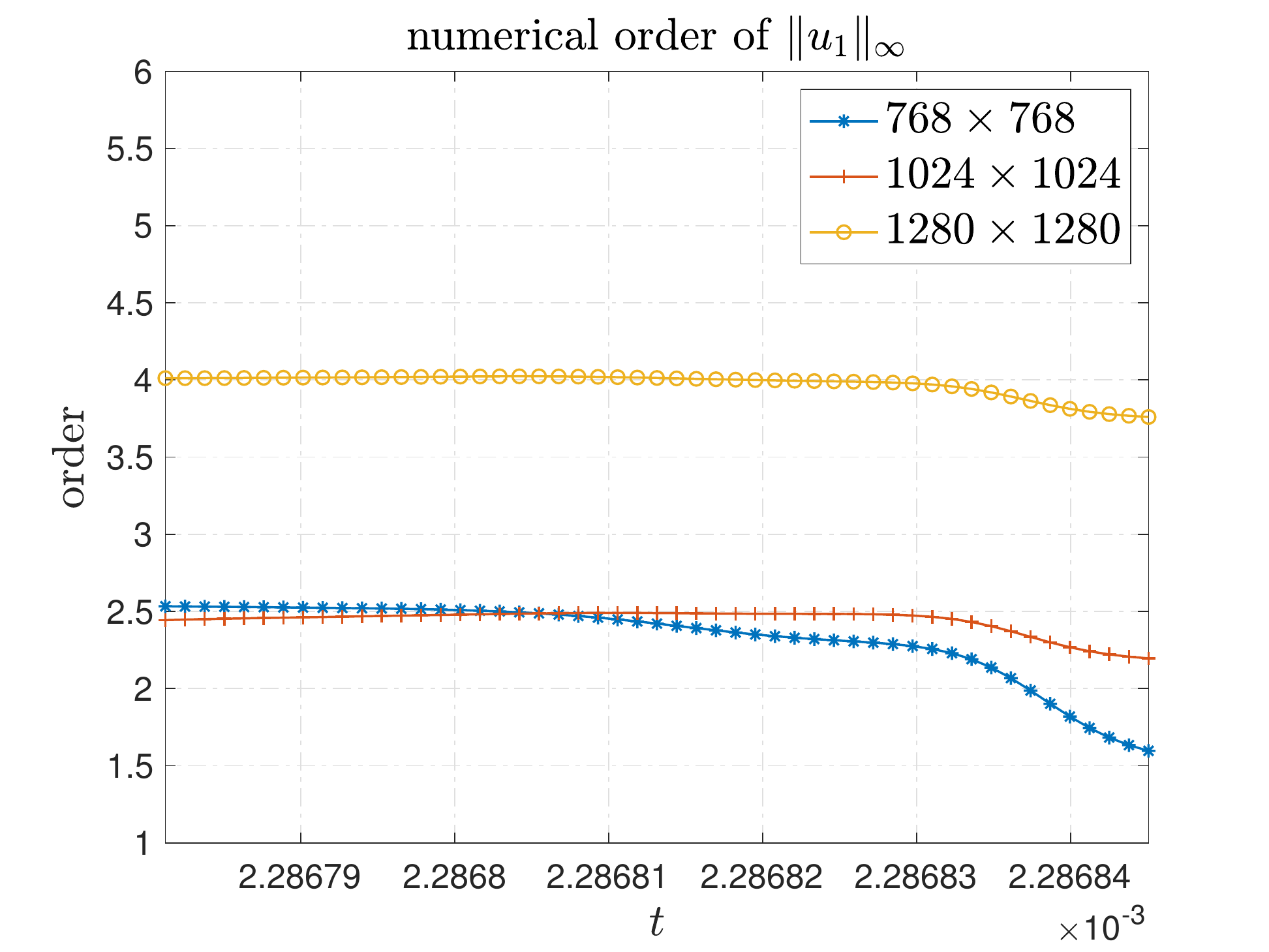}
            \includegraphics[width=0.35\textwidth]{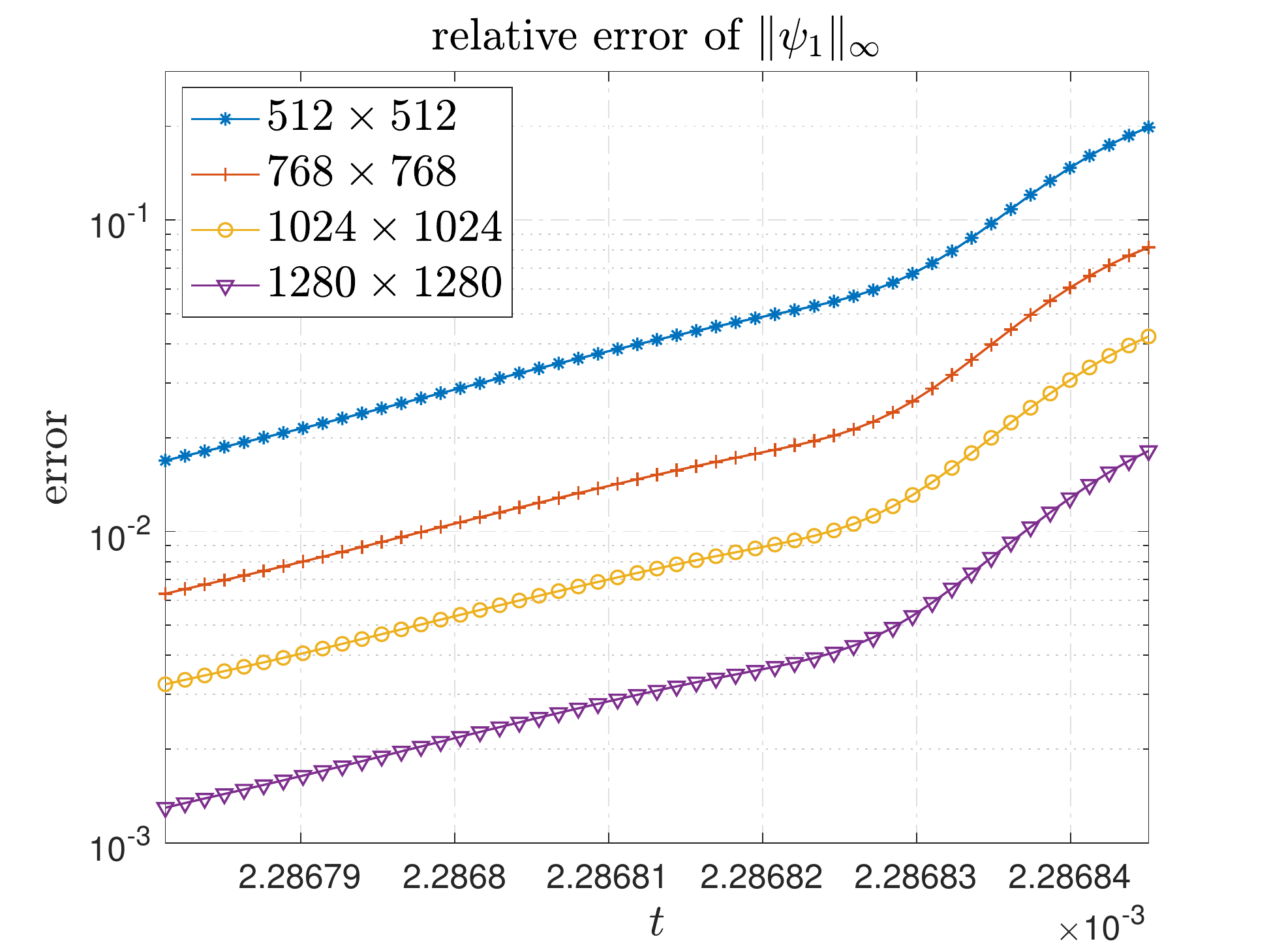}
    \includegraphics[width=0.35\textwidth]{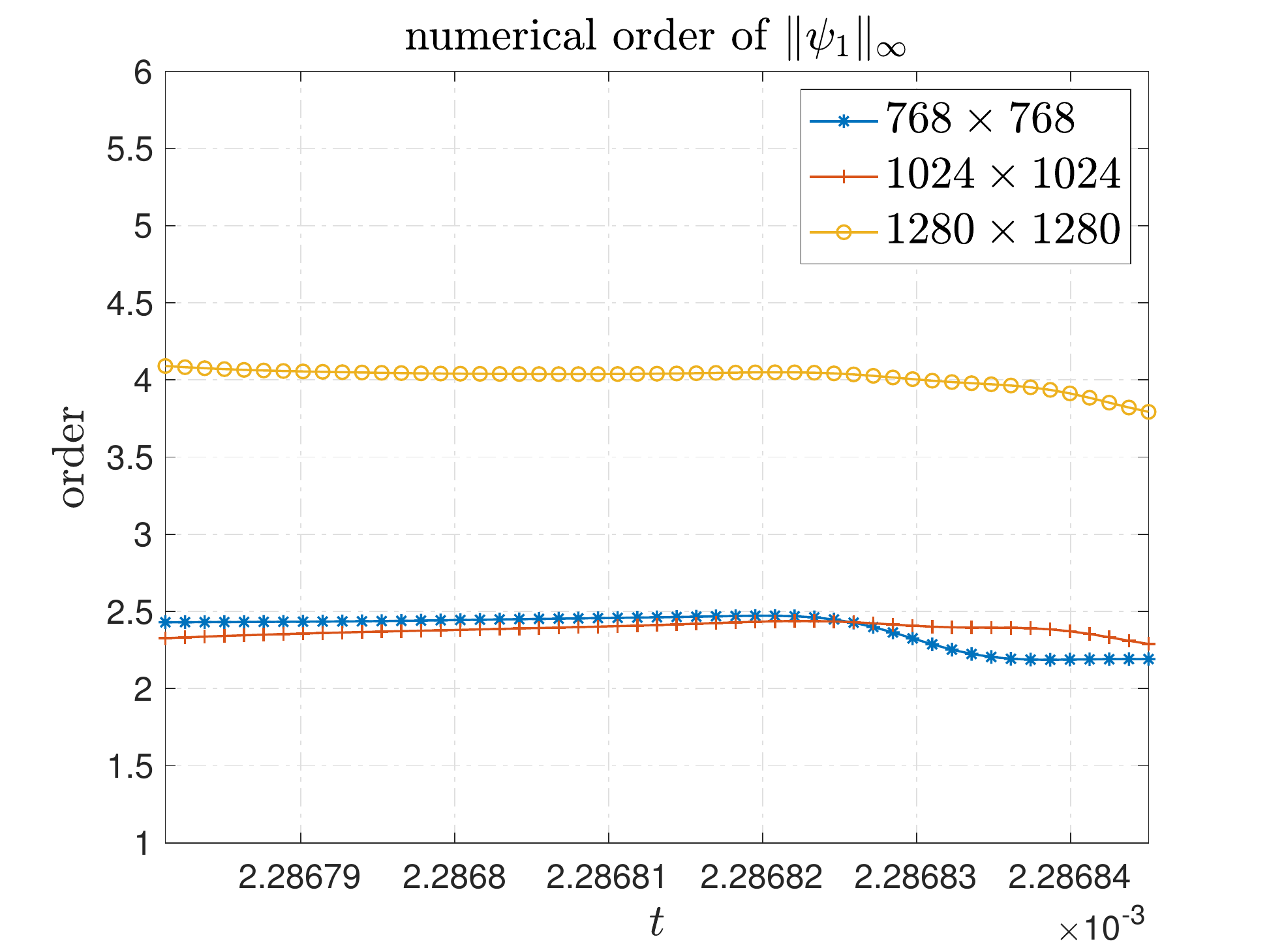}
    \caption[Relative error II]{First row: relative error and numerical order of $\|\vom(t)\|_{L^\infty}$. Second row: relative error and numerical order of $\|u_1(t)\|_{L^\infty}$. Third row: relative error and numerical order of $\psi_1(t)$. The time period shown in the figure is from $t_2=0.0022867812$ to $t_3= 0.002286845$.}  
    \label{fig:relative_error_4_nse}
       \vspace{-0.05in}
\end{figure}

%

We will also study the convergence of some variables as functions of time. In particular, we report the convergence of the quantities $\|\psi_1\|_{L^\infty}$, $\|u_1\|_{L^\infty}$, $\|\om_1\|_{L^\infty}$, $\|\vom\|_{L^\infty}$, and the kinetic energy $E$. Since we have already demonstrated the second order convergence in the first time interval $[0, t_0]$, we will focus  our study on the relative errors and the numerical orders in two different time intervals beyond $t=t_0$. 
In Figure \ref{fig:relative_error_2_nse}, we plot the relative errors and numerical orders of $\|\psi_{1z}\|_{L^\infty}$, $\|\psi_1\|_{L^\infty}$, and the kinetic energy $E$ as functions of time during the second time interval $[t_0,t_2]$. 
We observe that our method is at least $2$nd-order in $h_\rho,h_\eta$ during this time interval. 

In Figure \ref{fig:relative_error_4_nse}, we plot the relative errors and the numerical orders of $\|\vom\|_{L^\infty}$, $\|u_1\|_{L^\infty}$ and $\|\psi_1\|_{L^\infty}$ in
the third time interval between $t_2=0.0022867812$ and $t_3= 0.0022868453$. This corresponds to the late stage of our computation.
Due to the strong viscous regularization, we observe that the errors become quite stable in the late stage of our computation. This is very different from the situation for the $3$D Euler equations \cite{Hou-euler-2021}. It also shows that our adaptive mesh strategy is very effective.

\subsection{Scaling Properties of the solution}\label{sec:scaling_study_nse}
In this subsection, we will examine the scaling properties of the solution of the $3$D Navier--Stokes equations. 
We will provide some qualitative numerical evidences that the solution of the $3$D Navier--Stokes equations enjoys nearly self-similar scaling properties. 

\vspace{-0.15in}
\subsubsection{Fitting of the growth rate}\label{sec:growth_fitting_nse}
\begin{figure}[!ht]
\centering
	\begin{subfigure}[b]{0.40\textwidth}
    \includegraphics[width=1\textwidth]{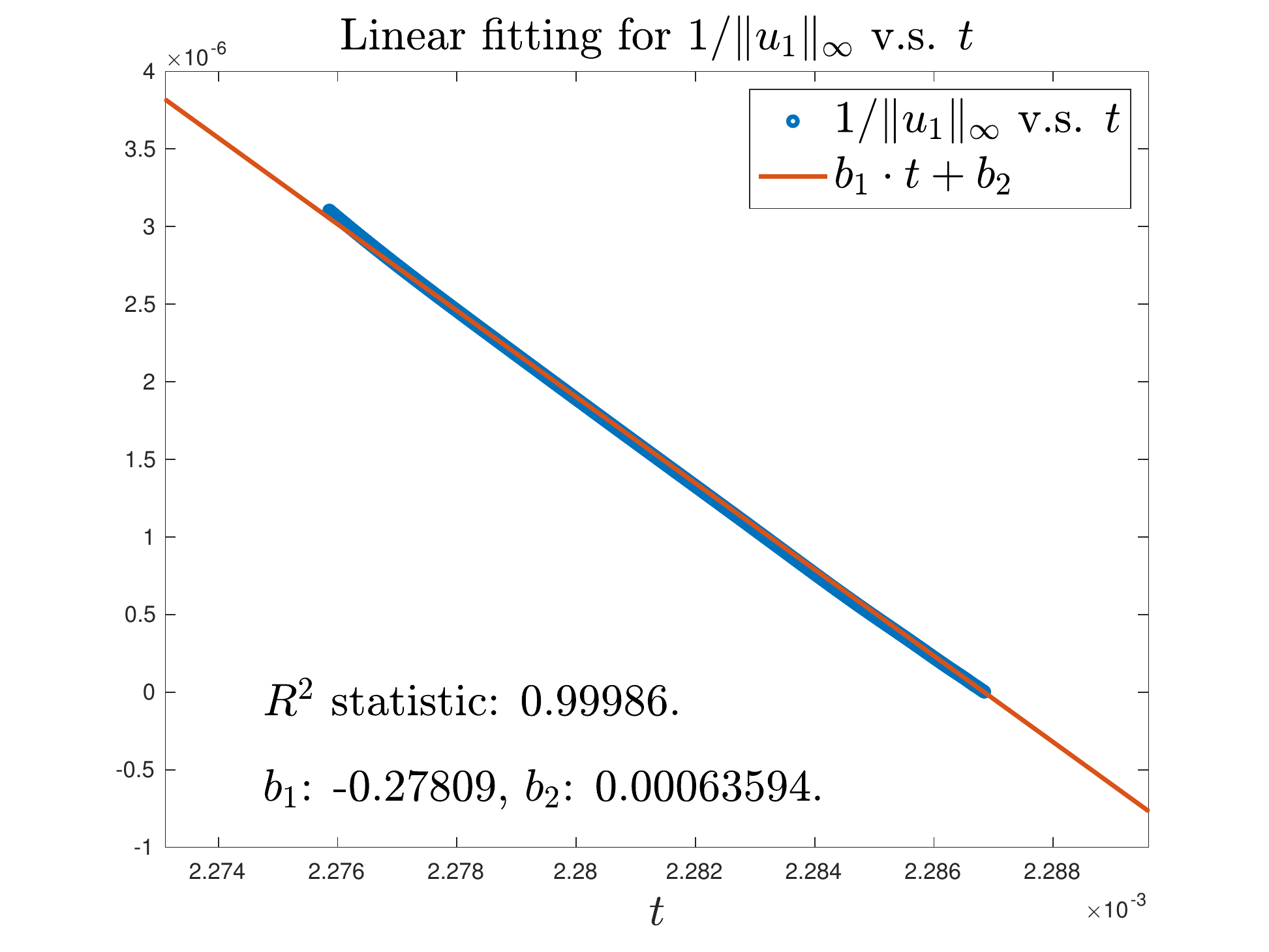}
    \caption{linear regression of $\|u_1(t)\|_{L^\infty}^{-1}$}
    \end{subfigure}
  	\begin{subfigure}[b]{0.40\textwidth} 
    \includegraphics[width=1\textwidth]{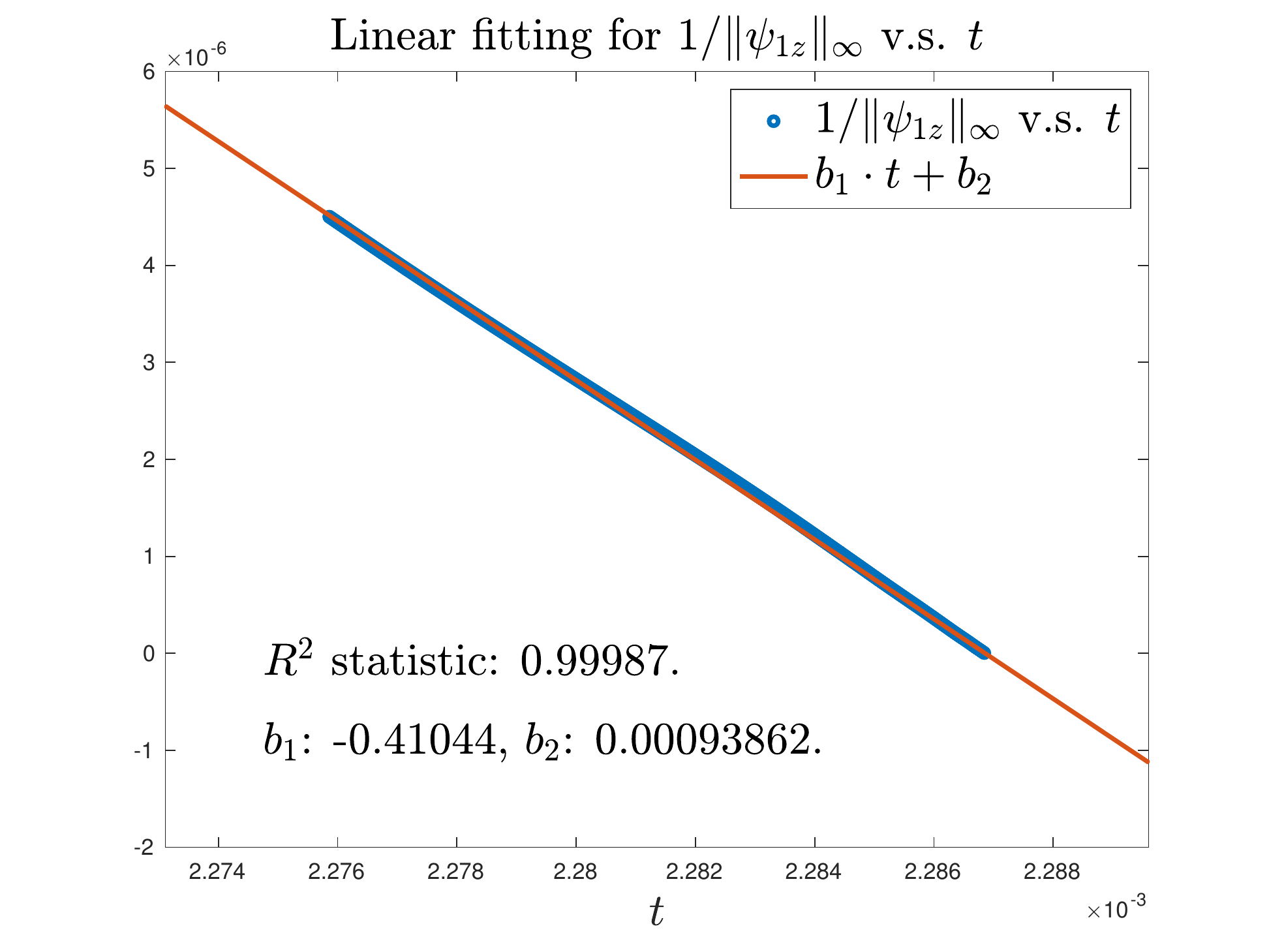}
    \caption{linear regression of $\|\psi_{1z}\|_{L^\infty}^{-1}$}
    \end{subfigure} 
    \caption[Linear regression $u_1$]{The linear regression of (a) $\|u_1\|_{L^\infty}^{-1}$ vs $t$, (b) $\|\psi_{1z}\|_{L^\infty}^{-1}$ vs $t$. 
    The blue points are the data points obtained from our computation, and the red lines are the linear models. The solution is computed using $1536\times 1536$ grid. The final time instant is $t_4=0.0022868502$.}   
    \label{fig:linear_regression_u1_nse}
       \vspace{-0.05in}
\end{figure}

We now study the scaling properties the numerical solutions obtained from $1536\times1536$ resolution. Figure\eqref{fig:linear_regression_u1_nse} shows the fitting results for the quantities $\|u_1(t)\|_{L^\infty}$ and $\|\psi_{1,z}\|_{L^\infty} $ on the time interval $[t_s,t_4] = [0.002275858596,0.0022868502]$. We would like to emphasize that the maximum vorticity relative to its initial maximum vorticity, i.e. $\|\vom (t_4) \|_{L^\infty}/\|\vom (0) \|_{L^\infty}$, has increased by a factor of $10^7$ while we only have $\|\vom (t_s) \|_{L^\infty}/\|\vom (0) \|_{L^\infty} = 514.58$. During this time interval, the viscosity remains the same and is equal to $\nu = 5 \cdot 10^{-3}$.  

We can see that 
$\|u_1(t)\|_{L^\infty} \sim (T-t)^{-1}$, 
$\|\psi_{1,z}\|_{L^\infty}  \sim (T-t)^{-1}$
have excellent linear fitness with $R$-Square values very close to $1$. Recall that we have observed a strong positive alignment between $\psi_{1,z}$ and $u_1$ around the maximum location $(R(t),Z(t))$ of $u_1$ after we increase the viscosity from $\nu=5\cdot 10^{-4}$ to $\nu=5\cdot 10^{-3}$ at $t_0=0.00227375$. Thus, we have $\psi_{1,z}(t,R(t),Z(t))\sim u_1(t,R(t),Z(t)) $ and the equation of $\|u_1(t)\|_{L^\infty}$ can be approximated by 
\[\frac{\diff \,}{\diff t} \|u_1(t)\|_{L^\infty} \approx 2\psi_{1,z}(R(t),Z(t),t)\cdot u_1(R(t),Z(t),t) \sim c_0 \|u_1(t)\|_{L^\infty}^2,\]
where we have neglected the viscous effect since vortex stretching dominates diffusion during this time period.
This would implies that 
$\|u_1(t)\|_{L^\infty} \sim  (T-t)^{-1} $
for some finite time $T$. This asymptotic analysis is consistent with our linear fitting results.

\begin{figure}[!ht]
\centering
    \begin{subfigure}[b]{0.40\textwidth}
    \includegraphics[width=1\textwidth]{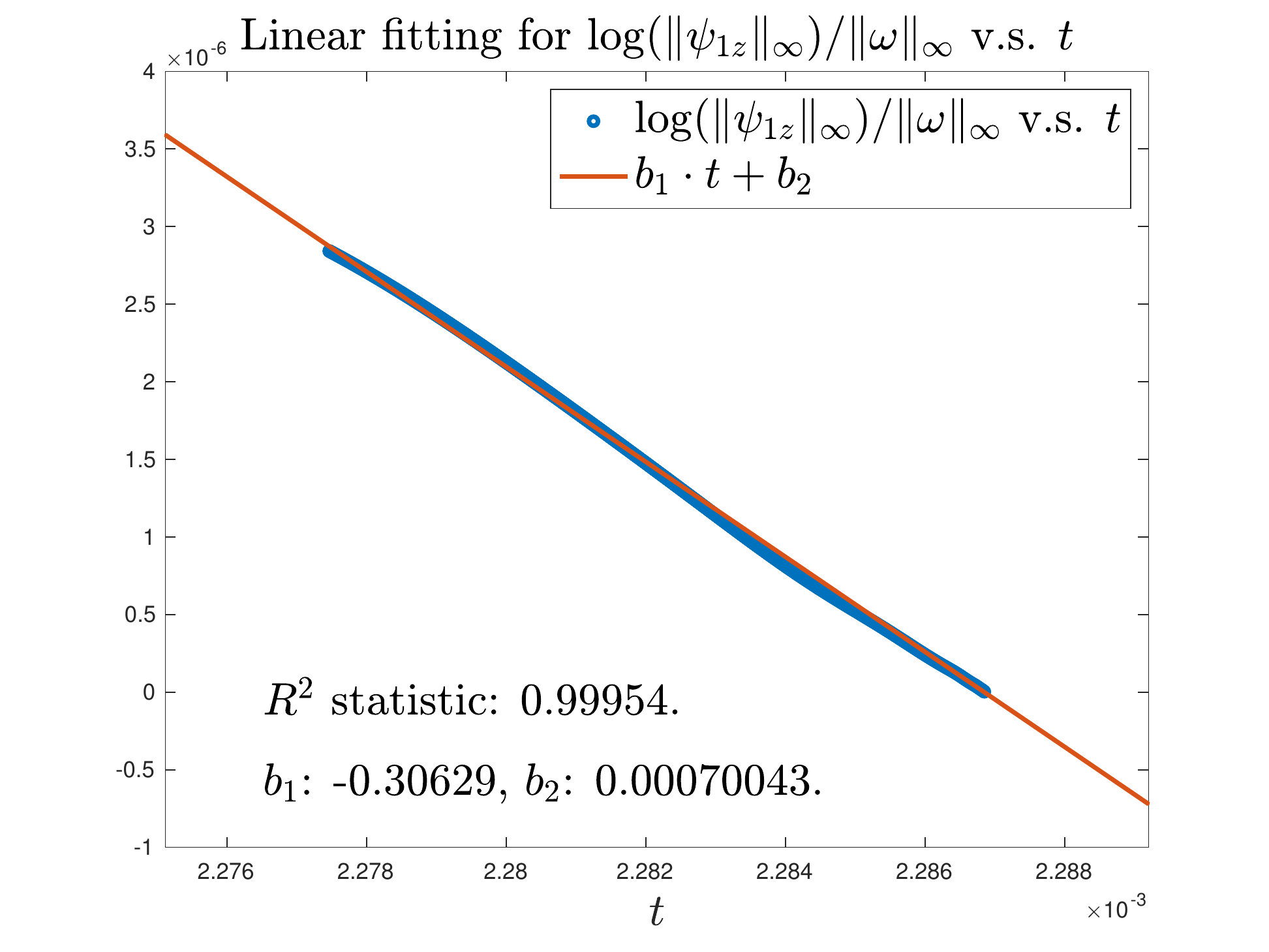}
    \caption{linear regression of $\|\omega (t)\|_{L^\infty}^{-1}$}
    \end{subfigure}
  	\begin{subfigure}[b]{0.40\textwidth} 
    \includegraphics[width=1\textwidth]{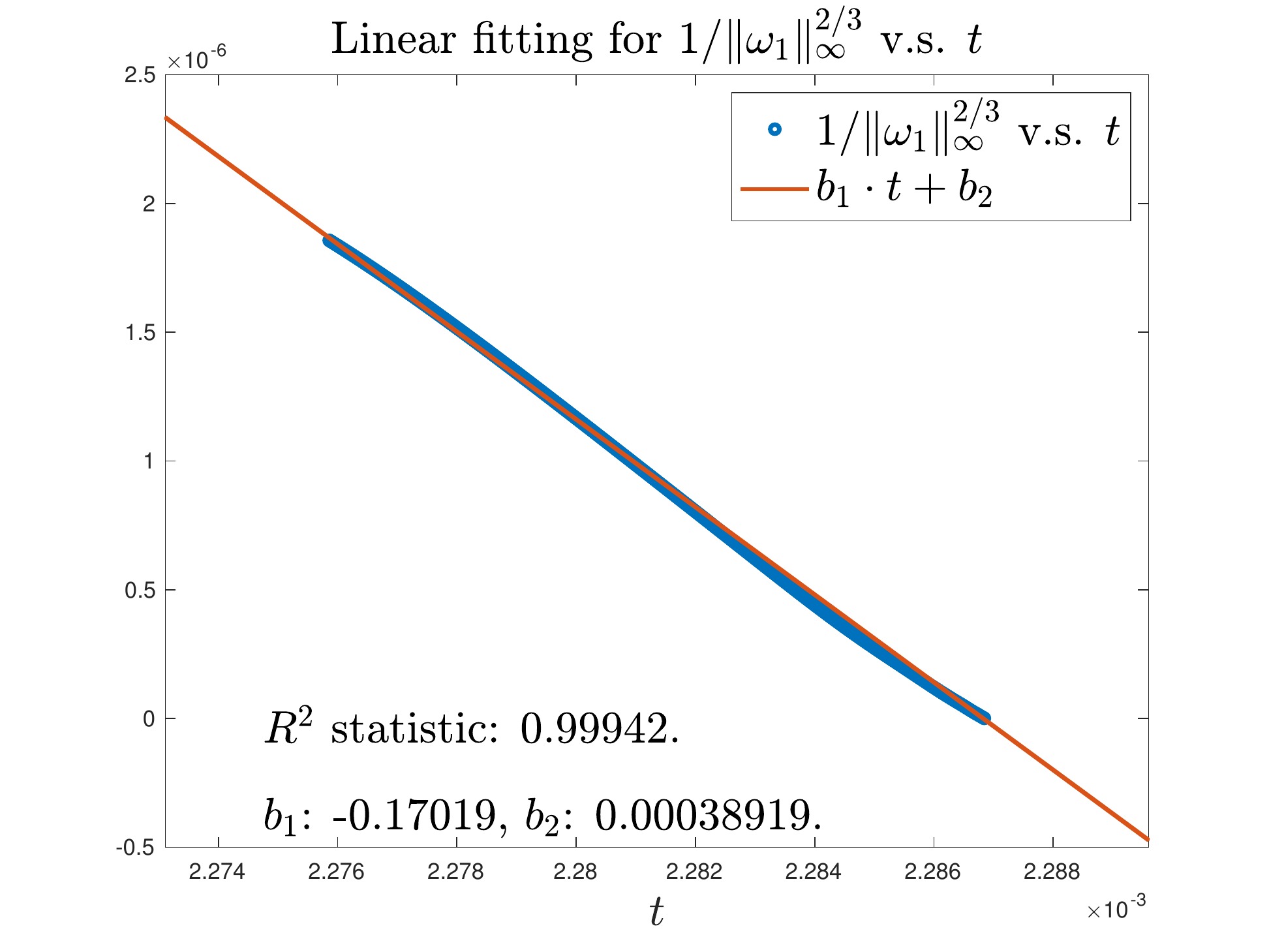}
    \caption{linear regression of $\|\omega_1(t)\|_{L^\infty}^{-2/3}$}
    \end{subfigure} 
    \caption[Linear regression $\omega$]{The linear regression of (a) $\log(\|\psi_{1z}\|_\infty)\|\omega(t)\|_{L^\infty}^{-1}$ vs $t$, (b) $\|\omega_1(t)\|_{L^\infty}^{-2/3}$ vs $t$.
    The blue points are the data points obtained from our computation, and the red lines are the linear models. The solution is computed using $1536\times 1536$ grid. The final time instant is $t_4=0.0022868502$.}   
    \label{fig:linear_regression_w1_nse}
       \vspace{-0.05in}
\end{figure}

\begin{figure}[!ht]
\centering
    \begin{subfigure}[b]{0.40\textwidth}
    \includegraphics[width=1\textwidth]{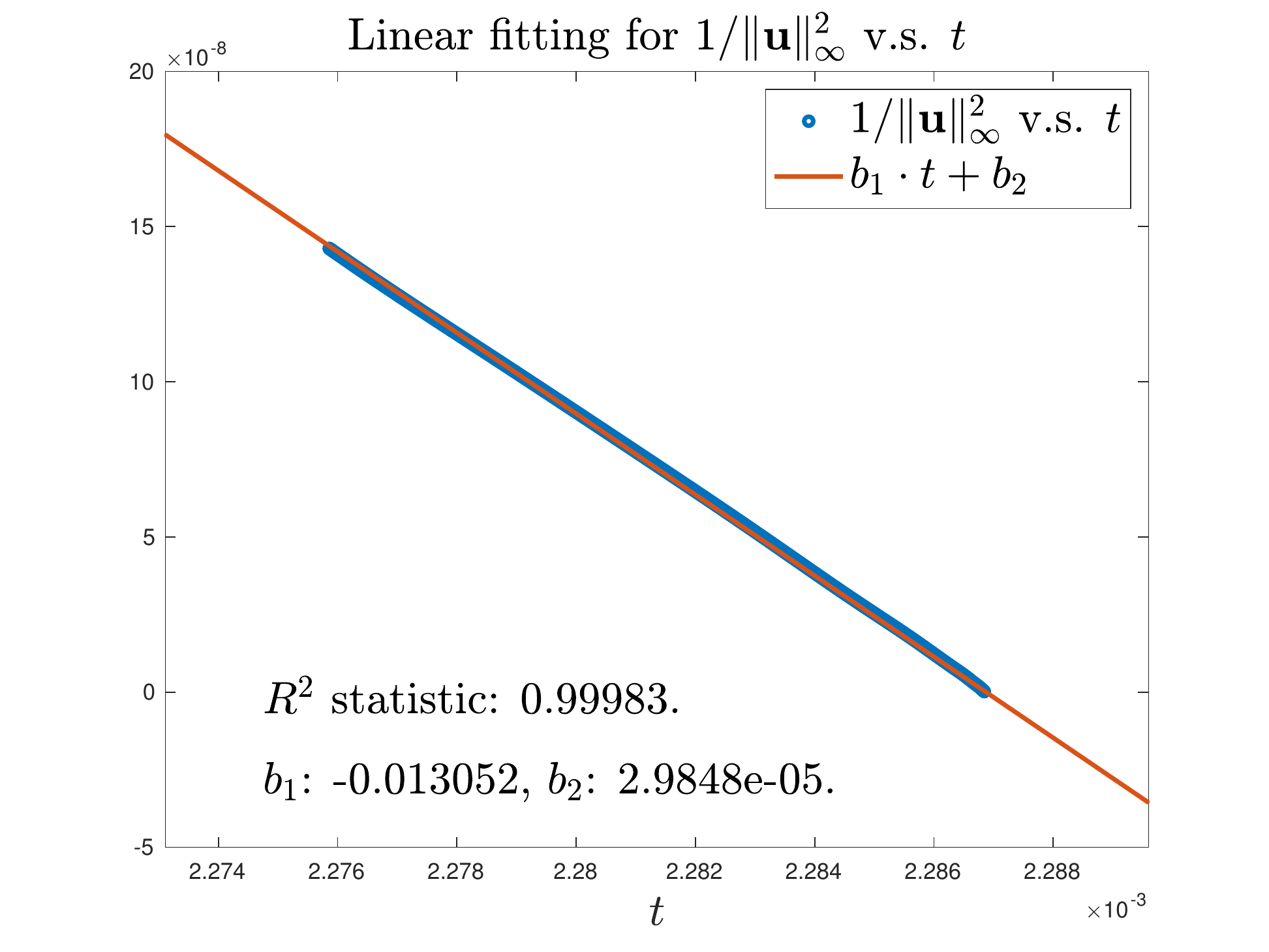}
    \caption{linear regression of $\|{\bf u} (t)\|_{L^\infty}^{-2}$}
    \end{subfigure}
  	\begin{subfigure}[b]{0.40\textwidth} 
    \includegraphics[width=1\textwidth]{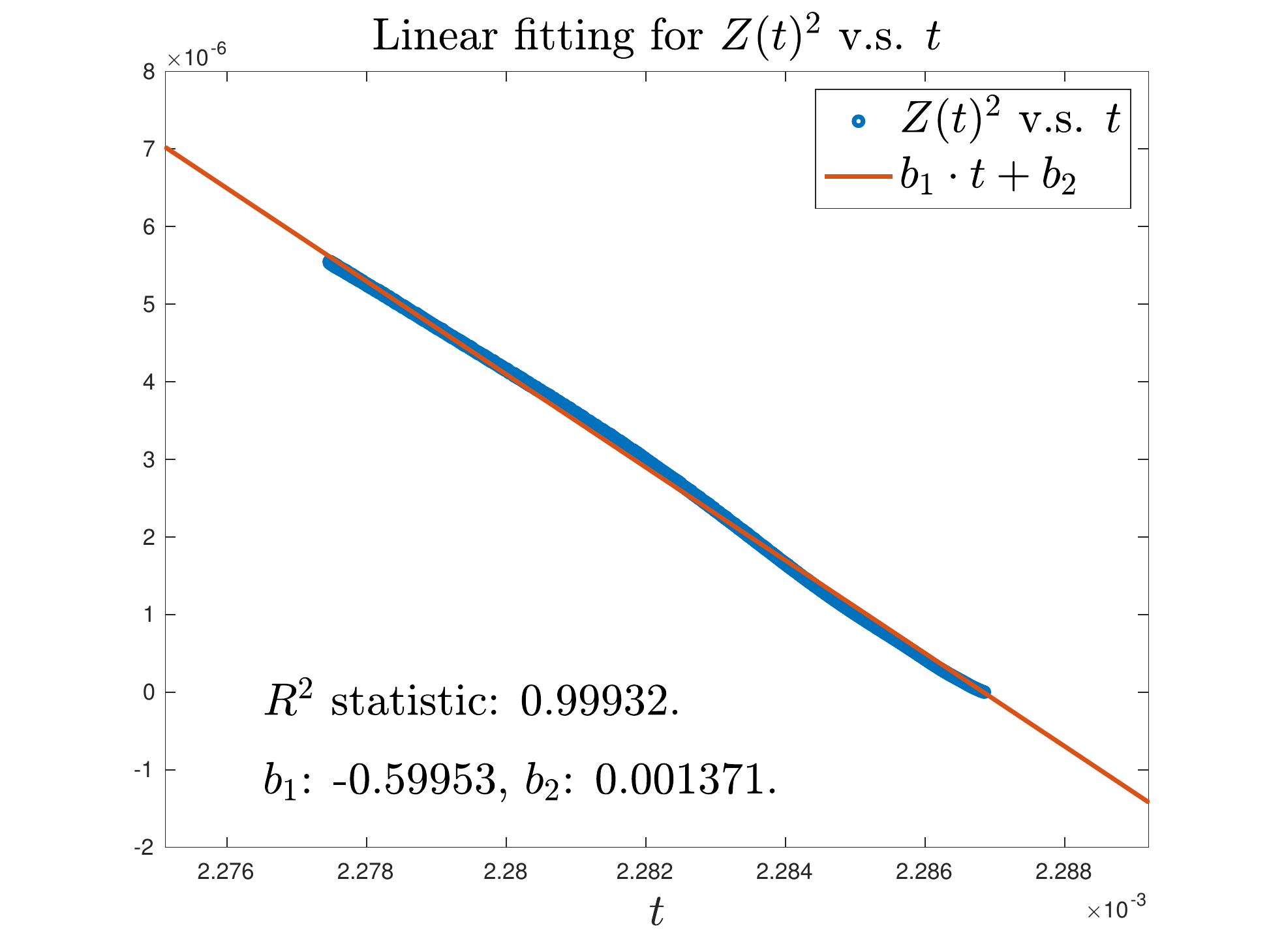}
    \caption{linear regression of $Z(t)^2$}
    \end{subfigure} 
    \caption[Linear regression ${\bf u}$]{The linear regression of (a) $\|{\bf u}(t)\|_{L^\infty}^{-2}$ vs $t$, (b) $Z(t)^2$ vs $t$.
    The blue points are the data points obtained from our computation, and the red lines are the linear models. The solution is computed using $1536\times 1536$ grid. The final time instant is $t_4=0.0022868502$.}   
    \label{fig:linear_regression_vel_nse}
       \vspace{-0.05in}
\end{figure}

Next, we study the growth of the maximum vorticity $\|\vom (t)\|_{L^\infty}$. In Figure \ref{fig:linear_regression_w1_nse} (a), we show the linear fitting of $\log(\|\psi_{1z}\|_\infty)\|\vom(t)\|_{L^\infty}^{-1}$ as a function of time on the time interval $[t_s^*,t_4] = [0.002277464739,0.0022868502]$. This $t_s^*$ is slightly larger than the $t_s$ that we use for the fitting of $u_1$ and $\psi_{1z}$ since $\vom$ enters the asymptotic self-similar regime slightly later than that for $u_1$ and $\psi_{1z}$. The end time $t_4$ is still the same as for $u_1$ and $\psi_{1z}$. We observe that $ \|\vom (t)\|_{L^\infty} = O(\log(\|\psi_{1z}\|_\infty)/(T-t))$ has good linear fitness with $R$-Square values very close to $1$. Since we observe $\|\psi_{1z}(t)\|_{L^\infty} = O(1/(T-t))$, this seems to imply that 
\[\|\vom(t)\|_{L^\infty} \sim \frac{|\log(T-t)|}{(T-t)}.\] 
We remark that this linear fitting of the asymptotic growth rate is qualitative in nature. Such asymptotic growth rate is consistent with the rapid dynamic growth of $\int_0^t \| \vom (s) \|_{L^\infty} ds $ presented in Figure \ref{fig:rapid_growth2_nse}(b). The Beale-Kato-Majda blowup criterion \cite{beale1984remarks} would then imply that the $3$D Navier--Stokes equations develop a finite time singularity.
In Figure \ref{fig:linear_regression_w1_nse} (b), we also plot the linear fitting of $\|\omega_1(t)\|_{L^\infty}^{-2/3}$ as a function of time on the same time interval for $u_1$ and $\psi_{1z}$. We see that 
 $\|\omega_1(t)\|_{L^\infty}  \sim \frac{1}{(T-t)^{3/2}} $
has good linear fitness with $R$-Square values very close to $1$.

To further illustrate the potentially singular behavior of the Navier--Stokes equations, we perform linear fitting for the maximum velocity and $Z(t)$. For the fitting of maximum velocity, the fitting time interval is the same as that for $u_1$ and $\psi_{1z}$. For $Z(t)$, we use the same time interval as that for $\vom$, i.e. $[t_s^*,t_4] = [0.002277464739,0.0022868502]$. In Figure \ref{fig:linear_regression_vel_nse} (a), we observe that $ \|{\bf u}(t)\|_{L^\infty}^{-2} = O((T-t))$ has good linear fitness with $R$-Square values very close to $1$, which seems to imply that $\|{\bf u} (t)\|_{L^\infty}$ has the form of an inverse power law
\[
\|{\bf u}(t)\|_{L^\infty} \sim \frac{1}{(T-t)^{1/2}}.
\] 

The scaling properties of the maximum vorticity and maximum velocity seem to suggest that the small scale of the solution, which is characterized by $Z(t)$, should have the form: $Z(t) \sim (T-t)^{1/2}$. Indeed, in Figure \ref{fig:linear_regression_vel_nse}(b), we observe that 
\[
Z(t) \sim (T-t)^{1/2}\;,
\]
has good linear fitness. 

\begin{figure}[!ht]
\centering
    \begin{subfigure}[b]{0.38\textwidth}
    \includegraphics[width=1\textwidth]{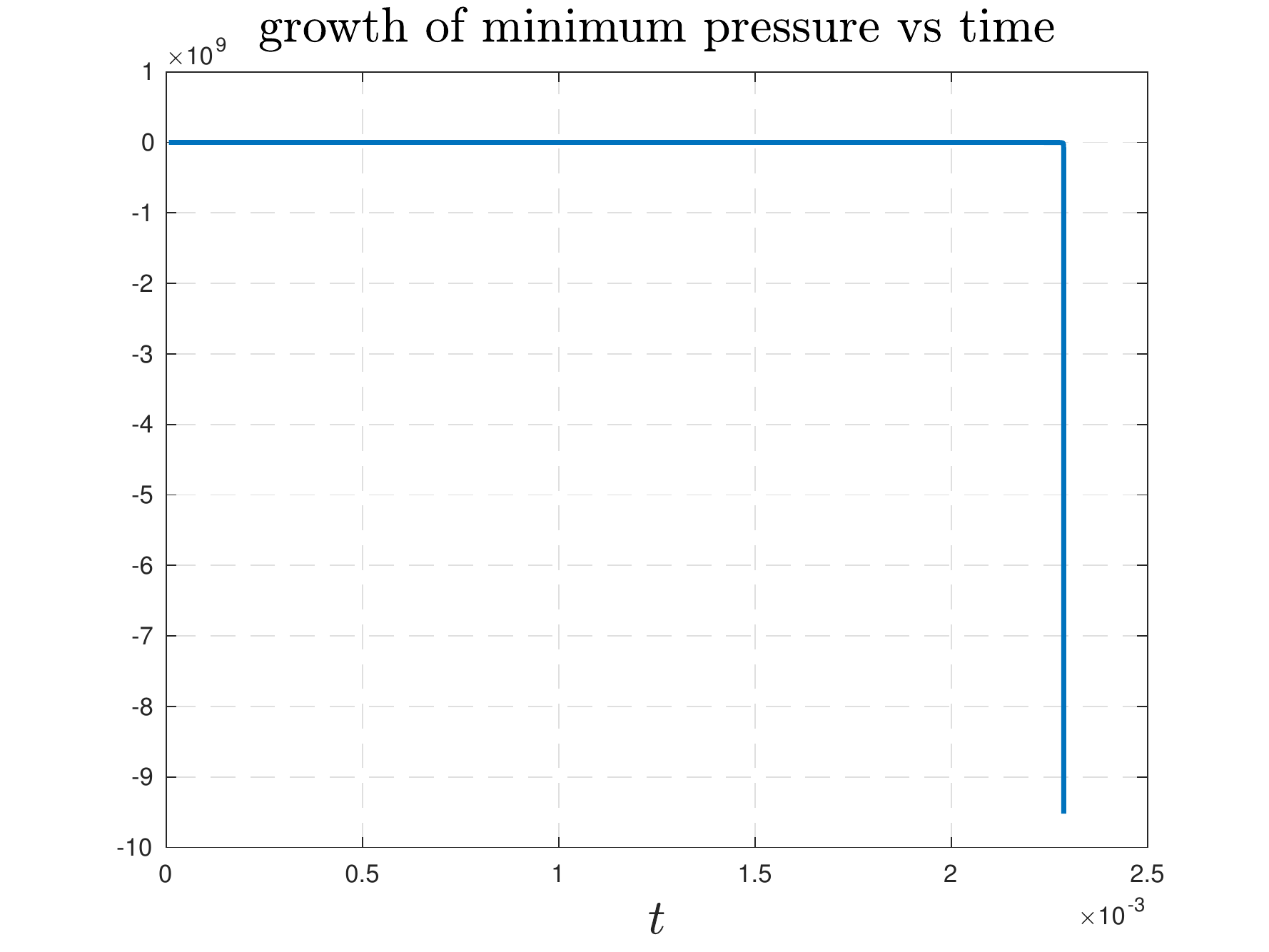}
    \caption{Minimum pressure vs time}
    \end{subfigure}
  	\begin{subfigure}[b]{0.38\textwidth} 
    \includegraphics[width=1\textwidth]
    {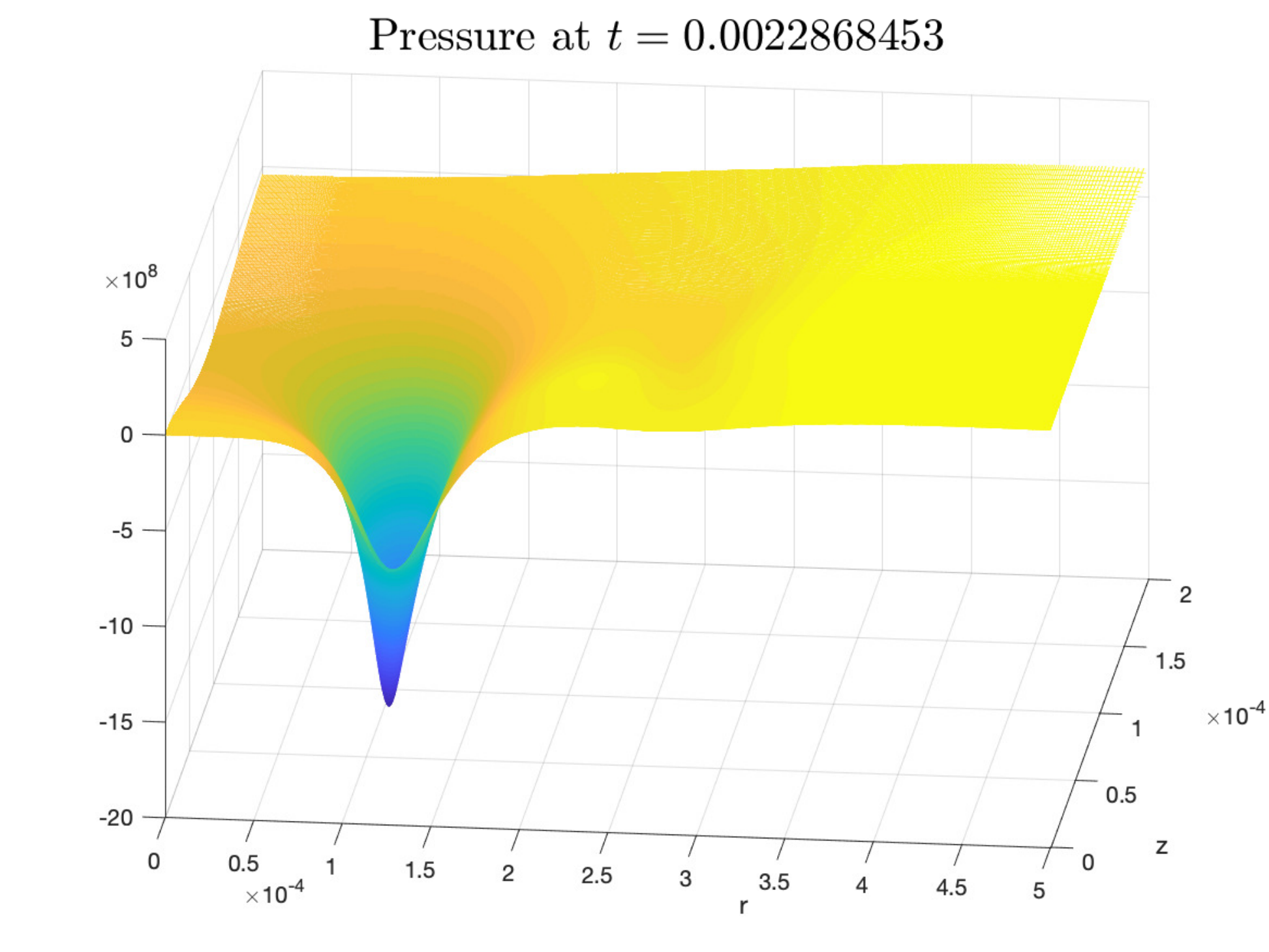}
    \caption{$3$D plot of pressure}
    \end{subfigure} 
    \caption[$3$D prerssure plot]{(a) Minimum pressure as a function of time. The solution is computed using $1536\times 1536$ grid. The final time instant is $t_4=0.0022868502$. (b) $3$D plot of pressure at $t_3=0.002286845$.}   
    \label{fig:pressure_3dpre_nse}
       \vspace{-0.05in}
\end{figure}

Another important indicator for a potential singular behavior of the $3$D Navier--Stokes equations is the growth of negative pressure, i.e. the pressure should approach to negative infinity at the singularity time, see \cite{sverak2002}. In Figure \ref{fig:pressure_3dpre_nse}(a), we plot the minimum pressure as a function of time. We observe that the minimum pressure grows rapidly in time and approaches $-10^{10}$ by $t_4=0.0022868502$. To give a better idea how the potentially singular pressure profile looks like, we plot the $3$D pressure profile at $t_3=0.0022868453$ in Figure \ref{fig:pressure_3dpre_nse}(b). By this time, the maximum vorticity has increased by a factor of $10^6$. We can see that the pressure profile is still very well resolved by our adaptive mesh. The pressure seems to develop a potential focusing singularity.

\begin{figure}[!ht]
\centering
    \begin{subfigure}[b]{0.40\textwidth}
    \includegraphics[width=1\textwidth]{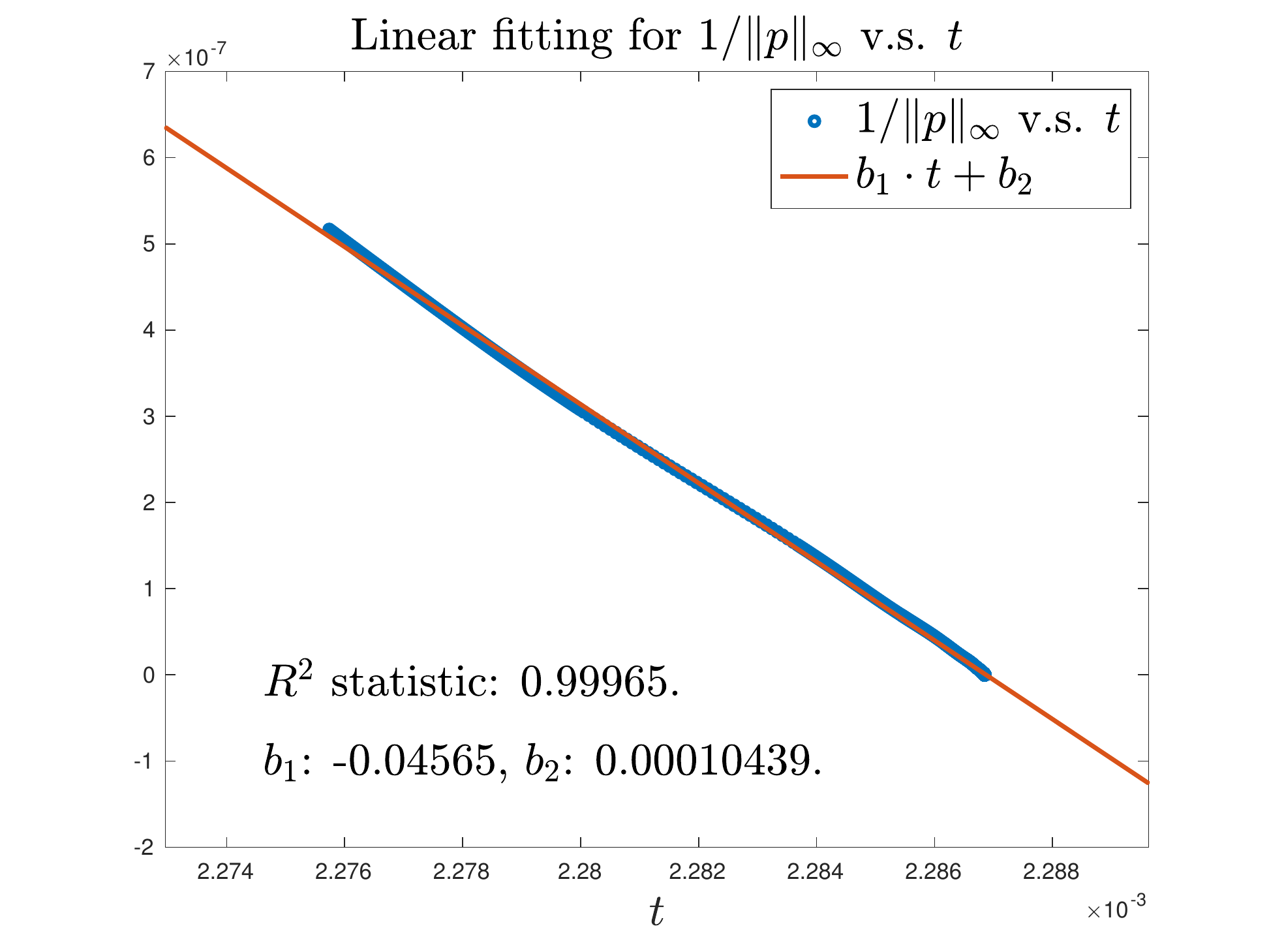}
    \caption{linear regression of $\| -p(t)\|_{L^\infty}^{-1}$}
    \end{subfigure}
  	\begin{subfigure}[b]{0.40\textwidth} 
    \includegraphics[width=1\textwidth]{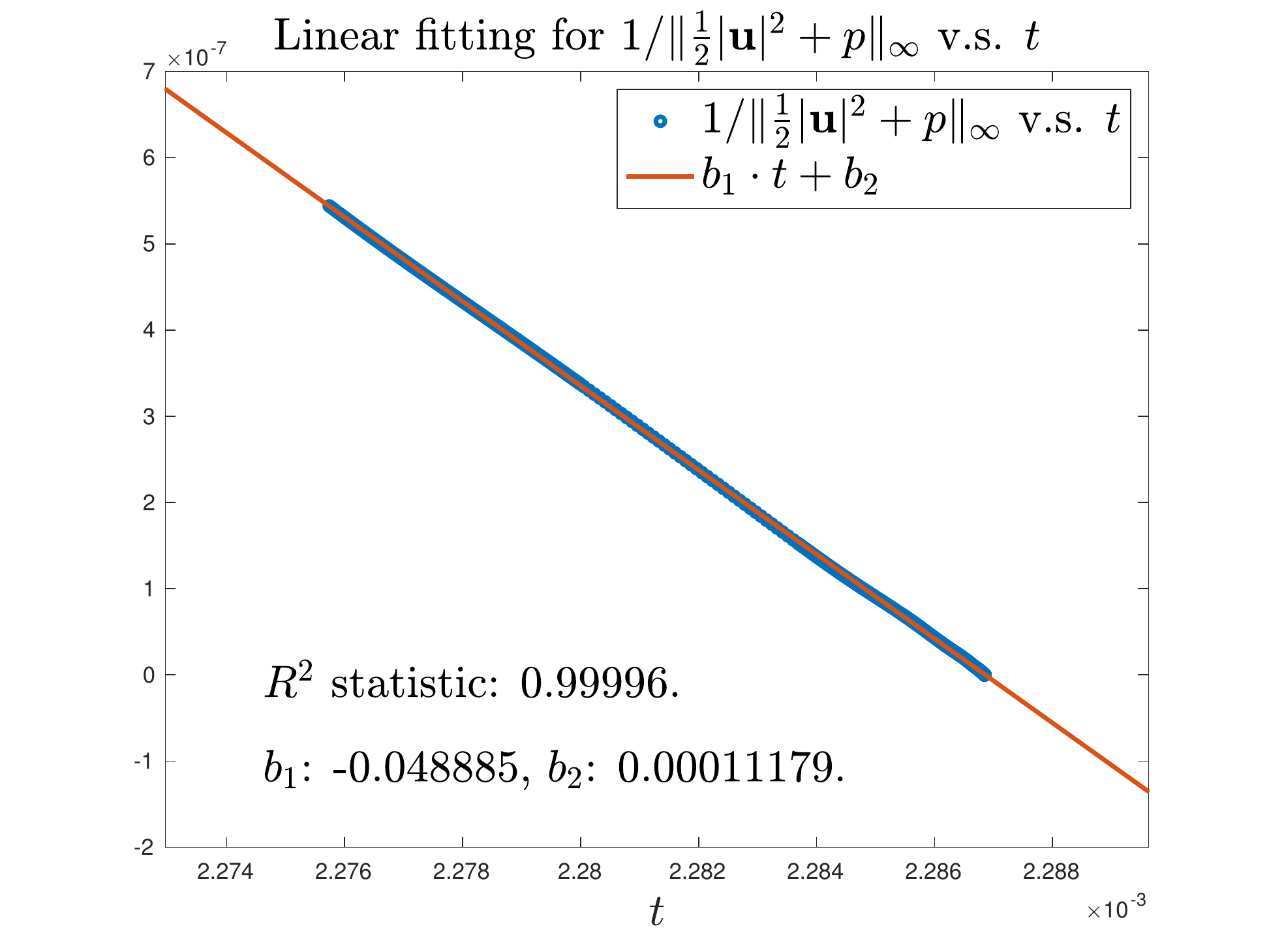}
    \caption{linear regression of $\|\frac{1}{2} |{\bf u}|^2 + p\|_\infty^{-1}$}
    \end{subfigure} 
    \caption[Linear regression ${\bf u}$]{The linear regression of (a) $\|- p(t)\|_{L^\infty}^{-1}$ vs $t$, (b) $\|\frac{1}{2} |{\bf u}|^2 + p\|_\infty^{-1}$ vs $t$.
    The blue points are the data points obtained from our computation, and the red lines are the linear models. The solution is computed using $1536\times 1536$ grid. The final time instant is $t_4=0.0022868502$.}   
    \label{fig:linear_regression_pre_nse}
       \vspace{-0.05in}
\end{figure}

In Figure \ref{fig:linear_regression_pre_nse} (a), we plot the linear fitting of the inverse of the maximum norm of negative pressure as a function of time. We obtain excellent linear fitting 
\[
\| -p(t) \|_{L^\infty}  \sim \frac{1}{(T-t)}\;,
\]
with $R$-Square values very close to $1$. We also perform linear fitting for another related quantity: $\|\frac{1}{2} |{\bf u}|^2 + p\|_{L^\infty}^{-1}$ as a function of time. 
We also obtain excellent linear fitting 
\[
\|\frac{1}{2} |{\bf u}|^2 + p\|_\infty \sim \frac{1}{(T-t)}\;,
\]
with $R$-Square values very close to $1$, see Figure \ref{fig:linear_regression_pre_nse} (b).

\noindent
{\bf Remark.} We would like to emphasize that the linear fitting results that we presented in this subsection are qualitative in nature. They shed light on the scaling properties of the potentially singular solution. Our current adaptive mesh strategy does not offer sufficient resolution in the late stage to enable us to obtain an accurate fitting for the potential singularity time $T$, the blow-up rate and the constant. Thus the qualitative fitting of $\|{\bf u}(t)\|_{L^\infty} \sim (T-t)^{-1/2} $ does not imply that we can obtain an upper bound on the growth rate of $\|{\bf u}(t)\|_{L^\infty}$ that is uniformly valid up to the singularity time $T$, i.e.
\begin{equation}
\label{MaxVelbound}
\| {\bf u } (t) \|_{L^\infty} \leq \frac{c}{(T - t)^{1/2}}, \quad \mbox{for}\; t < T,
\end{equation}
for some constant $c$. Such inequality would be almost impossible to verify numerically since it requires the exact value of $T$. If the numerically fitted value of $T$ is slightly larger than the exact value of $T$, the constant $c$ will be infinite. Moreover, the potentially singular solution of the Navier--Stokes equations cannot be asymptotically self-similar \cite{sverak1996,tsai1998}. Thus, the potential blow-up rate is most likely not exactly equal to $1/2$, which makes it extremely difficult to obtain an approximation of the potential blow-up time $T$ with high accuracy. Based on the reasons stated above, it is almost impossible to verify numerically the non-blowup conditions based on the lower bound of the blow-up rate for the axisymmetric Navier-Stokes equations \cite{chen2008lower,chen2009lower,sverak2009} (one of the conditions is given by \eqref{MaxVelbound}). In this paper, we will mainly use those blow-up criteria that do not involve the numerical fitting of an asymptotic blowup rate in our study of the potentially singular behavior of the Navier--Stokes equations. More precisely, we will mainly use the dynamic growth of $\int_0^t \|\vom(s)\|_{L^\infty} ds$, $\int_0^t \|{\bf u}(s)\|_{L^\infty}^2 ds$, $\int_0^t \| p(s)\|_{L^\infty} ds$ and the $L_t^q L_x^p$ norms of the velocity  to study the potentially singular behavior of the Navier--Stokes equations.

\vspace{-0.05in}
\subsubsection{Numerical evidence of nearly self-similar profiles}\label{sec:evidence_of_self-similar_nse}

To study the nearly self-similar scaling properties of the $3$D Navier--Stokes equations, we study the solution in the dynamically rescaled variables $(\xi, \zeta)$ as follows: 
\begin{equation}\label{eq:stretch_u1_nse}
u_1(t,r,z) = \max(u_1) \overline{U}(t, \xi,\zeta), \quad
\omega_1(t,r,z) = \max(\omega_1) \overline{\Omega}(t, \xi,\zeta)\;,
\end{equation}
where 
\[
\xi = \frac{r - R(t)}{Z(t)} ,\quad \zeta = \frac{z}{Z(t)} \;,
\]
are the dynamically rescaled variables.

\begin{figure}[!ht]
\centering
    \includegraphics[width=0.9\textwidth]{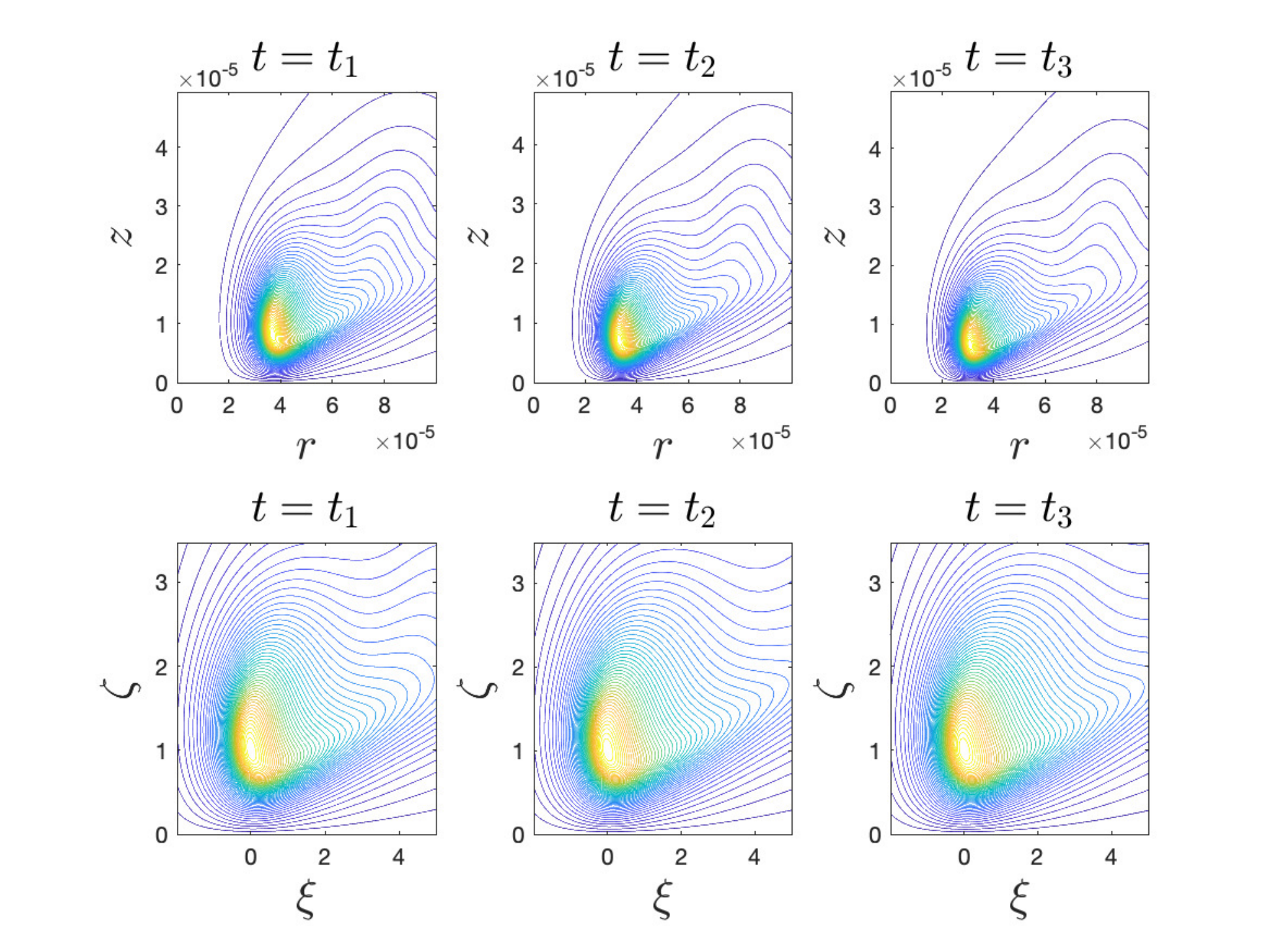}
    \caption{Comparison of the level sets of $u_1$ at different time instants. First row: original level sets of $u_1$ in the domain $(r,z)$ in different times. Second row: rescaled level sets of $u_1$ as a function of $(\xi,\zeta)$ in the domain $(\xi,\zeta)$. The computation is performed using the $1536\times1536$ grid. Here $t_1^*=0.002286850057$, 
    $t_2^*=0.002286850158$ and $t_3^*=0.002286850216$.} 
    \label{fig:levelset_compare_u1_nse} 
\end{figure}

\begin{figure}[!ht]
\centering
    \includegraphics[width=0.9\textwidth]{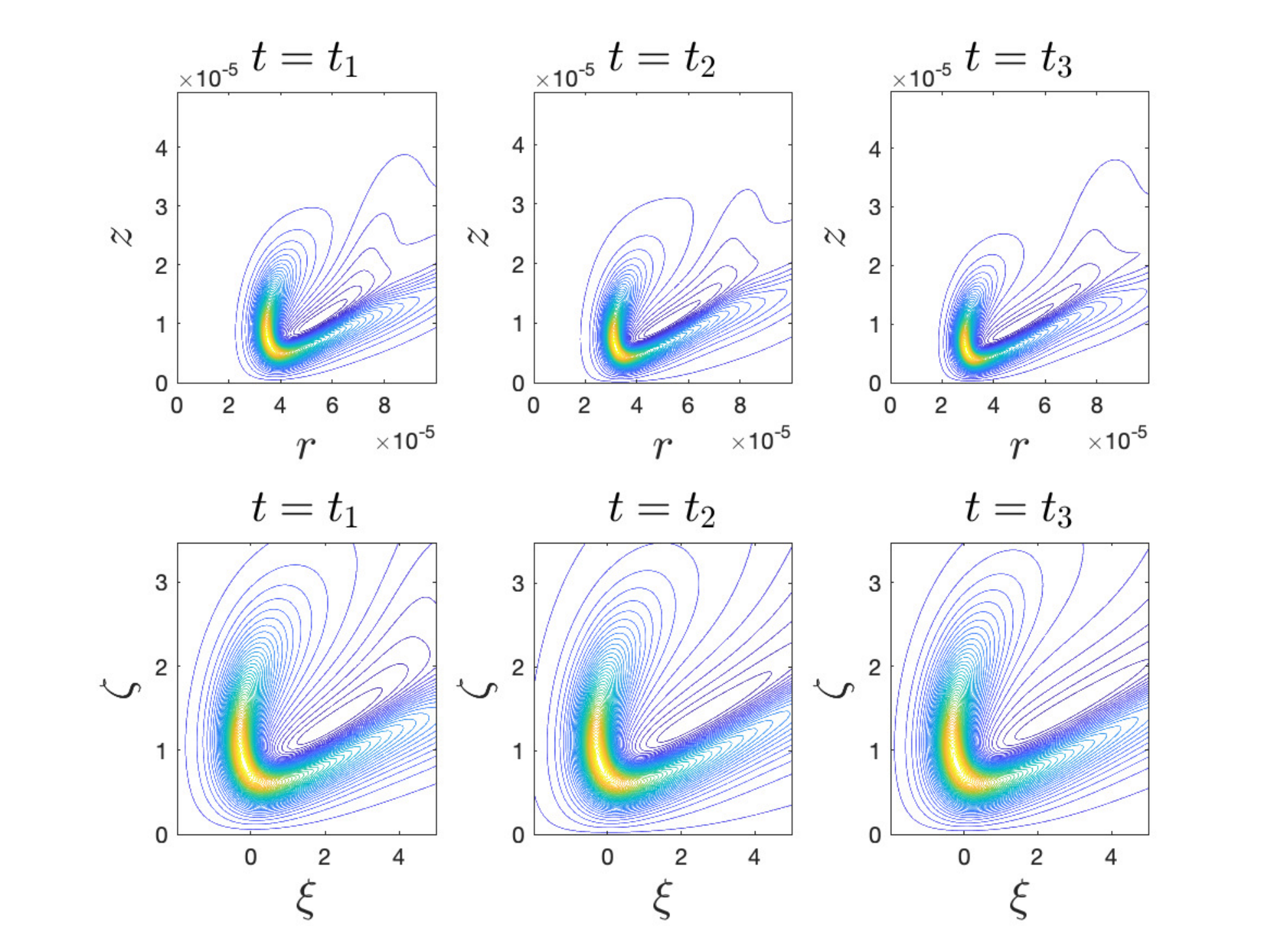}
    \caption{Comparison of the level sets of $\omega_1$ at different time instants. First row: original level sets of $u_1$ in the domain $(r,z)$ in different times. Second row: rescaled level sets of $\omega_1$ as a function of $(\xi,\zeta)$ in the domain $(\xi,\zeta)$.  The computation is performed using the $1536\times1536$ grid. Here $t_1^*=0.002286850057$, 
    $t_2^*=0.002286850158$ and $t_3^*=0.002286850216$.} 
    \label{fig:levelset_compare_w1_nse} 
\end{figure} 

In Figure \ref{fig:levelset_compare_u1_nse}, we compare the level sets of $u_1$ at different time instants. In the first row of Figure \ref{fig:levelset_compare_u1_nse}, we plot the level sets of $u_1$ in a local domain. We plot the profiles in a short time interval at three different times with $t_1^*=0.002286850057$, 
    $t_2^*=0.002286850158$ and $t_3^*=0.002286850216$. By the time $t=t_3^*$, $\|\vom(t_3)\|_{L^\infty}/\|\vom(0)\|_{L^\infty}$ has grown by a factor of $10^7$. As we can see, the singular support of the profile shrinks in space and travels toward the origin. Compared with the solution of the $3$D Euler equations, the sharp front has been regularized by the relatively large viscosity. In the second row of Figure \ref{fig:levelset_compare_u1_nse}, we can see that the rescaled profile $\overline{U}$ (in the $\xi\zeta$-plane) is almost static during this time interval. This observation suggests that there exists an approximate self-similar profile $U(\xi,\zeta)$ locally.

In Figure \ref{fig:levelset_compare_w1_nse}, we compare the level sets of $\om_1$ and the level sets of the spatially rescaled function $\overline{\Omega}$
in a similar manner. Again, we can see that although the profile of $\om_1$ has a noticeable change during this time interval, there seems to exist an approximate self-similar profile $\Omega(\xi, \zeta)$ locally.

\begin{figure}[!ht]
\centering
	\begin{subfigure}[b]{0.36\textwidth}
    \includegraphics[width=1\textwidth]{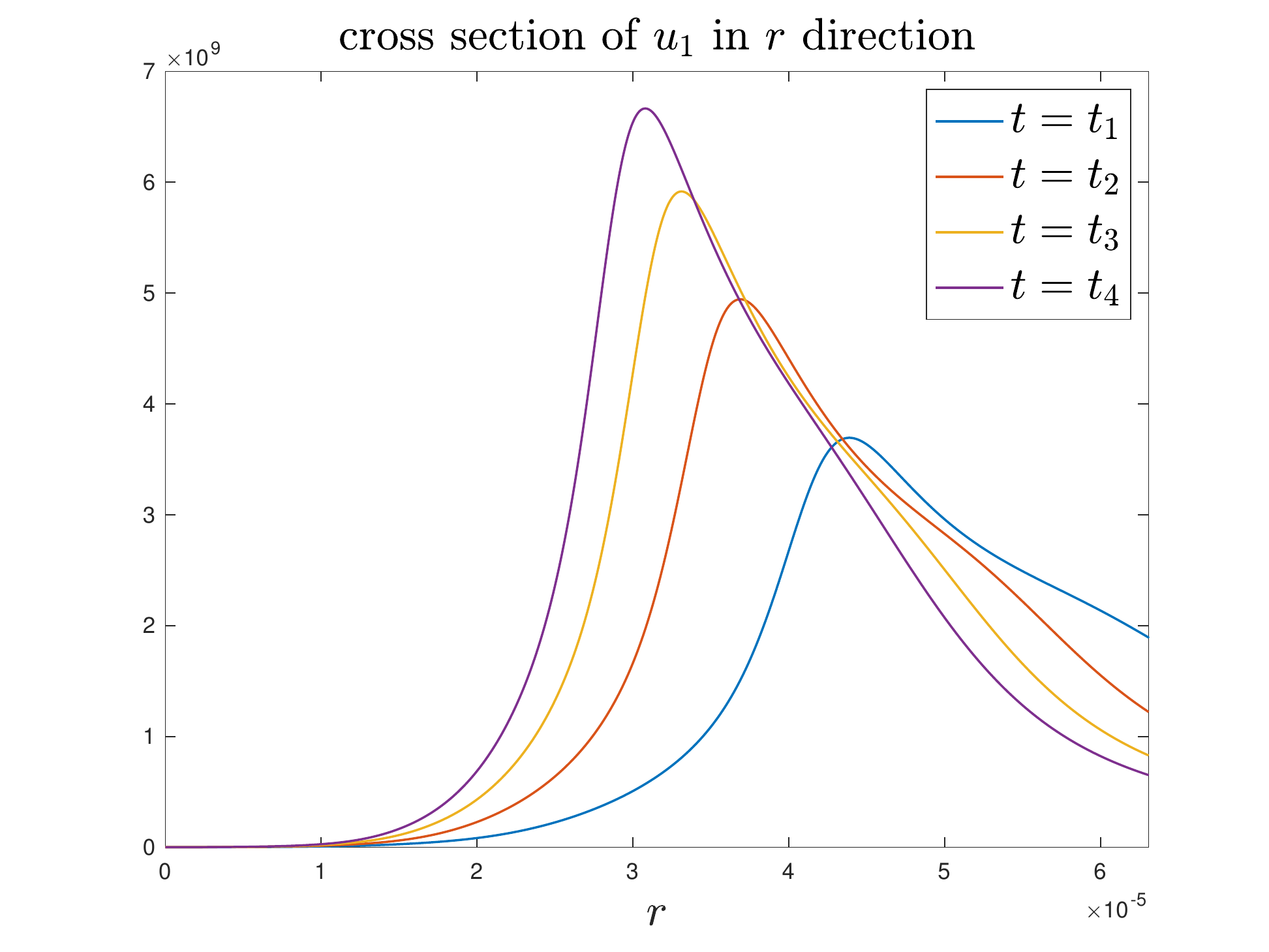}
    \caption{cross sections of $u_1$ in $r$}
    \end{subfigure}
  	\begin{subfigure}[b]{0.36\textwidth} 
    \includegraphics[width=1\textwidth]{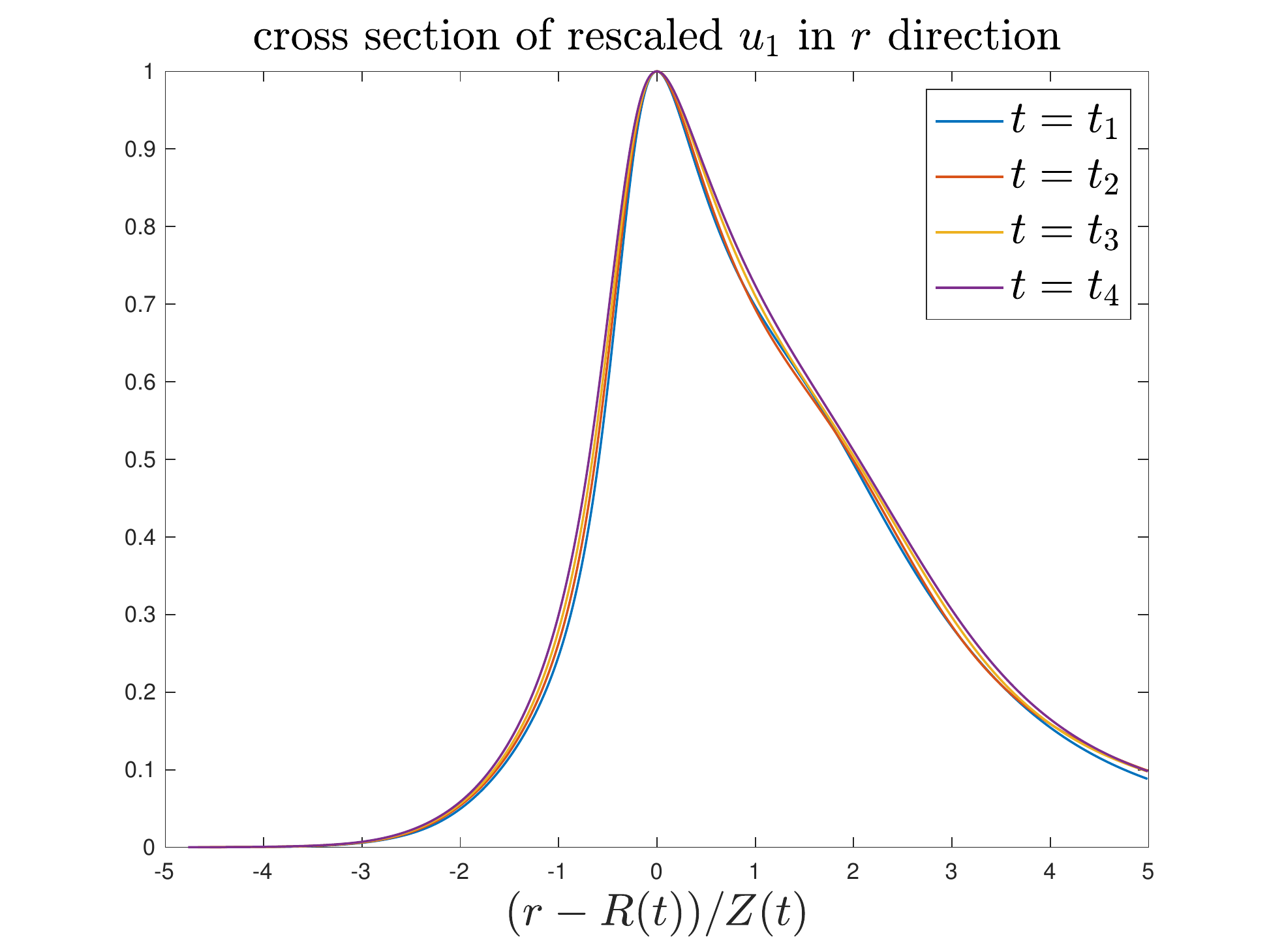}
    \caption{rescaled cross sections of $u_1$ in $r$}
    \end{subfigure} 
    \begin{subfigure}[b]{0.36\textwidth} 
    \includegraphics[width=1\textwidth]{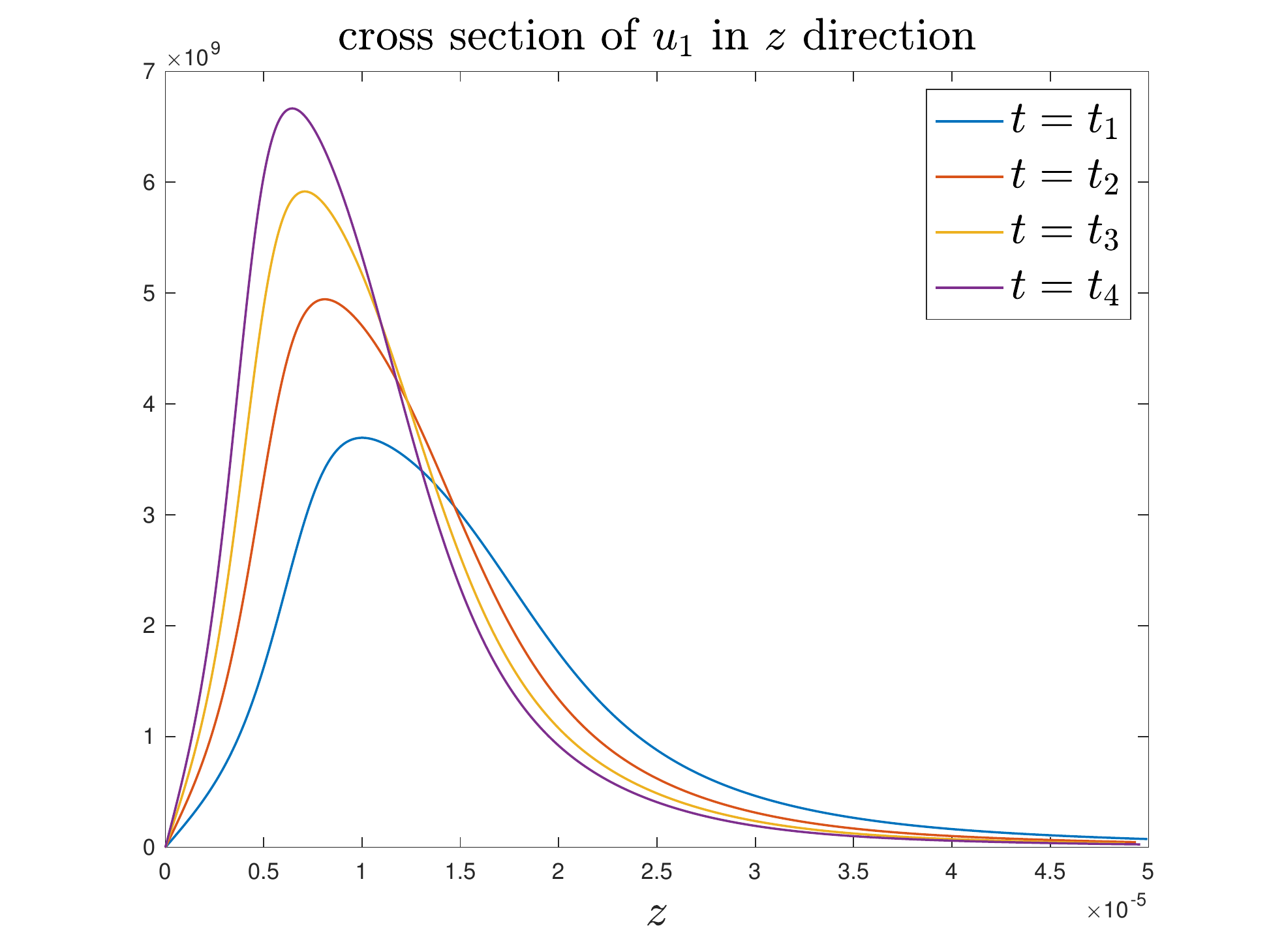}
    \caption{cross sections of $u_1$ in $z$}
    \end{subfigure}
  	\begin{subfigure}[b]{0.36\textwidth} 
    \includegraphics[width=1\textwidth]{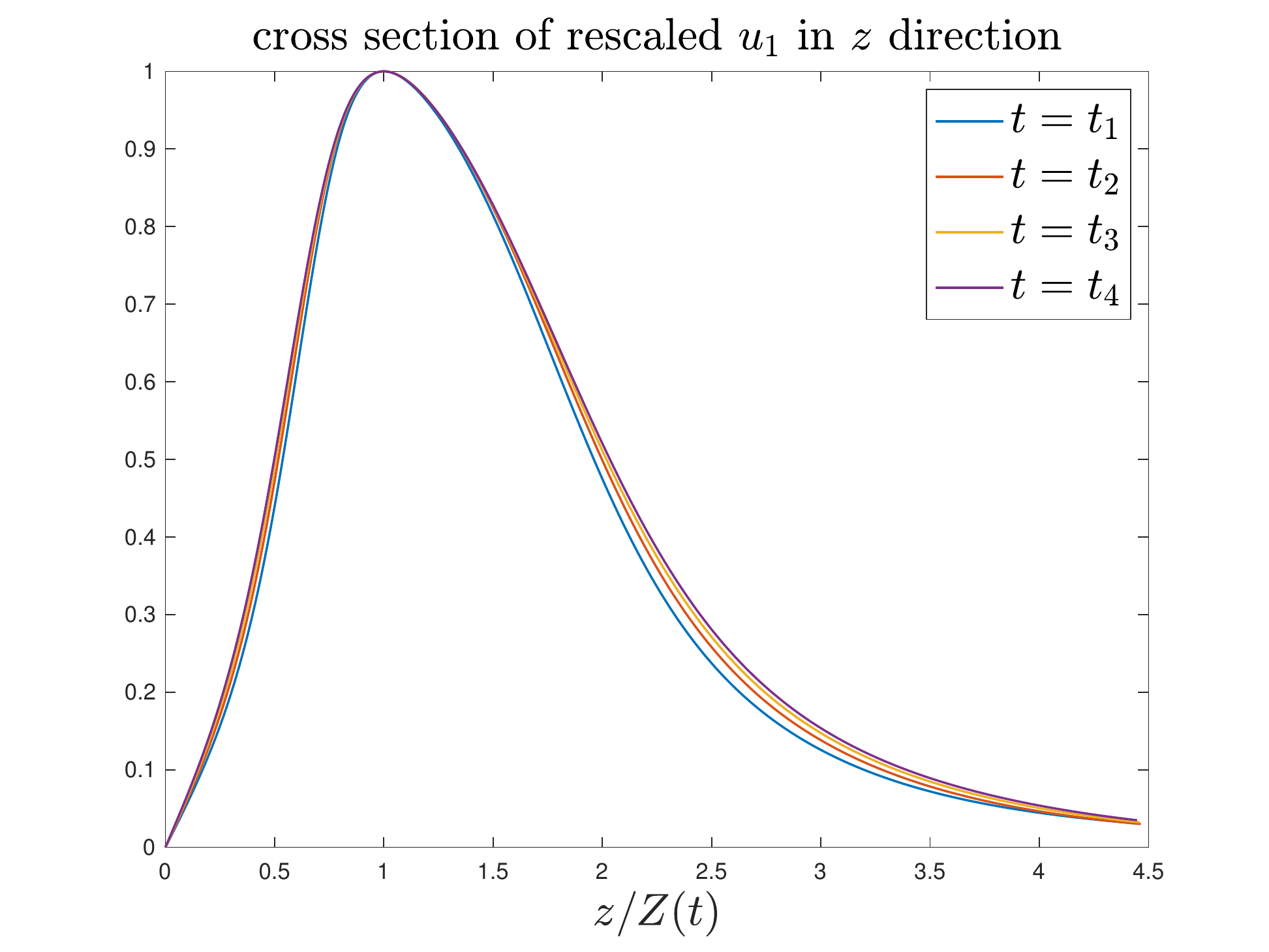}
    \caption{rescaled cross sections of $u_1$ in $z$}
    \end{subfigure} 
    \caption[Cross section compare]{Cross sections and rescaled cross sections of $u_1$ through the point $R(t),Z(t)$ in both directions at different time instants instants. (a) Cross sections in the $r$ direction. (b) Rescaled cross sections in the $r$ directions. (c) Cross sections in the $z$ direction. (d) Rescaled cross sections in the $z$ directions.
    The computation is performed using the $1536\times1536$ grid. Here $\tilde{t}_1^*=0.002286849845$, $\tilde{t}_2^*=0.002286850057$, 
    $\tilde{t}_3^*=0.002286850158$ and $\tilde{t}_4^*=0.002286850216$.}   
    \label{fig:cross_section_compare_nse}
       \vspace{-0.05in}
\end{figure}

Next, we compare the cross sections of the solution at different time instants to study the nearly self-similar scaling properties. In Figure \ref{fig:cross_section_compare_nse}(a) and (c), we present the evolution of the cross sections of $u_1$ through the point $(R(t),Z(t))$ in both directions. The length scale of the profile shrinks in both directions, and the sharp front along the $r$-direction travels toward $r=0$. For comparison, Figure \ref{fig:cross_section_compare_nse}(b) and (d) plot the corresponding cross sections of the rescaled profile $\overline{U}$ in terms of the dynamically rescaled variables $(\xi, \zeta)$. We can see that the rescaled profiles seem to be very close to a limiting profile. These results support the existence of an approximate self-similar profile locally.

\subsection{Further evidence of potentially singular behavior of the Navier--Stokes equations}

In this section, we will provide further numerical evidence on the potentially  singular behavior of the Navier--Stokes equations by using several local and global non-blowup criteria. 

\subsubsection{Non-blowup criteria based on enstrophy growth}
\begin{figure}[!ht]
\centering
    \begin{subfigure}[b]{0.38\textwidth}
     \includegraphics[width=1\textwidth]{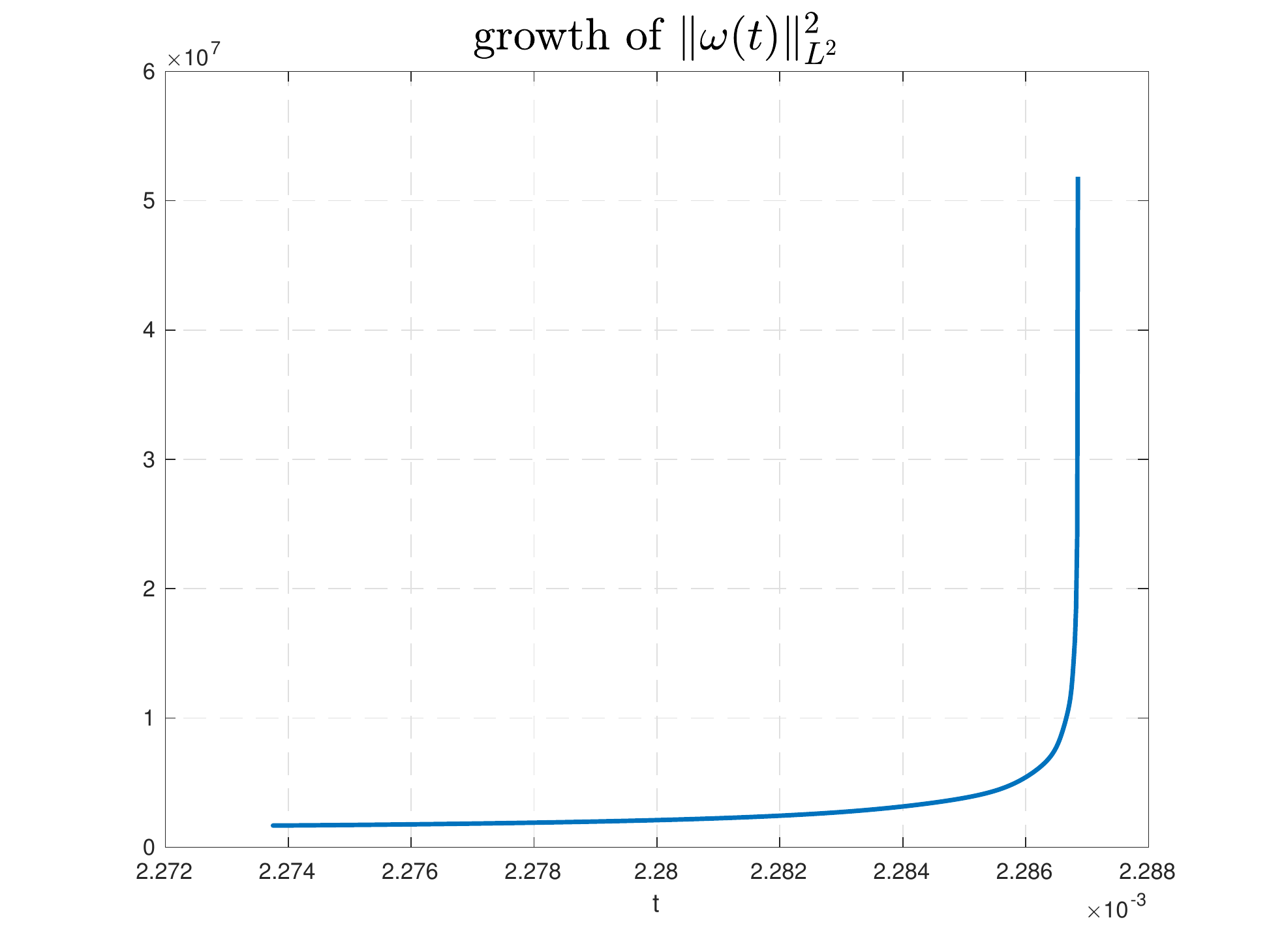}
    \caption{growth of $\|\vom (t)\|_{L^2}^2$ vs $t$}
    \end{subfigure}
  	\begin{subfigure}[b]{0.38\textwidth} 
    \includegraphics[width=1\textwidth]
    {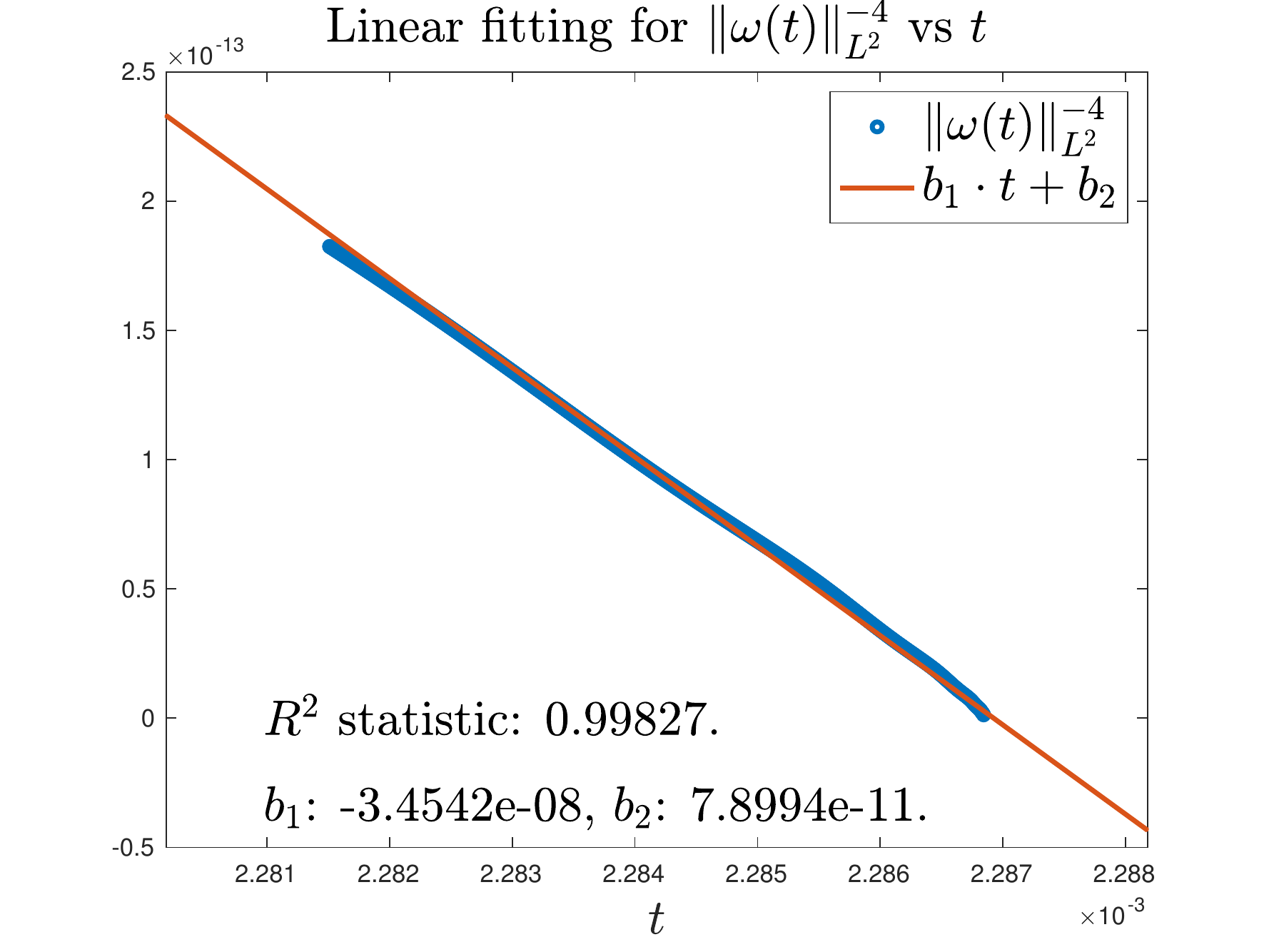}
    \caption{$\|\vom (t)\|_{L^2}^{-4}$ vs $t$}  
    \end{subfigure} 
    \caption[Enstrophy growth]{(a) growth of $\|\vom (t)\|_{L^2}^2$ vs $t$. The solution is computed using $1536\times 1536$ grid. The final time instant is $t_4=0.0022868502$. (b) Linear fitting of $\|\vom (t)\|_{L^2}^{-4}$ vs $t$, which implies $\|\vom (t)\|_{L^2}^{2} \sim (T-t)^{-1/2}$. }   
    \label{fig:enstrophy_growth_nse}
       \vspace{-0.05in}
\end{figure}

\begin{figure}[!ht]
\centering
  	\begin{subfigure}[b]{0.38\textwidth} 
    \includegraphics[width=1\textwidth]{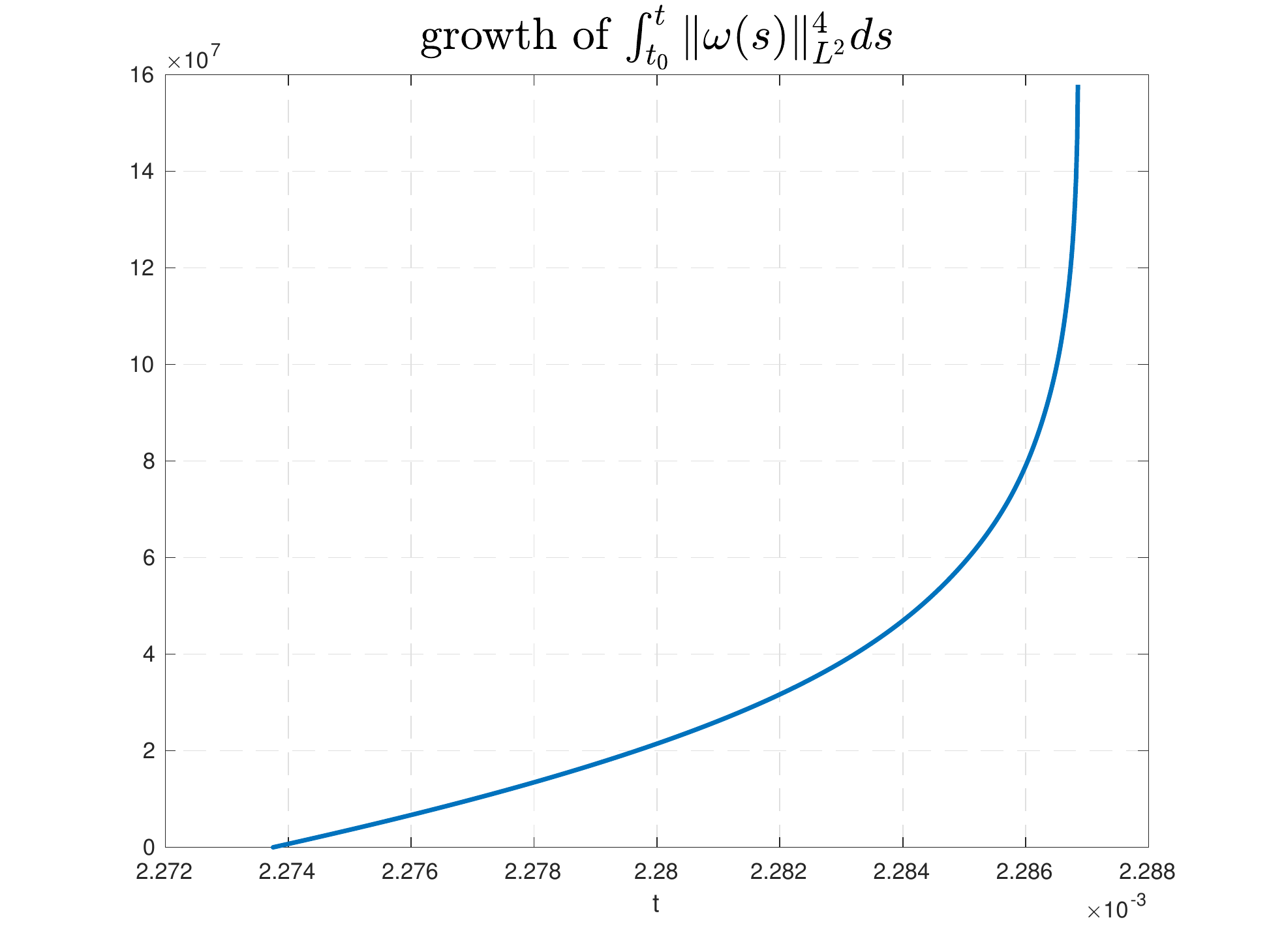}
    \caption{$\int_{t_0}^t\|\omega (s)\|_{L^2}^4 ds$ vs $t$}
    \end{subfigure} 
    \begin{subfigure}[b]{0.38\textwidth}
        \includegraphics[width=1\textwidth]{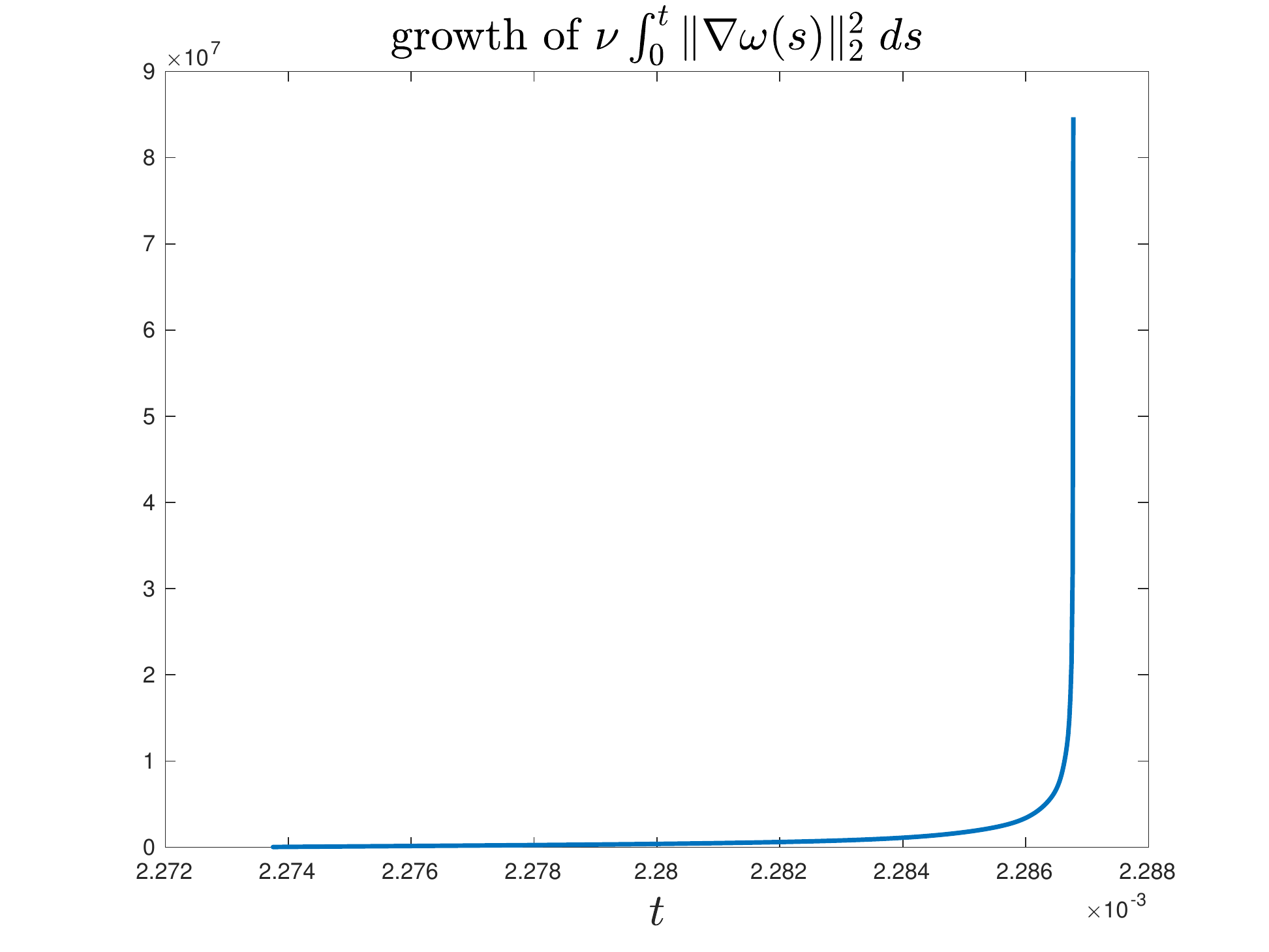}
    \caption{growth of $\nu \int_0^t \| \nabla \vom (s) \|_{L^2}^2 ds$ vs $t$}
    \end{subfigure} 
    \caption[Vorticity $L^2$ norm]{(a) $\int_{t_0}^t\|\vom (s)\|_{L^2}^4 ds$ vs $t$. The solution is computed using $1536\times 1536$ grid. Here $t_0= 0.00227375$. The final time instant is $t_4=0.0022868502$. (b) growth of $\nu \int_0^t \| \nabla \vom (s) \|_{L^2}^2 ds$ vs $t$. The solution is computed using $1024\times 1024$ grid. By the end of computation, we have $\|\vom(t)\|_{L^\infty}/\vom(0)\|_{L^\infty} \approx 1.6*10^7$.}   
    \label{fig:OmegaL2_growth_nse}
       \vspace{-0.05in}
\end{figure}

We first study the growth rate of the enstrophy, $\|\vom (t)\|_{L^2}^2$. A simple energy estimate would imply that if $\int_0^T\|\vom (t)\|_{L^2}^4 dt$ is bounded up to $T$, then the solution of the $3$D Navier--Stokes equations would remain smooth up to time $T$. In Figure 
\ref{fig:enstrophy_growth_nse} (a), we plot the growth of $\|\vom (t)\|_{L^2}^2$ as a function of time. We observe rapid growth of the enstrophy. In Figure 
\ref{fig:enstrophy_growth_nse} (b), we plot the linear fitting of $\|\vom (t)\|_{L^2}^{-4}$ vs time and observe very good linear fitness with $R$-Square values very close to $1$, which suggests that $\|\vom (t)\|_{L^2}^2 \sim (T-t)^{-1/2}$. In Figure 
\ref{fig:OmegaL2_growth_nse} (a), we plot the growth of $\int_{t_0}^t\|\vom (s)\|_{L^2}^4 ds$ as a function of time. Here $t_0= 0.00227375$ is the time when we switch the viscosity from $\nu = 5\cdot 10^{-4}$ to $\nu=5\cdot 10^{-3}$ and we use the same viscosity $\nu=5\cdot 10^{-3}$ for $t \geq t_0$. We observe that $\int_{t_0}^t\|\vom (s)\|_{L^2}^4 ds$ seems to grow without bound. 
This provides additional support for the potentially singular behavior of the Navier--Stokes equations.

In Figure 
\ref{fig:OmegaL2_growth_nse} (b), we plot the growth of $\nu \int_0^t \| \nabla \vom (s) \|_{L^2}^2 ds $ as a function of time. This quantity records how much enstrophy is depleted over time. As we can see, $\nu \int_0^t \| \nabla \vom (s) \|_{L^2}^2 ds $ grows rapidly in the late stage. This shows that the depletion of the enstrophy will be large near the potential singularity time, possibly due to the hyperbolic
nature of the flow.

%

\subsubsection{The Ladyzhenskaya-Prodi-Serrin regularity criteria} 

Next, we study the Ladyzhenskaya-Prodi-Serrin regularity criteria \cite{ladyzhenskaya1957,prodi1959,serrin1962}, which state that if a Leray-Hopf weak solution ${\bf u}$ \cite{leray1934,hopf1951} also lies in $L_t^q L_x^p$, with $3/p + 2/q \leq 1$, then the solution is unique and smooth in positive time. The endpoint result with $p=3$, $q=\infty$ has been proved in the work of Escauriaza-Seregin-Sverak in \cite{sverak2003}.

\begin{figure}[!ht]
\centering
	\begin{subfigure}[b]{0.38\textwidth}
    \includegraphics[width=1\textwidth]{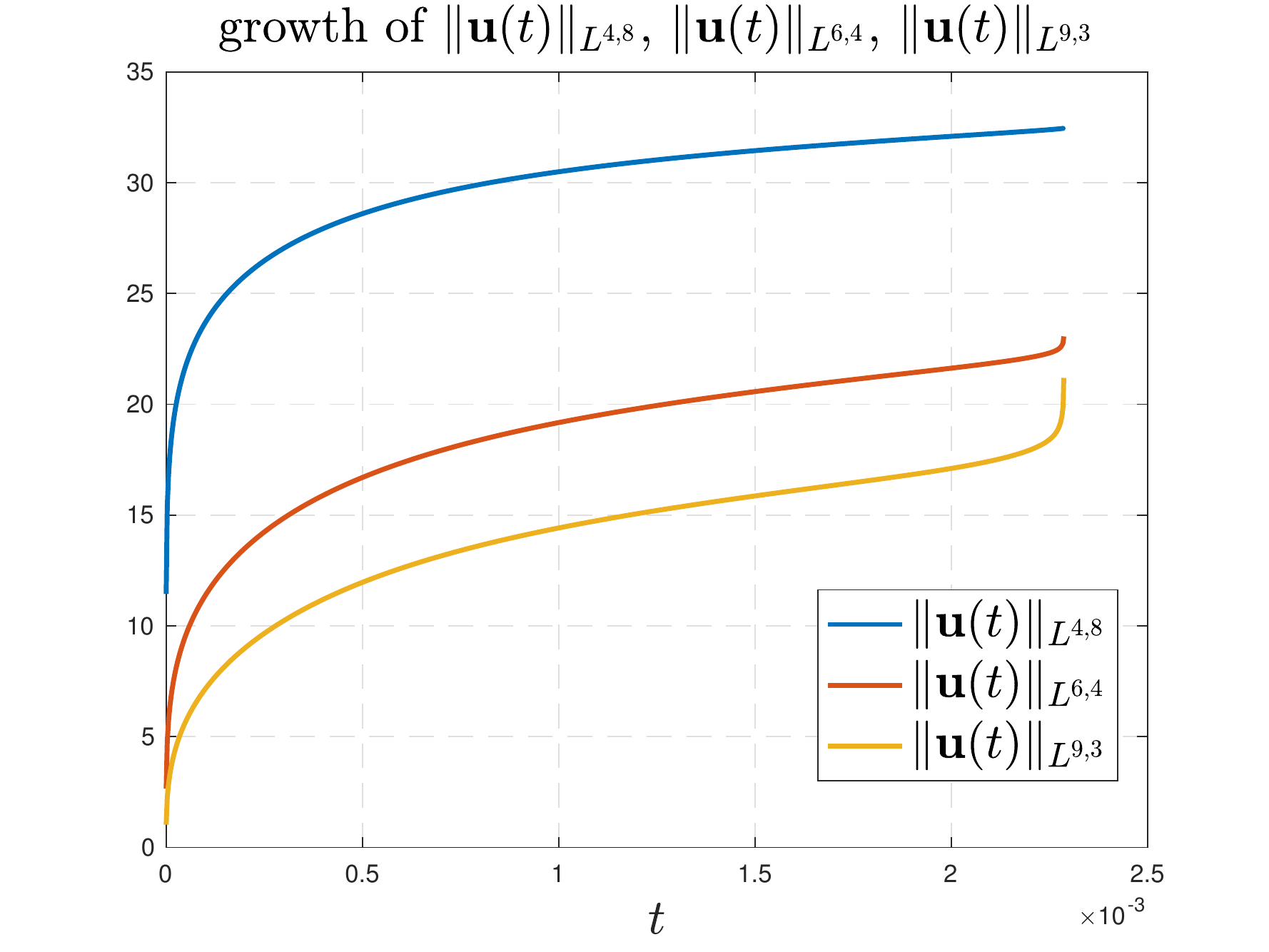}
    \caption{$\|{\bf u}\|_{L^{4,8}}$, $\|{\bf u}\|_{L^{6,4}}$, $\|{\bf u}\|_{L^{9,3}}$}
     \vspace{0.1in}
    \end{subfigure}
  	\begin{subfigure}[b]{0.38\textwidth} 
    \includegraphics[width=1\textwidth]{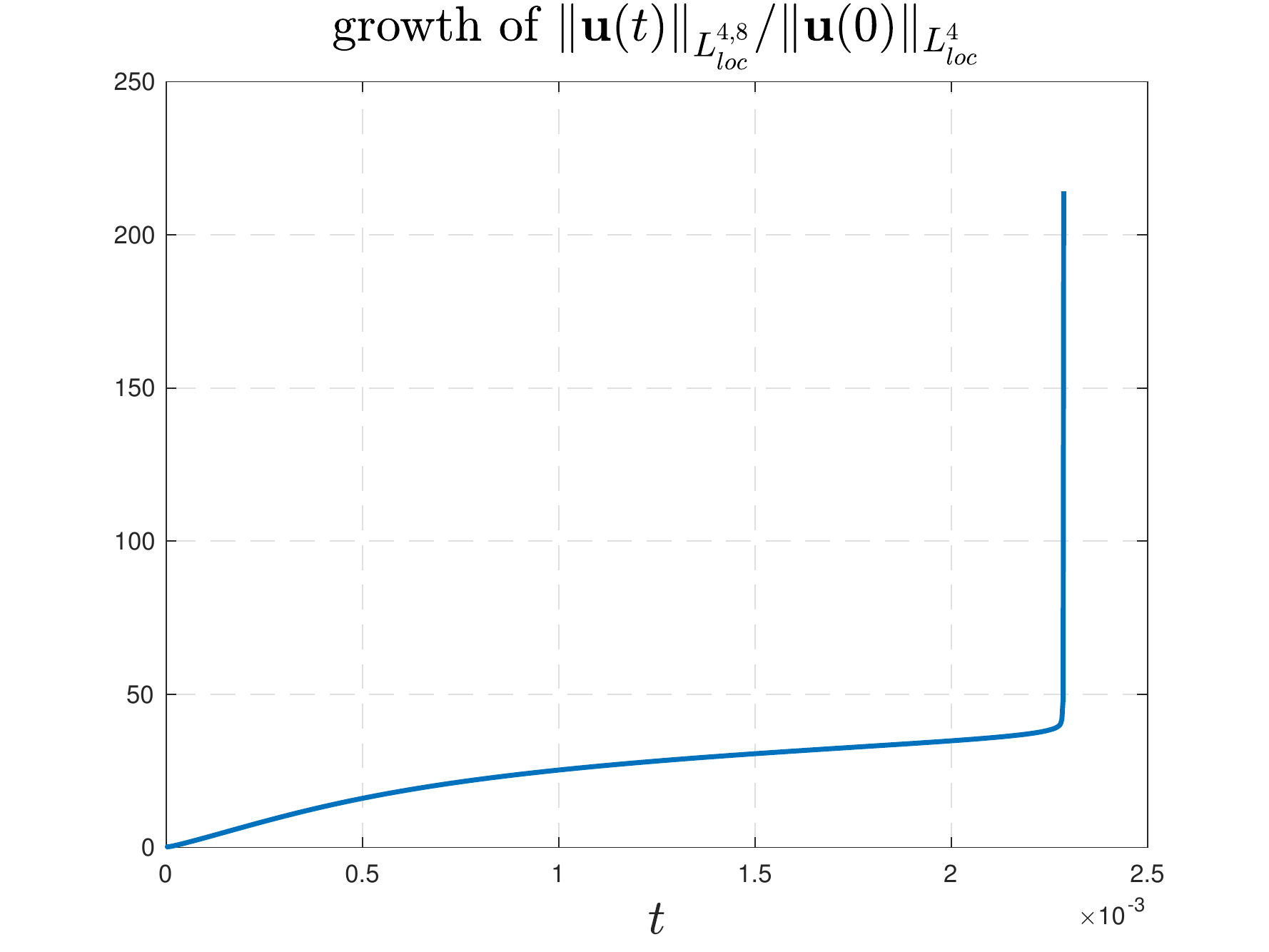}
    \caption{$\|{\bf u}\|_{L^{4,8}_{loc}}$}
    \vspace{0.1in}
    \end{subfigure} 
    \begin{subfigure}[b]{0.38\textwidth} 
    \includegraphics[width=1\textwidth]{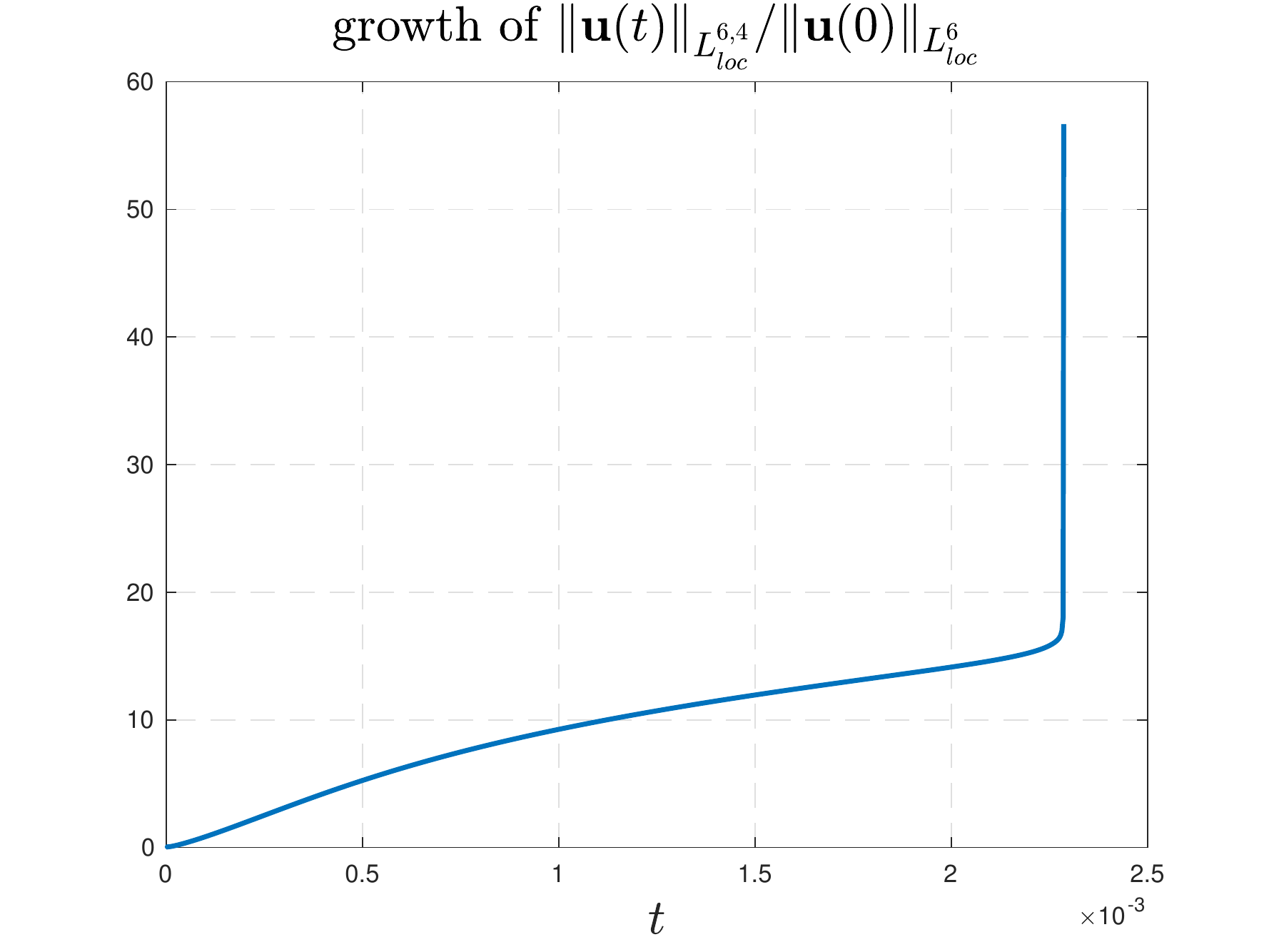}
    \caption{$\|{\bf u}\|_{L^{6,4}_{loc}}$}
    \end{subfigure}
  	\begin{subfigure}[b]{0.38\textwidth} 
    \includegraphics[width=1\textwidth]{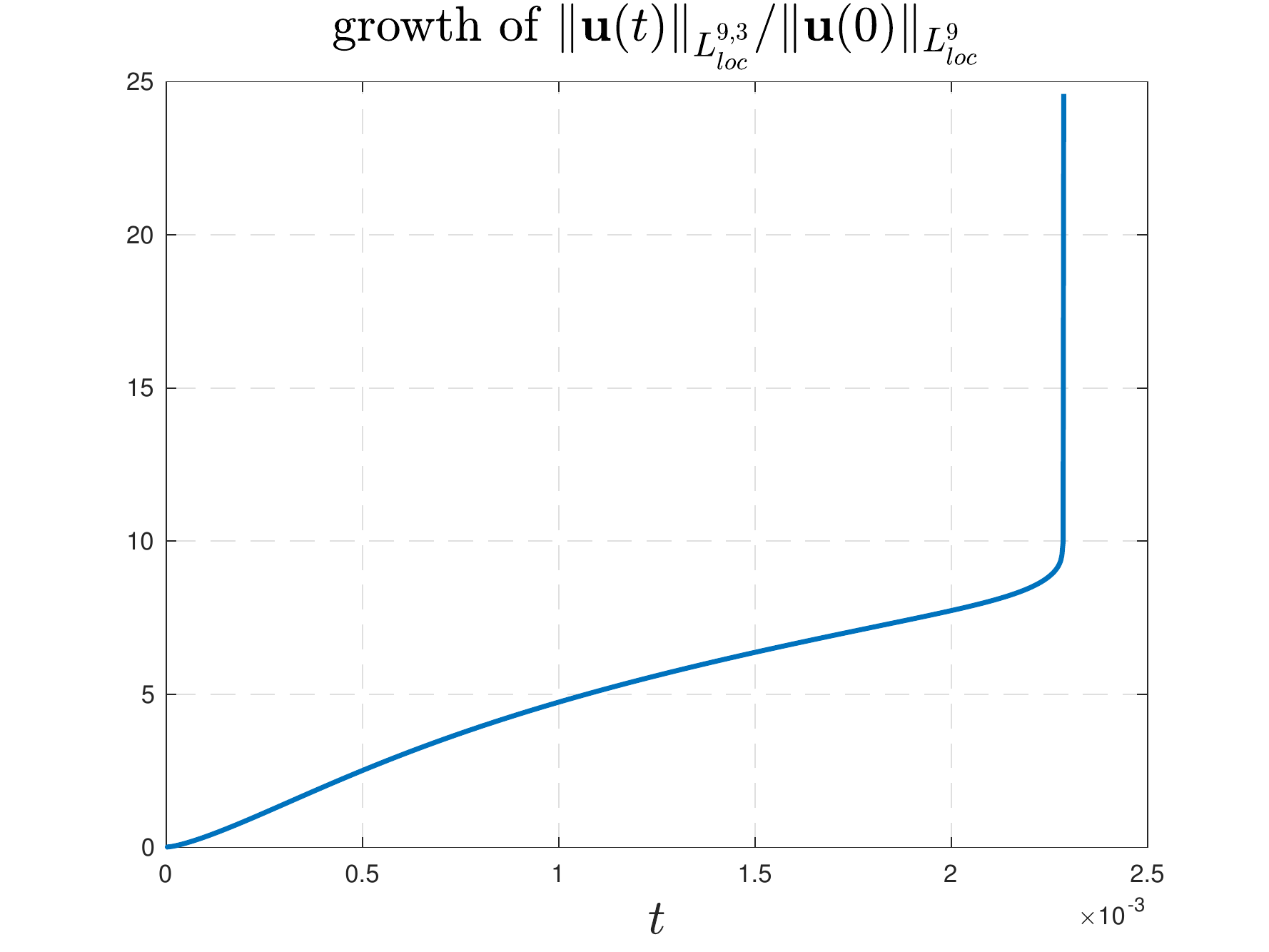}
    \caption{$\|{\bf u}\|_{L^{9,3}_{loc}}$}
    \end{subfigure} 
    \caption[Lpq compare]{Comparison of growth of $\|{\bf u}(t)\|_{L^{p,q}}$. (a) The growth of $\|{\bf u}(t)\|_{L^{4,8}}$, $\|{\bf u}(t)\|_{L^{6,4}}$, $\|{\bf u}(t)\|_{L^{9,3}}$ in the whole domain. (b) The relative growth of $\|{\bf u}(t)\|_{L^{4,8}_{loc}}/\|{\bf u}(0)\|_{L^{4}_{loc}}$ in the local domain $r \leq r_0 = 0.001$. (c) The relative growth of $\|{\bf u}(t)\|_{L^{6,4}_{loc}}/\|{\bf u}(0)\|_{L^{6}_{loc}}$ in the local domain $r \leq r_0$. (d) The relative growth of $\|{\bf u}(t)\|_{L^{9,3}_{loc}}/\|{\bf u}(0)\|_{L^{9}_{loc}}$ in the local domain $r \leq r_0$. The computation is performed using $1024\times 1024$ grid.}   
    \label{fig:Lpq-criteria}
       \vspace{-0.05in}
\end{figure}

Denote the $L^{p,q}$ norm of the velocity ${\bf u}$ as follows:

\[
\|{\bf u}(t)\|_{L^{p,q}} = \left (\int_0^t \|{\bf u(s)}\|_{L^p(\Omega)}^q ds \right )^{1/q}.
\]
Let $\Omega_{loc} = \{ (r,z) \; | \; r \leq r_0\}$ be a local cylindrical domain with radius $r_0 > 0$. We define a localized version of the $L^{p,q}$ norm as follows. 
\[
\|{\bf u}(t)\|_{L^{p,q}_{loc}} = \left (\int_0^t \|{\bf u(s)}\|_{L^p(\Omega_{loc})}^q ds \right )^{1/q}.
\]
By the partial regularity results due to Caffarelli-Kohn-Nirenberg \cite{caffarelli1982partial}, we know that the solution of the axisymmetric Navier--Stokes equations with smooth initial data of finite energy can only develop a potential singularity at $r=0$. This implies that the solution of the axisymmetric Navier--Stokes equations with smooth initial data will remain regular for $r > r_0 >0$. Since our adaptive mesh strategy allocates more and more grid points to the near field (the most singular region) dynamically, the adaptive mesh in the far field is relatively coarse in the late stage of the computation. Thus, it makes sense to compute the localized version of the $L^{p,q}$ norm of the velocity field with a small $r_0 = 0.001$.

In Figure \ref{fig:Lpq-criteria}(a), we plot the  dynamic growth of $\|{\bf u}\|_{L^{4,8}}$, $\|{\bf u}\|_{L^{6,4}}$, $\|{\bf u}\|_{L^{9,3}}$ in the whole domain. We can see that all three norms grow dynamically throughout the computation. By the end of the computation, the maximum vorticity has grown more than a factor of $10^6$. The growth rate for $\|{\bf u}\|_{L^{9,3}}$ is the fastest among the three cases while the growth rate for $\|{\bf u}\|_{L^{4,8}}$ is the slowest among the three cases. This trend is quite consistent. The larger the value of $p$ is, the faster the growth rate. As we will see next, the growth rate of the critical $L^3$ norm of the velocity is the slowest among all $ p \geq 3$.

In order to capture the rapid dynamic growth driven by the near field, we plot the relative growth of the localized version of $\|{\bf u}\|_{L^{4,8}_{loc}}$, $\|{\bf u}\|_{L^{6,4}_{loc}}$, $\|{\bf u}\|_{L^{9,3}_{loc}}$ with $r_0 = 0.001$ in Figure \ref{fig:Lpq-criteria} (b)-(d), respectively. We observe that all three quantities grow rapidly in time after a mild transient growth period. We also observe that $\|{\bf u}\|_{L^{4,8}_{loc}}$ grows the fastest among all three cases. This is due to the fact that $\|{\bf u}(0)\|_{L^{4}_{loc}} < \|{\bf u}(0)\|_{L^{6}_{loc}}  < \|{\bf u}(0)\|_{L^{9}_{loc}}$.
These results provide strong evidence that the Navier--Stokes equations with our initial data seem to develop a potential finite time singularity.

\begin{figure}[!ht]
\centering
    \begin{subfigure}[b]{0.38\textwidth}
    \includegraphics[width=1\textwidth]{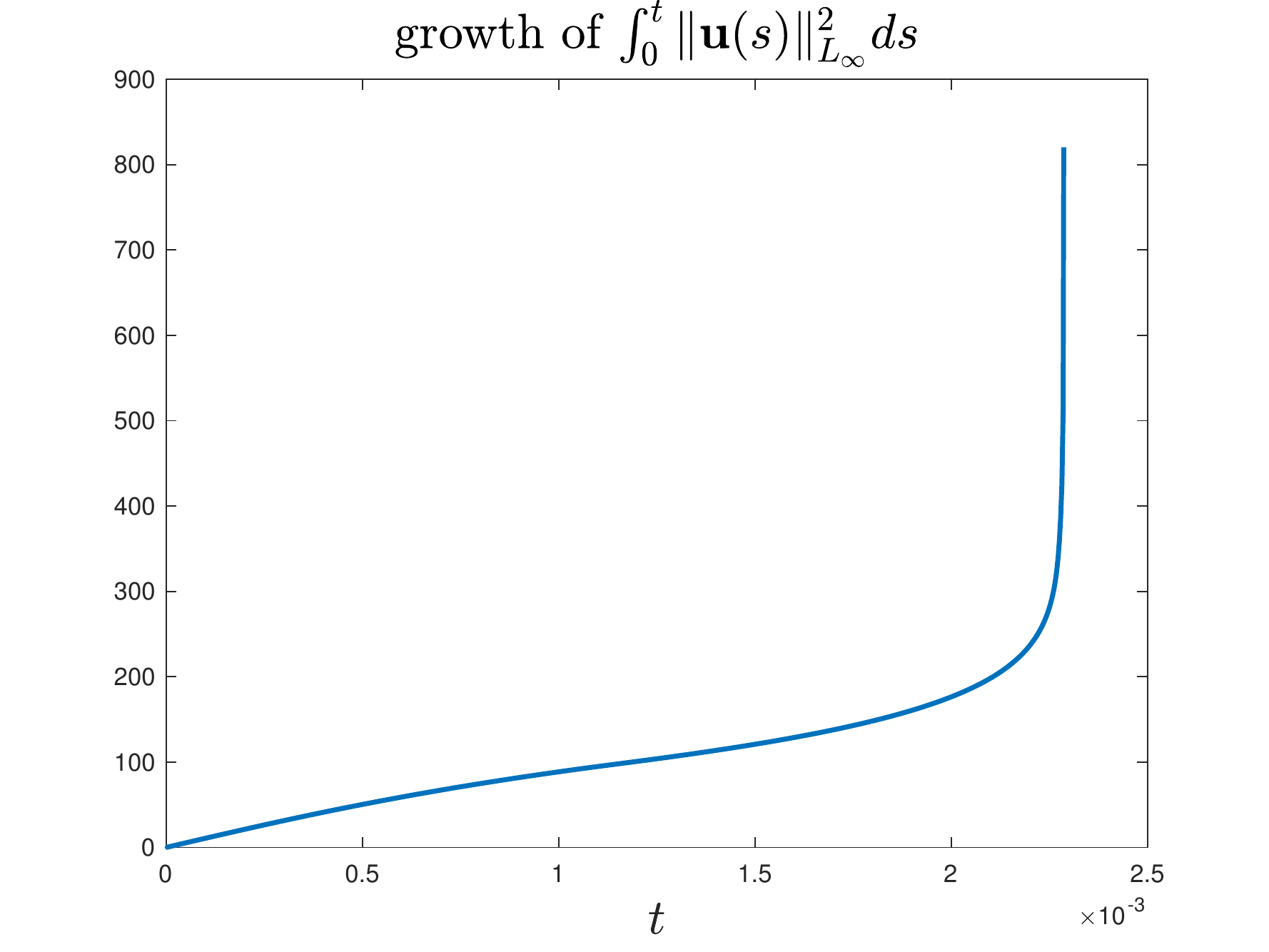}
      \end{subfigure}
    \begin{subfigure}[b]{0.38\textwidth}
    \includegraphics[width=1\textwidth]{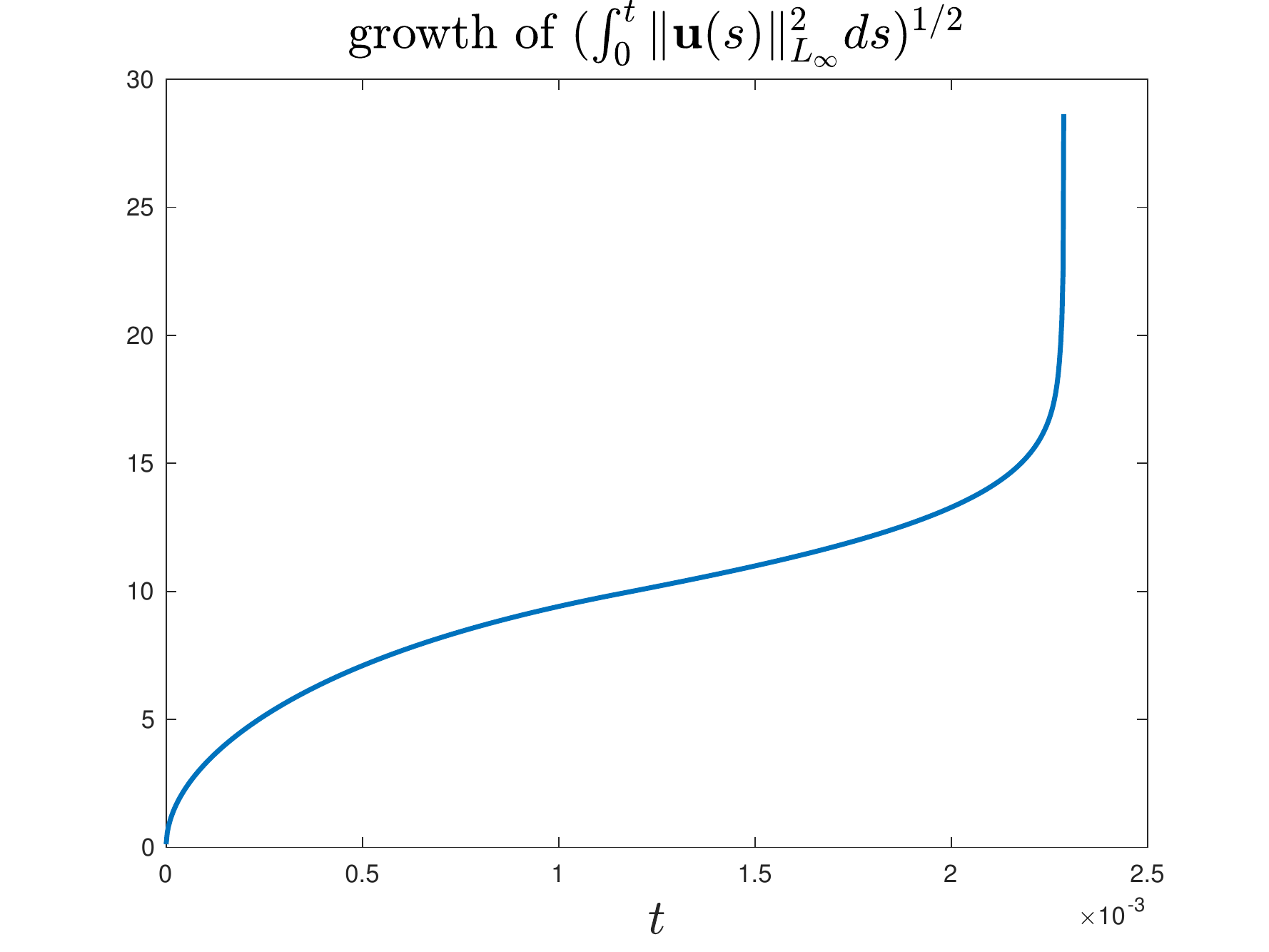}
      \end{subfigure}    
    \caption[The max Velocity test]{Left subplot. The growth of $\int_0^t \| | {\bf u}(s)\|_{L^\infty}^2 ds$. Right subplot. The growth of $\left ( \int_0^t \| | {\bf u}(s)\|_{L^\infty}^2 ds \right )^{1/2}$. The final time is $t=0.00228676968$. The computation is performed $1024\times 1024$ grid. }  
    \label{fig:maxVel_growth}
       \vspace{-0.05in}
\end{figure}

In Figure \ref{fig:maxVel_growth}(a)-(b), we plot the dynamic growth of $\int_0^t \| | {\bf u}(s)\|_{L^\infty}^2$ and $\left ( \int_0^t \| | {\bf u}(s)\|_{L^\infty}^2\right )^{1/2}$, respectively. The second quantity is the $L^{\infty,2}$ norm of the velocity over the whole domain, which is one of the endpoint cases in the the Ladyzhenskaya-Prodi-Serrin regularity criteria with $p=\infty$ and $q = 2$. We observe that both quantities develop rapid growth dynamically. This provides further evidence for the potentially singular behavior of the Navier--Stokes equations.

\begin{figure}[!ht]
\centering
    \begin{subfigure}[b]{0.38\textwidth}
    \includegraphics[width=1\textwidth]{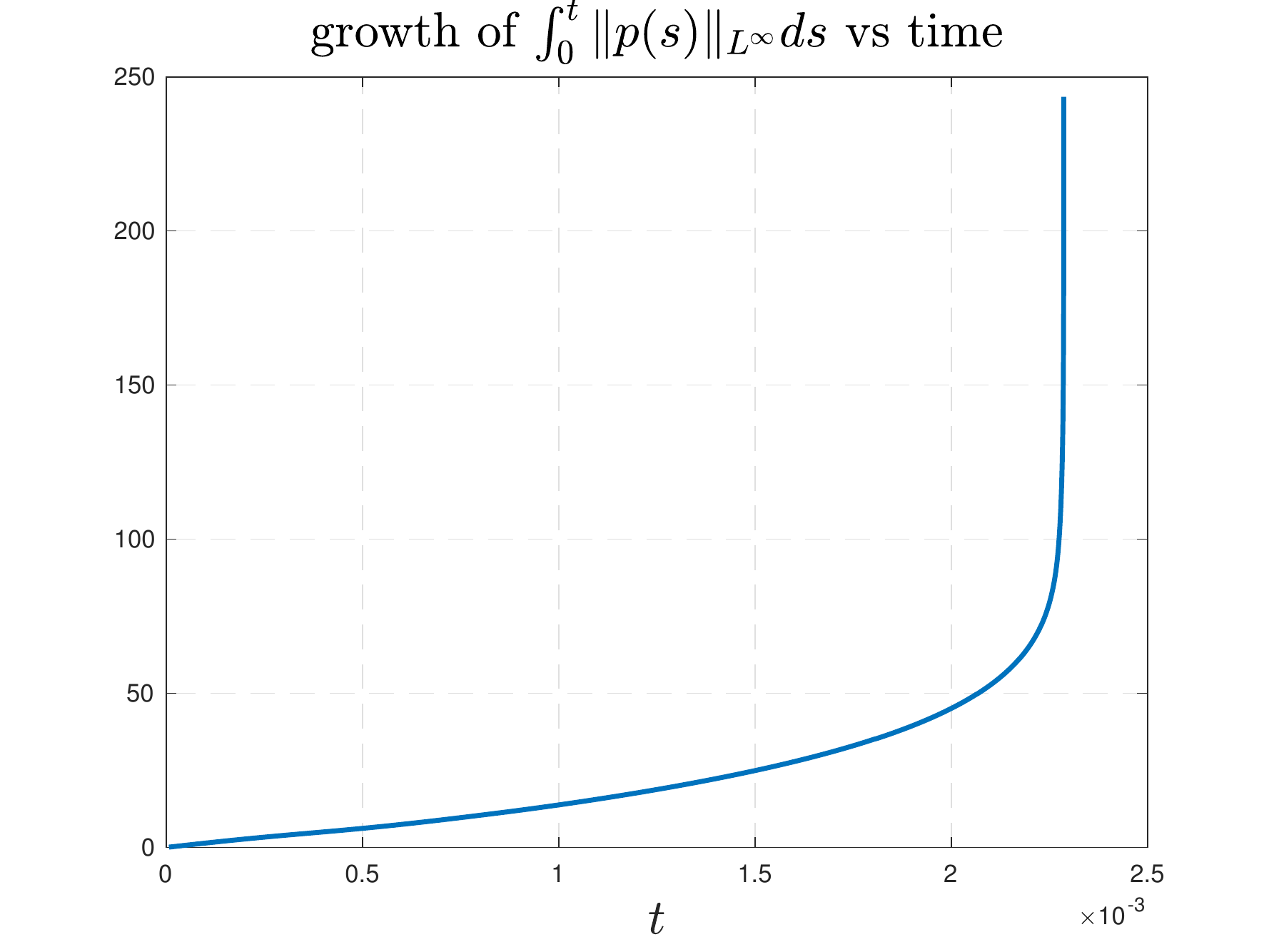}
      \end{subfigure}
    \begin{subfigure}[b]{0.38\textwidth}
    \includegraphics[width=1\textwidth]{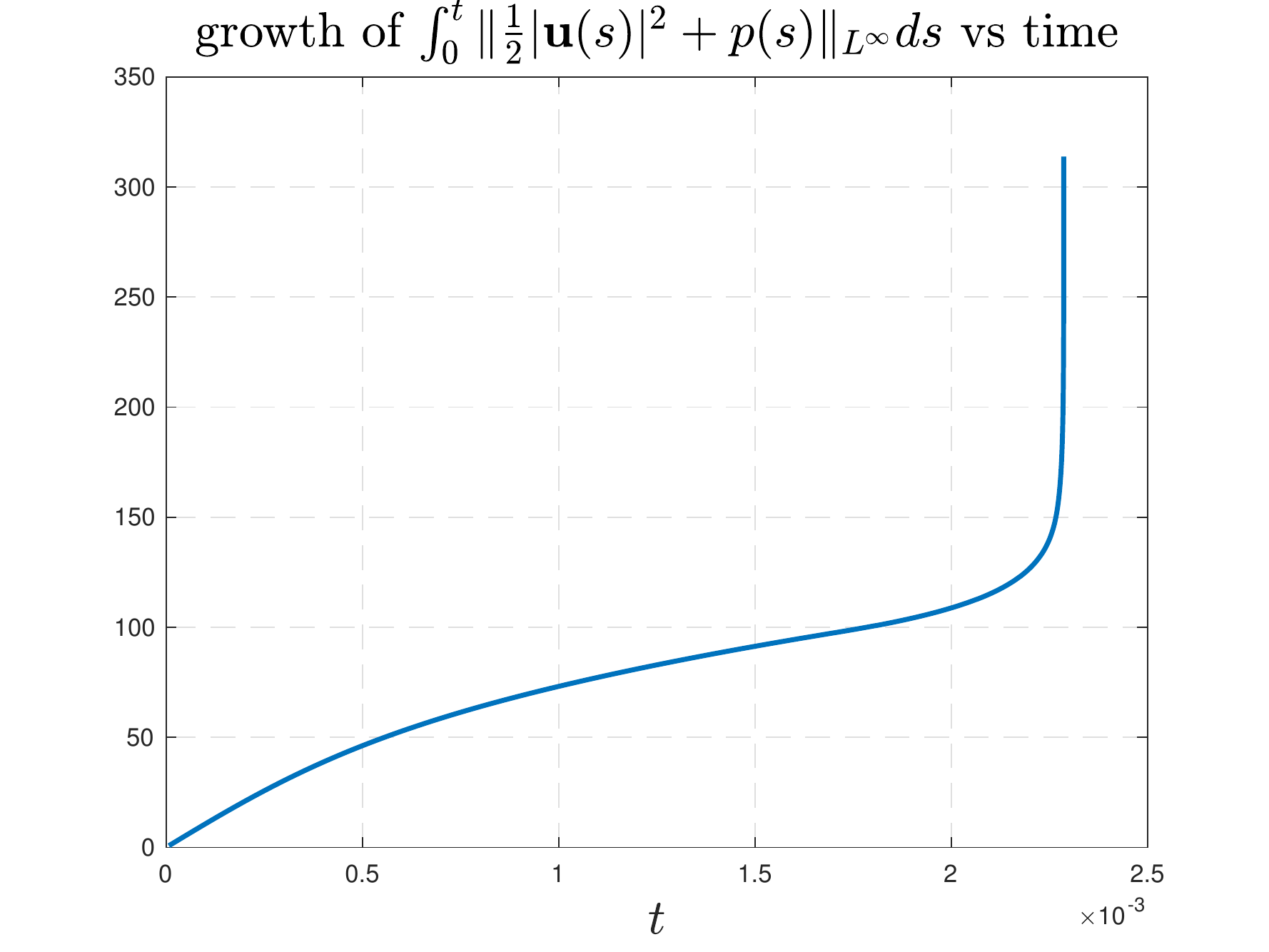}
      \end{subfigure}    
    \caption[The min pressure test]{Left subplot. The growth of $\int_0^t \| | p(s)\|_{L^\infty} ds$. Right subplot. The growth of $ \int_0^t \| | \frac{1}{2} |{\bf u}(s)|^2 + p(s) \|_{L^\infty} ds $. The final time is $t_4=0.0022868502$. The computation is performed $1536\times 1536$ grid. }  
    \label{fig:maxpre_growth}
       \vspace{-0.05in}
\end{figure}

In Figure \ref{fig:maxpre_growth} (a)-(b), we plot the dynamic growth of $\int_0^t \| | p(s)\|_{L^\infty} ds$ and $ \int_0^t \| | \frac{1}{2} |{\bf u}(s)|^2 + p(s) \|_{L^\infty} ds $, respectively. By the end of the computation, the maximum vorticity has grown by a factor of $10^7$. The rapid growth of these two quantities provides additional evidence for the development of potentially singular solutions of the Navier-Stokes equations with our initial data \cite{sverak2002}.

\subsubsection{The growth of the critical $L^3$ norm of the velocity}
\begin{figure}[!ht]
\centering
    \begin{subfigure}[b]{0.40\textwidth}
    \includegraphics[width=1\textwidth]{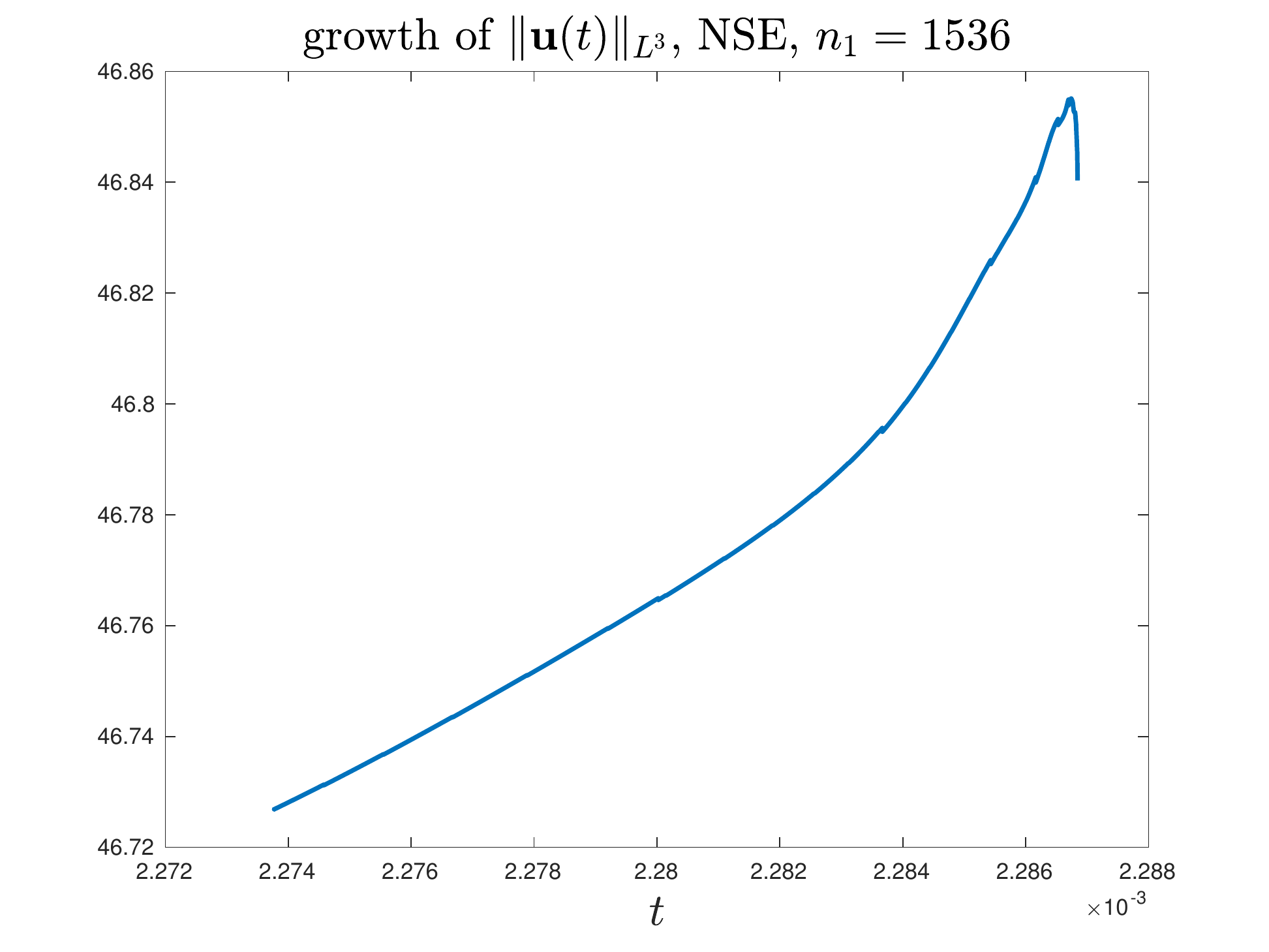}
    \caption{growth of $\| {\bf u} (t)\|_{L^3} $ vs $t$}
      \end{subfigure}
     \begin{subfigure}[b]{0.40\textwidth}
    \includegraphics[width=1\textwidth]{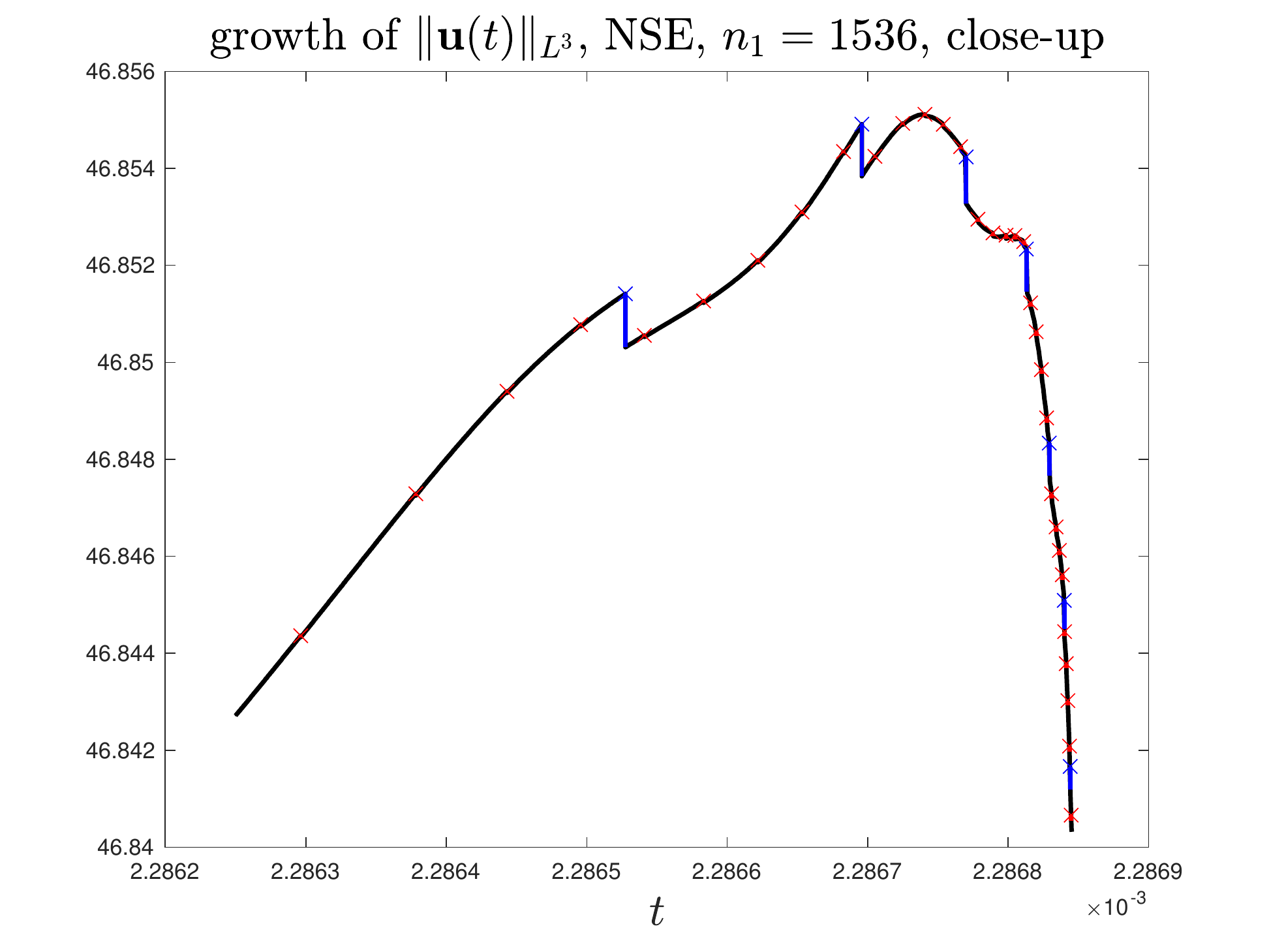}
    \caption{growth of $\| {\bf u} (t)\|_{L^3} $,  close-up view}
    \end{subfigure}
    \caption[$L^3$ norm of velocity test]{ (a) growth of $\| {\bf u} (t)\|_{L^3} $ vs $t$. (b)  
    growth of $\| {\bf u} (t)\|_{L^3} $, a close-up view. The red cross marks the time when the mesh $r(\rho)$ is updated, and the blue cross marks the time when $z(\eta)$ is updated.
  We observe that  $\| {\bf u} (t)\|_{L^3} $ drops every time we refine the mesh. The drop is larger when $z(\eta)$ is updated. The solution is computed using $1536\times 1536$ grid. The final time instant is $t_3=0.0022868453$.}   
    \label{fig:VelL3_growth_nse}
       \vspace{-0.05in}
\end{figure}

We now study the $L^{3}$ norm of the velocity field. As shown in \cite{sverak2003}, the Navier--Stokes equations cannot blow up at time $T$ if $\|{\bf u}(t)\|_{L^3}$ is bounded up to time $T$. In Figure \ref{fig:VelL3_growth_nse} (a), we plot the dynamic growth of $\|{\bf u} (t)\|_{L^3}$ as a function of time. We observe that $\|{\bf u} (t)\|_{L^3}$ first grows in time and then has a mild decrease in the late stage.  

We remark that the non-blowup criterion using the  $\|{\bf u}\|_{L^{3}}$ estimate is based on a compactness argument. As a result, the bound on $\max_{0\leq t \leq T}\|{\bf u}(t)\|_{L^3}$ does not provide a direct estimate on the dynamic growth rate of the Navier--Stokes solution up to $T$.
In a recent paper \cite{tao2020}, Tao further examined the role of the $L^3$ norm of the velocity on the potential blow-up of the Navier-Stokes equations. He showed that as one approaches a finite blow-up time $T$, the critical $L^3$ norm of the velocity must blow up at least at a rate $\left (\log \log \log \frac{1}{T-t}\right )^c$ for some absolute constant $c$. This implies that even for a potential finite time blow-up of the Navier--Stokes equations, $\|{\bf u}(t)\|_{L^3}$ may blow up extremely slowly. If $\|{\bf u}(t)\|_{L^3}$ indeed grows at a rate like $\left (\log \log \log \frac{1}{T-t}\right )^c$ for some absolute constant $c$, it would be almost impossible to capture such slow growth rate numerically with our current computational capacity.

\begin{figure}[!ht]
\centering
    \begin{subfigure}[b]{0.38\textwidth}
    \includegraphics[width=1\textwidth]{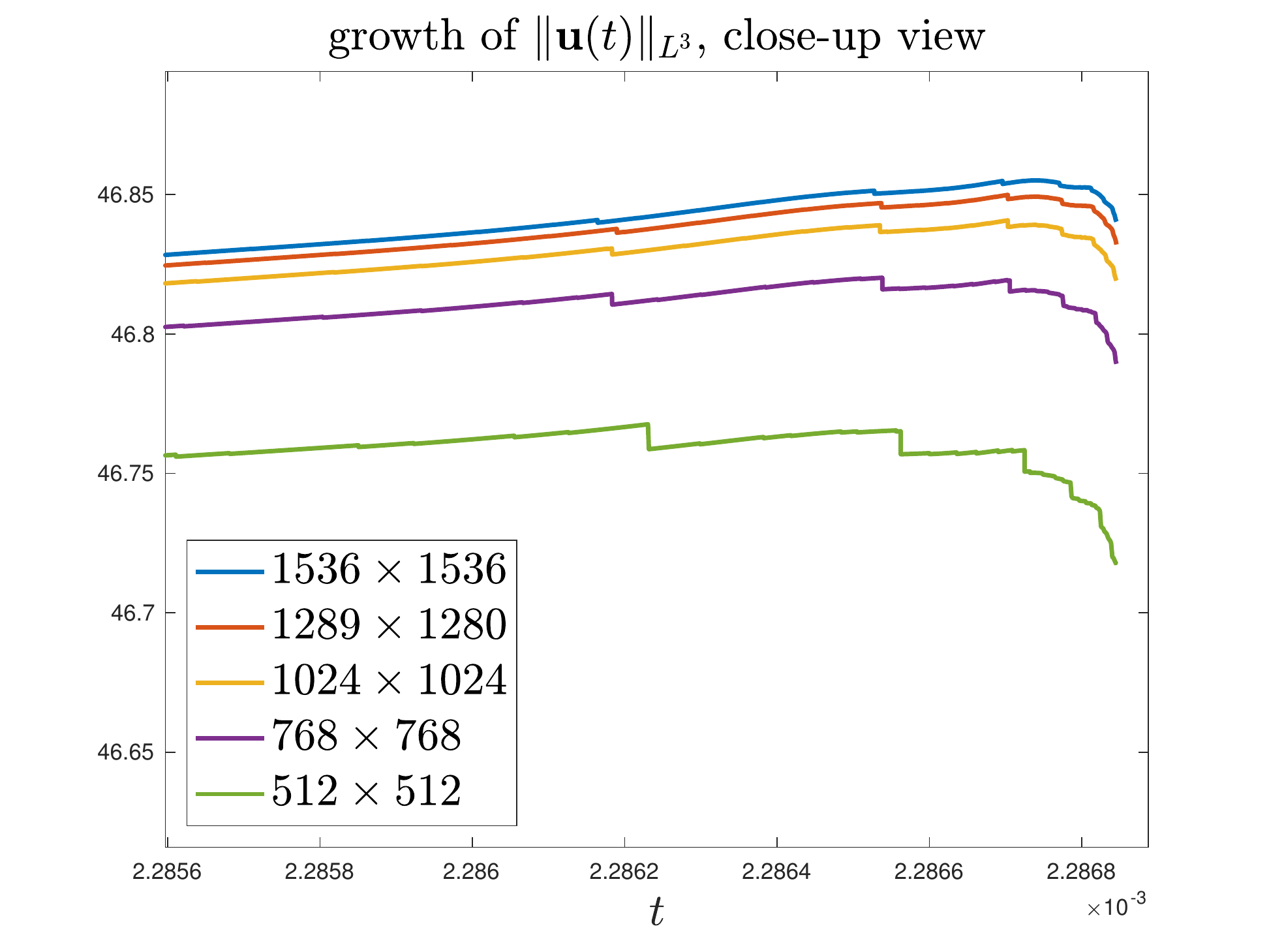}
      \end{subfigure}
    \begin{subfigure}[b]{0.38\textwidth}
    \includegraphics[width=1\textwidth]{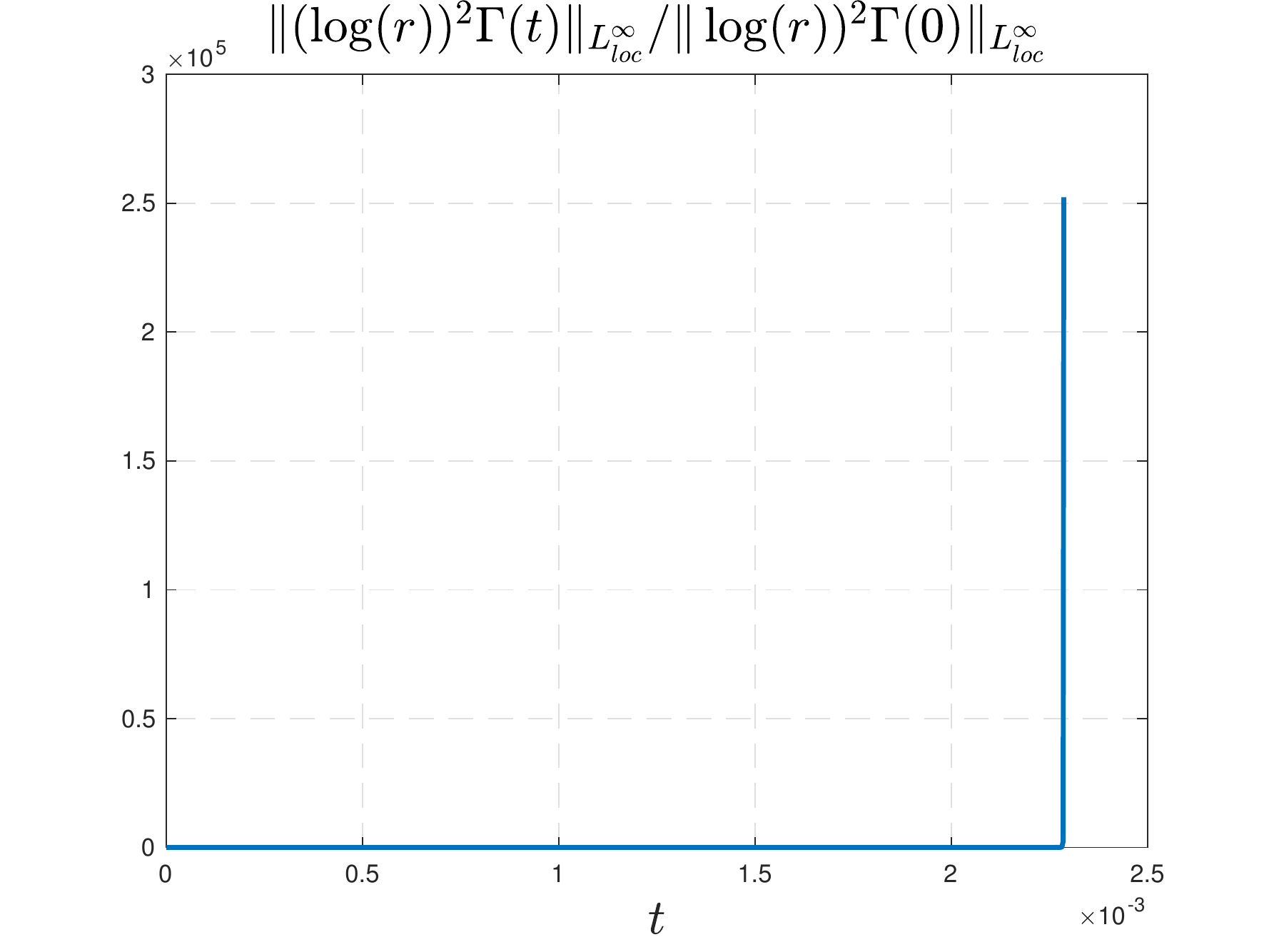}
      \end{subfigure}    
    \caption[$L^3$ norm of velocity test]{Left subplot. Comparison of $\| {\bf u} (t)\|_{L^3} $ vs $t$ for different resolution using $1536\times1536$, $1289\times1280$, $1024\times1024$, $768\times768$ and $512\times512$ grids, a close-up view. The final time instant is $t_3=0.0022868453$. 
      Right subplot. 
    The growth of $\| (\log(r))^2 \Gamma(t)\|_{L^\infty (\Omega_{loc}^*)}$ over a small local domain $\Omega_{loc}^{*} = \{(r,z) \;|\; (r,z) \in [0,0.001]^2\}$ using $1024\times 1024$ grid.  The final time is $t=0.00228691317$.}  
    \label{fig:VelL3_growth_nse2}
       \vspace{-0.05in}
\end{figure}

\begin{figure}[!ht]
\centering
    \begin{subfigure}[b]{0.38\textwidth}
    \includegraphics[width=1\textwidth]{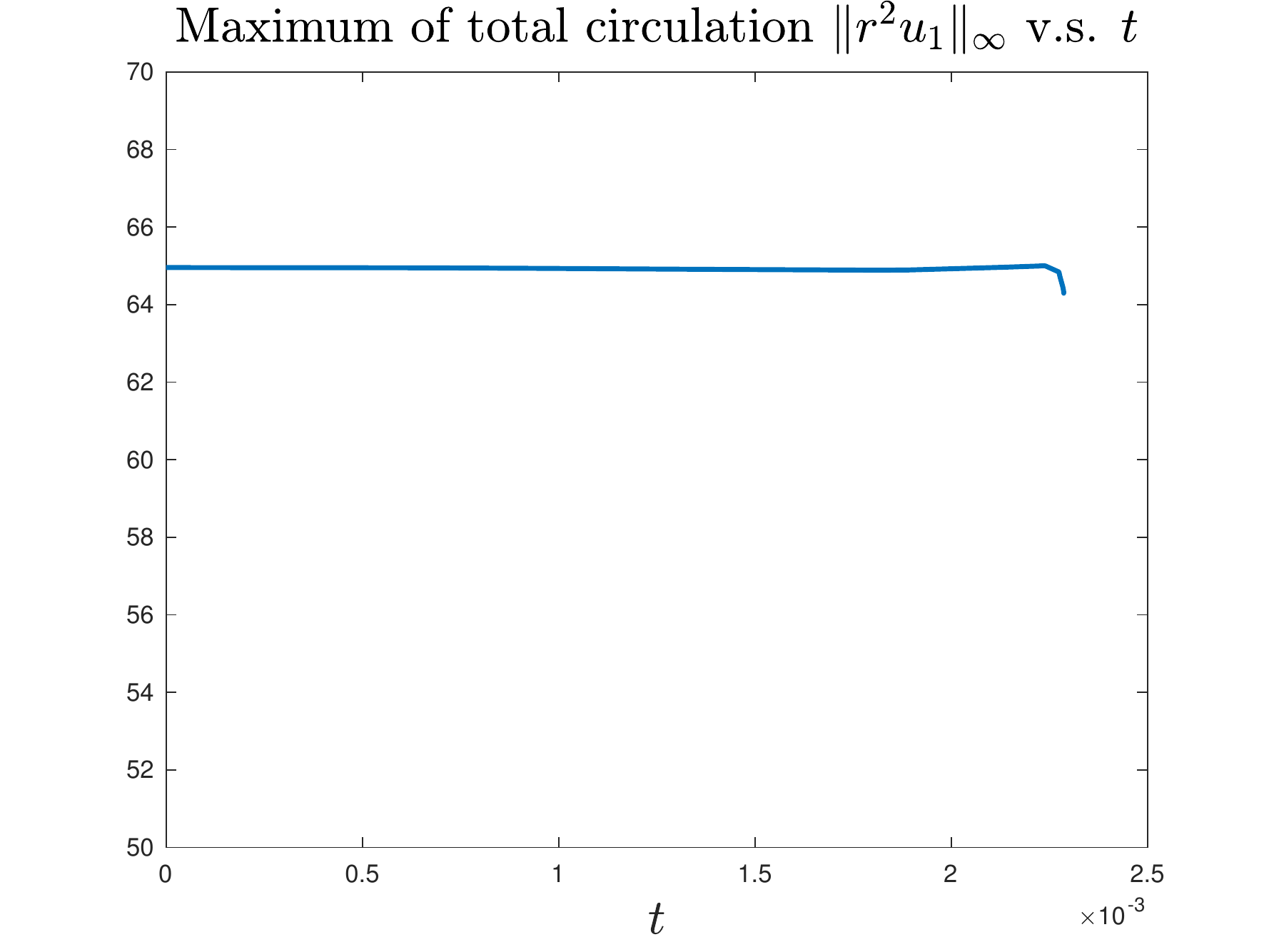}
      \end{subfigure}  
          \begin{subfigure}[b]{0.38\textwidth}
    \includegraphics[width=1\textwidth]{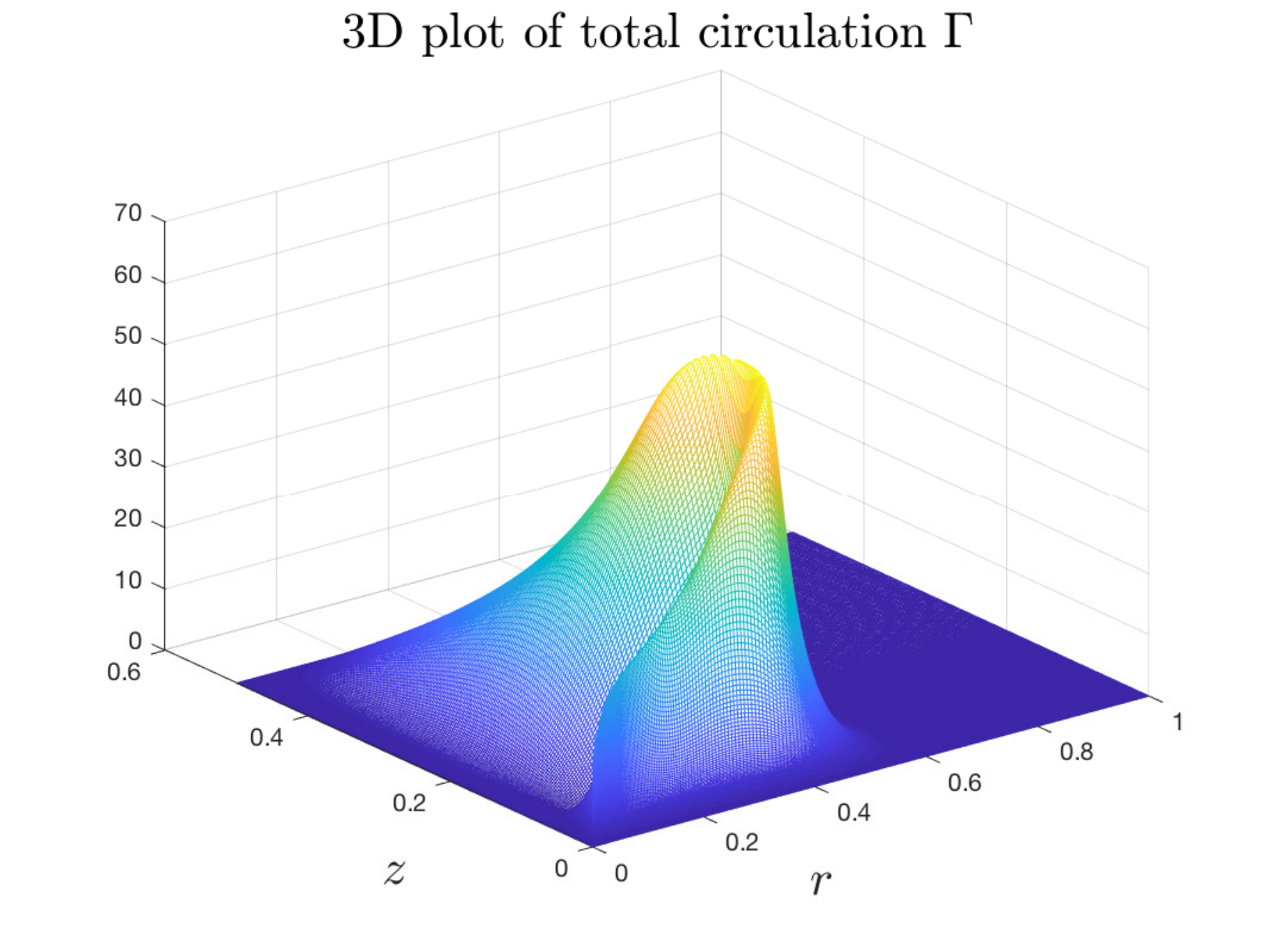}
      \end{subfigure}  
    \caption[total circulation]{
    Left plot: The maximum of total circulation $\Gamma(t) = r u^\theta = r^2 u_1$ in time. Right plot: The global $3$D plot of $\Gamma$ at the final time $t=0.00228685059$ using $1280\times 1280$ grid.}  
    \label{fig:total_circulationh_nse}
       \vspace{-0.05in}
\end{figure}

The total circulation $\Gamma = r u^\theta = r^2 u_1$ for the axisymmetric Navier--Stokes equations satisfies
\[
\Gamma_t + u^r \Gamma_r + u^z \Gamma_z = \nu (\Delta - \frac{2}{r}\partial_r ) \Gamma.
\]
Thus, we have $\|\Gamma (t)\|_{L^\infty}\leq \|\Gamma_0 \|_{L^\infty}$ for smooth solutions \cite{majda2002vorticity,chen2008lower,sverak2009}. 
In Figure \ref{fig:total_circulationh_nse}(a), we plot $\|\Gamma (t)\|_{L^\infty}$ as a function of time using a $1280\times 1280$ grid. We observe that $\|\Gamma (t)\|_{L^\infty}$ remains bounded throughout the computation. Moreover, $\| \Gamma (t)\|_{L^\infty}$ decays relatively fast in the late stage due to the strong viscous regularization of the nearly singular solution. We also observe that the maximum of $\Gamma$ is achieved in the far field in the late stage where we have a relatively coarse grid, see Figure \ref{fig:total_circulationh_nse}(b). The frequent changes of adaptive mesh also contribute to the relatively fast decay of $\| \Gamma (t)\|_{L^\infty}$ in the late stage.


Currently, our adaptive mesh strategy introduces some numerical dissipation in the late stage due to the frequent changes of adaptive meshes and the interpolation from the old adaptive mesh to the new adaptive mesh. A careful study reveals that such numerical dissipation contributes to the mild decrease in $\|{\bf u}(t)\|_{L^3}$ that we observed in the late stage. 
In Figure \ref{fig:VelL3_growth_nse}(b), we mark the time position by a red cross when we change the mesh map $r(\rho)$ and by a blue cross when we update the mesh map $z(\eta)$. We observe that the changes of adaptive mesh are much more frequent in the late stage. Moreover, every time we change the adaptive mesh, $\|{\bf u}(t)\|_{L^3}$ experiences a drop. The drop in $\|{\bf u}(t)\|_{L^3}$ is more significant when we update the mesh map $z(\eta)$ (marked by blue color in Figure \ref{fig:VelL3_growth_nse}(b)). 

In Figure \ref{fig:VelL3_growth_nse2}(a), we plot $\|{\bf u}(t)\|_{L^3}$ using different resolutions. We observe that the drop of $\|{\bf u}(t)\|_{L^3}$ due to a change of adaptive mesh is smaller for a higher resolution. For a fixed time, $\| {\bf u} (t)\|_{L^3} $ increases as we increase the resolution. Moreover, we observe that the time interval during which $\| {\bf u} (t)\|_{L^3} $ remains increasing is slightly longer if we use a higher resolution to solve the Navier--Stokes equations. This seems to suggest that $\| {\bf u} (t)\|_{L^3} $ may still have an extremely mild growth dynamically if we can afford sufficient resolution in both the near field and the far field.

Our numerical study shows that the far field velocity has a significant contribution to the $L^3$ norm of the velocity (about $83\%$) in the late stage. 
Our current adaptive mesh strategy only provides sufficient resolution in the near field, but the adaptive grid in the far field is relatively coarse. Thus, the numerical dissipation is relatively large in the far field, which contributes to the mild decrease of $\| {\bf u} (t)\|_{L^3} $ due to the frequent changes of adaptive mesh in the late stage.

\begin{figure}[!ht]
\centering
    \begin{subfigure}[b]{0.38\textwidth}
    \includegraphics[width=1\textwidth]
    {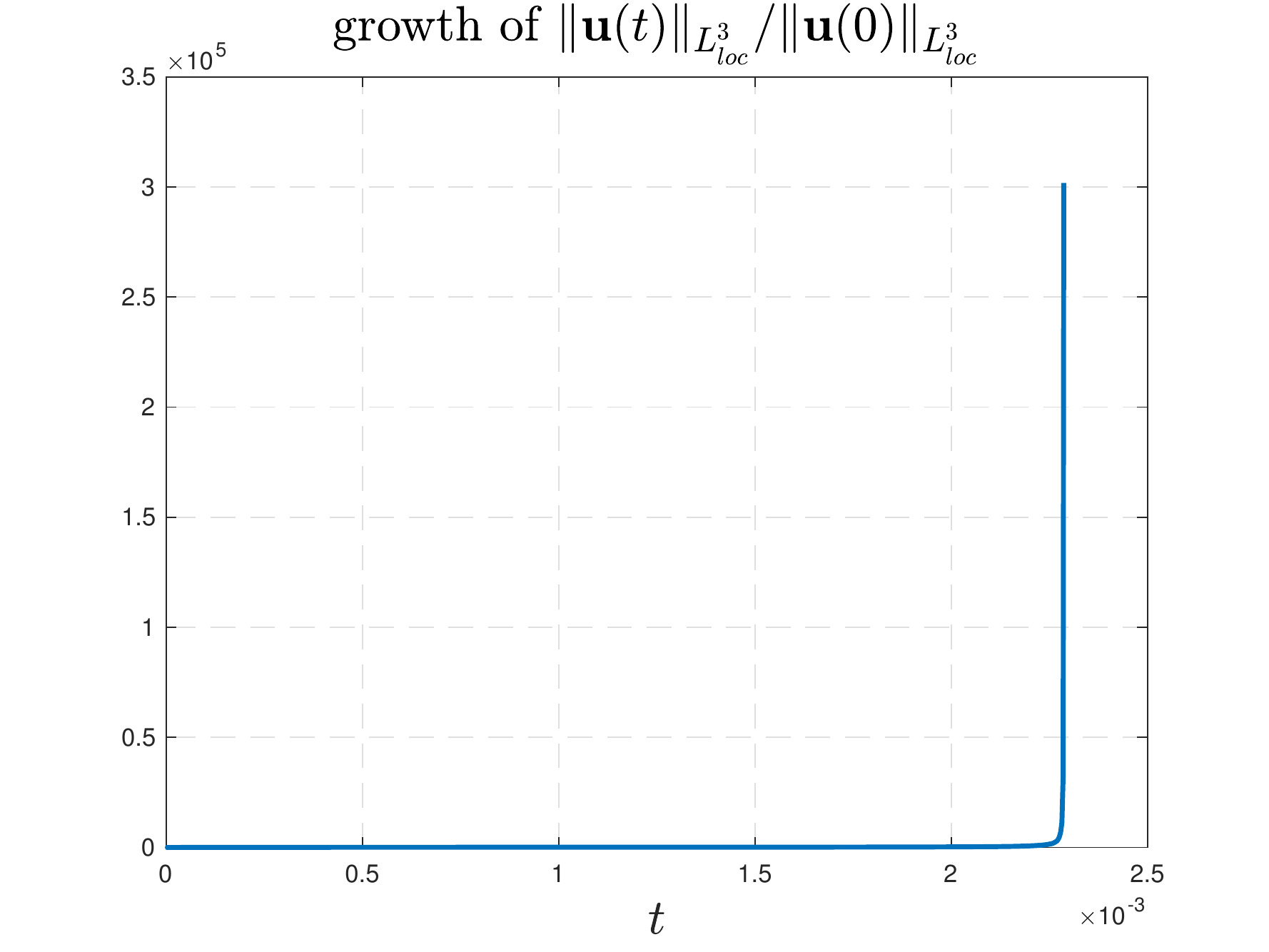}
      \end{subfigure}
    \begin{subfigure}[b]{0.38\textwidth}
    \includegraphics[width=1\textwidth]{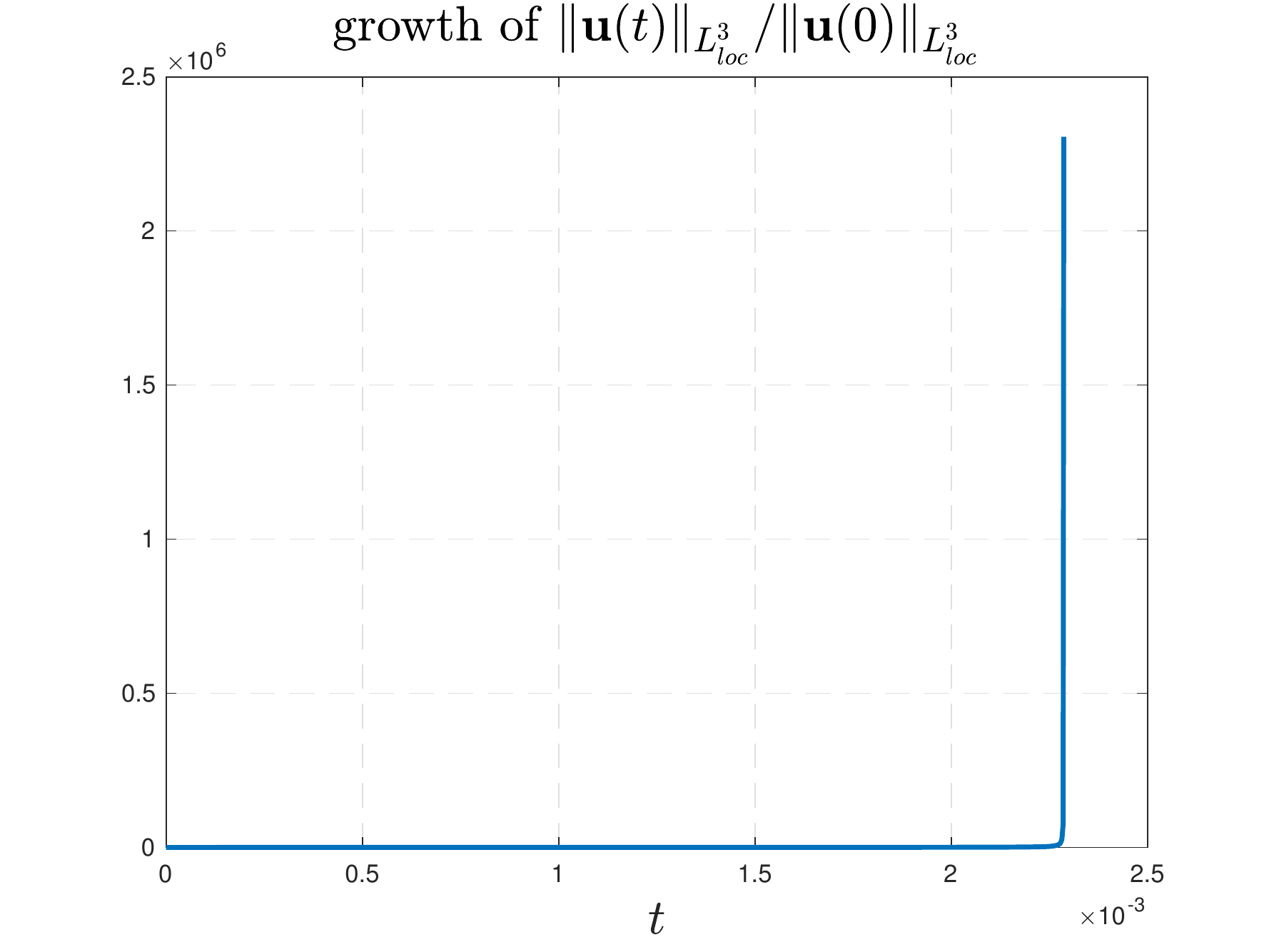}
      \end{subfigure}    
    \caption[The u2r test]{Left subplot. The relative growth of $\| {\bf u} (t)\|_{L^3_{loc}}/\| {\bf u} (0)\|_{L^3_{loc}}$ over a local domain $\Omega_{loc}^* = \{(r,z) \;|\; (r,z) \in [0,0.001]^2\}$ using $1024\times 1024$ grid. The final time is $t=0.00228689295$. Right subplot. The relative growth of $\| {\bf u} (t)\|_{L^3_{loc}}/\| {\bf u} (0)\|_{L^3_{loc}}$ over a smaller local domain $\Omega_{loc}^{**} = \{(r,z) \;|\; (r,z) \in [0,0.0005]^2\}$ using $1024\times 1024$ grid. The final time is $t=0.00228689798$.    
   }  
    \label{fig:VelL3_loc_growth}
       \vspace{-0.05in}
\end{figure}

Since the axisymmetric Navier--Stokes equations cannot develop a finite time singularity away from the symmetry axis $r=0$ and the potential singularity seems to occur at the origin, we compute the relative dynamic growth of $\|{\bf u}(t)\|_{L^3(\Omega_{loc}^*)}$ in a localized domain $\Omega_{loc}^* = \{(r,z) \;|\; (r,z) \in [0,0.001]^2\}$. In Figure \ref{fig:VelL3_loc_growth} (a), we plot $\| {\bf u} (t)\|_{L^3(\Omega_{loc}^*)}/\| {\bf u} (0)\|_{L^3(\Omega_{loc}^*)}$ as a function of time. In this computation, we allocate more grid points to the far field by using the adaptive strategy in the second time period along the $z$ direction (see Appendix) for all time. We have sufficient resolution in this localized domain to capture the  dynamic growth of $\| {\bf u} (t)\|_{L^3(\Omega_{loc}^*)}/\| {\bf u} (0)\|_{L^3(\Omega_{loc}^*)}$ up to $t_3^*=0.00228689295$. The maximum vorticity has grown more than a factor of $2 *10^5$. Beyond $t_3^*$, we will not be able to resolve the far field velocity accurately with $n_1 = 1024$.  
If we further localize the domain, we can capture a faster dynamic growth further in time, see Figure \ref{fig:VelL3_loc_growth} (b). Since the localized $L^3$ norm of the velocity provides a lower bound on the global $L^3$ norm of velocity, the rapid growth of the localized $L^3$ norm of velocity provides additional support of the potentially singular behavior of the Navier--Stokes equations. 


We also examine another non-blowup criteria based on the bound of $\| (\log(r))^2 \Gamma (t)\|_{L^\infty}(r \leq r_0)$ by Lei and Zhang in a recent paper \cite{lei2017criticality}. 
In order to obtain sufficient resolution, we restrict the computation of $\| (\log(r))^2 \Gamma (t)\|_{L^\infty}$ to the local domain $\Omega_{loc}^*$. In Figure \ref{fig:VelL3_growth_nse2}(b), we plot $\| (\log(r))^2 \Gamma(t)\|_{L^\infty(\Omega_{loc}^*)}/\| (\log(r))^2 \Gamma(0)\|_{L^\infty(\Omega_{loc}^*)}$ over the localized domain $\Omega_{loc}^* = \{(r,z) \;|\; (r,z) \in [0,0.001]^2\}$. We observe that this localized quantity develops rapid growth dynamically up to $t=0.00228673493$. By this time, the maximum vorticity has grown by a factor of roughly $5*10^5$. This seems to imply that the non-blowup condition stated in \cite{lei2017criticality} is to be violated.

\vspace{-0.05in}
\subsection{Stability of the nearly self-similar profile to small perturbation of initial data} 
\label{IC:stability}

In this subsection, we study whether the observed nearly self-similar profile is stable with respect to a small perturbation of the initial condition. We will solve the $3$D Navier--Stokes equations using four different initial data defined in \eqref{IC:case1}-\eqref{IC:case4}. Our study shows that the approximate self-similar profile seems to be very stable to small perturbation of the initial data. 

In our stability study, we consider the four different initial data given below. 

\vspace{0.05in}
\noindent
{\bf Case $1$.} We choose the same initial condition given in \eqref{eq:initial-data}, i.e.

\begin{equation}
\label{IC:case1}
u_1 (0,r,z) =\frac{12000(1-r^2)^{18}
\sin(2 \pi z)}{1+12.5(\sin(\pi z))^2}, \quad \om_1(0,r,z)=0.
\end{equation}
 
 \vspace{0.05in}
 \noindent
{\bf Case $2$.} We choose the initial condition as a small perturbation to the initial condition defined in \eqref{IC:case1}

 \begin{equation}
\label{IC:case2}
u_1 (0,r,z) =\frac{12000(1-r^2)^{18}
\sin(2 \pi z)}{1+12.5(\sin(\pi z))^2} + 
\frac{(1-r^2)^{10}
\sin(6 \pi z)}{1+12.5(\sin(\pi z))^2} 
, \quad \om_1(0,r,z)=0.
\end{equation}
 
 \vspace{0.05in}
 \noindent
{\bf Case $3$.} We choose the initial condition as a larger perturbation to the initial condition defined in \eqref{IC:case1}

 \begin{equation}
\label{IC:case3}
u_1 (0,r,z) =\frac{12000(1-r^2)^{18}
\sin(2 \pi z)}{1+12.5(\sin(\pi z))^2} + 
\frac{42(1-r^2)^{6}
\sin(10 \pi z)}{1+12.5(\sin(\pi z))^2} 
, \quad \om_1(0,r,z)=0.
\end{equation}
 
\vspace{0.05in}
 \noindent
{\bf Case $4$.} We choose the initial condition that is $O(1)$ perturbation to the initial condition defined in \eqref{IC:case1}, i.e.

 \begin{equation}
\label{IC:case4}
u_1 (0,r,z) =\frac{12000(1-r^2)^{18}
\sin(4 \pi z)}{1+12.5(\sin(\pi z))^2}
, \quad \om_1(0,r,z)=0.
\end{equation}

 \vspace{0.05in}
The relative size of the perturbation in Case $2$ is approximately $1.928\cdot 10^{-4}$ while the relative size of the perturbation in Case $3$ is about $10^{-2}$. We remark that the decay of the perturbation along the $r$-direction is slower than the original unperturbed initial condition and the perturbation along the $z$-direction is more oscillatory. 
In Case $4$, we just change $\sin(2 \pi z)$ in the numerator in the original initial condition to $\sin(4 \pi z)$. Everything else is the same. With this change, $u_1$ becomes negative for $z$ near $0.5$. This introduces an $O(1)$ structural change to $u_1$.
 
For the Case $1$ and Case $2$ initial data, we solve the Navier--Stokes equations by using viscosity $\nu=5\cdot 10^{-4}$ from $t=0$ to $t_0=0.00227375$, and then switch to $\nu=5\cdot 10^{-3}$.  For the Case $3$ initial data, we solve the Navier--Stokes equations by using viscosity $\nu=5\cdot 10^{-4}$ from $t=0$ to $t_0^*=0.0022266053$ and then switch to $\nu=5\cdot 10^{-3}$. We choose this time $t_0^*$ so that $\|\vom (t_0^*)\|_{L^\infty}/\|\vom (0)\|_{L^\infty}$ matches exactly $\|\vom (t_0)\|_{L^\infty}/\|\vom (0)\|_{L^\infty} \approx 493.08619$ obtained from the first initial data. Due to the relative large perturbation in the Case $3$ initial data, the growth rate of the solution is quite different from that for the first two initial data. So we cannot switch the viscosity at the same time for all three initial data. 

We solve the Navier--Stokes equations for these three different initial data for a very large number of time steps using a $1024\times 1024$ grid. For the Case $1$ initial data, we solve the Navier--Stokes equations for $370,000$ time steps and stop at $T_1=0.002286851153$. By this time, the maximum vorticity has increased by a factor of $1.03 \cdot 10^7$. For the Case $2$ initial data, we solve the the Navier--Stokes equations for $320,000$ time steps and stop at $T_2=0.002286395676$. By this time, the maximum vorticity has increased by a factor of $9.66\cdot 10^6$. For the Case $3$ initial data, we solve the the Navier--Stokes equations for $280,000$ time steps and stop at $T_3=0.002239211579$. The maximum vorticity has increased by a factor of $3.188\cdot 10^6$. One may expect that the potentially singular solution for these three initial data would behave very differently after solving the Navier--Stokes equations for so many time steps. To our surprise, the potentially singular profile seems to be quite stable with respect to the small perturbation of the initial data. 

 \begin{figure}[!ht]
\centering
	\begin{subfigure}[b]{0.37\textwidth}
    \includegraphics[width=1\textwidth]{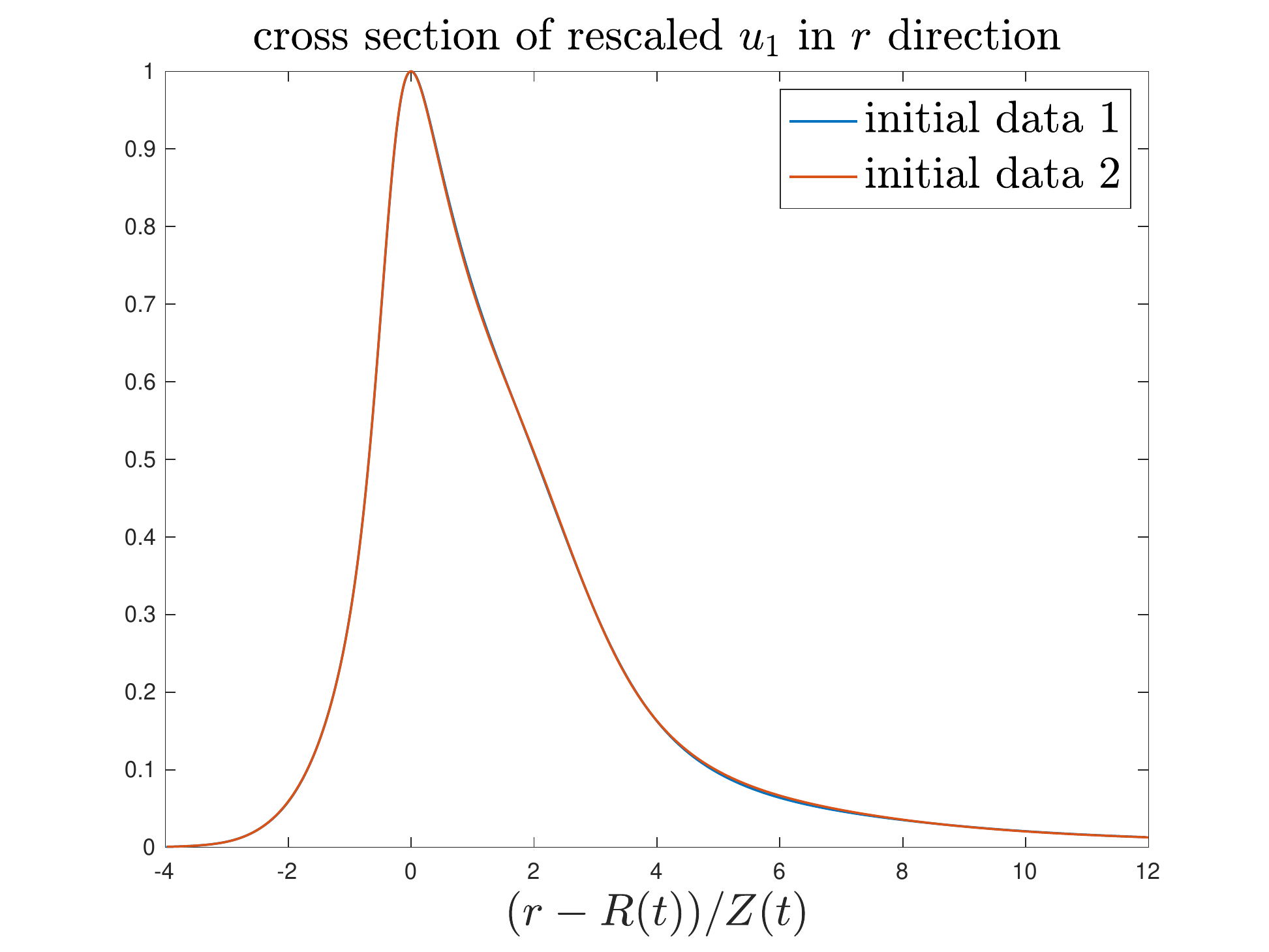}
    \caption{un-normalized $\xi$-cross section of $u_1$}
    \end{subfigure}
  	\begin{subfigure}[b]{0.37\textwidth} 
    \includegraphics[width=1\textwidth]{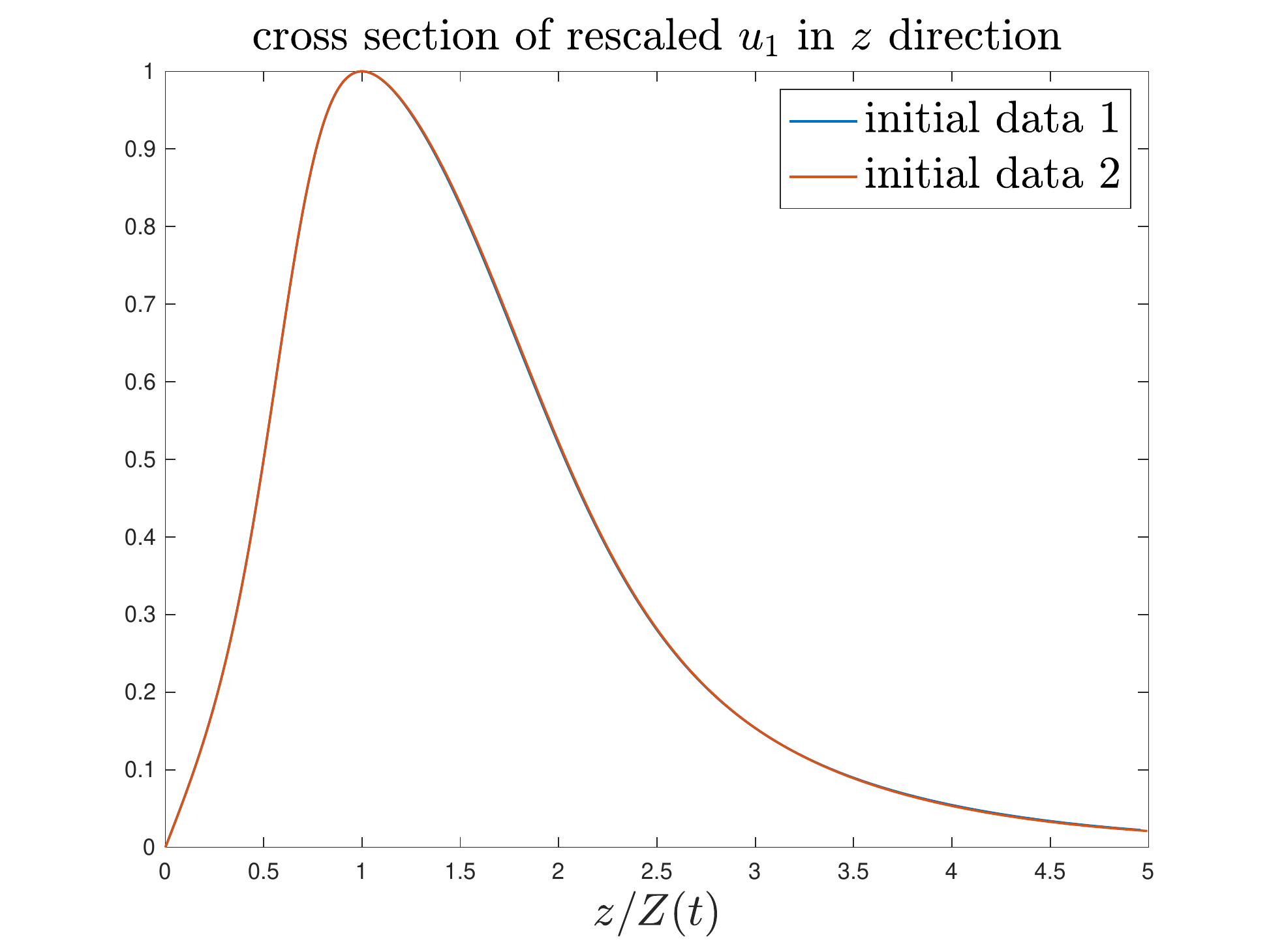}
    \caption{rescaled $\zeta$-cross section of $u_1$}
    \end{subfigure} 
    \begin{subfigure}[b]{0.37\textwidth} 
    \includegraphics[width=1\textwidth]{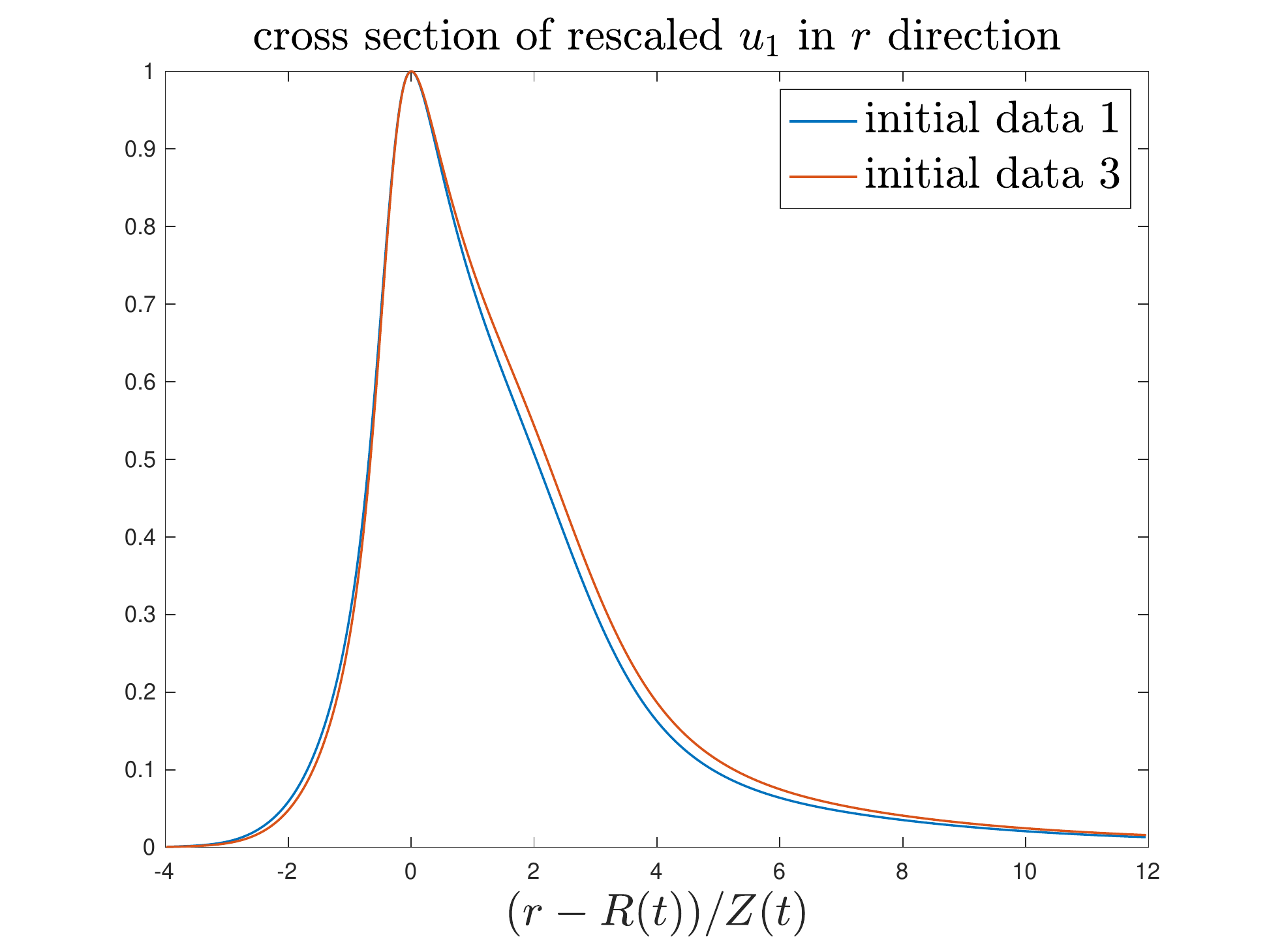}  
    \caption{rescaled $\xi$-cross section of $u_1$}
    \end{subfigure}
  	\begin{subfigure}[b]{0.37\textwidth} 
    \includegraphics[width=1\textwidth]{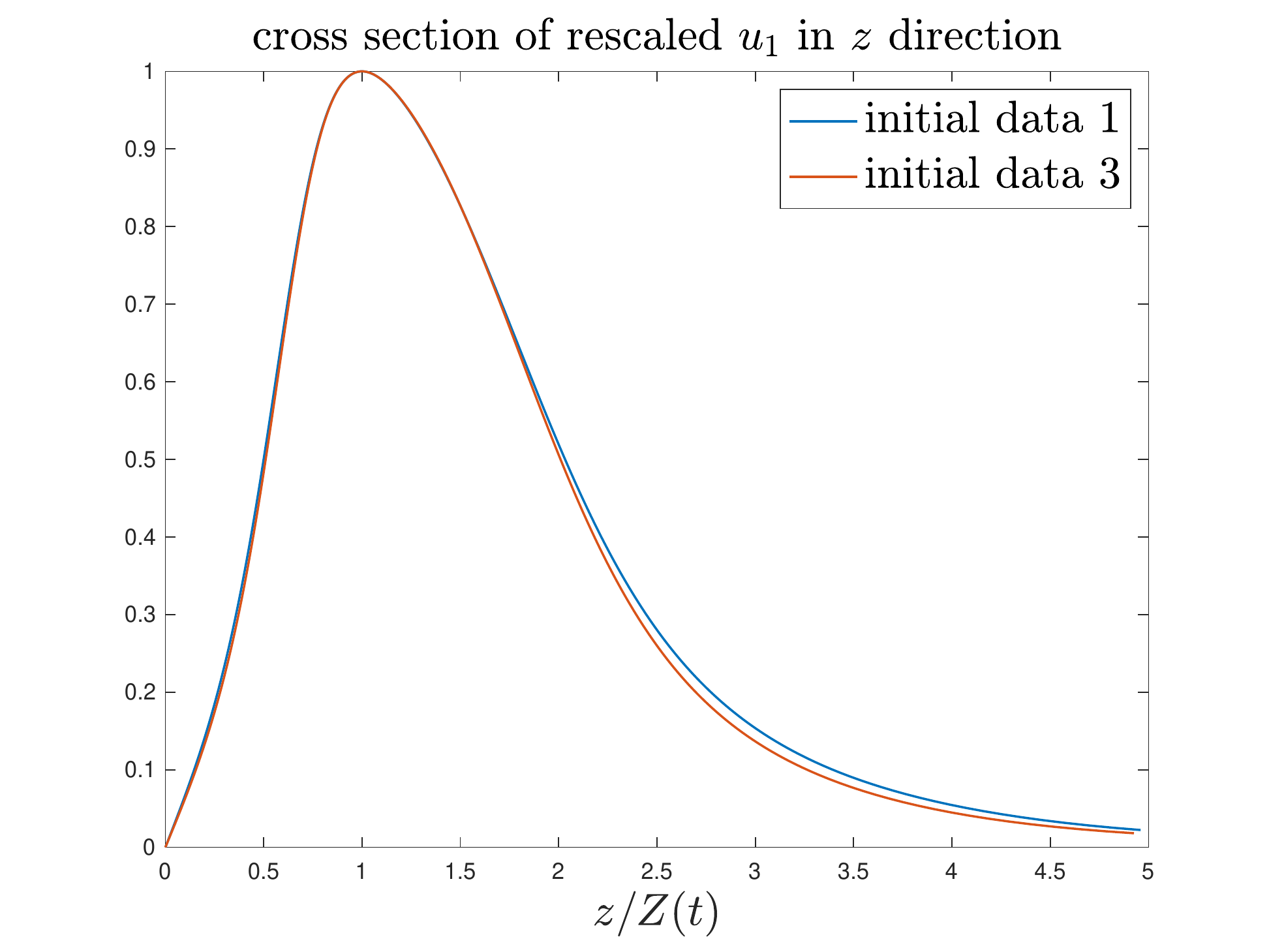}
    \caption{rescaled $\zeta$-cross section of $u_1$}
    \end{subfigure} 
    \caption[compare u1-profile]{(a) Rescaled $\xi$-cross section of $u_1$ at $z=Z(t)$ for the Case $1$ and Case $2$ initial data. (b) Rescaled $\zeta$-cross section of $u_1$ at $r=R(t)$ for the Case $1$ and Case $2$ initial data. (c) Rescaled $\xi$-cross section of $u_1$ at $z=Z(t)$ for the Case $1$ and Case $3$ initial data. (d) Rescaled $\zeta$-cross section of $u_1$ at $r=R(t)$ for the Case $1$ and Case $3$ initial data..}   
    \label{fig:stability_data_nse}
       \vspace{-0.05in}
\end{figure}

 \begin{figure}[!ht]
\centering
	\begin{subfigure}[b]{0.37\textwidth}
    \includegraphics[width=1\textwidth]{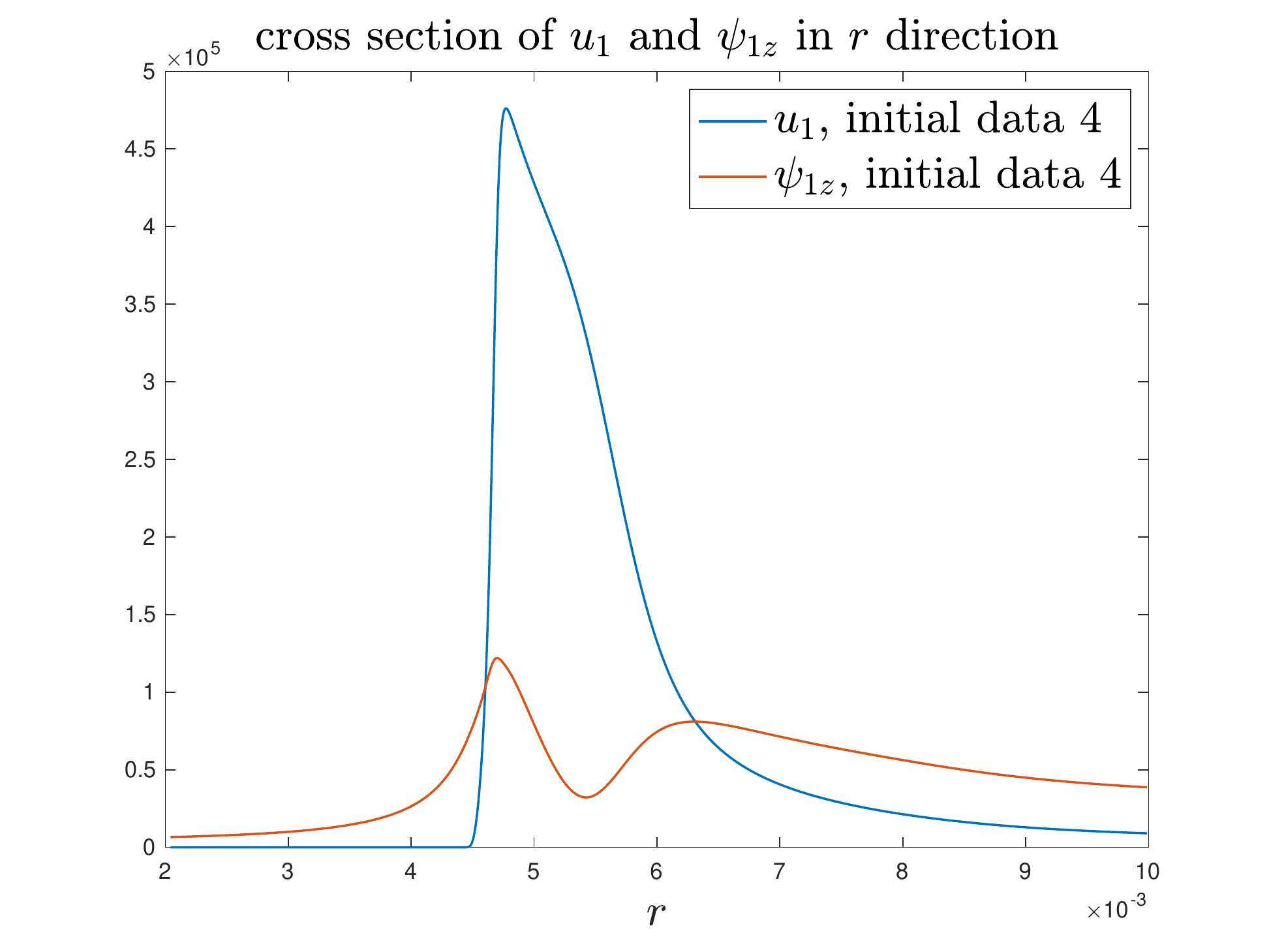}
    \caption{$r$-cross section, case 4 data}
    \end{subfigure}
    \begin{subfigure}[b]{0.37\textwidth} 
    \includegraphics[width=1\textwidth]{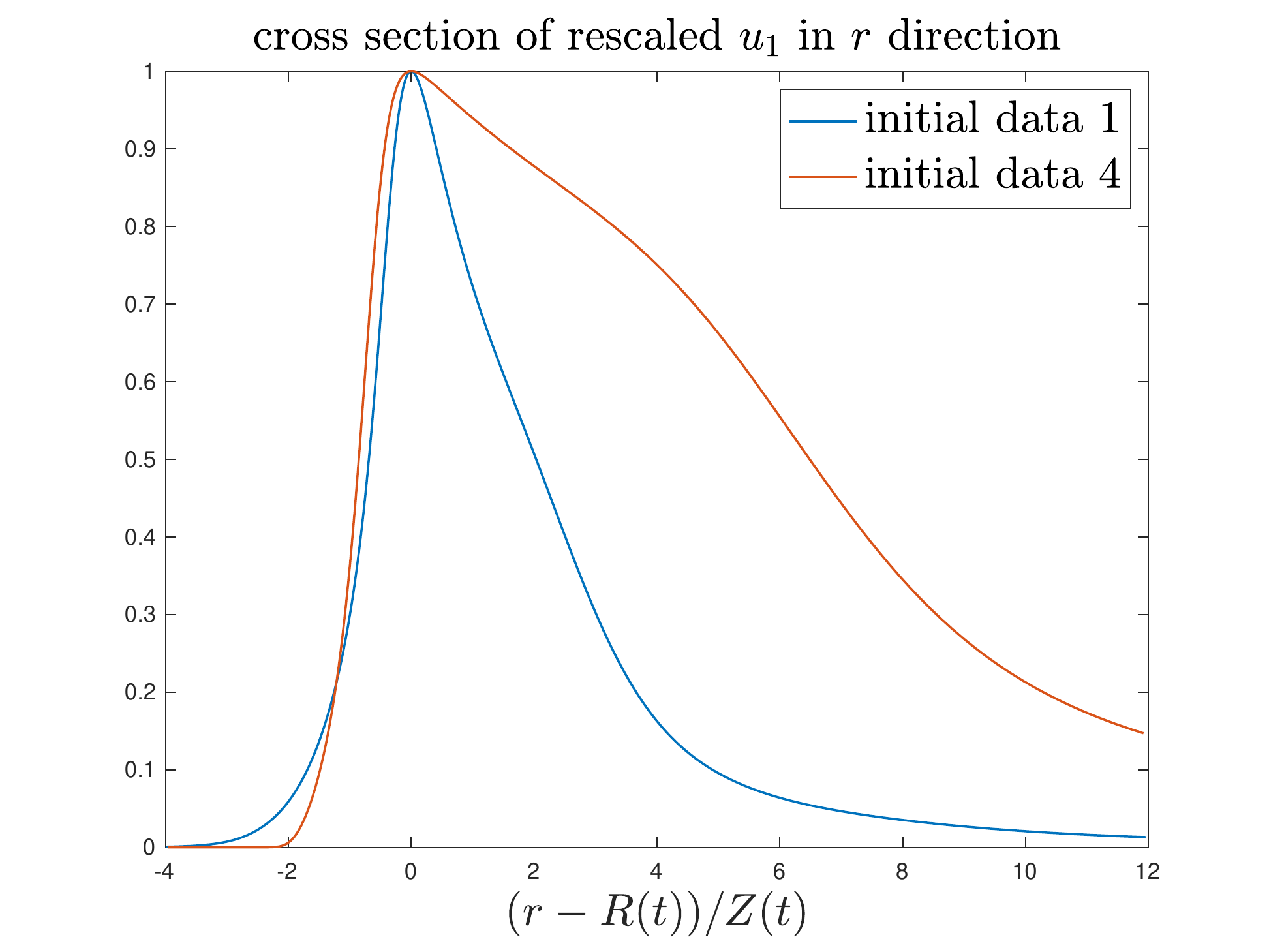}
    \caption{rescaled $\xi$-cross section of $u_1$}
    \end{subfigure}
    \caption[compare u1-profile]{(a) The $r$-cross section of $u_1$ and $\psi_{1z}$ at $z=Z(t)$ for the Case $4$ initial data in the original physical space. 
    (b) Rescaled $\xi$-cross section of $u_1$ for the Case $1$ and Case $4$ initial data. }
    \label{fig:stability_data_nse2}
       \vspace{-0.05in}
\end{figure}

In Figure \ref{fig:stability_data_nse} (a)-(b), we plot the rescaled solution $u_1$ using the Case $1$ and Case $2$ initial conditions. The rescaled solution $U$ is defined below:
$$u_1 = \max(u_1) U (t,\xi, \zeta ), \quad
\xi = (r-R(t))/Z(t), \; \zeta = z/Z(t).$$
We observe that the rescaled profiles $U$ as a function of $(\xi,\zeta)$ for the first two initial conditions almost collapse on each other. This shows that the nearly self-similar profile is very stable with respect to a small perturbation. In Figure \ref{fig:stability_data_nse} (c)-(d), we compare the rescaled profiles obtained using the first and the third initial data. We observe a small difference between the rescaled profile $U$ obtained from the first initial data and that obtained from the third initial data. As we mentioned earlier, the difference between the first and the third initial data is relatively large, about $1 \%$. The time at which we switch the viscosity is also quite different. The small difference that we observe in the rescaled profile is  compatible with the size of the perturbation. This suggests that the nearly singular profile of the solution is relatively stable to this moderate size ($1 \%$) perturbation of the initial condition.

We also solve the Case $4$ initial data using the same viscosity $\nu=5\cdot 10^{-4}$ for all time. 
We see a completely different behavior. The solution behaves qualitatively similar to the two-scale traveling wave solution reported in \cite{Hou-Huang-2021,Hou-Huang-2022}. In particular, there is a vacuum region for $u_1$ developed dynamically between the sharp front and $r=0$ and a compact support of $u_1$ along the $r$-direction, see Figure \ref{fig:stability_data_nse2} (a). The maximum vorticity grows much slower than that for the first three initial data. The maximum vorticity relative to its initial maximum voroticity grows only by a factor of $2078$ after solving the NSE for $300,000$ time steps. For the same number of time steps, the maximum vorticity for the Case $1$ initial data has grown by a factor of $4.186 \cdot 10^{6}$. The slow growth rate of the maximum vorticity for the Case $4$ initial data is due to the two-scale structure and the viscous dominance in the late stage. Moreover, as we can see from Figure \ref{fig:stability_data_nse2}(b), the rescaled profile of the Case $4$ initial data looks completely different from that of the Case $1$ initial data.

\subsection{Competition between vortex stretching and diffusion in the late stage}
\label{stretching-diffusion}

\begin{figure}[!ht]
\centering
    \begin{subfigure}[b]{0.38\textwidth}
    \includegraphics[width=1\textwidth]{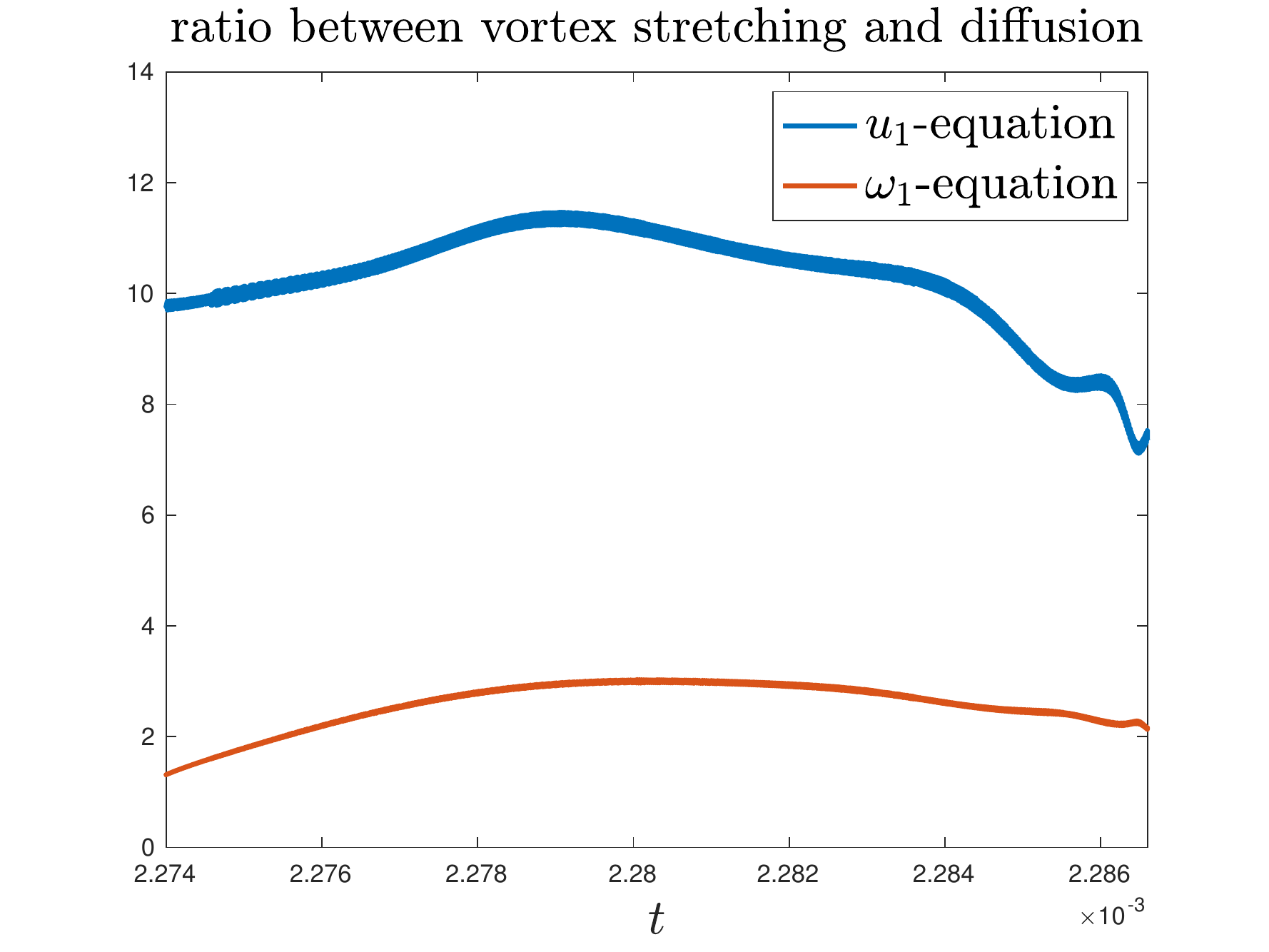}
    \end{subfigure}
  	\begin{subfigure}[b]{0.38\textwidth} 
    \includegraphics[width=1\textwidth]{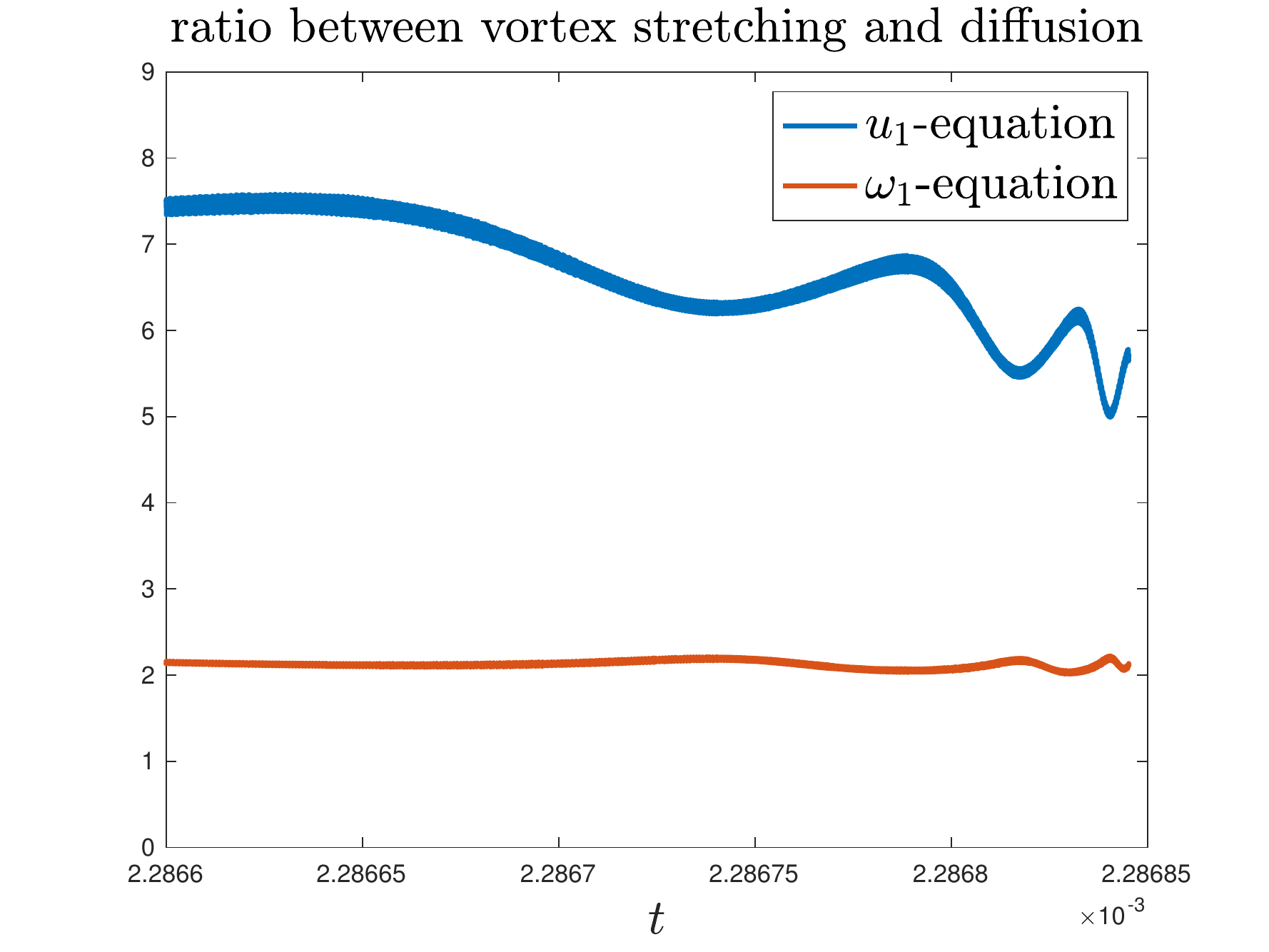}
    \end{subfigure} 
    \caption[vortex stretching vs diffusion]{(a) The ratio between vortex stretching and diffusion for the $3$D Navier--Stokes equations with a constant viscosity $\nu = 5\cdot 10^{-3}$, early stage (b) The same as in (a) but for the late stage. The solution is computed using $1536\times 1536$ grid. The final time of the computation is $t_3$.}   
    \label{fig:vort_diffus_nse}
       \vspace{-0.05in}
\end{figure}

To gain further understanding of the mechanism leading to a potential blow-up of the Navier--Stokes equations, we study the competition between the vortex stretching and the diffusion terms.
In Figure \ref{fig:vort_diffus_nse}, we plot the ratio between the vortex stretching term $2\psi_{1z} u_1$ and $\nu\Delta u_1$ evaluated at $(R(t), Z(t))$ (see the blue curve) for the Navier--Stokes equations with viscosity $\nu = 5\cdot 10^{-3}$. We also plot the ratio between the vortex stretching term $2 u_1 u_{1z}$ and $\nu \Delta \omega_1$ evaluated at the location where $\omega_1$ achieves its maximum (see the red curve). We can see that the vortex stretching term for the $u_1$-equation dominates the diffusion term throughout the computation. Although the ratio between vortex stretching and diffusion seems to develop a downward trend in the late stage, we observe that the vortex stretching term maintain a factor of $5$ larger than the diffusion term. On the other hand, the ratio between vortex stretching and diffusion for the $\omega_1$-equation seems to stabilize in the late stage.

\begin{figure}[!ht]
\centering
    \begin{subfigure}[b]{0.32\textwidth}
    \includegraphics[width=1\textwidth]{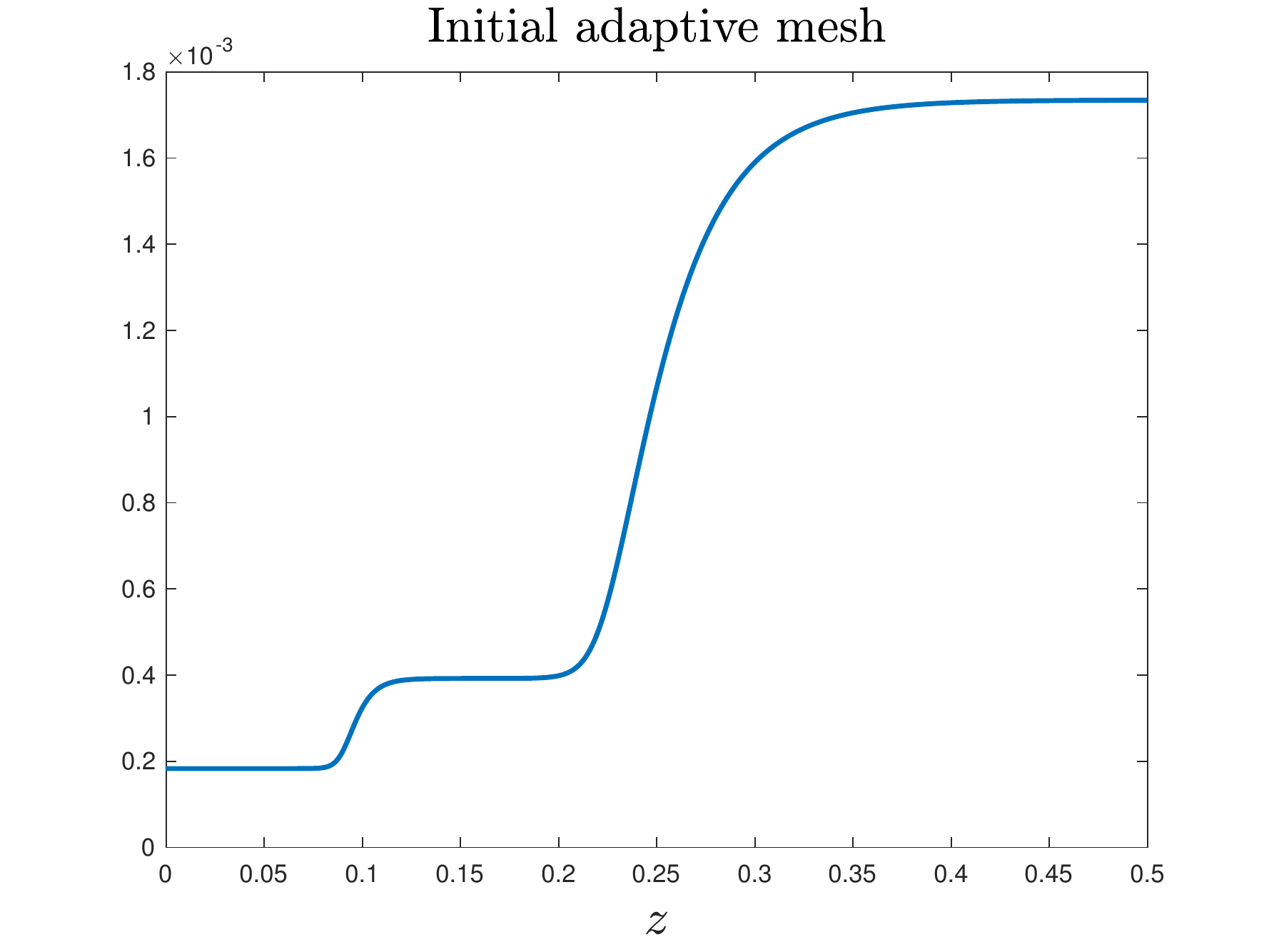}
    \end{subfigure}
  	\begin{subfigure}[b]{0.32\textwidth} 
    \includegraphics[width=1\textwidth]{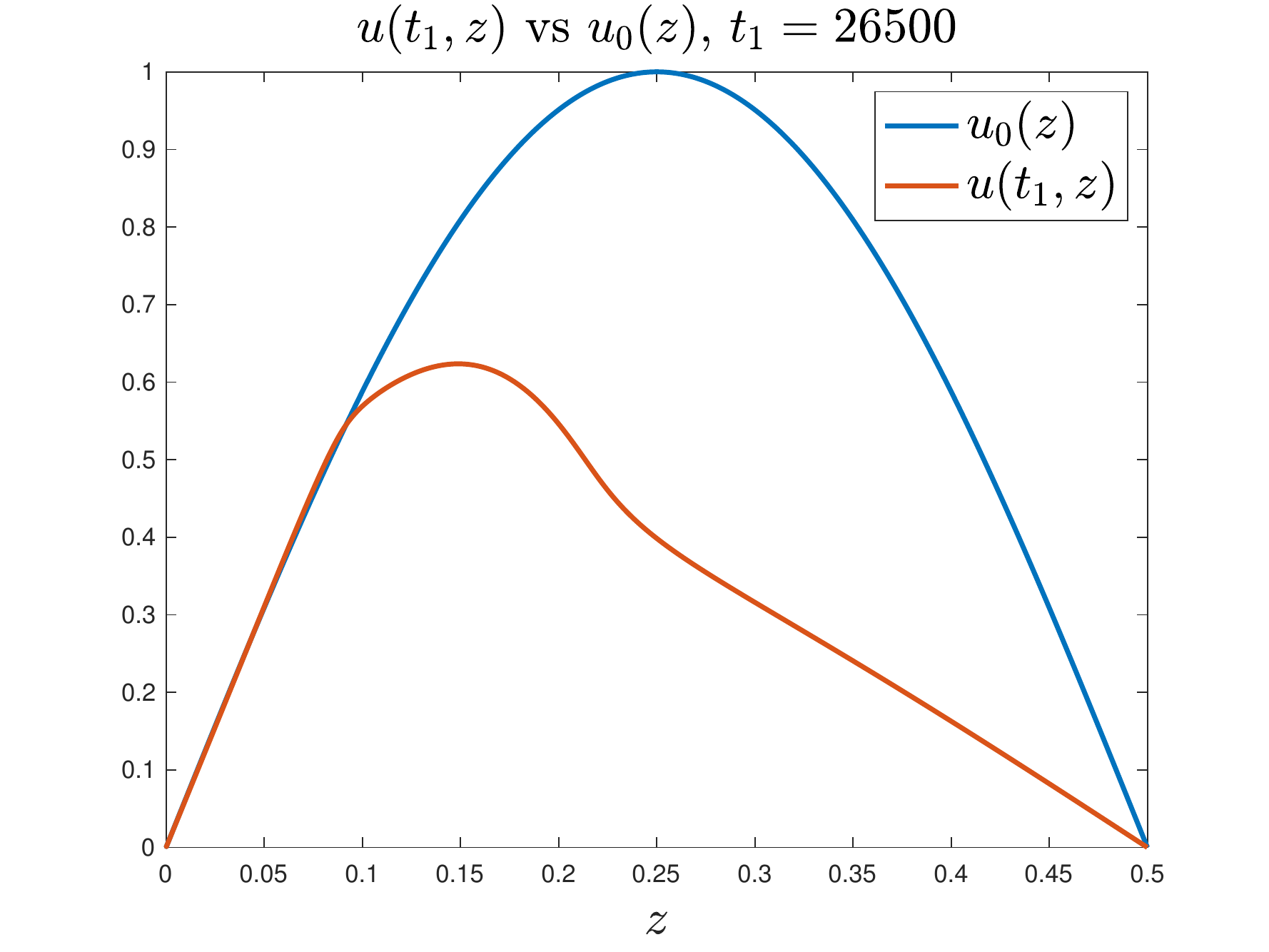}
    \end{subfigure} 
    \begin{subfigure}[b]{0.32\textwidth} 
    \includegraphics[width=1\textwidth]{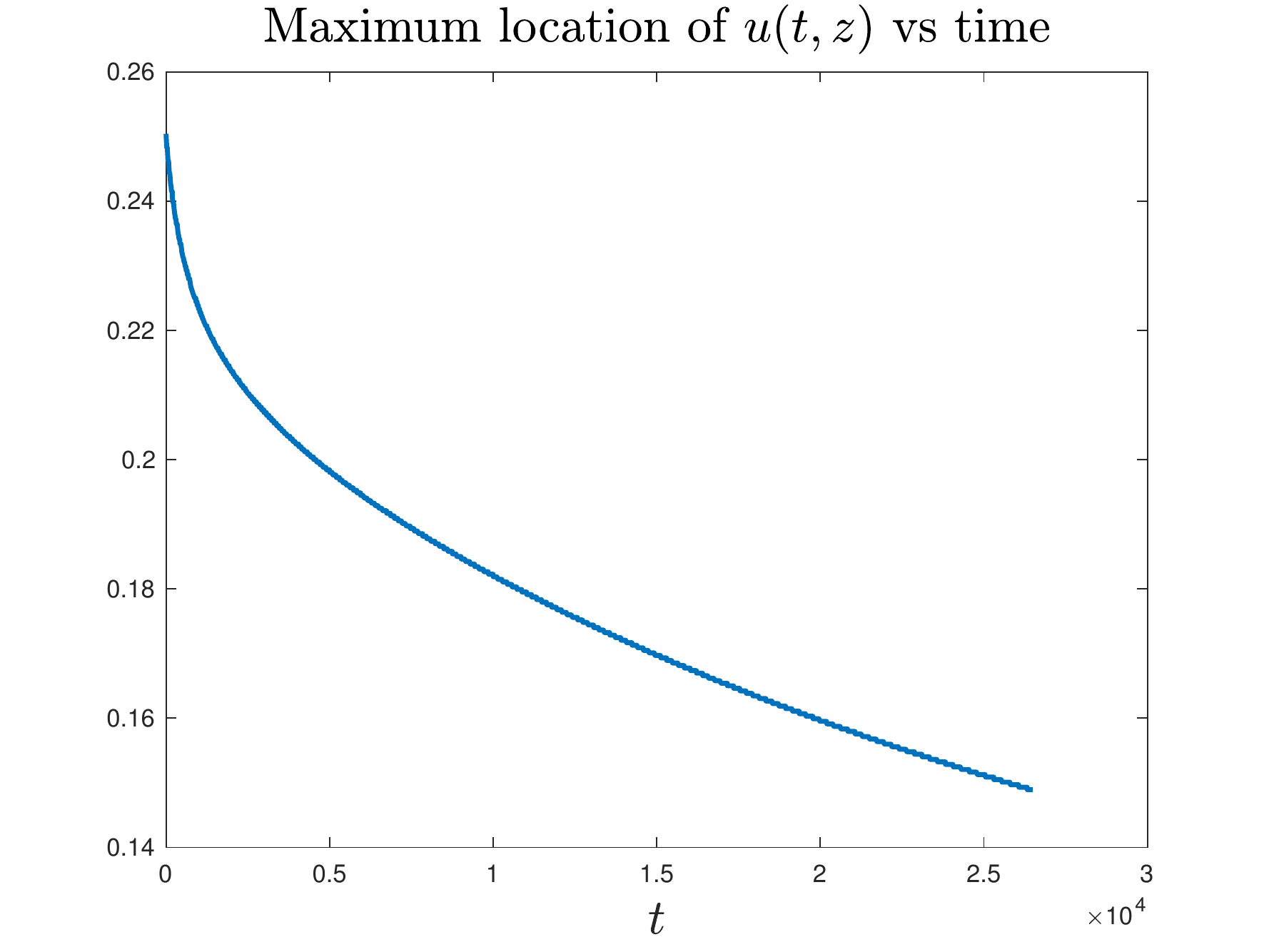}
    \end{subfigure} 
    \caption[error in Z and diffusion]{(a) The initial stage of the adaptive mesh in $z(\eta)$. (b) Comparison of the initial profile and the profile at the late stage. (c) The maximum position of $u(t,z)$ as a function of time.}   
    \label{fig:error_Z_diffv_toy}
       \vspace{-0.05in}
\end{figure}

The downward trend in the ratio between vortex stretching and diffusion for the $u_1$-equation described in Figure \ref{fig:vort_diffus_nse} seems to be driven by a slightly faster decay rate of $Z(t)$ in the late stage, possibly with a logarithmic correction. We tend to believe that the development of this mild logarithmic correction in the decay rate of $Z(t)$ is due to the frequent changes of adaptive mesh in the late stage. Since the potential singularity propagates toward the origin, we need to shift the boundary of the mesh map $z(\eta)$ in phase $1$ closer to $z=0$ every time we change the adaptive mesh. The interpolation from the old mesh to the new mesh introduces a numerical dissipation that depends on the local adaptive mesh. This mesh dependent numerical diffusion induces a traveling wave propagating toward $z=0$. 

The numerical dissipation introduced by the frequent changes of adaptive mesh in the late stage can be modeled by a $1$D diffusion equation with a second order mesh dependent numerical viscosity given by $\nu (z) = (h_1 z_\eta)^2$ ($h_1=1/n_1$). Here $h_1 z_\eta$ is the adaptive mesh along the $z$ direction and $z(\eta)$ is the adaptive mesh map generated by our adaptive mesh strategy at $t=0$ defined in the Appendix. In our computation, the adaptive mesh map will be updated frequently in the late stage. Here we just use the initial adaptive mesh to illustrate the main point. Note that $z=z(\eta)$ is a monotonically increasing function of $\eta$ that maps $[0,1]$ in $\eta$ to $[0,0.5]$ in $z$. The toy model diffusion equation is given by
\[
u_t =  \nu(z) u_{zz}, \quad u(0,z) = \sin(2 \pi z), \quad 0 \leq z \leq 0.5,
\]
where $\nu (z) = (h_1 z_\eta)^2$.
Let $v=u_z$. We can derive an equation for $v$ as follows;
\[
v_t - \nu_z v_z = \nu(z) v_{zz}, \quad v(0,z) = 2 \pi \cos(2\pi z)\;, \quad 0 \leq z \leq 0.5.
\]
Since the Jacobian of the adaptive mesh map $z_\eta$ is positive and monotonically increasing by construction, we have $\nu_z > 0$. Thus, the advection term $- \nu_z v_z$ in the $v$-equation, which is introdued by differentiating the $u$-equation, induces a traveling traveling wave that propagates toward $z=0$.

In Figure \ref{fig:error_Z_diffv_toy}(a), we plot the adaptive mesh $\nu(z)$ as a function of $z$. We observe a monotonically increasing mesh size as a function of $z$ with the finest mesh in the near field ($z \approx 0$) and the coarsest mesh in the far field  ($z \approx 0.5$). In Figure \ref{fig:error_Z_diffv_toy}(b), we plot the solution obtained after solving the above $1$D diffusion equation with $n_1=1024$ up to a relatively large time $t_1^*=26500$ (the red curve). Compared with the initial condition (the blue curve), we can see that the solution has traveled toward $z=0$ by a distance of roughly $0.1$. In Figure \ref{fig:error_Z_diffv_toy}(c), we plot the trajectory of the maximum location of $u(t,z)$ as a function of time. We can see the decay of the maximum location in time, confirming that the solution travels toward $z=0$. The decay slows down after the maximum location enters the plateau region of the adaptive mesh where the adaptive mesh does not vary too much. This may provide a partial explanation why $Z(t)$ seems to decay slightly faster than $(T-t)^{1/2}$ in the late stage.

For the axisymmetric Navier--Stokes equations, the diffusion term for the total circulation $\Gamma = r u^\theta$ has the form $\Delta - \frac{2}{r}\partial_r$. Thus, the diffusion term introduces an additional transport term $\frac{2\nu}{r} \Gamma_r$ in the $\Gamma$-equation, which transports $\Gamma$ away from $r=0$. This may slow down the decay of $R(t)$ as $R(t)$ approaches to $0$ in the late stage, which may contribute the mild increase of $R(t)/Z(t)$ in the late stage. In addition, the Jacobian $r_\rho$ is not monotonically increasing due to the decreasing of $r_\rho$ in phase $0$ near $r=0$. Therefore the numerical dissipation due to the frequent changes of mesh in the late stage does not introduce as strong a bias for $R(t)$ as for $Z(t)$.

\vspace{-0.1in}
\section{Concluding Remarks}
\label{sec:conclude}

In this paper, we presented numerical evidences that the $3$D axisymmetric Navier--Stokes equations with our initial condition seem to develop potentially singular behavior at the origin with maximum vorticity increased by a factor of $10^7$. The potentially singular solutions are nearly self-similar and preserve many essential features of the Euler solution using the same initial condition. 
We have applied several blow-up criteria to study the potentially singular behavior of the Navier--Stokes equations using our initial data. 
The Beale-Kato-Majda blow-up criterion based on the growth rate of  $\int_0^t \| \vom (s)\|_{L^\infty}ds$ seems to imply that the Navier--Stokes equations using our initial data would develop a potential finite time singularity at the origin. Moreover,
the blow-up criteria based on the enstrophy growth $\int_0^t \| \vom (s)\|_{L^2}^4 ds$ and the growth of negative pressure $\int_0^t \|- p (s)\|_{L^\infty}ds$ also provide support for the potentially singular behavior of the Navier--Stokes equations.

We have also examined the growth of $\| {\bf u}\|_{L^{p,q}}$ for $(p,q) = (4,8),\; (6,4),\; (9,3)$ and $(p,q)=(\infty,2)$, respectively. Our numerical results suggest that the localized $L^{p,q}$ norms of the velocity computed over $r \leq 0.001$ develop rapid growth dynamically. This provides further evidence for the development of a potential finite time singularity of the Navier--Stokes equations using our initial data.

The non-blowup criterion based on the critical $L^3$ norm of the velocity with $(p,q)=(3,\infty)$ is more difficult to capture numerically due to the extremely slow growth of $\|{\bf u}(t)\|_{L^3}$ in the late stage. Another difficulty is that there is a significant contribution to the $L^3$ norm of the velocity from the far field where we have a relatively coarse grid. According to a recent result by Tao \cite{tao2020}, as one approaches a finite blowup time $T$, the critical $L^3$ norm of the velocity may blow up as slowly as $\left (\log \log \log \frac{1}{T-t}\right )^c$ for some absolute constant $c$. In order to capture such mild dynamic growth rate, we computed the relative growth rate of the localized version of the $L^3$ norm of the velocity over a localized domain $\Omega_{loc}^* = \{ (r,z) \in [0,0.001]^2\}$. We observed rapid dynamic growth of $\| {\bf u}(t)\|_{L^3(\Omega_{loc})}/\| {\bf u}(0)\|_{L^3(\Omega_{loc})}$. This provides additional support for the potentially singular behavior of the Navier--Stokes equations.

Although all non-blowup criteria are equivalent in theory, the local blow-up criteria such as the Beale-Kato-Majda blow-up criterion are easier to verify numerically than the global blow-up criteria that are based on the critical $L^3$ norm of the velocity. Moreover, it is more reliable to use blow-up criteria such as $\int_0^t \| \vom (s)\|_{L^\infty}ds$ that do not require any numerical fitting of the asymptotic blow-up rate. In order to obtain an accurate numerical fitting of an asymptotic blow-up rate, one has to compute extremely close to the potential singularity time with sufficient resolution in both the near field and the far field, which requires tremendous computational resources. When the potential blow-up solution is not asymptotically self-similar, such asymptotic fitting is even more difficult to obtain.

In our future work, we plan to use the dynamic rescaling formulation \cite{mclaughlin1986focusing,chen2019finite2,chen2020finite,chen2021finite3} to further investigate the potentially singular behavior of the Navier--Stokes equations. One important advantage of using the dynamic rescaling approach is that we can use a fixed mesh to solve the dynamically rescaled Navier--Stokes equations, which avoids the frequent changes of adaptive mesh in the physical domain in the late stage. This will reduce the numerical dissipation in the far field and enable us to compute the nearly self-similar profile in a much larger rescaled domain. 


\vspace{0.1in}
{\bf Acknowledgments.} The research was in part supported by NSF Grants DMS-$1907977$ and DMS-$1912654$, and the Choi Family Gift Fund. I would like to thank Professor Vladimir Sverak, Dr. De Huang and Jiajie Chen for a number of stimulating discussions. I would also like to thank the referees and Jiajie Chen for their very constructive comments and suggestions, which significantly improves the quality of this paper. Finally, I have benefited a lot from the AIM SQarRE ``Towards a $3$D Euler singularity'', which has generated many stimulating discussions related to the $3$D Euler singularity.

\appendix

\section{Construction of the adaptive mesh}\label{apdx:adaptive_mesh}
In this appendix, we describe our adaptive mesh strategy that we use to study the potentially singular behavior of the Naveri--Stokes equations. 
We will use the method described in Appendix B of \cite{Hou-Huang-2021} to construct our adaptive mesh maps $r=r(\rho)$ and $z=z(\eta)$. Our adaptive mesh strategy does not require that the solution has a bell-shaped structure in the most singular region, see more discussion in the Appendix of \cite{Hou-euler-2021}. We will discretize the equations in the transformed variables $(\rho,\eta)$ with $n_1$ grid points along the $z$ direction and $n_2$ grid points along the $r$-direction. 

\vspace{-0.05in}
\subsection{The adaptive (moving) mesh algorithm}
To effectively and accurately compute the potential blowup, we have carefully designed a special meshing strategy that is dynamically adaptive to the more and more singular structure of the solution. The adaptive mesh covering the half-period computational domain $\mathcal{D}_1 = \{(r,z):0\leq r\leq 1,0\leq z\leq 1/2\}$ is characterized by a pair of analytic mesh maps
\[r = r(\rho),\quad \rho\in [0,1];\quad z = z(\eta),\quad \eta\in[0,1].\]
These mesh mapping functions are both monotonically increasing and infinitely differentiable on $[0,1]$, and satisfy 
$r(0) = 0,\; r(1) = 1,\; z(0) = 0,\; z(1) = 1/2.$
In particular, we construct these mapping functions by carefully designing their Jacobians/densities
\[r_\rho = r'(\rho),\quad z_\eta = z'(\eta),\] 
using analytic functions that are even functions at $0$. The even symmetries ensure that the resulting mesh can be smoothly extended to the full-period cylinder $ \{(r,z):0\leq r\leq 1,-1/2\leq z\leq 1/2\}$. The density functions contain a small number of parameters, which are dynamically adjusted to the solution. Once the mesh mapping functions are constructed, the computational domain is covered with a tensor-product mesh:
\begin{equation}\label{eq:mesh}
\mathcal{G} = \{(r_i,z_j): 0\leq i\leq n_2,\ 0\leq j\leq n_1\},
\end{equation}
where $r_i = r(ih_\rho),\quad h_\rho = 1/n_2;\quad z_j = z(jh_\eta),\quad h_\eta = 1/n_1.$
The precise definition and construction of the mesh mapping functions are described in Appendix B of \cite{Hou-Huang-2021}. 

We design the densities $r_\rho,z_\eta$ to have three phases: 
\begin{itemize}
\item Phase $1$ covers the inner profile of the smaller scale near the sharp front;
\item Phase $2$ covers the outer profile of the larger scale of the solution;
\item Phase $3$ covers the (far-field) solution away from the symmetry axis $r=0$.
\end{itemize} 
We add a phase $0$ in the density $r_\rho$ to cover the region near $r=0$ and also add a phase $0$ in the density $z_\eta$ to cover the region near $z=0$ in the late stage. In our computation, the number (percentage) of mesh points in each phase are fixed, but the physical location of each phase will change in time, dynamically adaptive to the structure of the solution. Between every two neighboring phases, there is also a smooth transition region that occupies a fixed percentage of mesh points. 
%

\vspace{-0.075in}
\subsection{Adaptive mesh for the 3D Navier--Stokes equations}
We also use three different adaptive mesh strategies for three different time periods. The first time period corresponds to the time interval between $t=0$ and $T_1=0.002191729$ with $\|\omega(T_1)\|_{L^\infty}/\|\omega(0)\|_{L^\infty} \approx 20.5235$ for the $1536\times 1536$ grid and the number of time steps equal to $45000$. The second time period corresponds to the time interval between $T_1=0.002191729$ and $T_2 = 0.002261605$ with $\|\omega(T_2)\|_{L^\infty}/\|\omega(0)\|_{L^\infty} \approx 139.5777$ for the $1536\times 1536$ grid and the number of time steps equal to $60000$. The third time period is for $ t \geq T_2$. 

For the first time period, since we use a very smooth initial condition whose support covers the whole domain, we use the following parameters $r_1=0.001,\; r_2=0.05, \; r_3=0.2$ and $s_{\rho_1}=0.001$, $s_{\rho_2}=0.5$, $s_{\rho_3}=0.85$ to construct the mapping $r=r(\rho)$ using a four-phase map. Similarly, we use the following parameters $z_1=0.1,\; z_2=0.25$ and 
$s_{\eta_1}=0.5$, $s_{\eta_2}=0.85$ to construct the mapping $z=z(\eta)$ using a three-phase map. We then update the mesh $z=z(\eta)$ dynamically using $z_1=2 z(I_w)$ and $z_2=10 z(I_w)$ with $s_{\eta_1}=0.6$, $s_{\eta_2}=0.9$ when $I < 0.2n_1$, but keep $r=r(\rho)$ unchanged during this early stage. Here $I_w$ is the grid point index along the $z$-direction at which $\omega_1$ achieves its maximum. 

In the second time period, we use the following parameters 
$s_{\rho_1}=0.05$, $s_{\rho_2}=0.6$, $s_{\rho_3}=0.9$, 
$r_2 = r(J) + 2dr$,
$r_1=\max((s_{\rho_1}/s_{\rho_2})r_2,r(J_r) - 5dr)$, and 
$r_3 = \max(3r(J),(r_2 - r_1)(s_{\rho_3}-s_{\rho_2})/(s_{\rho_2}-s_{\rho_1}) + r_2)$, where $J$ is the grid index at which $u_1$ achieves its maximum along the $r$-direction, $J_r$ is the grid index at which $u_{1,r}$ achieves its maximum along the $r$-direction, and $dr = r(J) - r(J_r)$. We update the mapping $r(\rho)$ dynamically when $J_r < 0.2n_2$. The adaptive mesh map for $z(\eta)$ in the second time period remains the same as in the first time period.

In the third time period, due to the viscous regularization, we do not need to allocate as many grid points to resolve the sharp front. Instead we allocate more grid points to cover the intermediate region in phase 2. We use the following parameters 
$s_{\rho_1}=0.05$, $s_{\rho_2}=0.5$, $s_{\rho_3}=0.9$, 
$r_2 = r(J) + 8dr$,
$r_1=\max((s_{\rho_1}/s_{\rho_2})r_2,r(J_r) - 3dr)$, and 
$r_3 = \max(4.5r(J),(r_2 - r_1)(s_{\rho_3}-s_{\rho_2})/(s_{\rho_2}-s_{\rho_1}) + r_2)$.
To construct the mesh map $z(\eta )$, we use the following parameters 
$s_{\eta_1}=0.05$, $s_{\eta_2}=0.5$, $s_{\eta_3}=0.85$, 
$z_2 = z(I_w) + 2dz$,
$z_1=\max((s_{\eta_1}/s_{\eta_2})z_2,z(I_{wz}) - 6dz)$, and 
$z_3 = \max(4.3z(I_w),(z_2 - z_1)(s_{\eta_3}-s_{\eta_2})/(s_{\eta_2}-s_{\eta_1}) + z_2)$, where $I_w$ is the grid index at which $\omega_1$ achieves its maximum along the $z$-direction, $I_{wz}$ is the grid index at which $\omega_{1,z}$ achieves its maximum along the $z$-direction, and $dz = z(I_w) - r(I_{wz})$. We will update $r(\rho)$ dynamically when $J_r < 0.2 n_2$ and update $z(\eta)$ when $I_z < 0.2 n_1$.

\bibliographystyle{abbrv}
\bibliography{reference}

\end{document}